\documentclass[fleqn,10pt]{wlscirep}
\usepackage[utf8]{inputenc}
\usepackage[T1]{fontenc}
\usepackage{graphics}
\usepackage{subfig}
\usepackage{csquotes}
\usepackage{float}
\usepackage{makecell}
\usepackage{tabularx}
\usepackage{booktabs}
\usepackage{url}
\usepackage{xr}
\usepackage{lineno}
\usepackage{bm}

\makeatletter
\newcommand\footnoteref[1]{\protected@xdef\@thefnmark{\ref{#1}}\@footnotemark}
\makeatother

\makeatletter
\newcommand*{\addFileDependency}[1]{
  \typeout{(#1)}
  \@addtofilelist{#1}
  \IfFileExists{#1}{}{\typeout{No file #1.}}
}
\makeatother

\title{The shapes of an epidemic: using Functional Data Analysis to characterize COVID-19 in Italy}

\author[1]{Tobia Boschi}
\author[2]{Jacopo Di Iorio}
\author[2]{Lorenzo Testa}
\author[1,3,4,*]{Marzia A. Cremona}
\author[1,2,*]{Francesca Chiaromonte}
\affil[1]{Penn State University, Dept. of Statistics and Huck Institutes of the Life Sciences, University Park, PA 16802, USA}
\affil[2]{Sant'Anna School of Advanced Studies, Institute of Economics and EMbeDS, Pisa, 56127, Italy}
\affil[3]{Université Laval, Dept. of Operations and Decision Systems, Québec, G1V 0A6, Canada}
\affil[4]{CHU de Québec – Université Laval Research Center, Québec, G1V 4G2, Canada}

\affil[*]{co-corresponding authors; marzia.cremona@fsa.ulaval.ca, fxc11@psu.edu}


\begin{abstract}

We investigate patterns of COVID-19 mortality across 20 Italian regions and their association with mobility, positivity, and socio-demographic, infrastructural and environmental covariates. Notwithstanding limitations in accuracy and resolution of the data available from public sources, we pinpoint significant trends exploiting information in curves and shapes with Functional Data Analysis techniques. These depict two starkly different epidemics; an "exponential" one unfolding in Lombardia and the worst hit areas of the north, and a milder, "flat(tened)" one in the rest of the country -- including Veneto, where cases appeared concurrently with Lombardia but aggressive testing was implemented early on. We find that mobility and positivity can predict COVID-19 mortality, also when controlling for relevant covariates. Among the latter, primary care appears to mitigate mortality, and contacts in hospitals, schools and work places to aggravate it. The techniques we describe could capture additional and potentially sharper signals if applied to richer data.
\end{abstract}

\begin{document}
    
    \flushbottom
    \maketitle
    %
    %
    \thispagestyle{empty}
    
    
    \section*{Introduction}

    At the end of January 2020, two Chinese tourists were hospitalized in Rome and tested positive to SARS-CoV-2. At the beginning of February, a group of Italian citizens was repatriated from Wuhan –- among them, one tested positive. As the news media reported these headlines, neither the Italian public nor the Italian authorities appeared to perceive an imminent threat, though retrospective analyses
    now suggest that the virus may have been circulating in the north of the country as far back as December 2019 (e.g., detection of SARS-CoV-2 in the wastewater of Milan and Turin \cite{iss2020wastewater}). 
    The first recorded non-travel related COVID-19 case occurred in Codogno (Lombardia) –- where a 38 years old male visited the hospital first on February 17, and then again on February 19 with worsening respiratory symptoms; in this date, he was tested and diagnosed. On February 20, two individuals tested positive in Vo’ Euganeo (Veneto). Notably, the outbreaks in Lombardia and Veneto took two very different paths, something many observers attributed to the early response and aggressive testing strategy adopted by the regional authorities in Veneto\cite{mugnai2020, Lavezzo2020}. After some initial, much debated inconsistencies (e.g., hesitations in implementing local lock-downs in areas hosting major industrial production hubs, contested decisions to move patients between hospitals and nursing homes and to keep major sports events open to the public in Lombardia), starting in early March, local and central authorities took progressively more stringent measures to limit mobility and social gatherings –- culminating with a general nationwide lock-down on March 9 and the suspension of all nonessential production activities on March 23 (starting in early May, activities restarted and mobility and gathering restrictions were gradually loosened).
    
    Lock-down notwithstanding, based on official records, Italy saw a total of 
    $\approx 35,200$ COVID-19 deaths as of the beginning of August. 
    While other countries (e.g., the U.S.~and Brazil) have reached much higher death counts, 
    Italy’s relative death toll remains rather stark at 
    $58.25$ per $100,000$ inhabitants.
    This may be partially attributable to the fact that Italy’s population is very old (nationally, the median age is almost 46 years and the percentage of individuals over 65 almost 22\%), and that age itself correlates with conditions such as type II diabetes, hypertension and chronic respiratory ailments, which substantially worsen illness and increase the likelihood of death for individuals affected by the virus. But perhaps the most striking aspect of the COVID-19 epidemic in Italy has been its heterogeneity. Some parts of Lombardia and of other regions in the industrialized north were hit early and especially hard, yet other demographically and socio-economically similar areas fared better. Moreover, most of the central and southern regions of the country experienced a much milder epidemic –- notwithstanding waves of relocations from employment-related domiciles in the north back to family homes in the center and south around the time of the nationwide lock-down. Potential contributors to this heterogeneity discussed by both scientists and the media include human density characteristics; centralized, hospital-based vs distributed, primary health care systems; and pollution levels\cite{wu2020exposure,coccia2020factors,Binkin2020.04.10.20060707,sylos2020}.
    
    A broad and extremely sophisticated literature exists on epidemiological models, which many research groups are utilizing both to aid policy through forecasts 
    and to dissect what happened, in Italy and around the world. We did not utilize these models. Instead, we applied a mix of statistical tools from the field of Functional Data Analysis (FDA\cite{ramsay2005,kokoszka2017}), some well-established, and some recently developed by our group. FDA offers very powerful approaches to analyze data sets composed of curves or surfaces, exploiting information in their shapes. These techniques, which have been successfully applied in a variety of scientific domains \cite{ramsay2007applied,ullah2013applications,cremona2019}, can effectively complement traditional epidemiological analyses and provide useful
    insights\cite{carroll2020}. We used them to characterize patterns of COVID-19 deaths occurring around the country and analyze their statistical association with two {\em key predictors}; namely, mobility and positivity (the fraction of performed tests returning positive results). We also considered various socio-demographic, infrastructural and environmental {\em covariates}. We focused on the period from February 16, right before the first cases were recorded in Codogno and Vo’ Euganeo, to April 30, right before the first lock-down relaxations (restarting of manufacturing and construction activities at the beginning of May). Based on data availability, we performed our analyses at the spatial resolution of regions, which is suboptimal for several reasons. An epidemic is certainly better studied at a much finer resolution (municipalities, urban areas, perhaps the provinces within which Italian regions are further partitioned) – and so are its links to predictors and covariates whose signals may dilute when aggregated at the regional level. Moreover, operating with 20 observational units (the Italian regions) limits the size of the statistical models one can reliably fit on the data. The techniques we employed allowed us to pinpoint significant trends working with what we could retrieve from public data sources. Unquestionably though, access to data at higher resolution would allow more nuanced, in-depth analyses and likely produce sharper results. 
    
    \begin{figure}[!hbt]
    \centering
    \subfloat[\label{subfig:dpc_vs_istat}]{
    \fbox{
    \begin{minipage}[b]{0.62\linewidth}
        \includegraphics[page=3,trim={0.1cm 1cm 0.7cm 0.2cm},clip,height=4.3cm]{./img/regions_istat_vs_dpc_smooth_16feb-30avr.pdf}
        \includegraphics[page=4,trim={1.2cm 1cm 0.7cm 0.2cm},clip,height=4.3cm]{./img/regions_istat_vs_dpc_smooth_16feb-30avr.pdf} \\
        \includegraphics[page=2,trim={0.1cm 1cm 0.7cm 0.2cm},clip,height=4.3cm]{./img/regions_istat_vs_dpc_smooth_16feb-30avr.pdf}
        \includegraphics[page=1,trim={1.2cm 1cm 0.7cm 0.2cm},clip,height=4.3cm]{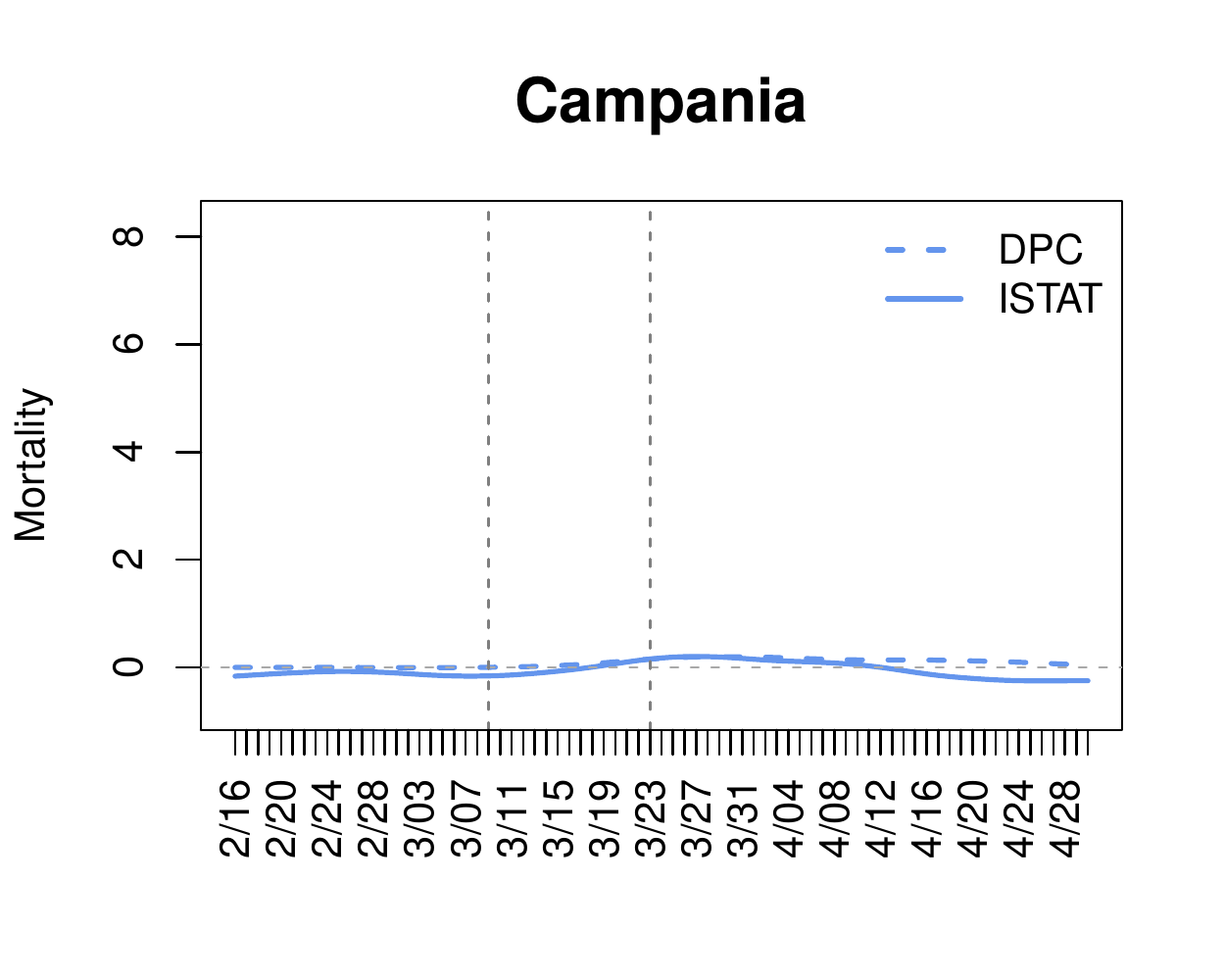}
    \end{minipage}}
    }
    \subfloat[\label{subfig:max}]{
    \fbox{
    \begin{minipage}[b]{0.32\linewidth}
        \includegraphics[page=1,trim={0.1cm 1cm 0.7cm 0.2cm},clip,height=4.3cm]{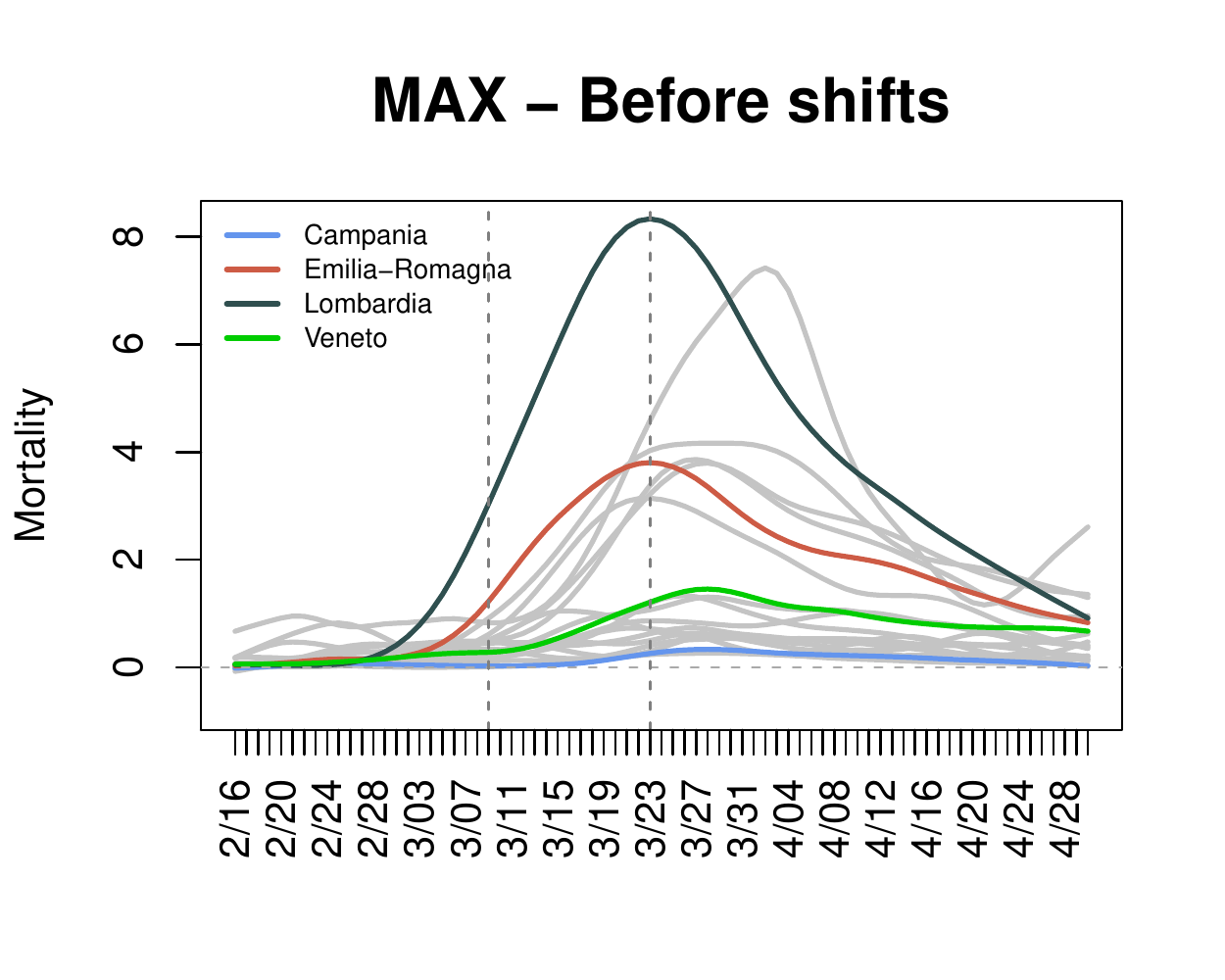}\\
        \includegraphics[page=2,trim={0.1cm 1cm 0.7cm 0.2cm},clip,height=4.3cm]{./img/regions_max_smooth_all.pdf}
    \end{minipage}}
    }
    \caption{
    {\bf Mortality curves}.
    {\bf \protect\subref{subfig:dpc_vs_istat}} DPC (dashed) and ISTAT (solid) differential mortality curves (per 100,000 inhabitants) in four example regions; Lombardia, Veneto, Emilia Romagna and Campania. 
    Curves are smoothed with splines, with degree of smoothing selected by generalized cross-validation (see Methods). ISTAT curves “take off” earlier and in some regions are as much as twice as high at their peak -- possibly due to many COVID-19 deaths happening at home and/or not being recorded as such in hospitals, especially in the early stages of the epidemic. 
    {\bf \protect\subref{subfig:max}} MAX mortality curves (per 100,000 inhabitants) in the 20 Italian regions, before (top) and after (bottom) the shifts produced by {\em probKMA} run with $K=2$. In the bottom panel, time is marked as a day number (as opposed to a date); this represents the region-specific time of the epidemic unfolding, and corresponds to actual time (starting on February 16 and ending on April  30) only for regions with no shifts, e.g., Lombardia). Curves are again smoothed with splines, with degree of smoothing selected by generalized cross-validation. Lombardia, Veneto, Emilia Romagna and Campania, also shown in \protect\subref{subfig:dpc_vs_istat}, are highlighted in color. 
    In all panels, vertical lines mark the dates of the national lock-down (March 9) and of the suspension of all nonessential production activities (March 23). In the bottom panel of \protect\subref{subfig:max} vertical lines still show these dates without shifts; stars on the curves mark the lock-down after the region specific shifts.
    }
    \label{fig:mortality}
    \end{figure}
    
    \section*{Results}
    %
    %
    Below we describe the salient outcomes of our analyses. After addressing some shortcomings in publicly available COVID-19 deaths records, we characterize two starkly different epidemic patterns and rank regional mortality curves. Next, we relate mortality to mobility and positivity, and to a number of socio-demographic, infrastructural and environmental factors.

    \subsection*{Under-counting deaths}
    
    Since February 24, the Italian Civil Protection agency (Dipartimento della Protezione Civile; DPC) has released daily counts of recorded COVID-19 deaths at the coarse resolution of regions (only the number of recorded cases are released at the finer resolution of provinces). In Italy and elsewhere, official death records have often been criticized as undercounts\cite{ciminelli2020covid}. Alternative data sources do exist, e.g., daily mortality rates -– which can be contrasted to those from prior years to gauge differential mortality. In Italy these are provided by the National Statistical Institute (ISTAT) at the resolution of municipalities. We aggregated the data over municipalities belonging to the same region and subtracted averages over the past 5 years (2015-19, see Methods)\cite{cremona2020}. Figure \ref{fig:mortality}\subref{subfig:dpc_vs_istat} shows smoothed DPC and ISTAT differential mortality curves (per 100,000 inhabitants) for some example regions (Lombardia, Veneto,  Emilia Romagna and Campania). The under-counting in the official DPC records was dramatic, especially in badly affected areas and in the initial stages of the epidemic. However, ISTAT differential mortality curves have themselves limitations,  especially in less affected areas, where they can fluctuate at small levels and even take negative values –- idiosyncratically or reflecting other COVID-19 related phenomena (e.g., increases in mortality due to untreated emergencies or reductions in mortality due to fewer accidents during the lock-down). We therefore formed maxima curves (MAX), where the largest between the DPC and the ISTAT datum is taken in each day and for each region, and then smoothed. These are shown in Fig.~\ref{fig:mortality}\subref{subfig:max} (DPC and ISTAT smoothed curves for all regions are shown in Figs.~\ref{fig:unshifted_curves_col_mortality} and \ref{fig:mortality_DPC_ISTAT}). 
    We repeated our analyses on all three data sets;
    given the small number of observational units at our disposal ($n=20$ regions), this allowed us to borrow strength replicating results across data sets, with their differences and limitations.
 
  \subsection*{Two different epidemics}
    
    \begin{figure}[!tb]
    \centering
    \subfloat[\label{subfig:probKMA_max}]{
    \fbox{
    \begin{minipage}[b]{0.63\linewidth}
        \includegraphics[page=1,trim={0.1cm 1cm 0.7cm 0.2cm},clip,height=4.3cm]{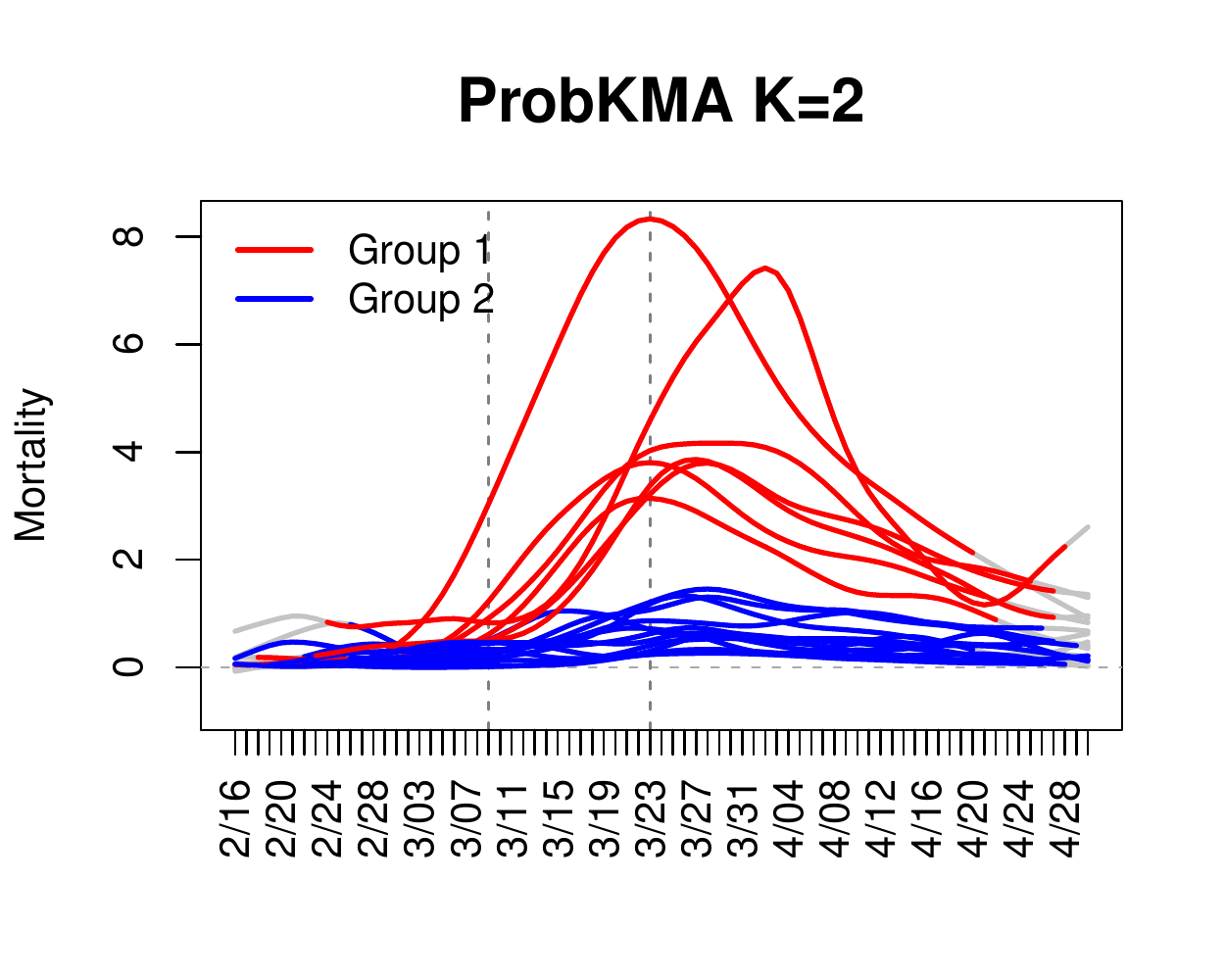}
        \includegraphics[page=4,trim={0.8cm 1cm 0.7cm 0.2cm},clip,height=4.3cm]{./img/regions_max_probKMA_d0_K2.pdf} \\
        \includegraphics[page=2,trim={0.1cm 1cm 0.7cm 0.2cm},clip,height=4.3cm]{./img/regions_max_probKMA_d0_K2.pdf}
        \includegraphics[page=3,trim={0.8cm 1cm 0.7cm 0.2cm},clip,height=4.3cm]{./img/regions_max_probKMA_d0_K2.pdf}
    \end{minipage}}
    }
    \subfloat[\label{subfig:IWTomics_max}]{
    \fbox{
    \begin{minipage}[b]{0.32\linewidth}
        \includegraphics[page=2,trim={0.3cm 0.2cm 0cm 0cm},clip,height=8.63cm]{./img/regions_max_IWTomics.pdf}
    \end{minipage}}
    }
    \caption{
    {\bf Characterizing two epidemics}.
    {\bf \protect\subref{subfig:probKMA_max}} MAX mortality curves are shown in the top left panel with 65-day portions identified by {\em probKMA} with  $K=2$ in red (Group 1; "exponential" pattern) and blue (Group 2; "flat(tened)" pattern). The curve portions are shown again, this time aligned with each other and separated by group, in the bottom panels. Black lines indicate group averages. The shifts produced by {\em probKMA} are shown in the top right panel (motifs, groups and shifts for Group 1 are stable across data sets; shifts for Group 2 are less stable and less interpretable -- see Fig. \ref{fig:two_epidemics_dpc_istat}). 
    {\bf \protect\subref{subfig:IWTomics_max}} Shifted Group 1 and Group 2 MAX mortality curves are tested against each other with {\em  IWTomics}. The heatmap at the top shows $p$-values 
    adjusted at all possible scales (from 1 to 65 days). 
    The middle panel shows in detail the top-most row of the heatmap; i.e.~the $p$-values adjusted across the whole 65-day interval. The bottom panel shows again the shifted curves.
    Gray areas in the middle and bottom panels mark days when the difference between the two groups is significant (adjusted $p$-value $<5\%$). Starting 
    a little over two weeks from the beginning of their epidemic, curves in the two groups differ significantly at all temporal scales.
    }
    \label{fig:two_epidemics}
    \end{figure}
    
    Italy saw the unfolding of two very different epidemics; a relatively mild one in the majority of the country, and a tragic, seemingly out of control one in its most hard-hit regions. These two epidemics can be effectively characterized with {\em probKMA}, an FDA technique designed to identify recurrent motifs within a set of curves, and group the curves based of the motifs they comprise\cite{cremona2020}. Here, the motifs are the temporal patterns of deaths that characterize alternative epidemic unfoldings, which may in fact start at different times in different curves (regions). Thus, the algorithm also produces the shifts required to align regions comprising the same motif to each other. 
    {\em ProbKMA} is similar to a $K$-mean algorithm; it requires the user to specify the number of motifs ($K$) at the outset, and to select a distance –- which can be defined on the curve levels, their derivatives, or a combination of both (see Methods). 
    
    The solution with $K=2$ and distance defined on curve levels depicts two starkly different epidemics, shown for the MAX curves in Fig.~\ref{fig:two_epidemics}\subref{subfig:probKMA_max}. Allowing for shifts, these are represented by 65-day long motifs.
    Group 1 undergoes a steep ascent (the “exponential” pattern) followed by a slower descent from the peak; it includes many northern regions. Based on the shifts, Lombardia was first, followed by Emilia Romagna, Marche, Liguria, Piemonte, Trento/Bolzano, and last Valle d’Aosta. Lombardia and Valle d’Aosta presented the most extreme peaks – but Valle d’Aosta's descent was 
    steeper (with a second late ascent likely due to data recording imprecisions; Valle d'Aosta is a very small region with only $\approx 125,000$ inhabitants).
    Group 2 follows a “flat(tened)” pattern; it includes all regions in southern and central Italy and, remarkably, Veneto -- where the curve was successfully curbed.
    The shifts produced for this group are less stable and less meaningful in terms of interpretation, as flatter profiles leave more leeway in aligning curves against each other. 
    All results (except for the shifts in Group 2) are rather consistent when using DPC and ISTAT curves (see Fig.~\ref{subfig:probKMA_dpc} and Fig.~\ref{subfig:probKMA_istat}), and when using distances defined on derivatives instead of curve levels.
    The solution with $K=3$ places Lombardia (ISTAT curves) or Lombardia and Valle d’Aosta (MAX and DPC curves) in a cluster of their own (see Fig.~\ref{fig:three_epidemics}). 
    We also validated our results using a modification of {\em  funBI\cite{funbi2019}}, a functional biclustering technique, and {\em IWTomics}\cite{cremona2018}, a functional testing technique which contrasts two sets of aligned curves pinpointing the locations and scales at which they differ (see Methods). Figure \ref{fig:two_epidemics}\subref{subfig:IWTomics_max} shows how, starting 
    a little over two weeks from the beginning of their motif (wherever that was in each curve), Group 1 and Group 2 differ significantly at all temporal scales (see also Fig.~\ref{subfig:IWTomics_dpc} and Fig.~\ref{subfig:IWTomics_istat}). 
    
    Why the two epidemics? The pattern of deaths characterizing Group 1 may be due, in large part, to the fact that the virus had circulated silently in the north of Italy for a long period of time before any kind of behavioral changes by the general public, medical protocols, or mitigation policies by local and central authorities were put in place. Mounting evidence suggests that a large share of COVID-19 cases are asymptomatic and yet contagious\cite{Lavezzo2020}; their numbers may have increased until a pent-up reservoir of virus found its way to vulnerable individuals 
    (some researchers also hypothesize Antibody-Dependent-Enhancement of SARS-CoV-2\cite{cegolon2020}, and  thus a role for re-infections). 
    But a variety of additional factors may have contributed to shaping the two epidemics; we explore some below.

   \subsection*{Ranking mortality curves}
    
\begin{figure}[!tb]
    \centering
        \subfloat[\label{subfig:boxplot}]{
    \fbox{
    \begin{minipage}[b]{0.407\linewidth}
        \includegraphics[width=0.99\linewidth]{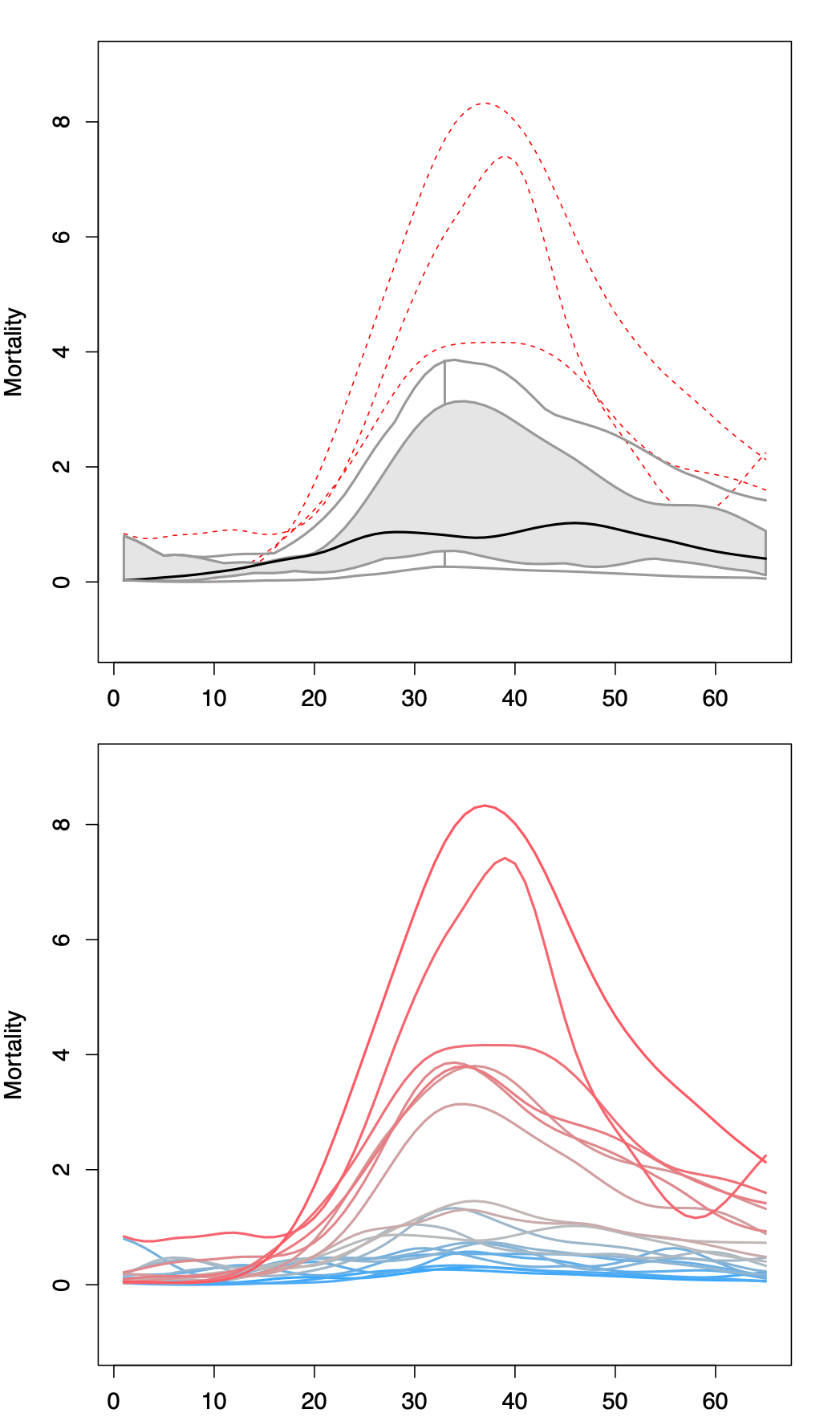}
    \end{minipage}}
    }
    \subfloat[\label{subfig:ranking}]{
    \fbox{
    \begin{minipage}[b]{0.53\linewidth}
        \includegraphics[width=0.99\linewidth]{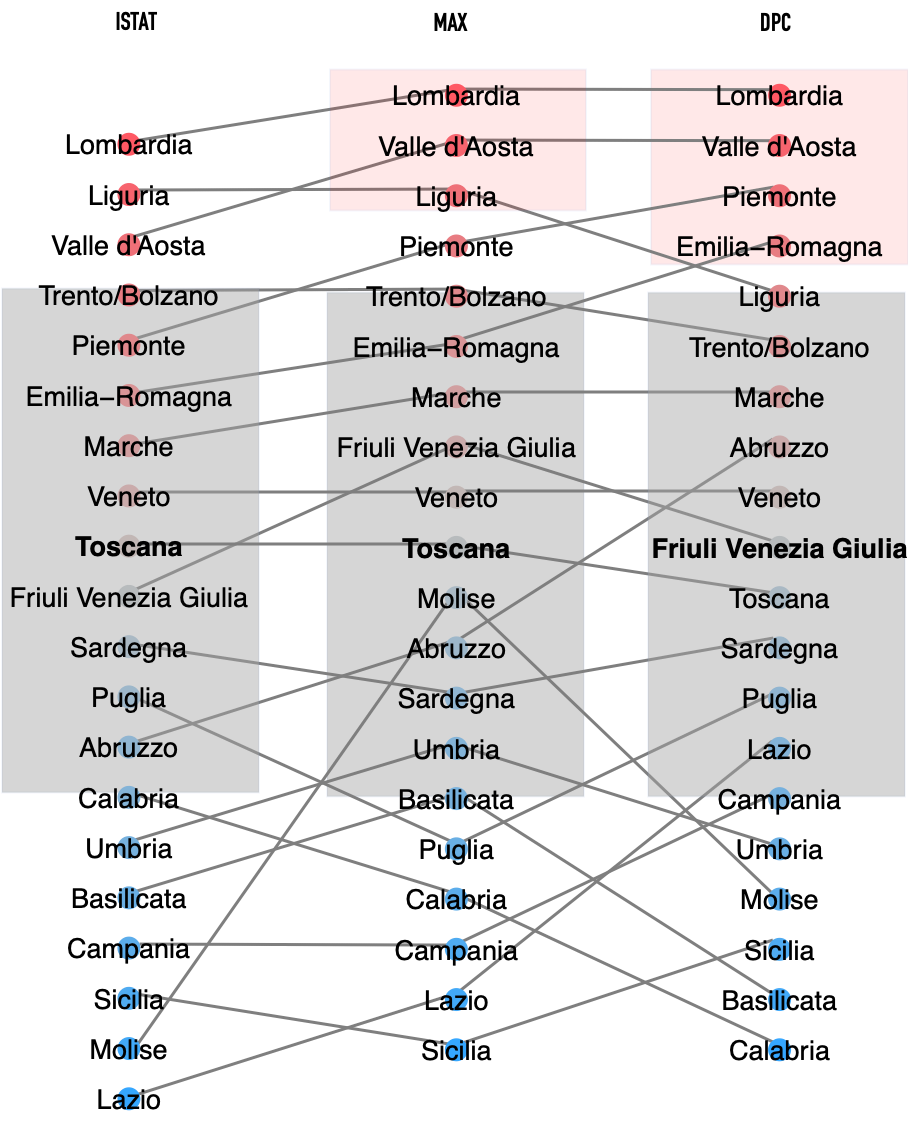}\\
    \end{minipage}}
    }
    \caption{
    {\bf Functional boxplot and ranking}.  
    {\bf \protect\subref{subfig:boxplot}} Functional boxplot of the MAX data set (top) and MAX mortality curves (bottom) color-coded according to their ranking, as shown in the MAX column of \protect\subref{subfig:ranking}. In the boxplot, Toscana is the median (black continuous line); Lombardia, Valle d'Aosta and Liguria are identified as outliers (red dashed lines); and the 50\% innermost "box" (grey area) include the curves for Trento/Bolzano, Emilia-Romagna, Marche, Friuli Venezia Giulia, Veneto, Toscana, Molise, Abruzzo, Sardegna, Umbria, and Basilicata. Note that the "box" is skewed upwardly. 
    {\bf \protect\subref{subfig:ranking}} Rankings of the ISTAT (left), MAX (center) and DPC (right) mortality curves. The median regions are in bold, gray rectangles mark the 50\% innermost boxes, and pale red rectangles mark outliers (no region is labeled as an outlier in the ISTAT data set; see Methods). The dots representing each region are color-coded (from intense red, through gray, to intense blue) according to their signed depth values (see Methods). In all three data sets, Lombardia's curve is the most extreme at the very top of the ranking and, in contrast, Veneto's curve is deep in the bulk close to the median (Toscana for ISTAT and MAX, Friuli Venezia Giulia for DPC). Segments joining the regions across the three rankings show how the top portion remains rather stable, while the mid- and bottom portions contain several crossings. Regions at the top are those characterized  by "exponential" epidemics (Group 1), while regions in the middle and at the bottom are those with "flat(tened)" epidemics (Group2), whose curves can more easily switch in their depth ranks.}
    \label{fig:quantiles}
    \end{figure}

    Non-parametric FDA methods can be used to rank curves based on the notion of depth –- from the innermost  to the most extreme, and to identify outliers\cite{sun2011functional,lopez2009concept}. Figure \ref{fig:quantiles} shows a functional box plot of the MAX mortality curves and a depth ranking of the curves in the DPC, ISTAT and MAX data sets -- shifted based on {\em probKMA} run with $K=2$ and restricted to their aligned 65-day portions. The ranking is directional; we attributed signs to the depth measurements, so that curves far over or under the median curve are at the top or bottom of the ranking, respectively (see Methods). The top portion of the ranking comprises regions with "exponential" epidemics (Group 1) and is rather stable across data sets; Lombardia and Valle  d'Aosta are consistently among the most extreme curves (they are also identified as outliers in the MAX and DPC data sets). The mid- and bottom portions of the ranking comprise regions with "flat(tened)" epidemics  (Group 2) and are less stable across data sets, as the flatter profiles can more easily switch in their depth ranks.
    However, Toscana (which is the median in the MAX and ISTAT data sets) and Veneto are consistently among the deepest, most central  curves. This analysis highlights again the tragic epidemic unfolding in Lombardia, and, by contrast, confirms how Veneto managed to “flatten” its curve back into the bulk.
    
    \subsection*{Local mobility and positivity as statistical predictors of mortality}
    
    \begin{figure}[!tb]
    \centering
    \subfloat[\label{subfig:mob_pos_shifted}]{
    \fbox{
    \begin{minipage}[b]{0.372\linewidth}
        \centering
        \includegraphics[width=0.99\linewidth]{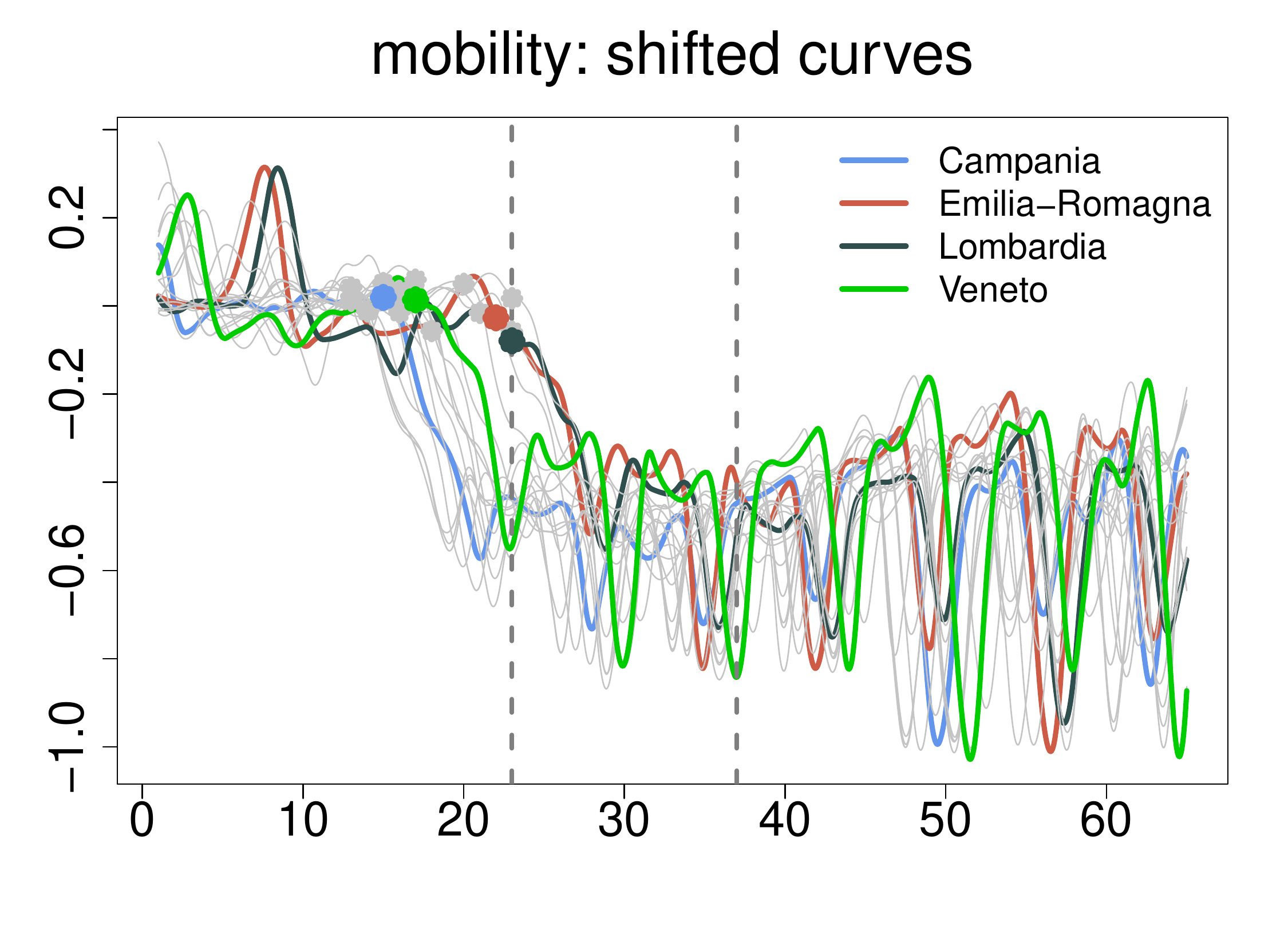} \\
        \includegraphics[width=0.99\linewidth]{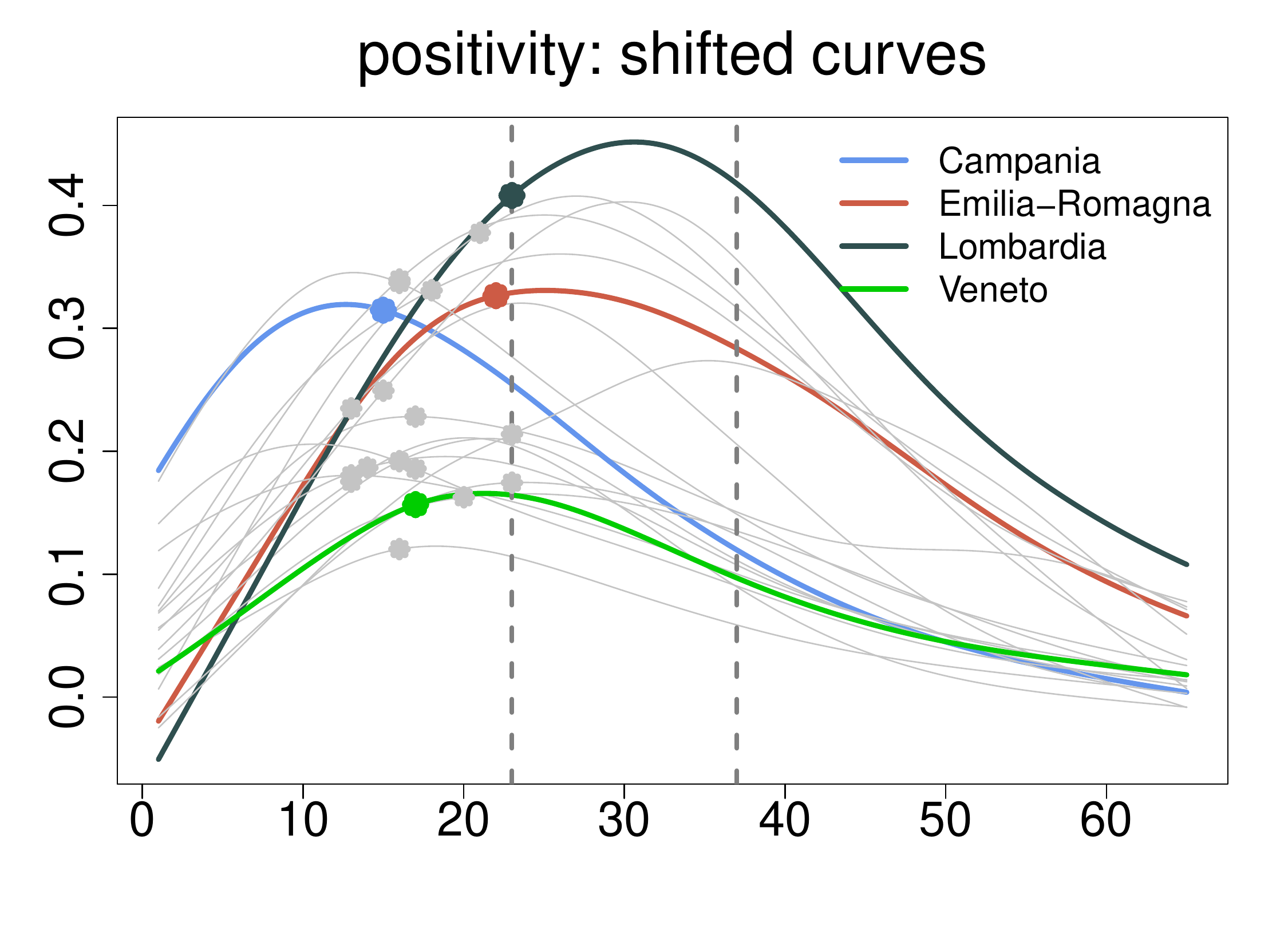}
    \end{minipage}
    }
    }
    \subfloat[\label{subfig:mob_pos_reg_res}]{
    \fbox{
    \begin{minipage}[b]{0.555\linewidth}
        \centering
        \includegraphics[width=0.48\linewidth]{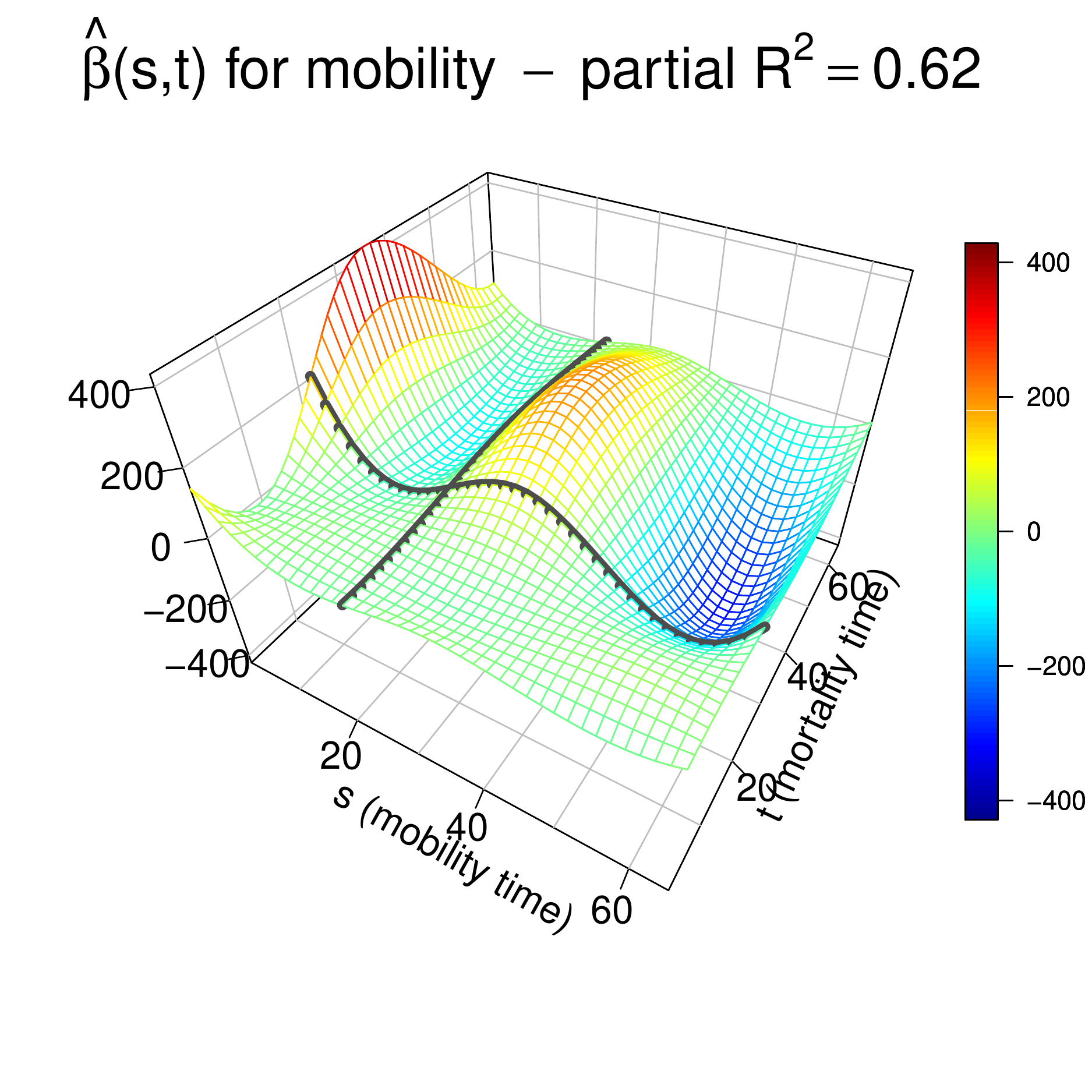}
        \hspace{0.2cm}
        \includegraphics[width=0.48\linewidth]{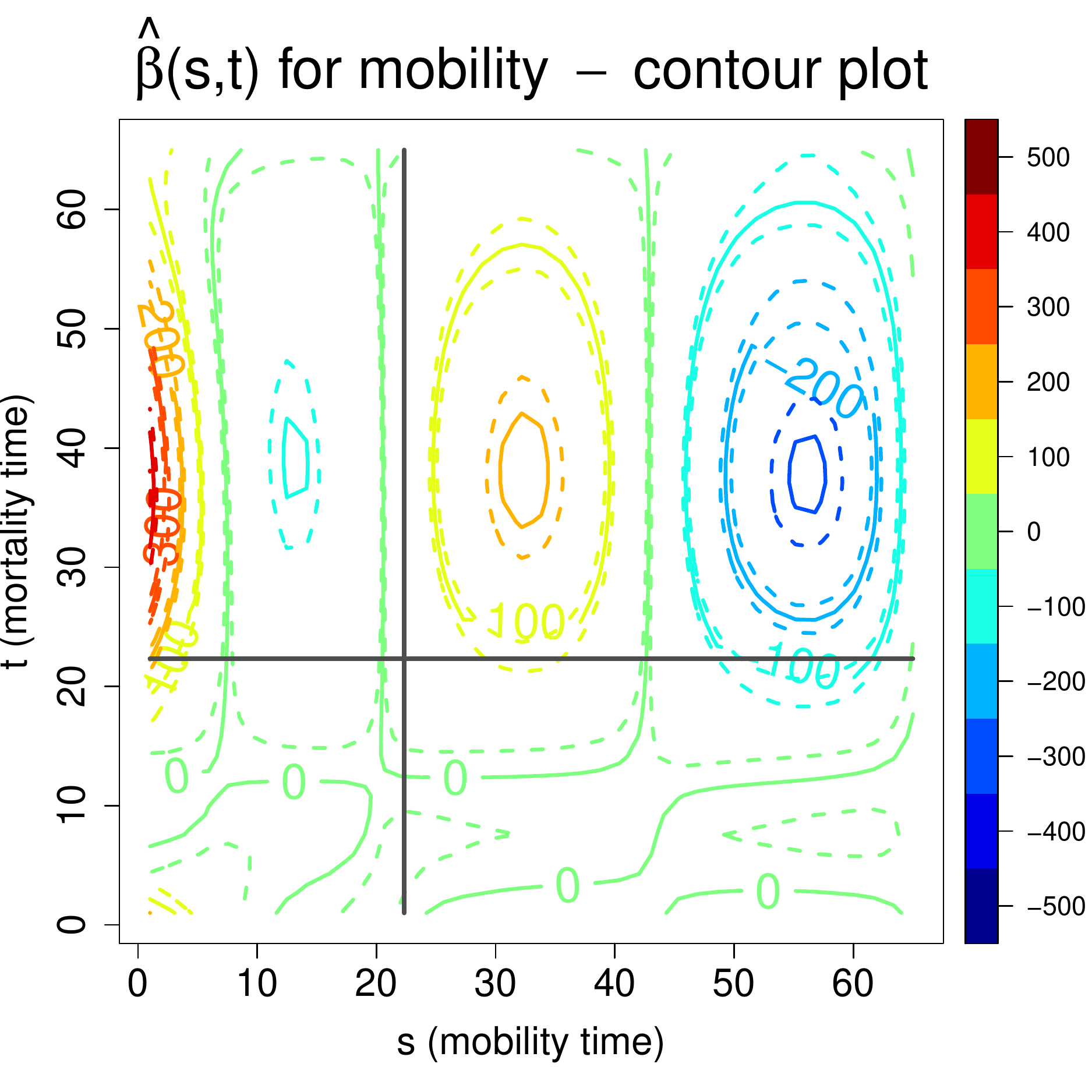} \\
      \includegraphics[width=0.48\linewidth]{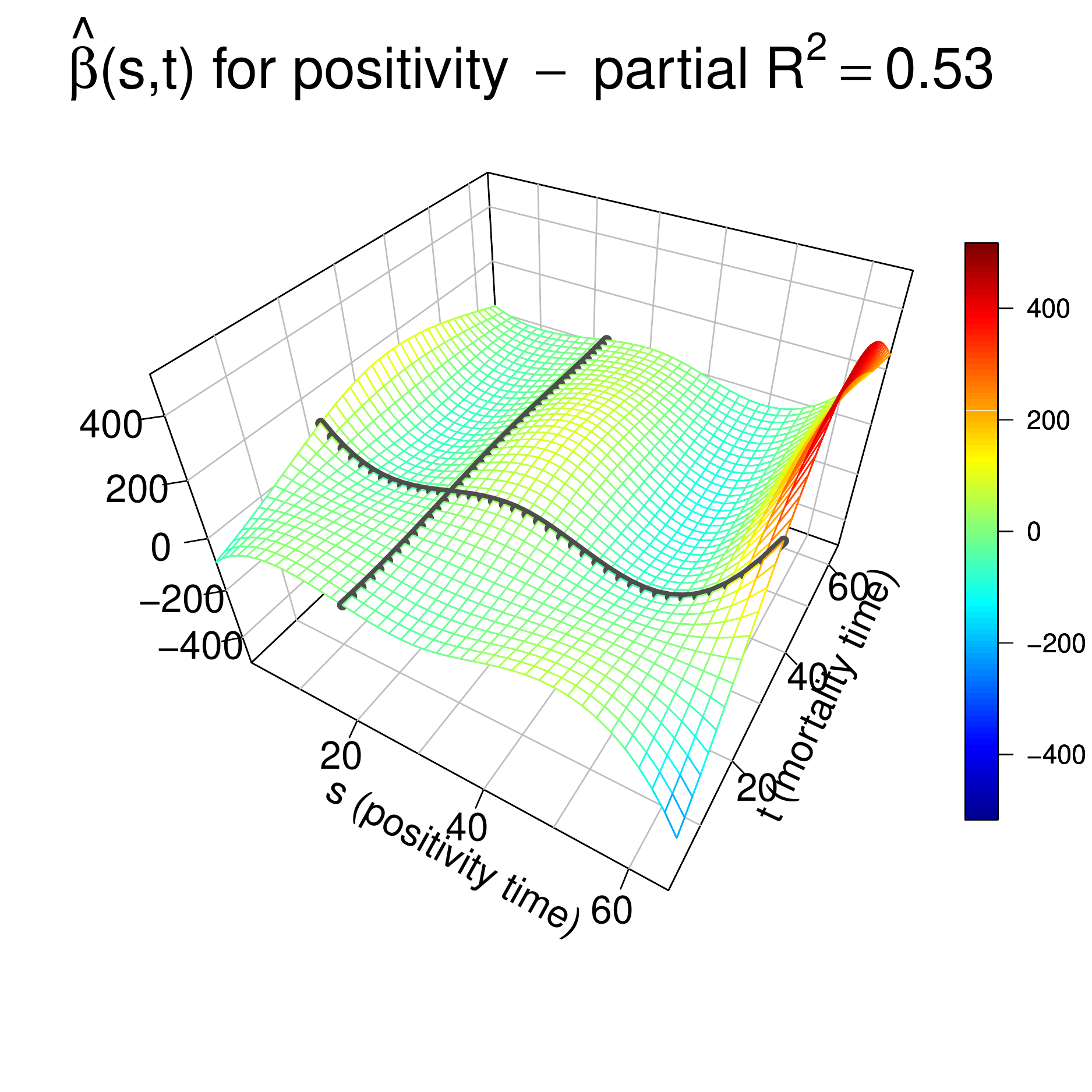}
      \hspace{0.2cm}
      \includegraphics[width=0.48\linewidth]{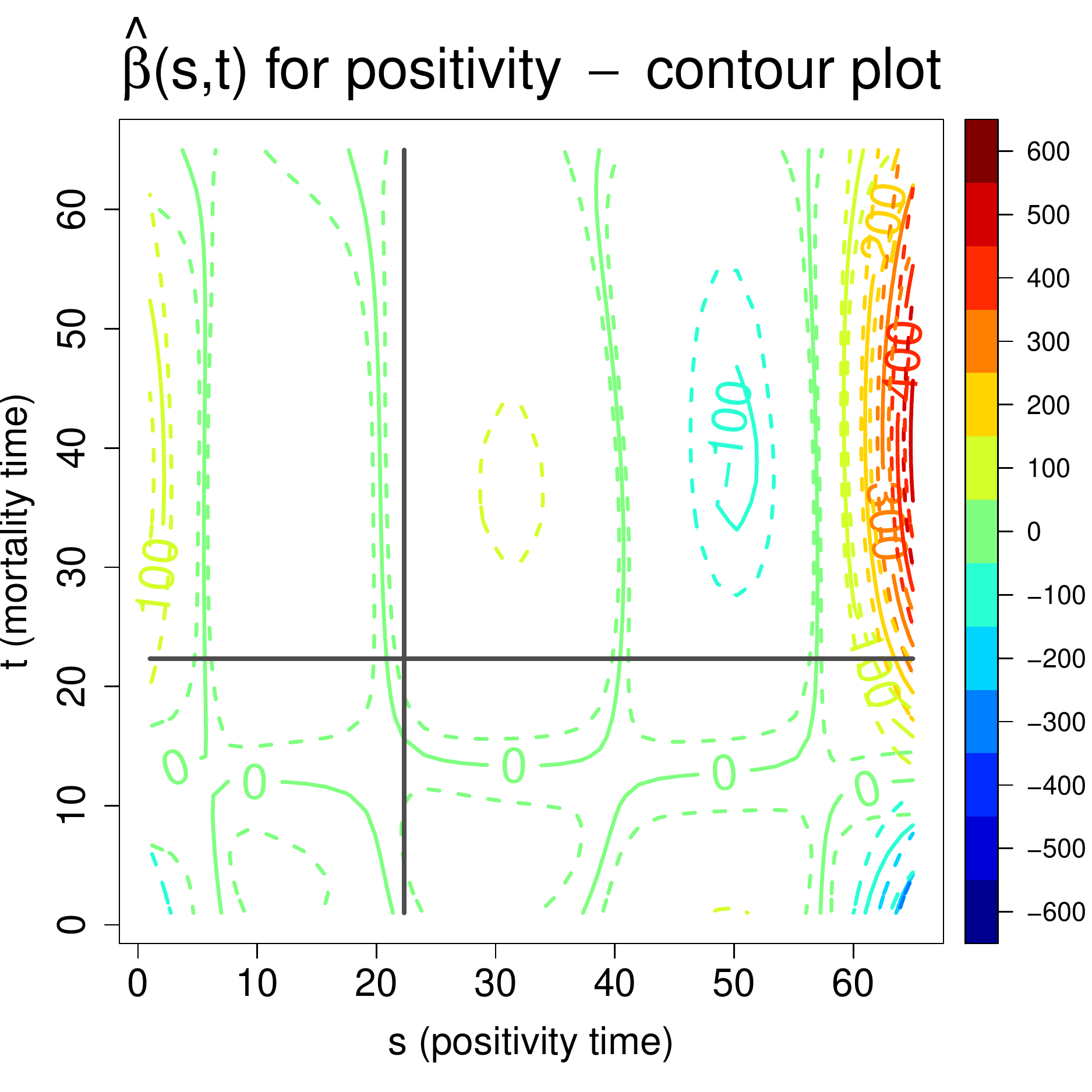}
    \end{minipage}
    }
    }
    \caption{
    {\bf Associating mortality to local mobility and positivity}. 
    {\bf \protect\subref{subfig:mob_pos_shifted}} Local mobility curves (Google's “Groceries \& pharmacy”) and positivity curves (regularized ratios of new cases to number of tests performed) in the 20 Italian regions. Curves are smoothed with splines, with degree of smoothing selected by generalized cross-validation, and shifted based on {\em probKMA} run on the MAX mortality curves with $K=2$; time is  marked as a day number representing the region-specific time of the epidemic unfolding, and corresponds to actual time (starting on February 16 and ending on April 30) only for regions with no shifts, e.g., Lombardia. Vertical lines show the days corresponding to the nationwide lock-down (March 9) and the suspension of all nonessential production activities (March 23)  without shifts, stars on the curves mark the lock-down after the region specific shifts. The example regions of Fig.~\ref{fig:mortality}\protect\subref{subfig:dpc_vs_istat} are highlighted in color. 
    {\bf \protect\subref{subfig:mob_pos_reg_res}} Estimated effect surfaces from the joint function-on-function regression of MAX mortality on local mobility and positivity shown in 3D and as contour plots (March 9, without shift, is again marked on both). Early and mid-period local mobility levels are strong positive predictors of mortality at its peak. Positivity has similar but much weaker predictive signals, likely because the effects are subsumed by mobility. Late local mobility has a negative association with mortality at its peak (mobility resumed faster in regions with milder epidemics), and late positivity a strong positive one (positivity remained elevated in regions with worse epidemics). The regression captures a large share of the variability in mortality curves (in-sample $R^2=0.90$, LOO-CV $R^2=0.52$), 
    with substantial and comparable contributions of the two predictors (partial $R^2$s $=0.62, 0.53$). 
    }
    \label{fig:mob_pos}
    \end{figure}
    
    Next, we focus on two key variables. The first is one of the most discussed policy-actionable variables, mobility, which has been curtailed to various degrees through lock-down measures in most of the countries affected by COVID-19. The second is one of the most discussed sentinel indicators, positivity, i.e.~the fraction
    of performed tests returning positive results. For both these variables daily values for the period February 16 – April 30 can be obtained from data in the public domain at regional resolution. 
    
    We considered differential mobility curves  provided by Google for the category “Grocery \& pharmacy”. These express the  fractional reduction with respect to January 2020 levels, and refer to mobility linked to first necessities –- such as buying food, medicine,  etc. For Italy,  they are provided at the resolution of regions. Even though individuals were allowed to leave their homes for these necessities also during the  most restrictive phase of the lock-down, the reduction captured by Google's “Grocery \& pharmacy” was substantial. 
    Mobility in weekdays fell by roughly $0.30$, i.e.~$30\%$, in the week after the lock-down (March 9), and further decreased in following weeks -- reaching the lowest levels (between approximately $-0.60$ and $-0.40$ depending on the region) in the week after the suspension of nonessential production activities (March 23). It then slowly increased,  getting back in a range between approximately $-0.40$ and $-0.20$ at the very end of April (see Fig.~\ref{fig:unshifted_curves_col_pos_mob}).
    In Lombardia, the peak MAX mortality was between March 20 and 25 -- i.e., roughly, simultaneous to the lowest mobility and two weeks after its first substantial drop. Notably, in most Italian regions mobility during lock-down weekends reached $-1.00$, i.e.~$-100\%$. For comparison, in the state of New York, which had among the strongest restriction  measures in the U.S., Google's  “Grocery \& pharmacy” never fell below $-0.40$.
    We refer to Google's “Grocery \& pharmacy” curves as {\em local} mobility because they measure how much individuals move around where they live, as opposed to how much individuals move from place to place –- e.g., to go from Wuhan to Milan, or from Milan to Palermo, or New York City. Obviously both types of mobility are relevant for the spread of a virus, and definitions depend on scale/resolution, but the first one is the one we analyzed.
    
    To construct positivity curves, we combined daily public records on number of tests performed and number of new cases, which are also provided by the Italian Civil Protection agency. Taking daily ratios of new cases on tests performed is clearly imperfect, because of (variable and unreported) delays in test results. But regularizing and smoothing these ratios (see Methods) produced a reasonable proxy. Smoothed positivity surpassed $0.1$, i.e.~$10\%$, as early as February 20 in some hard hit regions, peaked in a staggered fashion throughout March, and fell below $0.10$ for all regions by around April 22 (see Fig.~\ref{fig:unshifted_curves_col_pos_mob}). 
    Lombardia surpassed $0.10$ around February 22 and peaked around March 15-18; that is, roughly, about a month and about a week prior to the peak of MAX mortality,
    respectively. Though we cannot draw exact parallels (our positivity curves are approximate and smoothed), this is consistent with what was observed, e.g., in New York City -- where positivity  was above $0.10$ approximately from March 6-7 to May 12-13 and peaked at about $0.70$ around March 28, with deaths peaking between April 5 and 13. 

    To anchor local mobility and positivity curves to the epidemic unfolding in each region, we shifted them congruently with the mortality curves. Figure \ref{subfig:mob_pos_shifted} displays shifted curves based on {\em probKMA} run on MAX data with $K=2$ (Fig.~\ref{fig:shifted_curves_istat_dpc}
    displays shifted curves based on \textit{probKMA} run on DPC and ISTAT data).
    The horizontal axis now indicates again days in the region-specific epidemic unfolding, restricted to the 65-day portions where mortality curves align forming the two {\em probKMA} motifs.

    We then used function-on-function regressions\cite{ramsay2005,horvath2012inference} to model the statistical dependence of mortality on local mobility and positivity; in symbols, we fit the joint model 
    $ y(t) = \alpha(t) + \int \beta_{mob}(s,t)x_{mob}(s)ds + \int \beta_{pos}(s,t)x_{pos}(s)ds + \epsilon(t)$, where $y(t)$ is the response curve, i.e.~mortality, $\alpha(t)$ is the intercept, $\epsilon(t)$ is the model error, and $x_{mob}(s)$ and $x_{pos}(s)$ are the predictor curves  -- mobility and positivity, respectively. These predictors are integrated over time, 
    with “effects” represented by {\em surfaces}; $\beta_{mob}(t,s)$ is the association of mortality at time $t$ with local mobility at time $s$, and similarly $\beta_{pos}(t,s)$ for positivity (see Methods).
    
    Figure \ref{subfig:mob_pos_reg_res} shows the effect surfaces for local mobility and positivity estimated using the MAX curves as response. $\hat{\beta}_{mob}(t,s)$ suggests that local mobility levels early on and mid-way through the epidemic (e.g., around the March 9 lock-down date for Lombardia) are strong positive predictors of mortality at its peak, with the early predictive signal stronger than the mid-way one. In contrast, the local mobility level late in the epidemic 
    has a negative association with mortality at its peak, likely reflecting a faster resumption of mobility in regions with milder epidemics. $\hat{\beta}_{pos}(t,s)$ suggests that positivity levels early on and mid-way through the epidemic are also positive predictors of mortality at its peak -- though the predictive signals are substantially weaker than those of mobility, likely because they are 
    confounded with the latter. However, the positivity level late in the epidemic has a marked positive association with mortality at its peak. Here the signal is "detangled" from that of mobility, and one finds a sort of retrospective signature; regions which fared worse still had hightened positivity in the late stages of their epidemics.
    Estimated effect surfaces are remarkably similar across the three data sets (MAX, DPC and ISTAT), and the joint models all have in-sample $R^2$s above 90\% and leave-one-out cross-validated (LOO-CV) $R^2$s above 50\% (see Table \ref{tab:R2}), 
    with strong and comparable contributions of local mobility and positivity (e.g., for the MAX curves, the partial $R^2$s are 62\% and 53\%, respectively). Also, while this is not the case for all regions, residuals are rather consistent across data sets for Veneto, whose mortality is well predicted, and for Lombardia, whose mortality is always and sizably underestimated (see Fig.~\ref{subfig:mob_pos} and Fig.~\ref{fig:mob_pos_dpc_istat}).

     In order to further assess the roles of local mobility and positivity, we also considered {\em marginal} function-on-function regressions for mortality on each, separately; in symbols,
    $ y(t) = \alpha(t) + \int \beta_{mob}(s,t)x_{mob}(s)ds +  \epsilon(t)$ and $y(t) = \alpha(t) + \int \beta_{pos}(s,t)x_{pos}(s)ds + \epsilon(t)$.
     Effect surface estimates for local mobility are very similar to those in the joint models for all three data sets (see Fig.~\ref{fig:mob_marginal}). 
    Those for positivity confirm a strong association with mortality at its peak, but are less defined in terms of time profile (see Fig.~\ref{fig:pos_marginal}). 
    In summary, we find substantial 
    evidence that local mobility and positivity are associated with COVID-19 mortality, and can predict it with some lag-time. Though the data at our disposal does not allow us to pinpoint lag lengths with accuracy, our analysis does support their roles as policy-actionable and monitoring variables, respectively. We also find that, even when considered jointly, these  variables are not enough to fully account for the massive numbers of COVID-19 deaths recorded in Lombardia, the worst hit region in the country.
    
    \subsection*{The role of socio-demographic, infrastructural and environmental factors}
    
    We considered several scalar 
    (non-longitudinal) covariates proxying for socio-demographic, infrastructural and environmental factors debated by scientists and policy-makers during the epidemic (see Table \ref{tab:covariatelis}).
    Many of these are suboptimal proxies
    retrieved from  public sources. They refer to the closest times we could find data for (usually 2017 or 2018) and are, too, at the coarse resolution of regions. After an initial screen to guarantee data quality and limit the amount of collinearities, we focused on $12$ covariates capturing aging of the population; prevalence of pre-existing conditions believed to affect disease severity; quality of distributed primary health care vs.~centralized hospital-based health care; the potential of hospitals and nursing homes, but also schools, work places, households and public transport to act as contagion hubs; and pollution levels (see Table~\ref{tab:proxies}; Fig. \ref{fig:ggally} provides marginal densities, pair-wise scatter plots and correlations). 
    
    \begin{table}[htb]
  
        \begin{tabularx}{\textwidth}{lXl}
        \toprule
        {\bf Covariate} & {\bf Description [comment in legend]} & {\bf Year and Source} \\
       \midrule 
        \% Over 65  
        & Aging of the population [1]  
        &  2018, ISTAT\\
        
        \% Diabetics  
        & Prevalence of relevant pre-existing conditions [2]  
        &  2018, ISTAT \\
        
        \% Allergic  
        & Another potentially relevant pre-existing condition 
        &  2018, ISTAT \\
        
        Adults per family doctor 
        & Quality of distributed, primary health care 
        &  2017, Ministry of Health\\
        
        ICU beds per 100K inhabitants  
        & Quality of centralized, hospital-based health care [3] 
        &  2018, Ministry of Health \\
        
        Ave.~beds per hospital (whole) 
        & Ability of hospitals to act as contagion hubs 
        &  2018, Ministry of Health \\
        
        Ave.~beds per nursing home (ward) 
        & Ability of nursing homes to act as contagion hubs 
        &  2018, Ministry of Health \\
        
        Ave.~students per classroom  
        & Ability of schools to act as contagion hubs 
        &  2018, Ministry of Education \\
        
        Ave.~employees per firm  
        & Ability of work places to act as contagion hubs 
        &  2017, ISTAT \\
       
        Ave.~members per household   
        & Ability of households to act as contagion hubs [4] 
        &  2017, ASR Lombardia \\
        
        Public transport rides per capita  
        & Ability of public transport to act as contagion hub 
        &  2017, ISTAT \\
       
        PM10  
        & Pollution levels (particulates)
        &  2018, ISTAT \\
       \bottomrule
        \end{tabularx}
        \caption{
        {\bf Scalar covariates potentially affecting COVID-19 mortality}.
        [1] The percentages of over 65, 70, 75 and 80 are highly correlated at the resolution of regions; we took over 65 as representative.
        [2] The prevalence of diabetes, hypertension and chronic bronchitis are highly correlated at the resolution of regions; we took diabetes as representative (allergies are not as highly  collinear and were retained as separate).
        [3] Availability of ICU beds is also directly relevant for withstanding the impact of COVID-19 surges. 
        [4] Average members per household is {\em not} a direct proxi of inter-generational contacts, but it may capture some of its effects. 
        }
         \label{tab:proxies}
    \end{table}
    
\begin{figure}[!hbt]
    \centering
    \subfloat[\label{subfig:cluster}]{
    \fbox{
    \begin{minipage}[b]{0.58\linewidth}
        \includegraphics[width=0.99\linewidth]{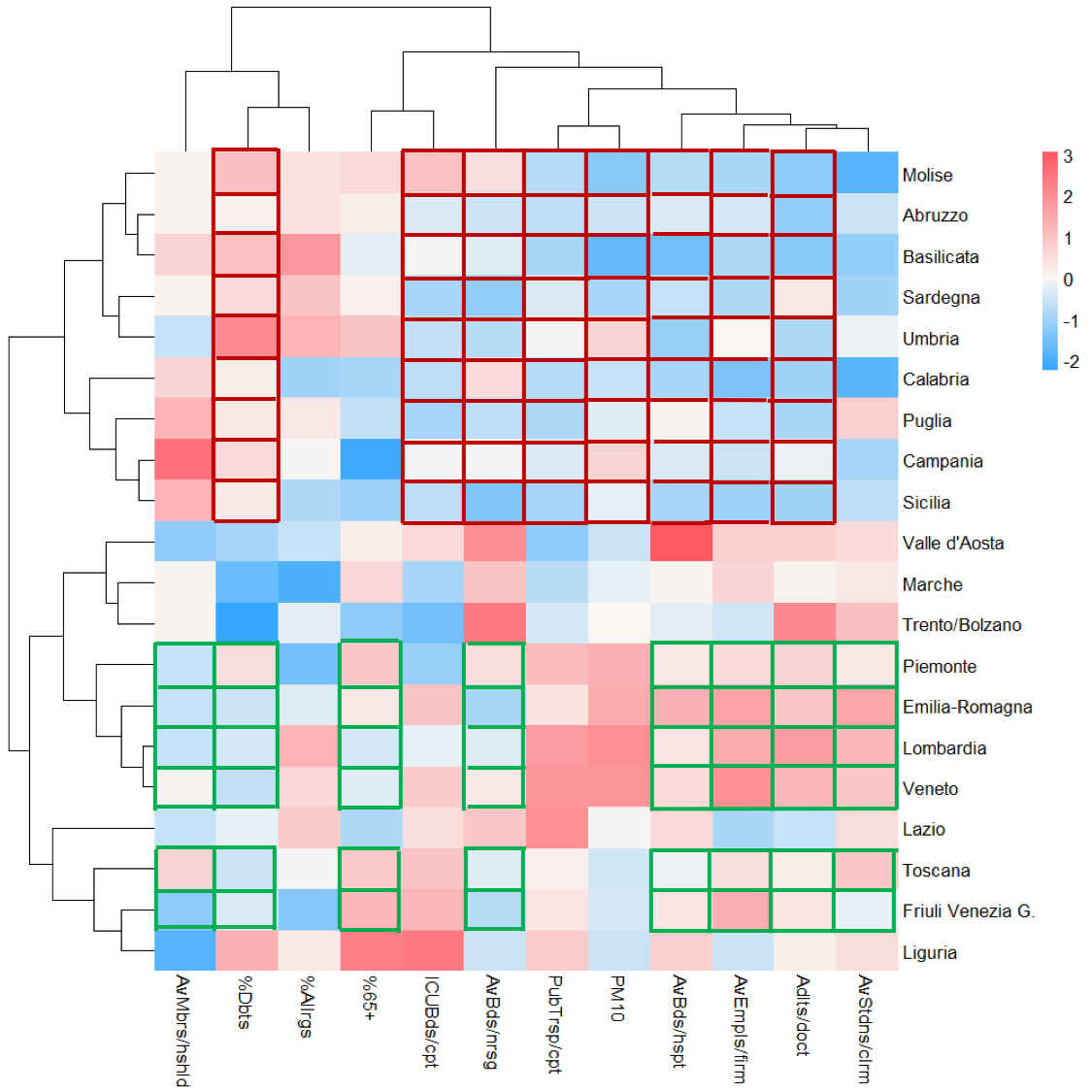}
    \end{minipage}}
    }
    \subfloat[\label{subfig:bicluster}]{
    \fbox{
    \begin{minipage}[b]{0.372\linewidth}
        \includegraphics[width=0.99\linewidth]{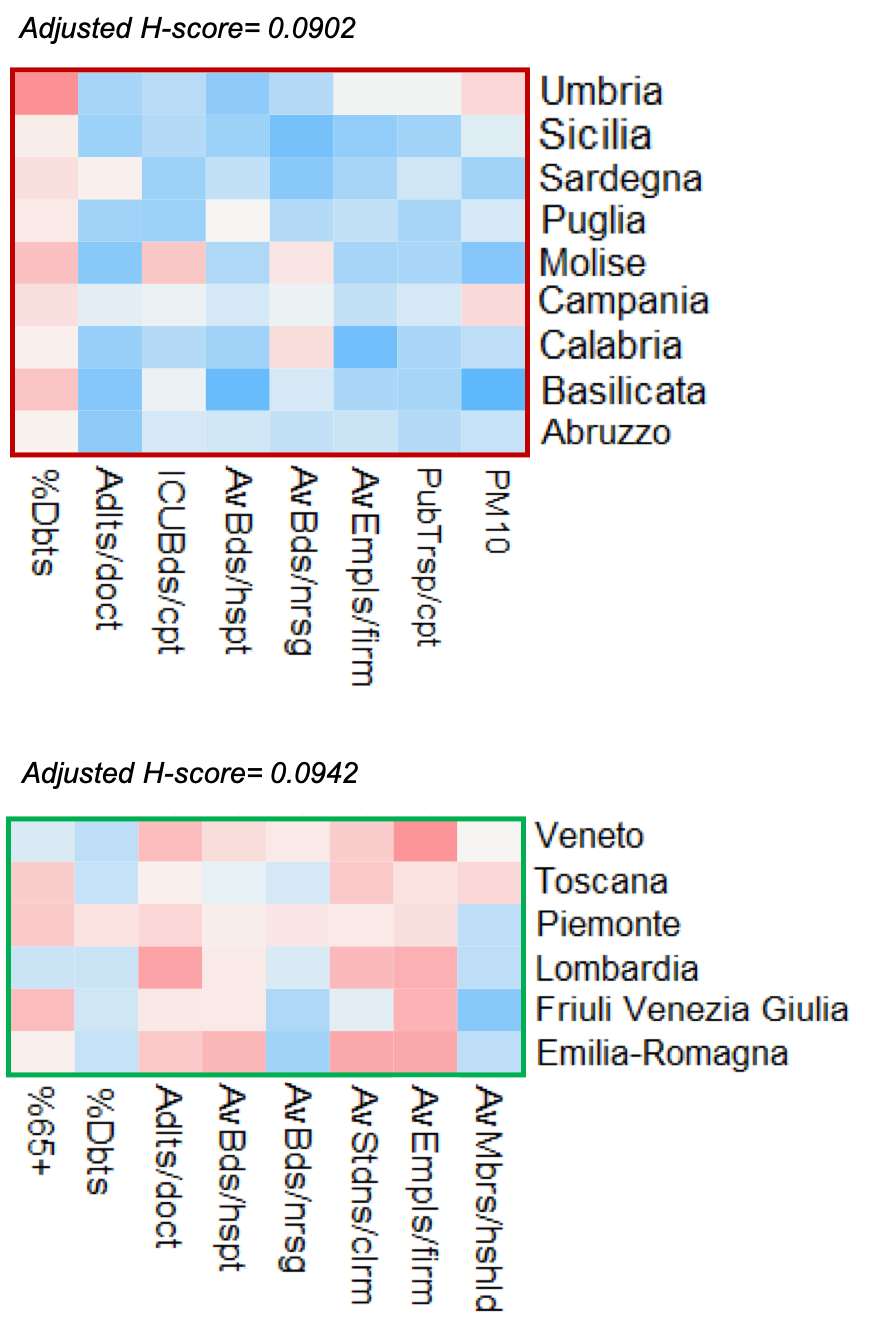}\\
    \end{minipage}}
    }
    \caption{
     {\bf Interdependencies among scalar covariates and regions}.
    {\bf \protect\subref{subfig:cluster}}: heatmap of the 20 (regions) x 12 (covariates) data matrix, with dendrograms from separate hierarchical clustering (correlation distance, complete linkage) of the regions (left) and the covariates (top). Color coding within cells represents values of the standardized covariates (centered and scaled to mean $0$ and standard deviation $1$). Color coding of some cell borders identifies the biclusters in \protect\subref{subfig:bicluster}.
    The dendrograms capture a distinct interdependence structure. For instance, there are marked similarities among Lombardia, Veneto, Emilia Romagna and Piemonte, as well as among some groups  of southern regions (Sicilia, Campania, Puglia and Calabria; Basilicata, Abruzzo and Molise). There are also marked associations among groups of covariates. The contagion hubs proxies for hospitals, schools and work places, and number of adults per family doctor, vary closely together. So do the contagion hub proxy for public transport and pollution levels; the percentages of individuals affected by diabetes and allergies; and ICU beds 
    and the percentage of individuals over 65. 
    {\bf \protect\subref{subfig:bicluster}}: restricted heat-maps further illustrating interdependencies through two biclusters of regions and covariates.
    Color-coding within cells corresponds to that in \protect\subref{subfig:cluster}, and each bicluster is identified by a border color and its adjusted H-score (an inverse measure of bicluster strength; see Methods). The first bicluster (adjusted H-score $=0.0902$) comprises central and southern regions with "flat(tened)" epidemics (Group 2). The second bicluster (adjusted H-score $=0.0942$) comprises northern regions with "exponential" epidemics (Group 1) but also  northern and central regions from Group 2.
    }
    \label{fig:cluster_bicluster}
    \end{figure}

    Even the restricted set of $12$ covariates presents a distinct interdependence structure (see covariates dendrogram in Fig.~\ref{fig:cluster_bicluster}\subref{subfig:bicluster} and Variance Inflation Factors in Table \ref{tab:vif}). For instance, our contagion hubs proxies for hospitals, schools and work places, and our (inverse) proxy for quality of distributed, primary health care (number of adults per family doctor), tend to vary closely together across regions.
    Also, our contagion hub proxy for public transport and pollution levels tend to vary together (this is not counter-intuitive, as both increase in more industrialized regions with large metropolitan areas), as do the percentages of individuals affected by diabetes and allergies, and our proxy for quality of centralized, hospital-based health care (ICU beds per $100,000$ inhabitants) and the percentage of individuals over 65.   
    Conversely, some regions show similar profiles across covariates (see regions dendrogram in Fig.~\ref{fig:cluster_bicluster}\subref{subfig:cluster}). For instance, Lombardia, Veneto, Emilia Romagna and Piemonte have strong similarities, as do groups of southern regions (e.g., Sicilia, Campania, Puglia and Calabria; Basilicata,  Abruzzo and Molise). An interesting characterization is produced using the {\em Cheng and Church's biclustering algorithm} \cite{cheng2000biclustering}, which we implement with an adjusted mean squared residue, or H-score \cite{di2020bias}. A bicluster is a subset of regions which exhibit similar behavior across a subset of covariates. Figure \ref{fig:cluster_bicluster}\subref{subfig:bicluster} shows two 
    biclusters with similar adjusted H-score values, obtained through the same run of the algorithm.
    The first bicluster
    comprises central and southern regions, all with "flat(tened)" epidemics (Group 2). Its regions have low ratios of adults to family doctors, limited concentrations in hospitals, nursing homes, work places and public transport, and low pollution levels. They also have high percentages of diabetic individuals and limited availability of ICU beds.
    The second bicluster
    comprises northern regions with "exponential" epidemics (Group 1), such as Lombardia, Emilia-Romagna and Piemonte, but also northern  and central regions with "flat(tened)" epidemics (Group 2), such as Veneto, Friuli Venezia Giulia and Toscana. Its regions have high ratios of adults to family doctors, high concentrations in hospitals, work places and classrooms, and tend to have large percentages of individuals over $65$. They also have low percentages of diabetic individuals and medium or small-sized households.

    Next, we used functional regressions with a two-fold aim: pursue a more direct, systematic assessment of the associations between the scalar covariates and COVID-19 mortality; and use the scalar covariates as controls in models comprising mobility and positivity to re-assess these key predictors. We stress again that the coarse resolution of the data poses serious limitations for these analyses, because it may dilute some predictive signals and because it bounds us to a small sample size. With only $n=20$ observational units (the regions), fitting functional regression models comprising many terms (e.g., several scalar covariates and possibly their interactions; mobility and positivity curves along with more than one scalar covariate) produces unstable, overfit outcomes. Thus, we evaluate only the marginal effects of the scalar predictors, and the effects of mobility and stability with one scalar control at a time. 
    The marginal function-on-scalar regressions of mortality curves on each of the $12$ covariates
    have in-sample $R^2$s ranging between $\approx 20$ and $65\%$.
    Here the "effects" are {\em curves}; $\beta_{x}(t)$ represents the association of mortality at time $t$ with the covariate $x$. For $8$ of the covariates the $\hat{\beta}_{x}(t)$'s show the expected signs throughout the peak period of the epidemic. 
    In particular, the (inverse) proxy for quality of distributed, primary health care is the strongest marginal predictor; adults per family doctor shows a very large positive association with mortality. Also hospital, school and work place contagion hub proxies show strong positive associations with mortality. Nursing homes and public transport contagion hub proxies, pollution and the percentage of individuals over $65$ are positive but comparatively weaker marginal predictors. For $4$ of the covariates the $\hat{\beta}_{x}(t)$'s show unexpected signs. The percentages of diabetics and allergic individuals show negative associations with  mortality, likely due to the fact that their prevalence is high(er) in areas which were spared the brunt of the epidemic. In fact, estimated effect curves become positive when a differential intercept is included in the model to account for different overall mortality levels in Group 1 and Group 2 regions (see Fig. \ref{fig:beta_signs_1_2_inter}).
    Also the average number of members per household shows a negative association with mortality. Its small range of variation across regions ($\approx 2.0 - 2.8$, mean $2.3$, s.d. $0.16$) may not allow it to properly proxy the effect of household contagions.  At the same time, a strong negative correlation with the percentage of individuals over $65$ may not allow it to properly proxy   inter-generational contacts; regions with more elderly people are in fact those with smaller households.
    The negative association of average number of members per household with mortality, which persists even when including a differential intercept for Group 1 and Group 2 in the model (see Fig. \ref{fig:beta_signs_1_2_inter}), 
    may simply be a "shadow" of this its negative correlation with the percentage of individuals over 65.  
    Finally, ICU beds per $100,000$ inhabitants shows a positive association with mortality which, too, persists when including a differential intercept for Group 1 and Group 2 in the model (see Fig.~\ref{fig:beta_signs_1_2_inter}), 
    and may be in part a "shadow" of positive correlations with percentage of individuals over $65$ and average number of beds per hospital. However, this proxy for quality of centralized,  hospital-based health care, so prominent to the public debate during the epidemic, is {\em not} a negative predictor of mortality in our analysis. 
    In conclusion, better proxies and finer resolution may reveal stronger aggravating roles for age, nursing homes, public transport and pollution 
    \cite{wu2020exposure,coccia2020factors} and better dissect the roles of chronic conditions, households and inter-generational contacts, and ICU availability\cite{dowd2020demographic,nepomuceno2020besides}. But our analysis, notwithstanding limitations in the data, suggests important roles of primary care in mitigating mortality, and of contacts in hospitals, schools and work places in aggravating it. 

    \begin{figure}[!tb]
    \hspace{-0.2cm}
    \subfloat[\label{subfig:marginal_reg_max}]{
    \fbox{
    \begin{minipage}[b]{0.349\linewidth}
        \centering
        \includegraphics[width=1\linewidth]{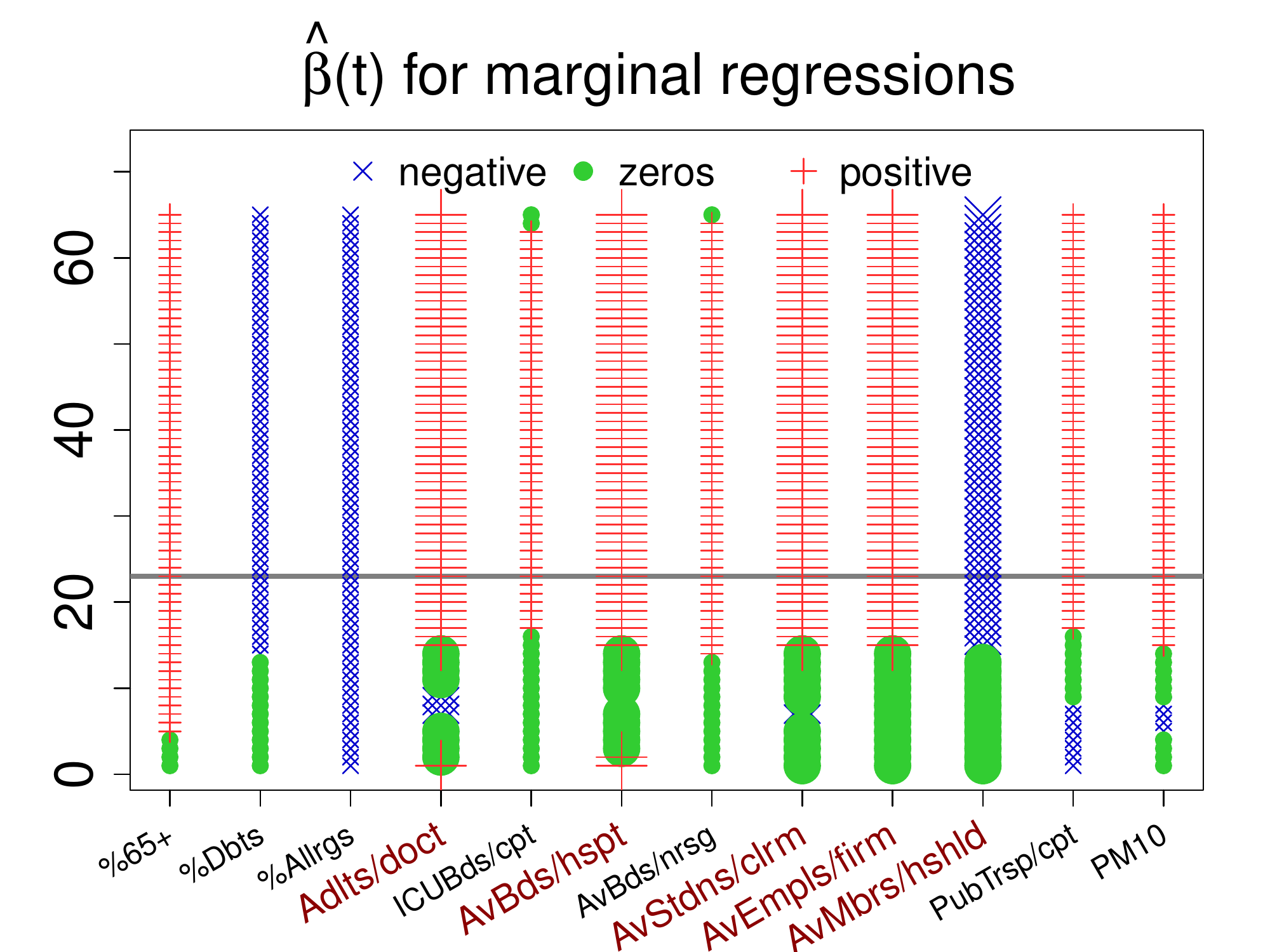} \\
        \includegraphics[width=1\linewidth]{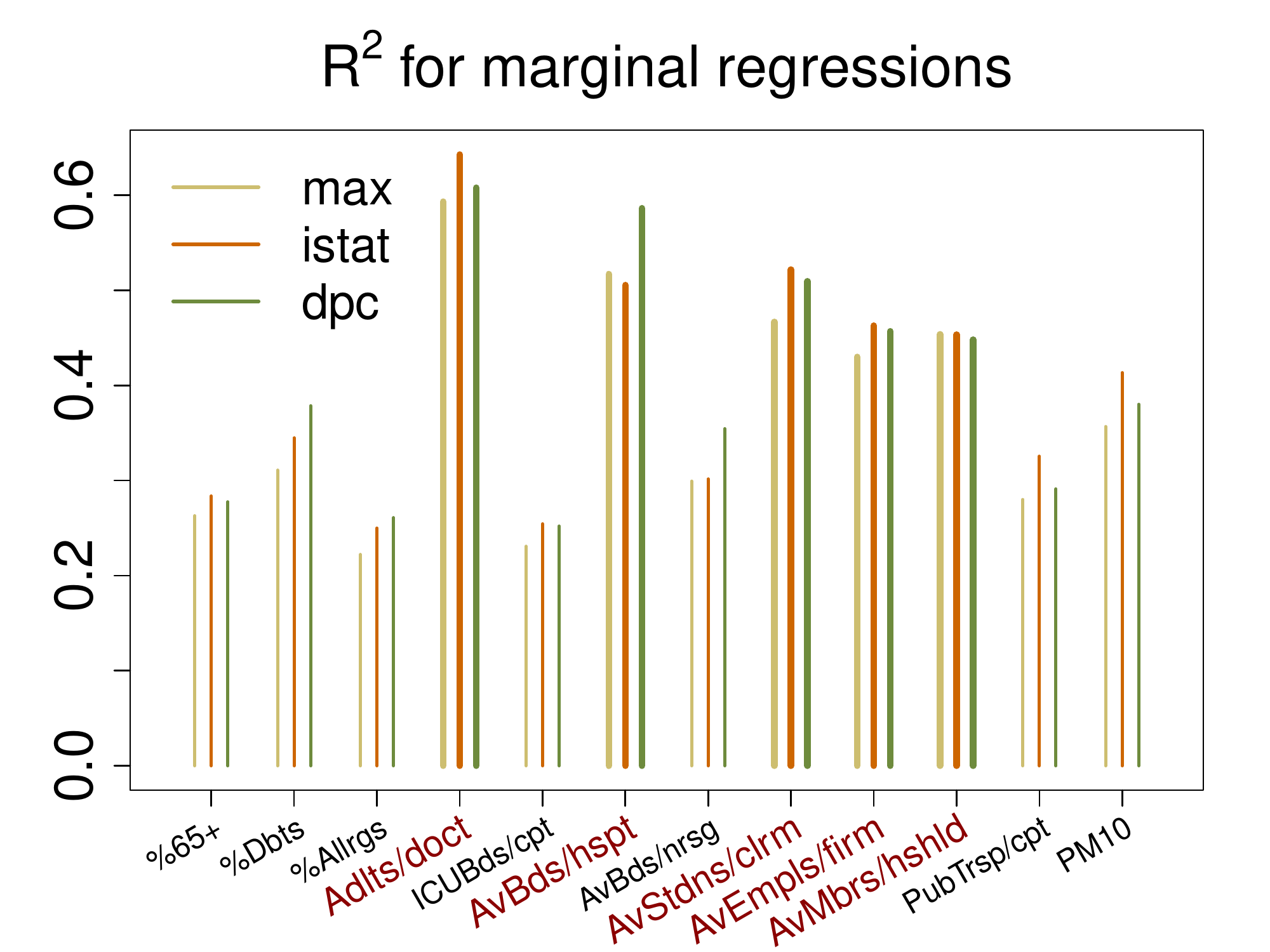}
    \end{minipage}
    }
    }
    \subfloat[\label{subfig:mob_pos_pc_reg}]{
    \fbox{
    \begin{minipage}[b]{0.6\linewidth}
        \centering
        \includegraphics[width=0.49\linewidth]{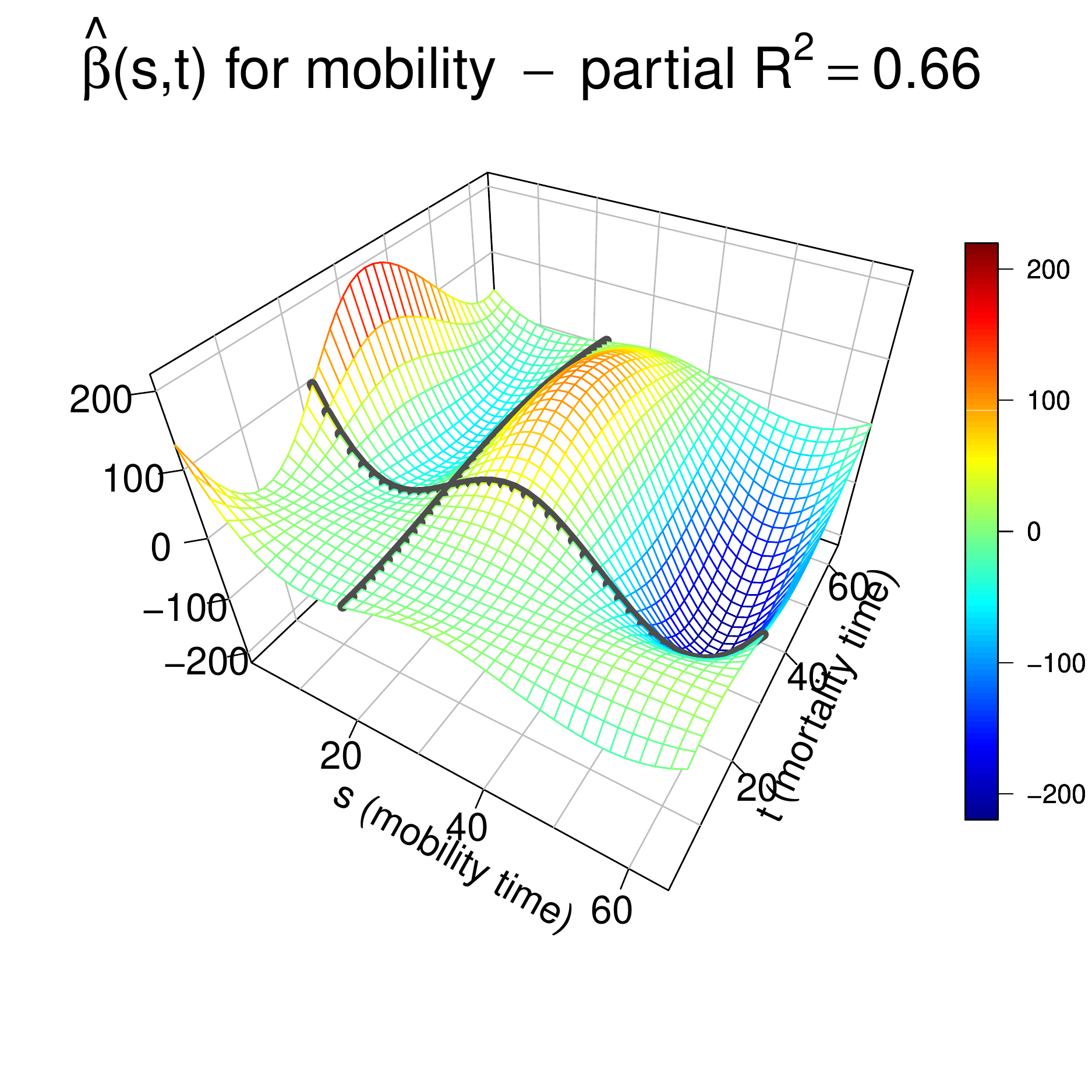}
        \includegraphics[width=0.49\linewidth]{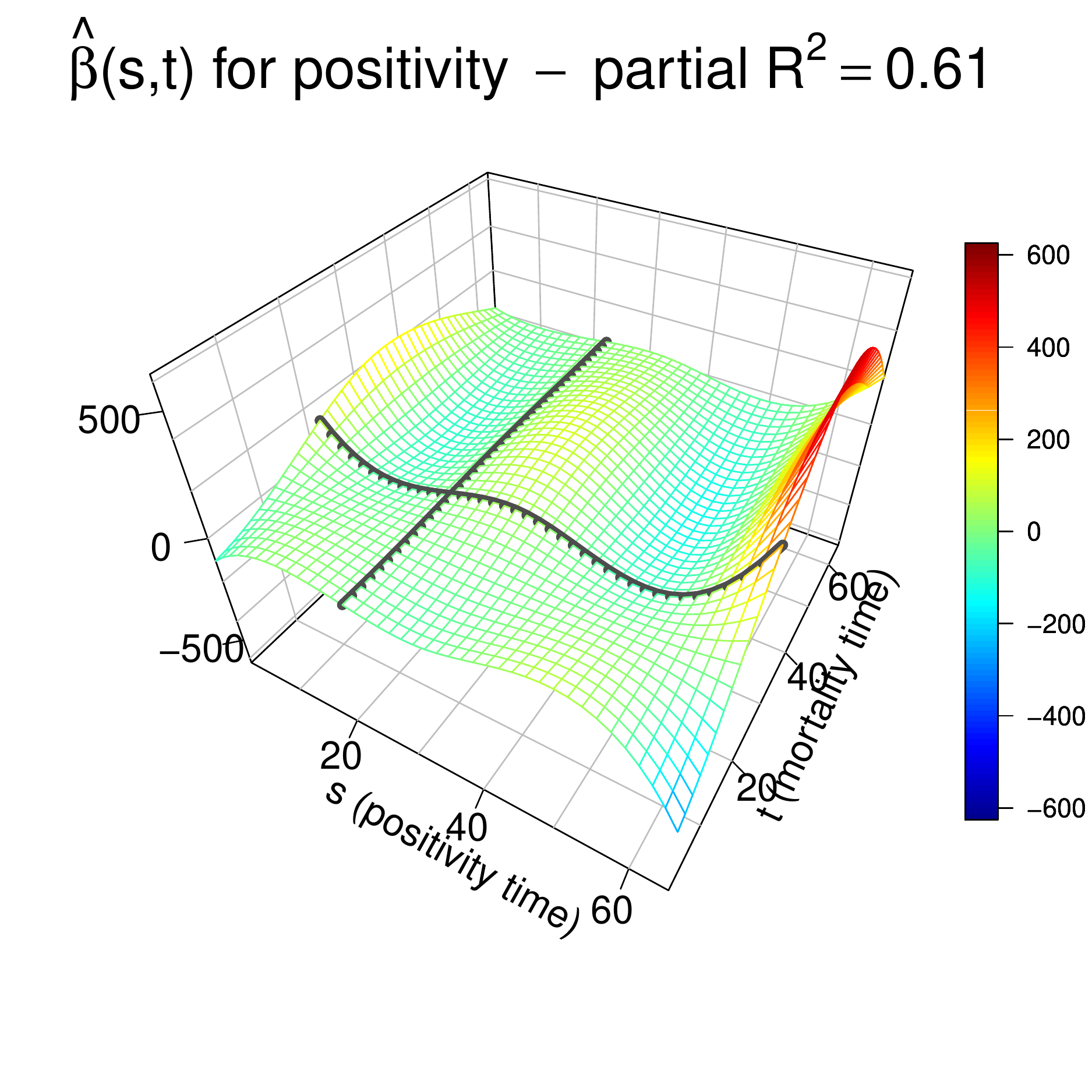} \\
      \vspace{-0.45cm}
      \includegraphics[width=0.39\linewidth]{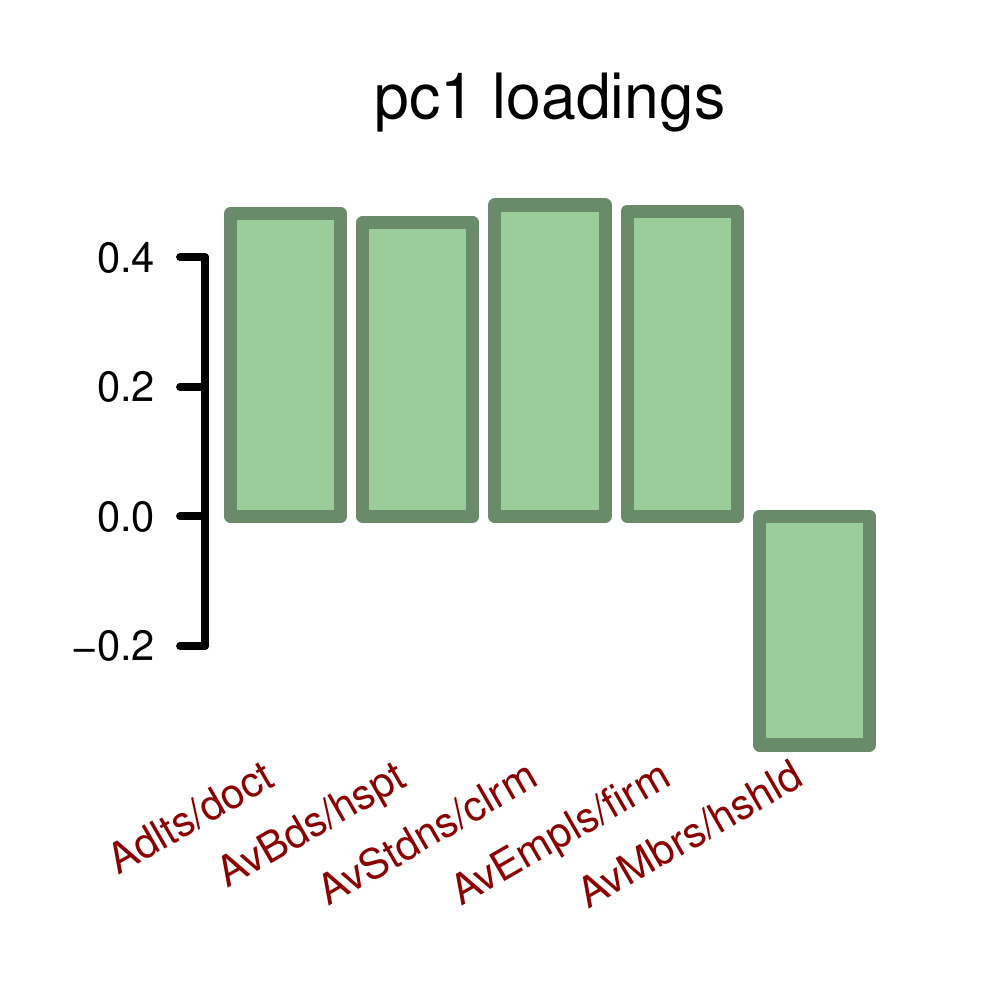}
      \includegraphics[width=0.59\linewidth]{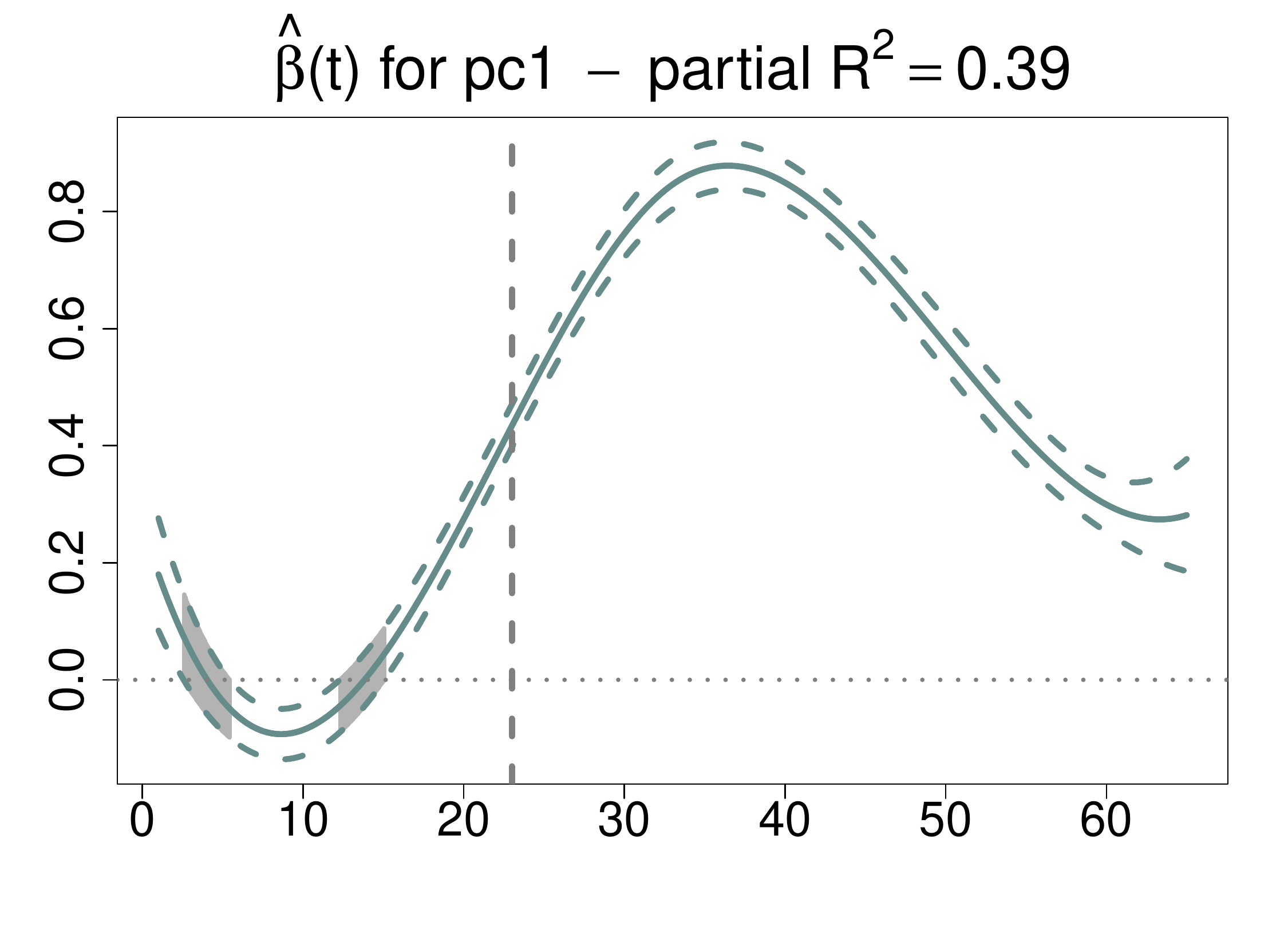}
    \end{minipage}
    }
    }
    \caption{
    {\bf Associating mortality to socio-demographic, infrastructural and environmental factors}.
    {\bf \protect\subref{subfig:marginal_reg_max}} Results from marginal function-on-scalar regressions. Mortality curves are regressed against each of the scalar covariates in Table~\ref{tab:proxies}. The top plot displays the signs of the effect curves 
    estimated on the MAX data. Time, marked as the 65 days of the region specific epidemic unfoldings, is on the vertical axis (the nationwide lock down on March 9, without shift, is marked by a horizontal line. Red, blue and green indicate,  respectively, positive, negative and non-significant portions
    (i.e., where $95\%$ confidence bands around the estimated effect curve are entirely above, entirely below, or contain $0$; see Methods).  
    The bottom plot displays in-sample $R^2$s for the regressions fitted on MAX, ISTAT and DPC data; these are remarkably consistent. 
    The names in red on  the horizontal axes indicate the top 5 covariates selected by SnNAL-EN on all three data sets (see Methods); these are also the ones with the largest $R^2$s. 
    {\bf \protect\subref{subfig:mob_pos_pc_reg}} Results from the joint function-on-function regression of MAX mortality on local mobility, positivity, and the first principal component (pc1) of the top 5 covariates, used as a "summary" control.
    This control does not modify the shapes of the estimated effect surfaces for mobility and positivity (shown on top) -- which are very similar to the ones in Fig.~\ref{fig:mob_pos}\protect\subref{subfig:mob_pos_reg_res}. 
    The estimated effect curve 
    for pc1 shows a positive and significant association with mortality at its peak (bottom right; $95\%$ confidence band in dashes, gray corresponds to non  significant portions, vertical dashed line corresponds to March 9, without shift). The sign of this effect is consistent with marginal findings, based on the loadings of the first principal component (bottom left; positive for adults per family doctor, average beds per hospital, average students per classroom and average employees per firm, and negative for average members per household).
     With the addition of pc1, the regression reaches an in-sample $R^2=0.94$ and a  LOO-CV $R^2=0.7$.
    The contributions of local mobility and positivity remain high (partial $R^2=0.66$ and $0.61$, respectively). That of our "summary" covariate is also substantial (partial $R^2=0.39$).
    }
    \label{fig:mob_pos_pc}
    \end{figure}

    The results of our marginal function-on-scalar regressions, which are summarized in Fig.~\ref{fig:mob_pos_pc}\subref{subfig:marginal_reg_max} for MAX mortality curves, are also consistent across data sets (see Fig. \ref{fig:beta_signs_1_2_inter})
    -- which lends them support, at least at the resolution of regions. To further validate their stability we ran a functional generalization  of {\em SsNAL-EN} \cite{boschi2020efficient} -- an Elastic Net-like algorithm that performs feature selection for
     regressions with many predictors, producing reasonably stable outcomes even with small sample sizes and collinear features. Reassuringly, the output of {\em SsNAL-EN} is consistent with the marginal analysis, and again consistent across data sets (see Table \ref{tab:fgen_selection}): 
    the top feature is always adults per family doctor, and the top 5 always include, in addition to it, average beds per hospital,  average students per classroom, average employees per firm, and average members per household. 
    
    Finally, we ran again the function-on-function regression of mortality on local mobility and positivity, and re-evaluated the effects of these predictors introducing in the model one of the top 5 scalar covariates at a time (see Fig.~\ref{fig:mob_pos_controls}
    for results on DPC, ISTAT and MAX data), as well as their first principal component, which explains $\approx 68\%$ of their variability and can act as a "summary" control (see Fig.~\ref{fig:mob_pos_pc}\subref{subfig:mob_pos_pc_reg} for MAX curves and Fig.~\ref{fig:mob_pos_pc1_dpc_istat}
    for DPC and ISTAT curves). Remarkably, while the control covariate "subsumes" some of the predictive power in each model, the estimated effect surfaces of local mobility and positivity retain the same shapes, and they remain very strong and comparable contributors (e.g., for MAX curves in Fig.~\ref{fig:mob_pos_pc}\subref{subfig:mob_pos_pc_reg}, the overall in-sample $R^2$ reaches $94\%$, the LOO-CV $R^2$ is 70\%, and the partial $R^2$s are $66, 61$ and $39\%$,  respectively, for local  mobility, positivity and the first principal component; see also Table~\ref{tab:R2}).
    Thus, with all the limitations of the data at our disposal, controlling for relevant covariates does not modify how the epidemic unfolding is associated to local mobility and positivity over time. Introducing socio-demographic, infrastructural and environmental factors in the modeling also does not change what we observed concerning residuals: mortality in Veneto is well predicted, and mortality in Lombardia remains sizably underestimated  (see Fig.~\ref{subfig:mob_pos_pc1})
    for MAX and Fig.~\ref{fig:mob_pos_pc1_dpc_istat}
    for DPC and ISTAT).

    \section*{Discussion}
    
    Notwithstanding the limitations of the data employed in this study, using FDA techniques we were able to characterize heterogeneous and staggered epidemics in different areas of Italy -– recapitulating and quantitating what scientists, policy makers and the public saw unfolding during the months of February, March and April 2020. In addition, we were able to document strong associations of COVID-19 mortality with local mobility and positivity, which persist in models that control for other relevant covariates. Investigating local mobility and positivity as, respectively, an actionable effector and a sentinel indicator of epidemic strength and progression, possibly to be used to adapt
    mitigation and containment efforts in real time, will require more and better data. In particular, accurate data on cases and hospitalizations in addition to deaths, and at a resolution much finer than that of Italian regions. Such data would also be critical to better capture predictive signals in a number of covariates -- which may weaken and/or become confounded when aggregating data over broad, internally heterogeneous areas. But our results, along with those of other recent studies\cite{cintia2020relationship}, do support a role for mobility as a key modulator of COVID-19 spread and for positivity as a monitoring variable. Moreover, they support a role for distributed, primary health care in mitigating mortality, and for hospitals, schools and work places as contagion hubs that may aggravate the epidemic. If confirmed and fine-tuned on higher resolution data, also these findings could inform 
    decision making -- e.g., on short- and medium-term investments to boost distributed health care,  or "pod" patients, students or employees. 
    Finally, an extension of the temporal span of the data would also be of great interest to properly characterize different phases of the Italian epidemic – including its evolution after the gradual weakening of lock-down measures in May 2020. 
    We believe that our work demonstrates the potential of FDA techniques for analyzing  epidemiological data. Our pipelines and the mix of FDA tools used in this study could be applied to COVID-19 data from other parts of the world.

    \section*{Methods}
    
    \subsection*{Data retrieval and pre-processing}
    
    \subsubsection*{Functional variables}
    Daily cumulative COVID-19 death counts per region were retrieved from the Italian Civil Protection agency (Dipartimento della Protezione Civile; DPC\footnote{\label{note_a}https://github.com/pcm-dpc/COVID-19/tree/master/dati-regioni}). {\em DPC mortality curves} from February 24 to April 30 were computed for each region as the daily increments in COVID-19 death counts, divided by the population of the region as of January 1, 2019 (as recorded by ISTAT\footnote{\label{note_b}http://asti.istat.it/asti}). DPC mortality curves were set to zero for the period February 16-23, before the Civil Protection agency started releasing data.
    Daily death counts from all causes in $7270$ Italian municipalities (about 93.5\% of the Italian population) for the years 2015-20 were downloaded from the Italian National Institute of Statistics (ISTAT\footnote{\label{note_c}https://www.istat.it/it/files/2020/03/Dataset-decessi-comunali-giornalieri-e-tracciato-record-4giugno.zip}) on June 4, 2020. Data were aggregated by region, and {\em ISTAT differential mortality curves} from February 16 to April 30 were computed for each region as the daily difference between 2020 deaths and the average daily deaths in 2015-19, divided by the total population of the municipalities included in the death counts as of January 1, 2019
    \footnote{\label{note_d}http://dati.istat.it/Index.aspx (Popolazione e famiglie/Popolazione/Popolazione residente al 1$^{\circ}$ gennaio/Tutti i comuni/2019)}). 
    {\em MAX mortality curves} were created taking, for each region and each day, the maximum between DPC mortality and ISTAT differential mortality.
    Daily measurements concerning “Grocery \& pharmacy” mobility from February 16 to April 30 were downloaded for each region from the Google Mobility Report\footnote{\label{note_e}https://www.google.com/covid19/mobility/} ({\em local mobility curves}). These measurements express percent changes with respect to the corresponding daily mobility levels in the first five weeks of 2020 (January 3 to February 6).
    {\em Positivity curves} were constructed using raw data from the Italian Civil Protection agency\footnoteref{note_a}. For each day from February 24  to  April 30 and each region, we took the ratio between the number of new positive cases and the number of new tests performed. The ratios were truncated at 0 and 1 to account for irregularities in the  row data (e.g., positive cases $=-1$, or positive cases exceeding tests performed, presumably due to delays in test results). Like DPC mortality, positivity curves were set to zero for the period before the Civil Protection agency started releasing data (February 16-23).
    For all functional data sets, the two self-governing provinces of Trento and Bolzano were considered together as the Trento/Bolzano region, since not all data were available for both provinces separately. 
    %
    The 20 curves in each functional data set were smoothed using cubic {\em smoothing B-splines} with knots at each day and roughness penalty on the curve second derivative\cite{ramsay2005}. For each functional data set the smoothing parameter was selected minimizing the average generalized cross-validation error (GCV\cite{craven1978smoothing}) across the 20 curves. All computations were performed using the \texttt{R package fda}\cite{ramsayfdapackage}.

    \subsubsection*{Scalar covariates}
    We considered a large number of scalar covariates of potential interest (see Table \ref{tab:covariatelis}), and focused on the $12$ listed in Table \ref{tab:proxies} and below. In retrieving and computing various measurements, as was done for the functional variables, the provinces of Trento and Bolzano were aggregated into the Trento/Bolzano region.
    {\em \% Over 65} was retrieved from ISTAT\footnoteref{note_b} at the regional level for the year 2018. 
    {\em \% Diabetics} and {\em \% Allergics} were retrieved from ISTAT\footnote{\label{note_f}http://dati.istat.it/Index.aspx?QueryId=15448} at the regional level for the year 2018.
    {\em Adults per family doctor} was retrieved from the Ministry of Health\footnote{http://www.salute.gov.it/imgs/C\_17\_pubblicazioni\_1203\_ulterioriallegati\_ulterioreallegato\_8\_alleg.pdf} at the regional level for the year 2017.
    To compute {\em ICU beds per $100,000$ inhabitants}, we collected the total number of ICU beds in each region in 2018 from the Ministry of Health\footnote{\label{note_g}http://www.dati.salute.gov.it/dati/dettaglioDataset.jsp?menu=dati\&idPag=96}, multiplied by 100,000 and divided by the population of the region as of January 1, 2019\footnoteref{note_b}.
    To compute {\em Ave.~beds per hospital (whole)} we used data from the Ministry of Health\footnote{http://www.salute.gov.it/imgs/C\_17\_bancheDati\_6\_0\_1\_file.xls}, which provides the number of beds per ward in each hospital in 2018. We first aggregated them over wards belonging to the same hospital, and then averaged over hospitals in each region. 
    {\em Ave.~beds per nursing home (ward)} was also obtained based on data for the year 2018 from the Ministry of Health\footnote{http://www.salute.gov.it/imgs/C\_17\_bancheDati\_6\_0\_0\_file.xls} -- here we considered regional averages at the level of wards, without  aggregating over wards inside the same nursing home (the ward-level covariate had a slightly higher association with mortality outcomes).   
    To compute {\em Ave.~students per classroom} we used data from the Ministry of Education\footnote{https://dati.istruzione.it/opendata/opendata/catalogo/elements1/leaf/?area=Studenti\&datasetId=DS0030ALUCORSOINDCLASTA}\footnote{https://dati.istruzione.it/opendata/opendata/catalogo/elements1/leaf/?area=Studenti\&datasetId=DS0030ALUCORSOINDCLAPAR}\footnote{https://dati.istruzione.it/opendata/opendata/catalogo/elements1/leaf/?area=Studenti\&datasetId=DS1114INFANZIACLASTA}\footnote{https://dati.istruzione.it/opendata/opendata/catalogo/elements1/leaf/?area=Studenti\&datasetId=DS1115INFANZIACLAPAR}, which provides the number of students in each classroom of each school in the country 
    (public or private, at every level of education), for the year 2018. We averaged them over schools in each region. Data for Trento/Bolzano and Valle d'Aosta were missing, and were imputed through random forest imputation\cite{stekhoven2012missforest}, with default parameters \texttt{maxiter=10} (maximum number of iterations to be performed given the stopping criterion is not met beforehand) and \texttt{ntree=100} (number of trees to grow in each forest).
    To compute {\em Ave.~employees per firm} we used data from ISTAT\footnote{http://asc.istat.it/asc\_BL/}, which provides number of employees per firm at the level of municipalities. We averaged them over firms in each region. Data for Valle d'Aosta were missing, and were again imputed through random forest imputation with default parameters. 
    {\em Ave.~members per household} was retrieved from ASR Lombardia\footnote{https://www.asr-lombardia.it/asrlomb/it/13740numero-di-famiglie-convivenze-e-numero-medio-di-componenti-famiglia-regionale} at the regional level for the year 2017.
    To compute {\em Public transport rides per capita} we used data  from ISTAT\footnote{\label{note_h}https://www.istat.it/it/archivio/236912}, which provides the number of rides per capita for each Italian province in 2017. We multiplied these by the provinces' population as of January 1, 2019\footnoteref{note_b},
    summed up over provinces in the same region, and divided by the region population as of January 1, 2019\footnoteref{note_b}.
    To compute {\em PM10} we used data from ISTAT \footnoteref{note_h}, which provides the average annual concentrations of PM10 (in $\mu g/m3$) detected by air quality meters distributed over the Italian territory. We averaged them over meters located in each region.

    \subsection*{Multivariate analysis tools}
    We used a number of standard multivariate techniques to analyze data on the $12$ scalar covariates -- including the extraction of {\em Principal Components}\cite{hastie2009elements}, the calculation of {\em Variance Inflation Factors}\cite{allison1999multiple} to evaluate multicollinearities, and clustering based on hierarchical agglomeration\cite{hastie2009elements}. The latter was used both to agglomerate covariates with similar behavior across regions and to agglomerate regions with similar behavior across covariates. 
    {\em Agglomerative hierarchical clustering} groups elements in a set with a bottom-up procedure that results in a dendrogram. 
    Each element starts in its own cluster, and pairs of clusters are merged iteratively with a chosen distance for  elements and linkage criterion for clusters. We employed the correlation distance, defined as $d(x_1,x_2)=1-corr(x_1,x_2)$ for two generic elements $x_1$ and $x_2$, and the complete linkage, defined as $D(X_1,X_2) = max_{x_1 \in X_1, x_2 \in X_2} d(x_1,x_2)$ for two generic clusters $X_1$ and $X_2$ (thus, the distance between two clusters is defined as the furthest distance between their elements). 
    We also used biclustering on the $20$ (regions) by $12$ (covariates) data matrix, to identify subsets of regions exhibiting similar behaviors across subsets of covariates. Following standard literature, we sought 
    sub-matrices of the data whose entries are consistent with the "ideal" additive model $x_{i,j} = \mu + \alpha_i + \tau_j$, where $\mu$ is the typical value within the bicluster, and $\alpha_i$ and $\tau_j$ are additive adjustments for row $i$ and column $j$, but we set all $\alpha_i$s to $0$ in order to find constant column biclusters, i.e., sub-matrices with constant columns (covariates). We employed the {\em Cheng and Church Biclustering Algorithm} \cite{cheng2000biclustering}, a greedy algorithm which finds the largest sub-matrices whose departure from the additive model is below a user-defined threshold. The departure is computed using the H-score (or mean squared residue score); in symbols, $H(I,J) = \frac{1}{\mid I \mid \mid J \mid} \sum_{i \in I, j \in J} \left( x_{i,j} - x_{I,j} \right)^{2}$, where $I$ and $J$ index the sets of rows and columns composing the bicluster, $x_{i,j}$ is a generic 
    cell in the bicluster and $x_{I,j}$ is the mean of column $j$. We implemented this algorithm with a recently proposed {\em adjustment to the H-score} \cite{di2020bias} that corrects a bias towards smaller biclusters in the original formulation. The adjusted H-score is defined as $H_{adj}(I,J) = (\prod_{r=2}^{I-1}\frac{r^2}{r^2-1}\prod_{q=2}^{J-1}\frac{q^2}{q^2-1})^{-1} H(I,J)$.

    \subsection*{Functional Data Analysis tools}
    
    \subsubsection*{Local clustering of curves and functional motif discovery}
    We performed local clustering of smoothed mortality curves  (DPC, ISTAT and MAX, separately) using {\em probabilistic $K$-mean with local alignment} (\emph{probKMA}\cite{cremona2020}). 
    {\em ProbKMA} is a $K$-mean-like algorithm for functional data that finds $K$ groups in a set of curves based on a local similarity among portions of the curves themselves. This allows the discovery of functional motifs, i.e.~of typical local shapes that recur within and across the curves. In symbols, the algorithm finds $K$ motifs $v_1,\dots,v_K$, membership probabilities $p_{k,i}$ and shifts $s_{k,i}$ (i.e.~the starting points of the motif instances) for each cluster-curve pair that minimize the generalized least-squares functional $J(v_1,\dots,v_K,p_{k,i},s_{k,i})=\sum_{i=1}^N \sum_{k=1}^K p_{k,i}^2 d^2(\tilde{x}_{i},v_k)$, where $\tilde{x}_{i}$ is the portion of the curve $i$ corresponding to the shift $s_{k,i}$, and
    $d$ is the distance used to  capture local similarity. 
    For each data set, we considered $K=2$ and $K=3$. {\em ProbKMA} is probabilistic; it returns as output a membership probability $p_{k,i}$ for each cluster-curve pair. However, such an output can be turned into a hard partition by assigning each curve to the group with highest membership probability -- which is what we did here. Notably, for $K=2$, membership probabilities showed that Lombardia's and Valle d'Aosta's extreme mortality patterns were not well accommodated even in the "exponential" group\cite{cremona2020}. 
    The algorithm can employ different definitions of similarity $d$ and thus capture different aspects of curve shapes. We used Euclidean ($L^2$) distance between curve levels for our main analysis -- in symbols, $d=\frac{1}{c} \int_0^c (x(t)-v(t))^2 dt$ for two generic curves $x$ and $v$ -- though using Euclidean distance between curve derivatives produced similar results (not shown). 
    {\em ProbKMA} allows the length of the motifs to be extended endogenously  starting from a minimal one fixed in input. However, to identify epidemic patterns we ran it with a fixed motif length of 65 days -- hence allowing a maximum shift of 10 days between curves (the mortality curves are 75 days long). The same clusters and very similar shifts were obtained with a fixed motif length of 50 days, which allows a maximum shift of 25 days (results not shown). 
    The shifts produced by {\em probKMA} with $K=2$ on the three mortality data sets (DPC, ISTAT and MAX) were employed to align, in addition to the mortality curves themselves, local mobility and positivity curves. All subsequent analyses employing shifted curves (tests contrasting groups of curves, functional boxplots and depth analyses, and functional regression models) were therefore restricted to the 65-day portions where mortality curves aligned following the two {\em probKMA} motifs. 
    We also validated the groups produced by {\em probKMA} with a modified version of {\em funBI }\cite{funbi2019}, an algorithm tipically used for finding functional biclusters. We used the modified {\em funBI} to identify groups of curves characterized by group-specific fixed length motifs, considering all possible sub-curves of a fixed length and clustering them with a divisive hierarchical algorithm (results not shown).

    \subsubsection*{Testing for differences between groups of curves}
    We employed an {\em Interval-Wise Testing} algorithm developed for omics data (\emph{IWTomics}\cite{cremona2018}) to test for differences between the two groups of shifted mortality curves produced by {\em probKMA} with $K=2$ (again, separately for DPC, ISTAT and MAX). 
    {\em IWTomics} is a non-parametric, permutation-based functional hypothesis test. It contrasts two sets of curves aligned on a common domain to detect locations where the two sets differ significantly, and scales at which such significant differences are displayed (scales correspond to varying degrees of adjustment for multiple testing on intervals of varying lengths). 
    Here locations are represented by the 65 days where the shifted mortality curves are defined, while scales vary from 1 day to the whole 65 days. 
    The test was performed with the \texttt{R package IWTomics} \cite{cremona2018}. The package allows the user to select among various possible test statistics; we employed the mean.

    \subsubsection*{Functional boxplots and depth analyses}
    The functional boxplot \cite{sun2011functional} is an exploratory tool used to visualize functional data. It is constructed after ordering a set of curves based on a depth measure, such as the modified band depth \cite{lopez2009concept}. The statistics employed to construct a functional boxplots are: the 50\% central region envelope, the median curve, and the maximum non-outlying envelope. The 50\% central region envelope corresponds to the box in a classical boxplot; it contains the 50\% deepest, most centrally located curves.
    The median, i.e.~the deepest curve, is inside this box and represents a robust "center" of the functional data set. The maximum non-outlying envelope is obtained by inflating the 50\% central region envelope by 1.5 times its range. All curves extending outside of this envelope are flagged as outliers (the fact that the ISTAT data set in Fig.~\ref{fig:quantiles}\subref{subfig:ranking} lacks outlying curves based on this definition is due to the width of its 50\% central region envelope). We ranked the curves based on their depth measurements, after attributing a sign to such measurements with an ad hoc procedure. We subtract the median from each curve, and consider the share of the domain on which the difference is positive. If this is larger than $50\%$, we attribute a positive sign to the curve's depth -- otherwise, we attribute a negative sign. Curves can thus be ranked from the most outlying above the median (labeled as positive), down to those close to the median, down to the most outlying below the median (labeled as negative) -- see Fig.~\ref{fig:quantiles}\subref{subfig:ranking}. While this is not a fully general procedure, it works well on the DPC, ISTAT and MAX mortality curves we considered, which are rather unambiguously above/below the median (the share of the domain where the difference from the median is positive is $\geq 70$ or $\leq 30\%$ for all curves in all three data sets). Note also that the median curve of a data set, defined as the deepest, does not necessarily have half of the curves above it and half of the curves below it in the signed ranking we created (e.g., Toscana is the median curve in both ISTAT and MAX data sets, but the number of curves above/below it differs). 

    \subsubsection*{Functional regression models}
    \label{subsec:fun_reg}
    
    We consider models where a functional response variable is regressed against functional predictors and/or scalar covariates \cite{ramsay2005, kokoszka2017}. All are special cases of the general equation \cite{horvath2012inference}
    \begin{equation*}
    \label{eq:fun_reg}
        y_i(t) = \alpha(t)  + \sum_{\ell=1}^L \int \beta_\ell(s,t) x_{i,\ell}(s) ds + \sum_{j=1}^J \beta_j(t) x_{i,j} + \epsilon_i(t) \quad i=1,\dots n \ .
    \end{equation*}
    $n$ is the number of observations, in our case $n=20$ regions. 
    $y_i(t)$,  $i=1,\ldots n$ are the aligned mortality curves (DPC, ISTAT or MAX, modeled separately),  $\alpha(t)$ is a functional intercept and
    $\epsilon_i(t)$,  $i=1,\ldots n$ are i.i.d. Gaussian model errors.
    $L$ is the number of functional predictors. $x_{i,\ell}(s)$, $i=1,\ldots n$, $\ell=1, \ldots L$,  are such predictors, measured on the $n$ observations. The regression coefficient of each functional predictor, $\beta_\ell(s,t)$, is a surface. $J$ is the number of scalar covariates. $x_{i,j}(s)$, $i=1,\ldots n$, $j=1, \ldots J$, are such covariates, measured on the $n$ observations. The regression coefficients of each scalar covariate, 
    $\beta_j(t)$, is a curve.
    For the marginal regression of mortality on local mobility and mortality on positivity, we have $L = 1$ and $J=0$. For the joint regression of mortality on local mobility and positivity, we have $L = 2$ and $J = 0$. For the marginal regressions of mortality on individual scalar covariates, we have $L=0$ and $J=1$. In Fig. \ref{fig:beta_signs_1_2_inter}
    we fit marginal regressions of this type allowing the estimation of two different intercepts: $\alpha_1(t)$ for curves in Group 1 and $\alpha_2(t)$ for curves in Group 2.
    Finally, for the joint regression of mortality on local mobility, positivity and one scalar control variable, we have $L=2$ and $J=1$.
    To fit all these functional regressions we used the \texttt{R package refund} \cite{goldsmith2016refund}, which estimates the functional coefficients as well as their standard errors. We used these standard errors to construct confidence bands around the estimated functional coefficients. 
    To gauge the explanatory power of each model, we computed the in-sample $R^2$ as well as the Leave-One-Out Cross-Validation (LOO-CV) $R^2$. The former is a functional generalization of the classical \emph{coefficient of determination} defined as $SS_{reg}/(SS_{reg} + SS_{res})$, where $SS_{reg}$ and $SS_{res}$ are the regression and the residual sum of squares, respectively.
    To compute the latter, for each observation $i$, one replaces the fitted response curve $\hat y_i(t)$ (from the model fitted on all observations) with the predicted response curve $\hat y_{pred, i}(t)$ obtained for $i$ from the model fitted withholding $i$ itself. 
    Finally, for models with multiple terms (predictor and/or covariate), the partial $R^2$ of each term is computed as $(R^2 - R^2_{red})/(1 -R^2_{red})$, where $R^2$ is the coefficient of determination of the complete model, and $R^2_{red}$ that of the model comprising all terms but the one being evaluated. 
    

    \subsubsection*{SsNAL-EN for feature selection}
    \label{subsec:ssnal_en}
    
    SsNAL-EN \cite{boschi2020efficient} is an efficient algorithm to perform Elastic Net \cite{zou2005regularization} feature selection in a standard regression framework, i.e.~when both response and features are scalars. The Elastic Net is a hybrid between LASSO and Ridge, which  penalizes both the $L_1$ and the $L_2$ (Euclidean) norm of the regression coefficients. The $L_1$ penalty induces sparsity selecting only the most predictive among 
    the features.  The $L_2$ penalty regularizes coefficient estimates mitigating variance inflation due to collinearity. To perform feature selection in the functional regression setting, we applied a generalization of  SsNAL-EN which incorporates a group structure in the Elastic Net objective function
    and uses the Functional Principal Components basis expansion to represent a functional response. 
    In particular, we performed feature selection for the regression of mortality against all $12$ scalar covariates in Table~\ref{tab:proxies}. Notably, we selected the same top 5 features across all three data sets (DPC, ISTAT and MAX) (see Table \ref{tab:fgen_selection}) -- 
    lending strong support to their association with mortality.

    \bibliography{references}
    
    %
   
    \section*{Acknowledgements 
    }
    
    M.A.~Cremona acknowledges support from the NSERC.
    F.~Chiaromonte and T.~Boschi acknowledge support from the Huck Institutes of the Life Sciences (Penn State University).  F.~Chiaromonte and L.~Testa acknowledge support from the Sant'Anna School of Advanced Studies. We are grateful to Paola  Cesari, Christian Esposito, Giovanni Felici, Daniele Licari, Andrea Mina and Flavia Petruso for useful feedback.
    
    \section*{Author contributions}
    
     All authors conceived ideas and analysis approaches. T.B., J.Di I., L.T.~and M.A.C.~retrieved and processed data from multiple public sources, implemented pipelines and performed statistical  analyses. All authors interpreted findings and participated to the writing of the manuscript. M.A.C.~and F.C. supervised the research. 
    
    \section*{Competing interests}
    %
    
    The authors declare no competing interests.

    
    %


\newpage
\appendix

\setcounter{table}{0}
\setcounter{figure}{0}
\renewcommand{\thesection}{S\arabic{section}}  
\renewcommand{\thetable}{S\arabic{table}}  
\renewcommand{\thefigure}{S\arabic{figure}}

\section*{\Huge Supplementary Material}

\begin{figure}[H]
\centering
\vspace{-0.2cm}
\hrule height 1pt
\vrule width 1pt
\hspace{0.15cm}
\includegraphics[width=0.48\linewidth]{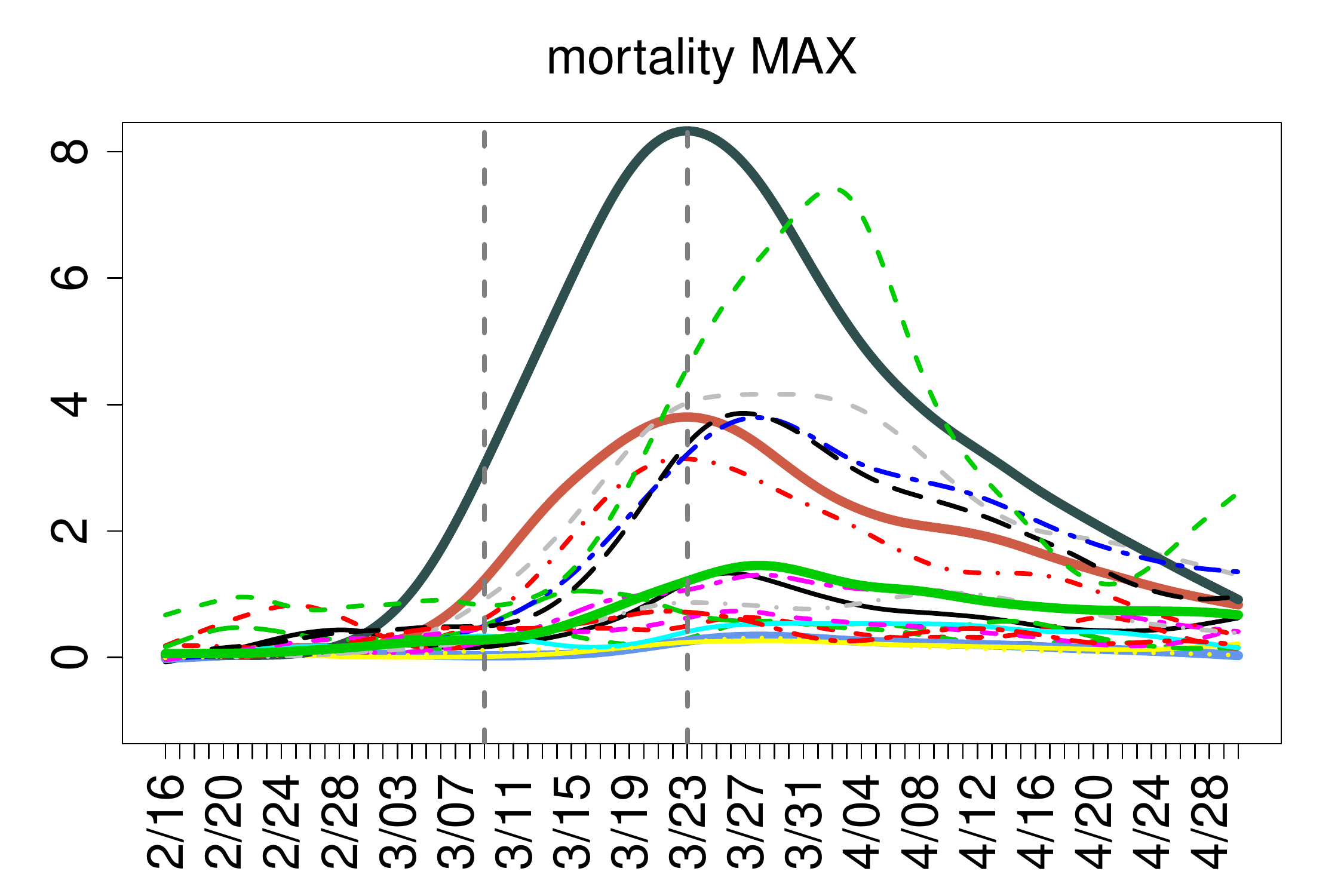}
\includegraphics[width=0.48\linewidth]{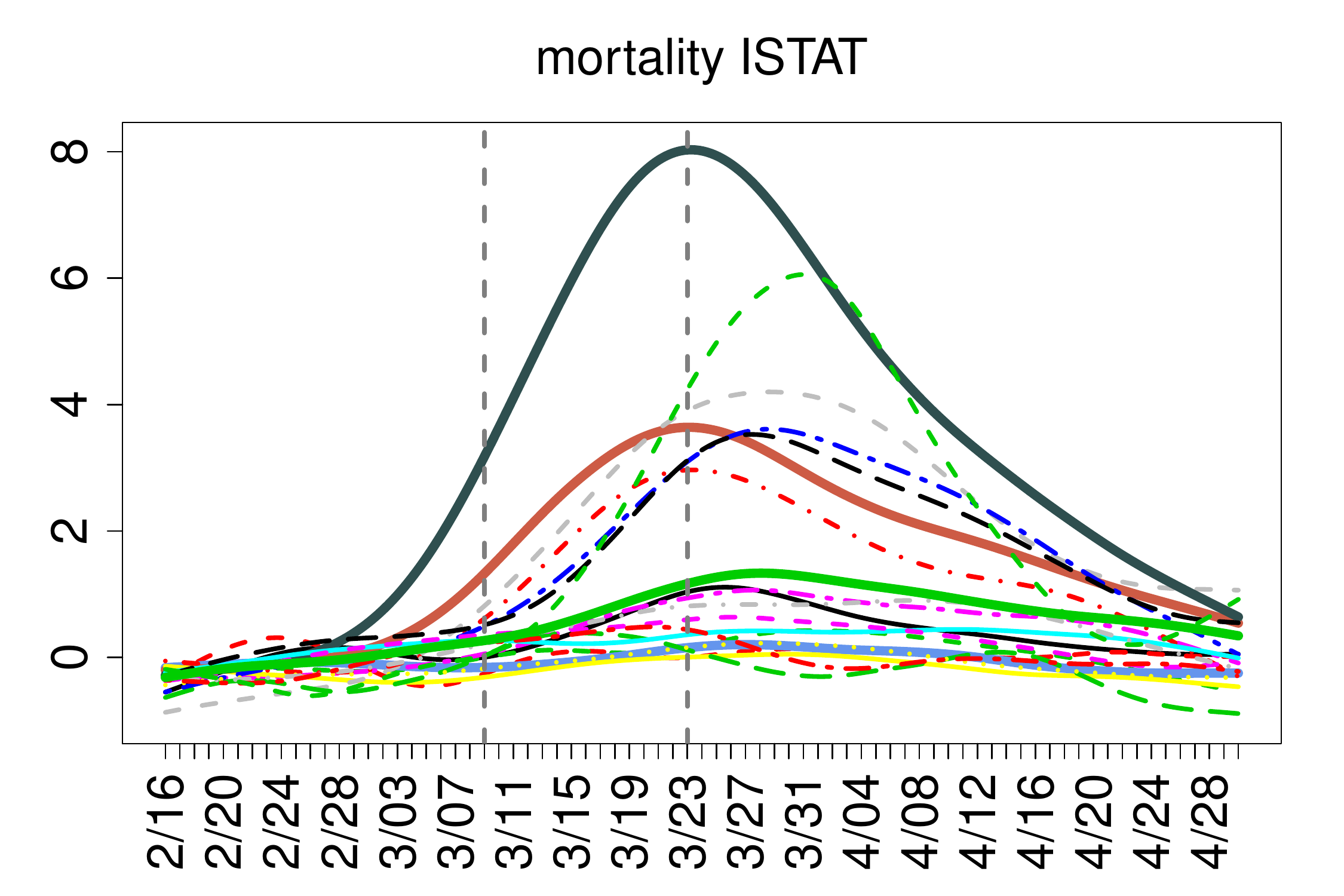}
\hspace{0.15cm}
\vrule width 1pt \\
\vspace{-0.1cm}
\vrule width 1pt
\hspace{0.15cm}
\includegraphics[width=0.48\linewidth]{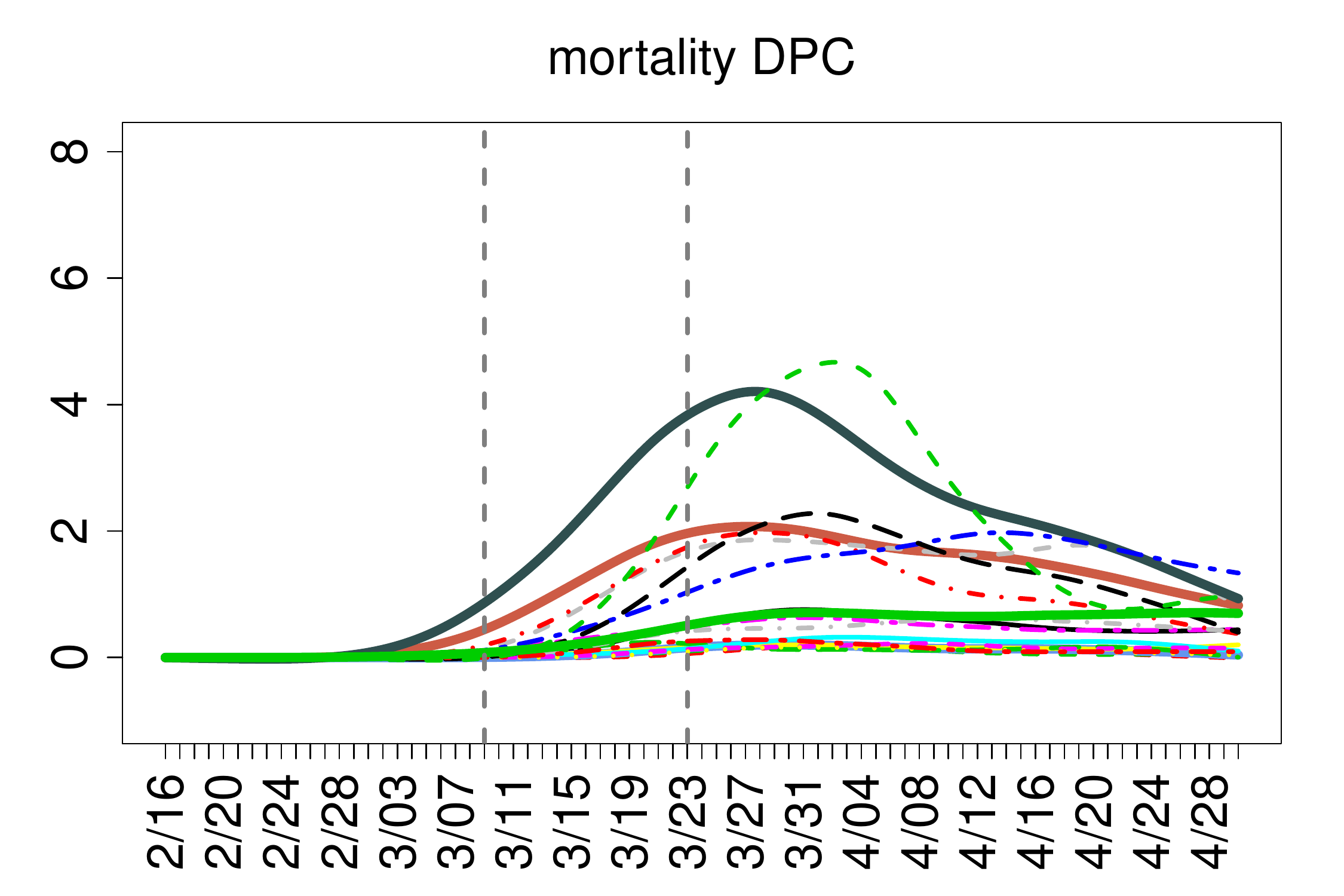}
\includegraphics[width=0.48\linewidth]{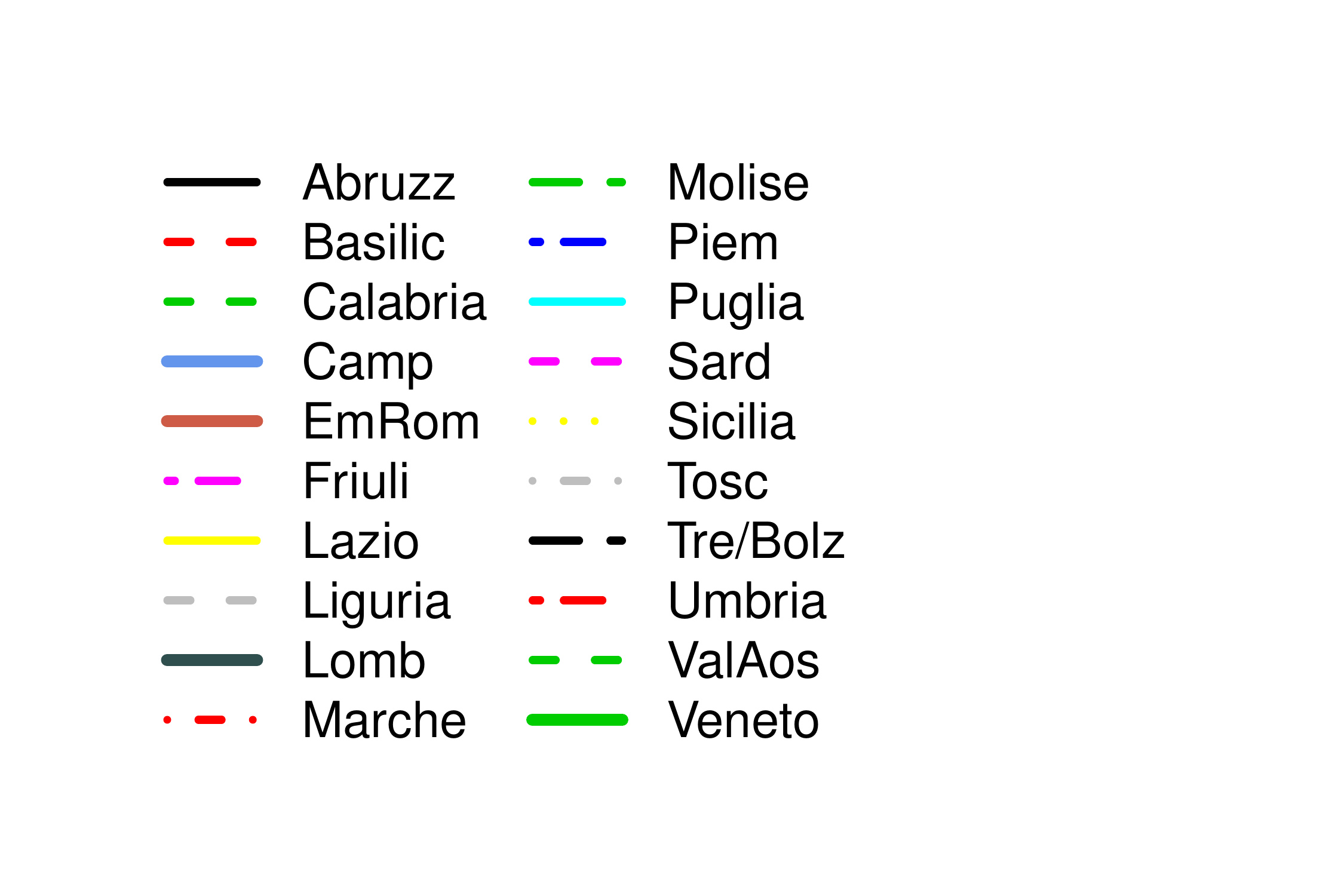}
\hspace{0.15cm}
\vrule width 1pt
\hrule height 1pt
\vspace{0.2cm}  
\caption{ {\bf Unshifted mortality curves.} MAX, ISTAT and DPC mortality curves (per 100,000 inhabitants) without shift. Vertical lines show the days corresponding to the national lock down (March 9) and the suspension of all non-essential production activities (March 23). 
}
\label{fig:unshifted_curves_col_mortality}
\end{figure}

\begin{figure}[!tb]
    \centering
    \subfloat[\label{subfig:dpc}]{
    \fbox{
    \begin{minipage}[b]{0.32\linewidth}
        \includegraphics[page=1,trim={0.1cm 1cm 0.7cm 0.2cm},clip,height=4.3cm]{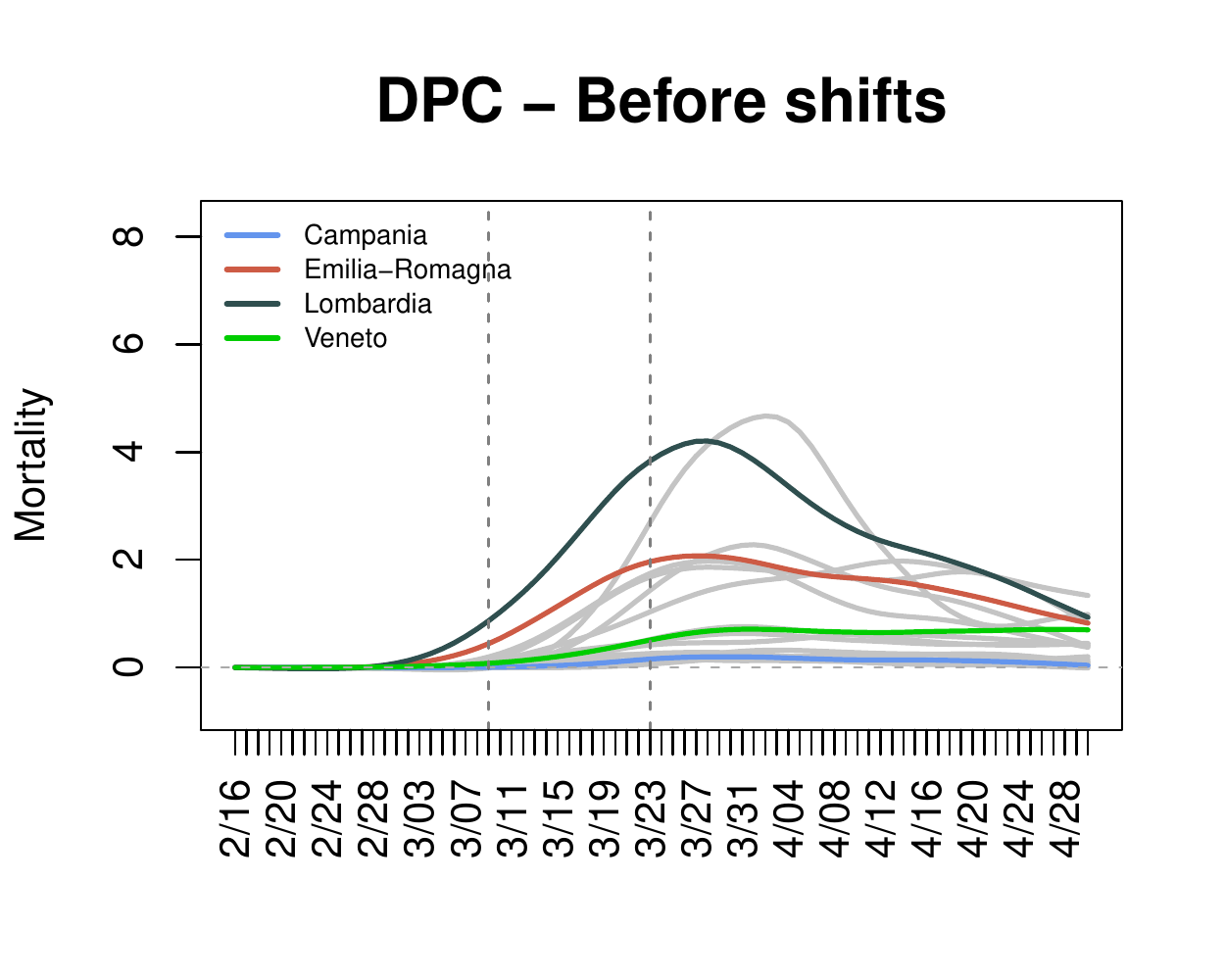}\\
        \includegraphics[page=2,trim={0.1cm 1cm 0.7cm 0.2cm},clip,height=4.3cm]{./img_suppl/regions_dpc_smooth_all.pdf}
    \end{minipage}}
    }
    \subfloat[\label{subfig:istat}]{
    \fbox{
    \begin{minipage}[b]{0.32\linewidth}
        \includegraphics[page=1,trim={0.1cm 1cm 0.7cm 0.2cm},clip,height=4.3cm]{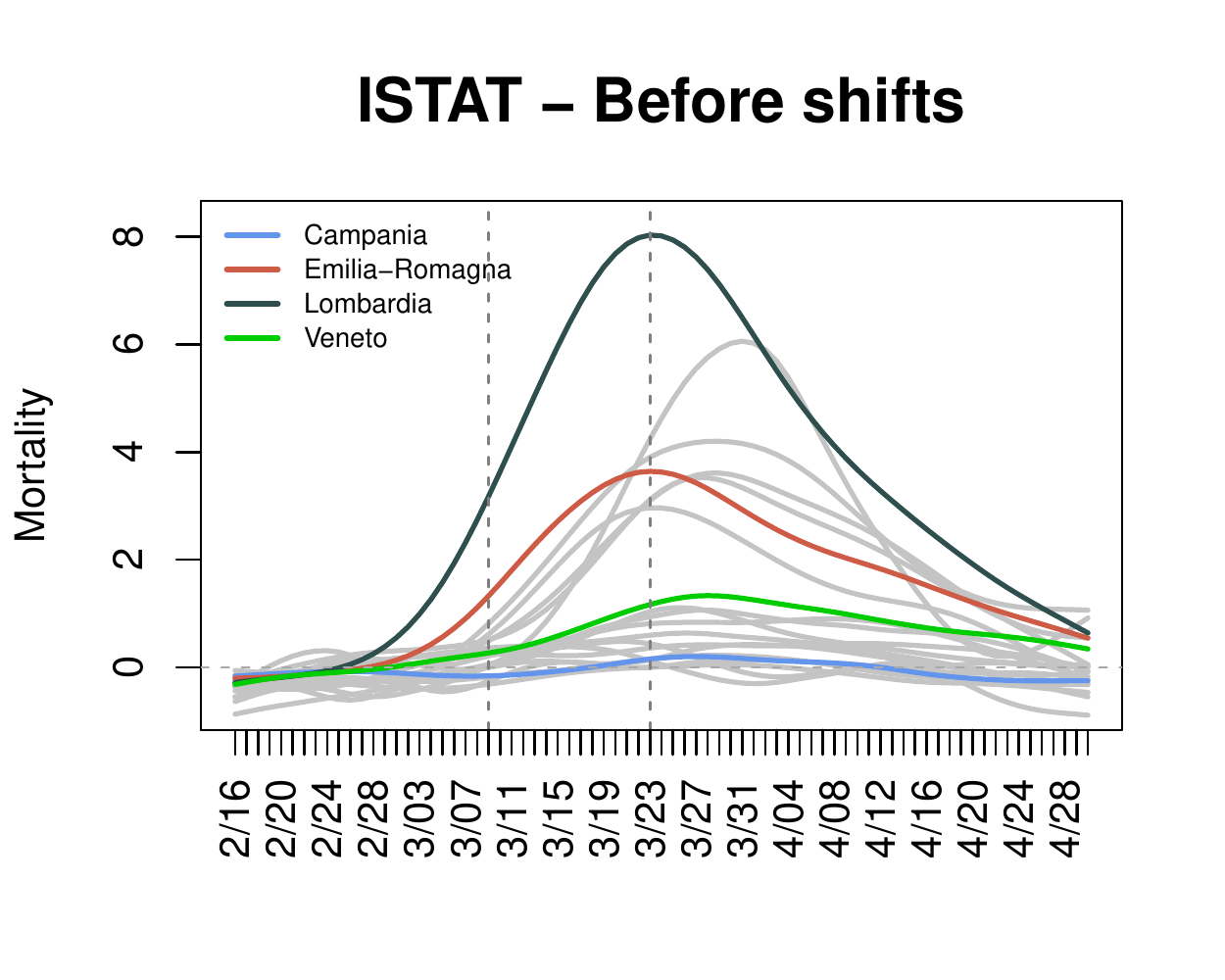}\\
        \includegraphics[page=2,trim={0.1cm 1cm 0.7cm 0.2cm},clip,height=4.3cm]{./img_suppl/regions_istat_smooth_all.pdf}
    \end{minipage}}
    }
    \caption{
    {\bf Mortality curves}.
    {\bf \protect\subref{subfig:dpc}}: DPC mortality curves (per 100,000 inhabitants) in the 20 Italian regions -- before (top) and after (bottom) the shifts produced by {\em probKMA} run with $K=2$. {\bf \protect\subref{subfig:istat}}: ISTAT mortality curves (per 100,000 inhabitants) in the 20 Italian regions -- before (top) and after (bottom) the shifts produced by {\em probKMA} run with $K=2$.
    In all panels, vertical lines mark the dates of the national lock-down (March 9) and the suspension of all non-essential production activities (March 23). In the bottom panels vertical lines still show these dates without shifts, stars on the curves mark the lock-down after the region-specific shifts.
    }
    \label{fig:mortality_DPC_ISTAT}
\end{figure}

\begin{figure}[!tb]
    \centering
    \subfloat[\label{subfig:probKMA_dpc}]{
    \fbox{
    \begin{minipage}[b]{0.63\linewidth}
        \includegraphics[page=1,trim={0.1cm 1cm 0.7cm 0.2cm},clip,height=4.3cm]{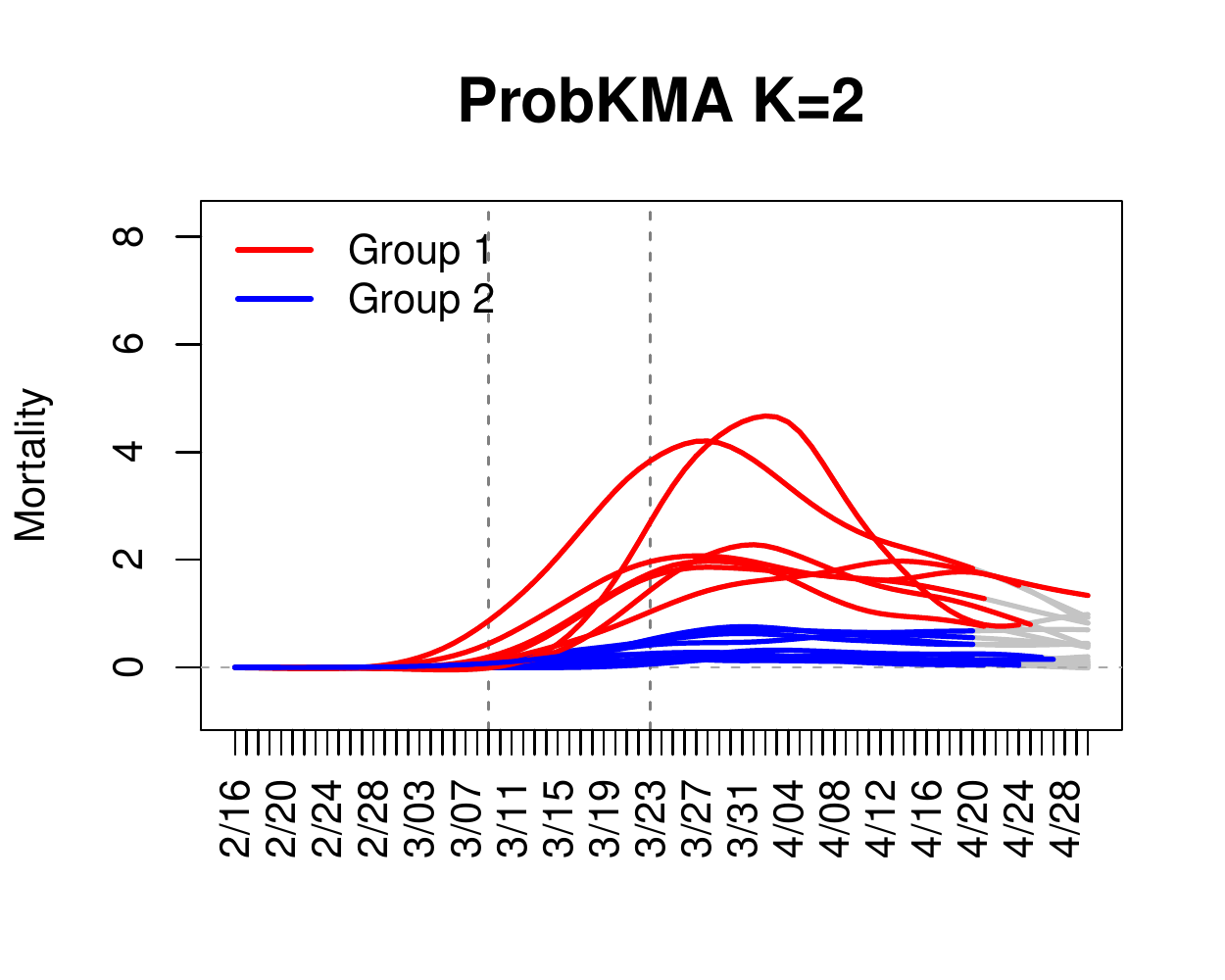}
        \includegraphics[page=4,trim={0.8cm 1cm 0.7cm 0.2cm},clip,height=4.3cm]{./img_suppl/regions_dpc_probKMA_d0_K2.pdf} \\
        \includegraphics[page=2,trim={0.1cm 1cm 0.7cm 0.2cm},clip,height=4.3cm]{./img_suppl/regions_dpc_probKMA_d0_K2.pdf}
        \includegraphics[page=3,trim={0.8cm 1cm 0.7cm 0.2cm},clip,height=4.3cm]{./img_suppl/regions_dpc_probKMA_d0_K2.pdf}
    \end{minipage}}
    }
    \subfloat[\label{subfig:IWTomics_dpc}]{
    \fbox{
    \begin{minipage}[b]{0.32\linewidth}
        \includegraphics[page=2,trim={0.3cm 0.2cm 0cm 0cm},clip,height=8.63cm]{./img_suppl/regions_dpc_IWTomics.pdf}
    \end{minipage}}
    } \\ 
        \subfloat[\label{subfig:probKMA_istat}]{
    \fbox{
    \begin{minipage}[b]{0.63\linewidth}
        \includegraphics[page=1,trim={0.1cm 1cm 0.7cm 0.2cm},clip,height=4.3cm]{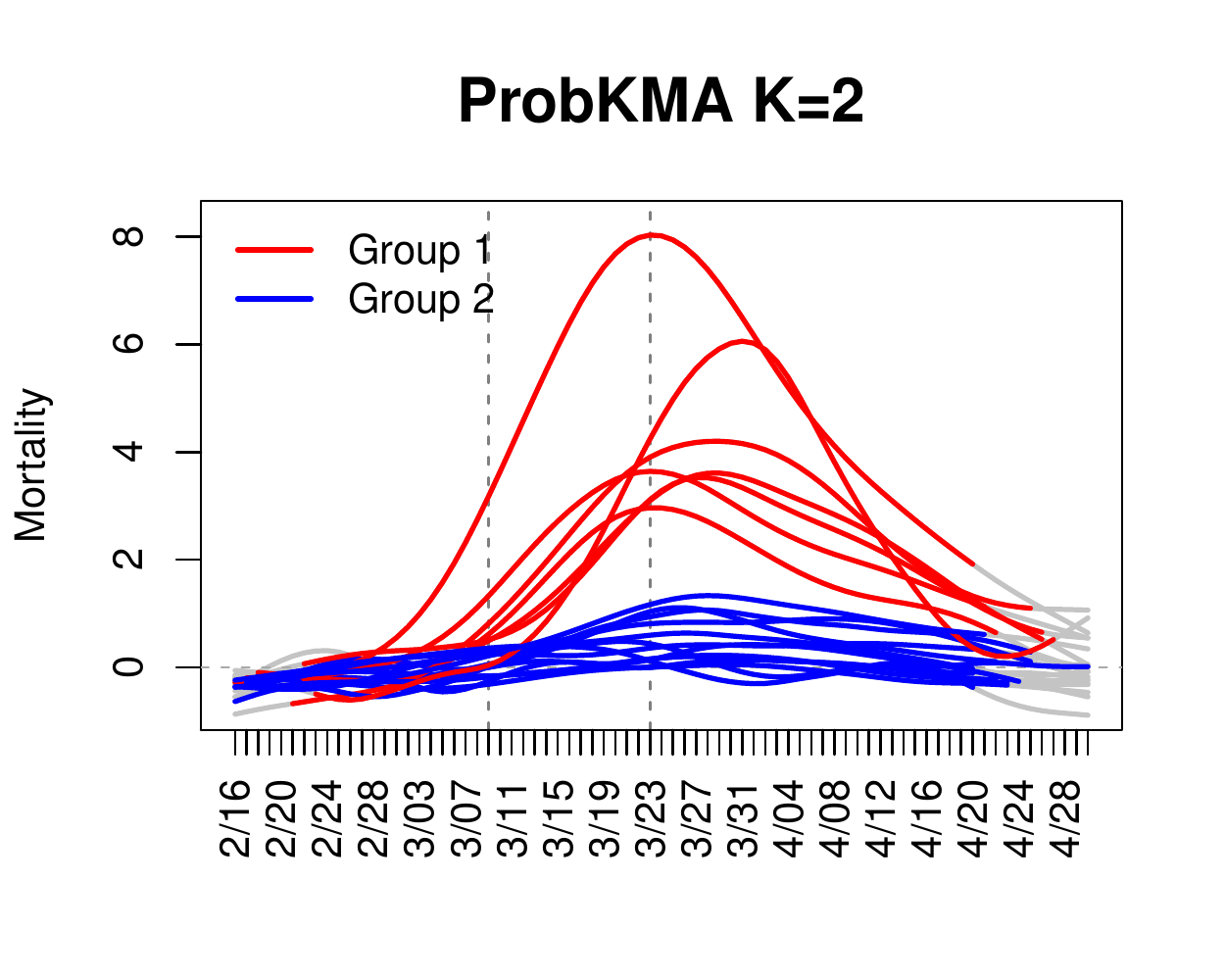}
        \includegraphics[page=4,trim={0.8cm 1cm 0.7cm 0.2cm},clip,height=4.3cm]{./img_suppl/regions_istat_probKMA_d0_K2.pdf} \\
        \includegraphics[page=2,trim={0.1cm 1cm 0.7cm 0.2cm},clip,height=4.3cm]{./img_suppl/regions_istat_probKMA_d0_K2.pdf}
        \includegraphics[page=3,trim={0.8cm 1cm 0.7cm 0.2cm},clip,height=4.3cm]{./img_suppl/regions_istat_probKMA_d0_K2.pdf}
    \end{minipage}}
    }
    \subfloat[\label{subfig:IWTomics_istat}]{
    \fbox{
    \begin{minipage}[b]{0.32\linewidth}
        \includegraphics[page=2,trim={0.3cm 0.2cm 0cm 0cm},clip,height=8.63cm]{./img_suppl/regions_istat_IWTomics.pdf}
    \end{minipage}}
    }
    \caption{
    {\bf Characterizing two epidemics}.
    Results of {\em probKMA} and {\em  IWTomics} on {\bf \protect\subref{subfig:probKMA_dpc}}-{\bf \protect\subref{subfig:IWTomics_dpc}} DPC mortality curves and {\bf \protect\subref{subfig:probKMA_istat}}-{\bf \protect\subref{subfig:IWTomics_istat}} ISTAT curves. 
    {\bf \protect\subref{subfig:probKMA_dpc}} and {\bf \protect\subref{subfig:probKMA_istat}}: Mortality curves are shown in the top left panel with portions identified by {\em probKMA} with  $K=2$ in red (Group 1; "exponential" pattern) and blue (Group 2; "flat(tened)" pattern). The curve portions are shown again, this time aligned with each other and separated by group, in the bottom panels. Black lines indicate group averages. The shifts produced by {\em probKMA} are shown in the top right panel. 
    {\bf \protect\subref{subfig:IWTomics_dpc}} and {\bf \protect\subref{subfig:IWTomics_istat}}: Shifted Group 1 and Group 2 mortality curves are tested against each other with {\em  IWTomics}. The heatmap at the top shows $p$-values adjusted at all possible scales (from 1 to 65 days). The middle panel shows in detail the top-most row of the heatmap; i.e.~the $p$-values adjusted across the whole 65-day interval. The bottom panel shows again the shifted curves. Gray areas in the middle and bottom panels mark days when the difference between the two groups is significant (adjusted $p$-value $<5\%$). }
    \label{fig:two_epidemics_dpc_istat}
\end{figure}

\begin{figure}[!tb]
    \centering
    \subfloat[\label{subfig:probKMA_max_K3}]{
    \fbox{
    \begin{minipage}[b]{0.94\linewidth}
        \includegraphics[page=1,trim={0.1cm 1cm 0.7cm 0.2cm},clip,height=4.3cm]{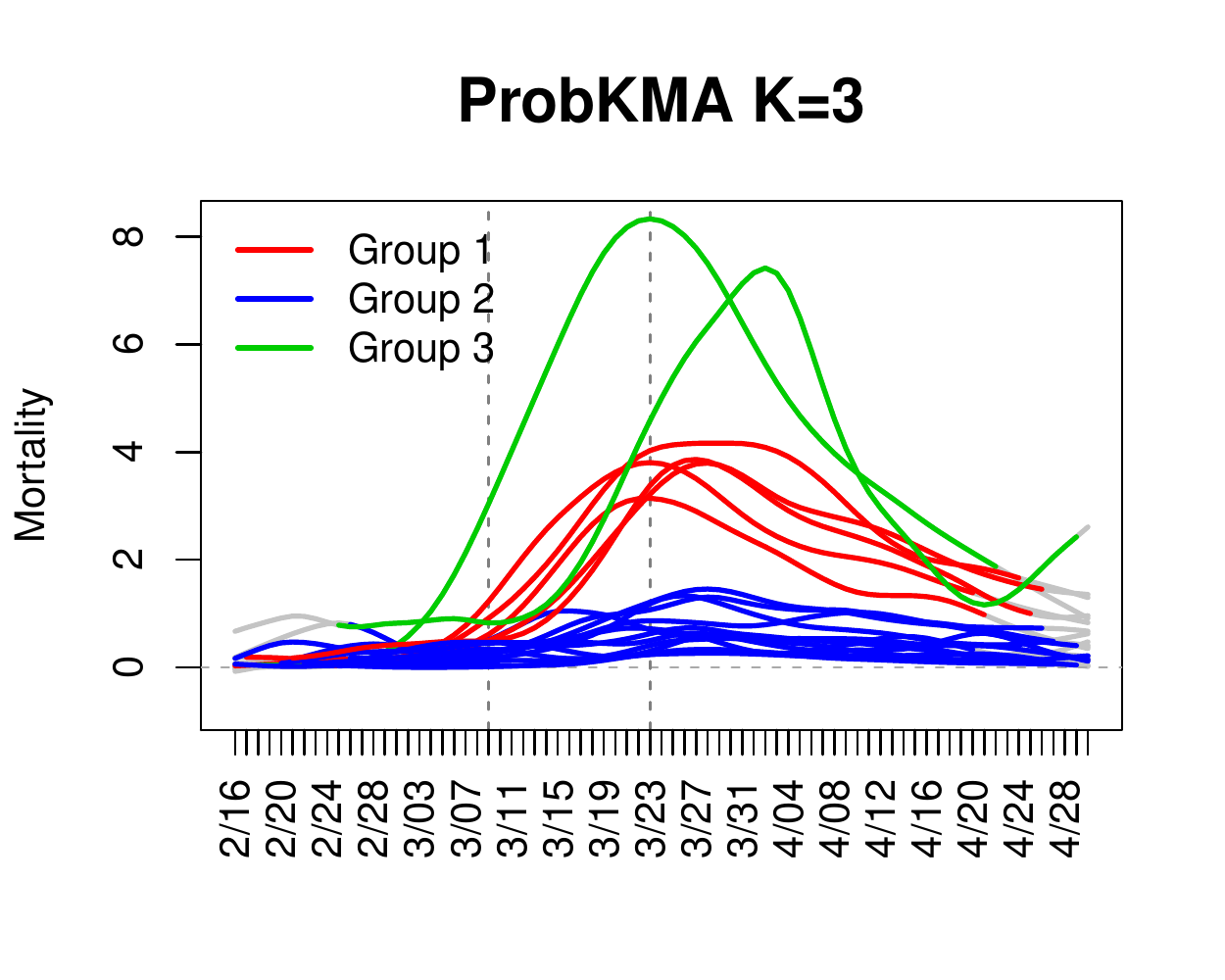}
        \includegraphics[page=5,trim={0.8cm 1cm 0.7cm 0.2cm},clip,height=4.3cm]{./img_suppl/regions_max_probKMA_d0_K3.pdf} \\
        \includegraphics[page=2,trim={0.1cm 1cm 0.7cm 0.2cm},clip,height=4.3cm]{./img_suppl/regions_max_probKMA_d0_K3.pdf}
        \includegraphics[page=3,trim={0.8cm 1cm 0.7cm 0.2cm},clip,height=4.3cm]{./img_suppl/regions_max_probKMA_d0_K3.pdf}
        \includegraphics[page=4,trim={0.8cm 1cm 0.7cm 0.2cm},clip,height=4.3cm]{./img_suppl/regions_max_probKMA_d0_K3.pdf}
    \end{minipage}}
    } \\ 
    \subfloat[\label{subfig:probKMA_dpc_K3}]{
    \fbox{
    \begin{minipage}[b]{0.94\linewidth}
        \includegraphics[page=1,trim={0.1cm 1cm 0.7cm 0.2cm},clip,height=4.3cm]{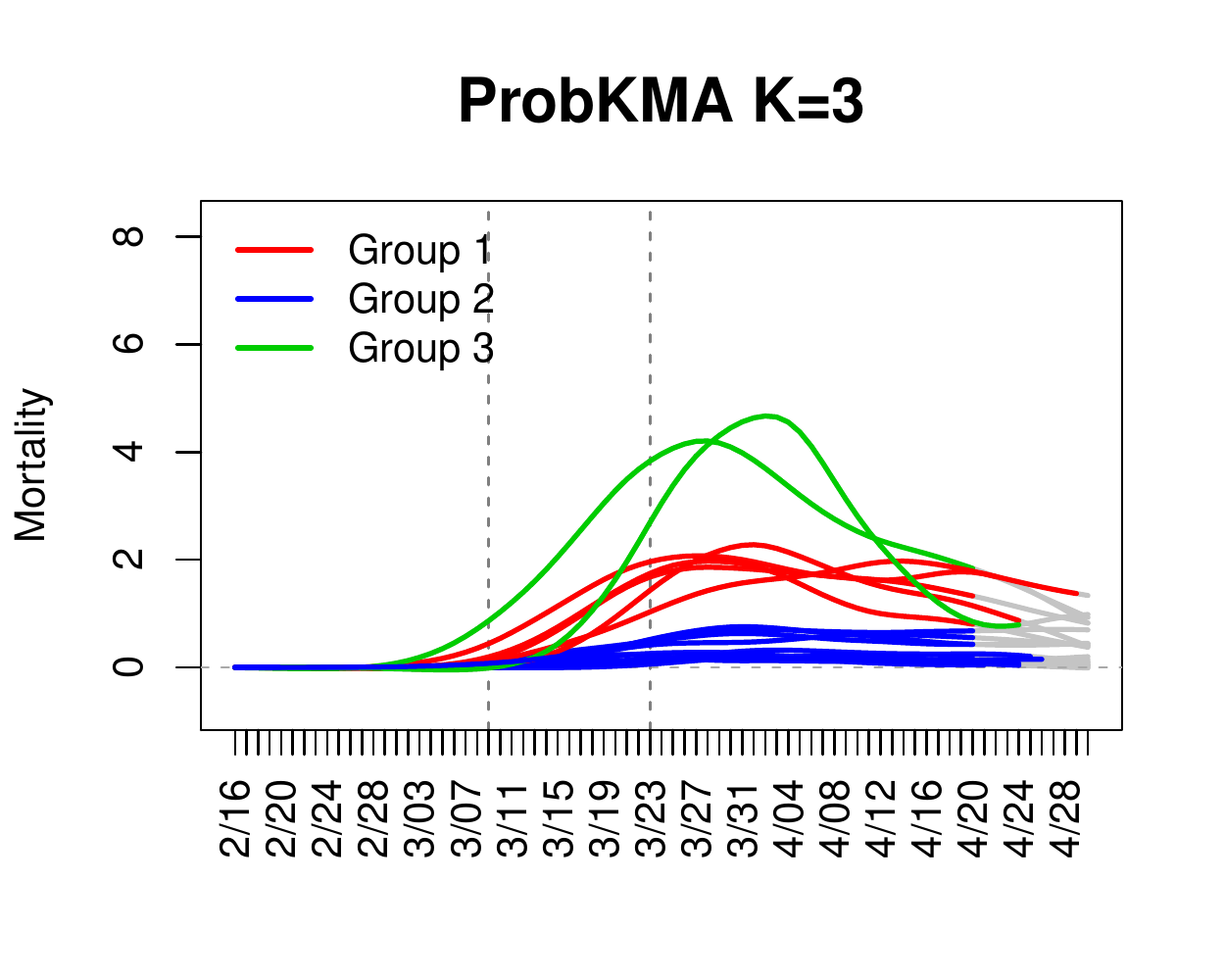}
        \includegraphics[page=5,trim={0.8cm 1cm 0.7cm 0.2cm},clip,height=4.3cm]{./img_suppl/regions_dpc_probKMA_d0_K3.pdf} \\
        \includegraphics[page=2,trim={0.1cm 1cm 0.7cm 0.2cm},clip,height=4.3cm]{./img_suppl/regions_dpc_probKMA_d0_K3.pdf}
        \includegraphics[page=3,trim={0.8cm 1cm 0.7cm 0.2cm},clip,height=4.3cm]{./img_suppl/regions_dpc_probKMA_d0_K3.pdf}
        \includegraphics[page=4,trim={0.8cm 1cm 0.7cm 0.2cm},clip,height=4.3cm]{./img_suppl/regions_dpc_probKMA_d0_K3.pdf}
    \end{minipage}}
    }
\end{figure}
\begin{figure}[!tb]
    \centering
    \subfloat[\label{subfig:probKMA_istat_K3}]{
    \fbox{
    \begin{minipage}[b]{0.94\linewidth}
        \includegraphics[page=1,trim={0.1cm 1cm 0.7cm 0.2cm},clip,height=4.3cm]{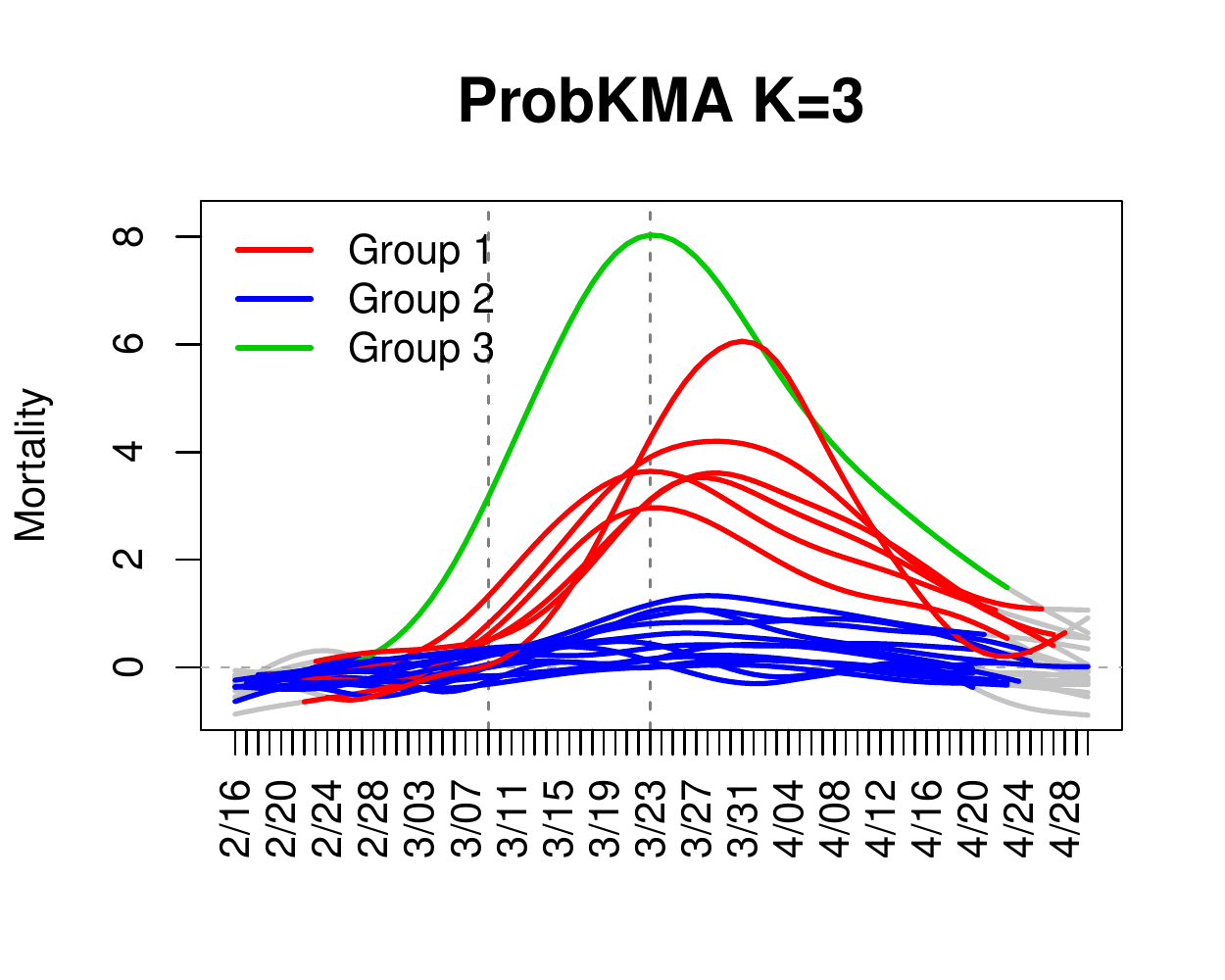}
        \includegraphics[page=5,trim={0.8cm 1cm 0.7cm 0.2cm},clip,height=4.3cm]{./img_suppl/regions_istat_probKMA_d0_K3.pdf} \\
        \includegraphics[page=2,trim={0.1cm 1cm 0.7cm 0.2cm},clip,height=4.3cm]{./img_suppl/regions_istat_probKMA_d0_K3.pdf}
        \includegraphics[page=3,trim={0.8cm 1cm 0.7cm 0.2cm},clip,height=4.3cm]{./img_suppl/regions_istat_probKMA_d0_K3.pdf}
        \includegraphics[page=4,trim={0.8cm 1cm 0.7cm 0.2cm},clip,height=4.3cm]{./img_suppl/regions_istat_probKMA_d0_K3.pdf}
    \end{minipage}}
    }
    \caption{
    {\bf Characterizing three epidemics}.
    Results of {\em probKMA} with $K=3$ on {\bf \protect\subref{subfig:probKMA_max_K3}} MAX mortality curves, {\bf \protect\subref{subfig:probKMA_dpc_K3}} DPC mortality curves and {\bf \protect\subref{subfig:probKMA_istat}} ISTAT curves. 
    Mortality curves are shown in the top left panel with portions identified by {\em probKMA} with  $K=3$ in red (Group 1; "exponential" pattern), blue (Group 2; "flat(tened)" pattern) and green (Group 3; "extreme" pattern). The curve portions are shown again, this time aligned with each other and separated by group, in the bottom panels. Black lines indicate the average curves of the group. The shifts produced by {\em probKMA} are shown in the top right panel. 
    }
    \label{fig:three_epidemics}
\end{figure}

\begin{figure}[!tb]
\centering
\vspace{-0.2cm}
\hrule height 1pt
\vrule width 1pt
\hspace{0.15cm}
\includegraphics[width=0.48\linewidth]{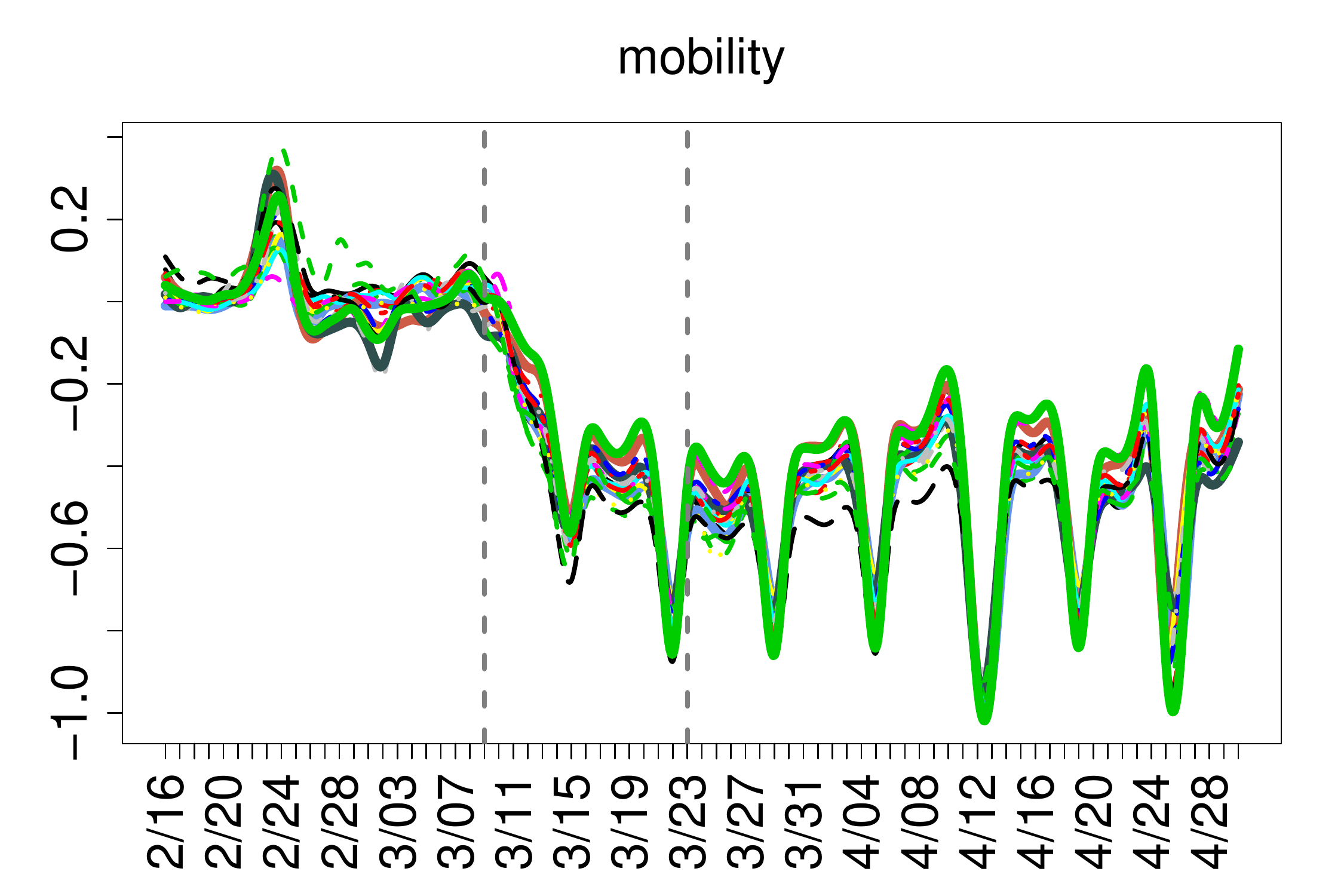}
\includegraphics[width=0.48\linewidth]{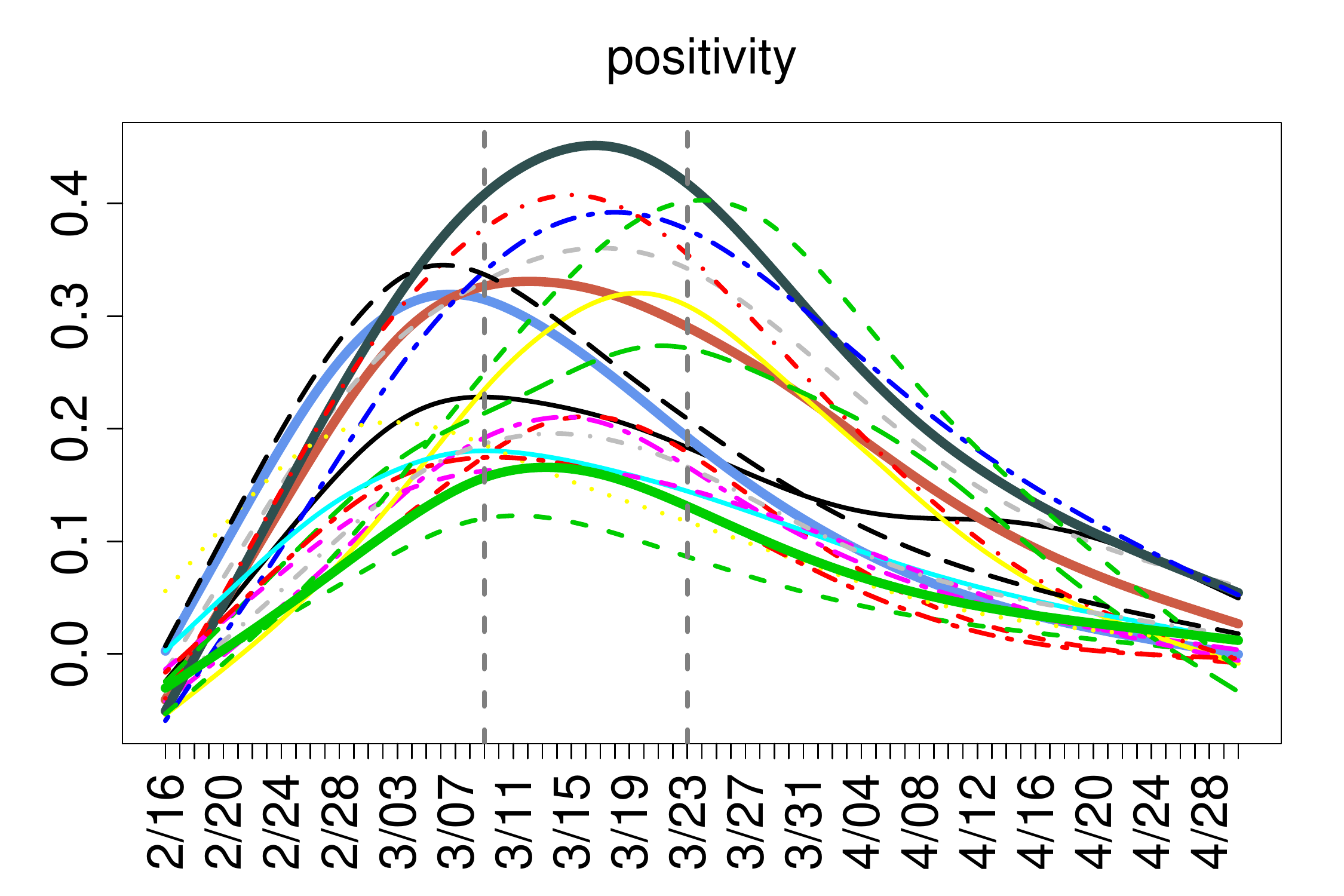}
\hspace{0.15cm}
\vrule width 1pt \\
\vspace{-0.7cm}
\vrule width 1pt
\includegraphics[width=0.97\linewidth]{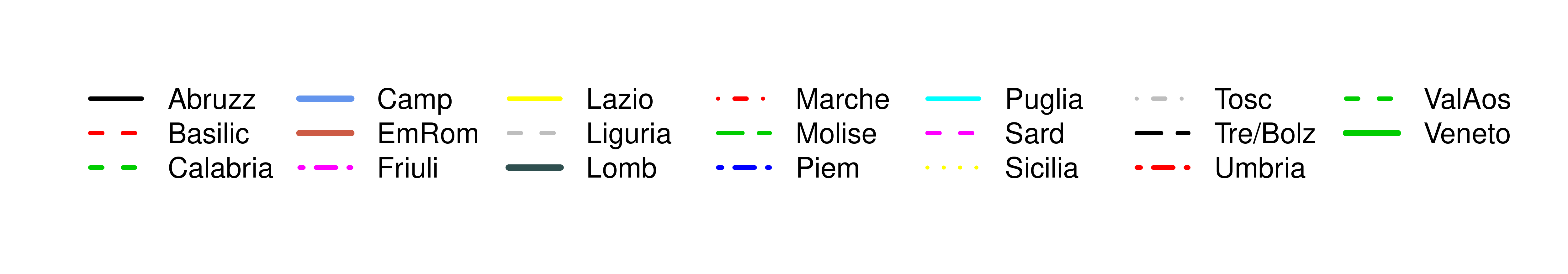}
\hspace{0.28cm}
\vrule width 1pt
\hrule height 1pt
\vspace{0.2cm}  
\caption{ {\bf Unshifted mobility and positivity curves.} Local mobility and positivity curves without shift in the 20 Italian regions. Vertical lines show the days corresponding to the national lock down (March 9) and the suspension of all non-essential production activities (March 23).
}
\label{fig:unshifted_curves_col_pos_mob}
\end{figure}

\begin{figure}[!tb]
\begin{center}
{\small \bf shifted curves ISTAT}
\end{center}
\centering
\vspace{-0.2cm}
\hrule height 1pt
\vrule width 1pt
\includegraphics[width=0.32\linewidth]{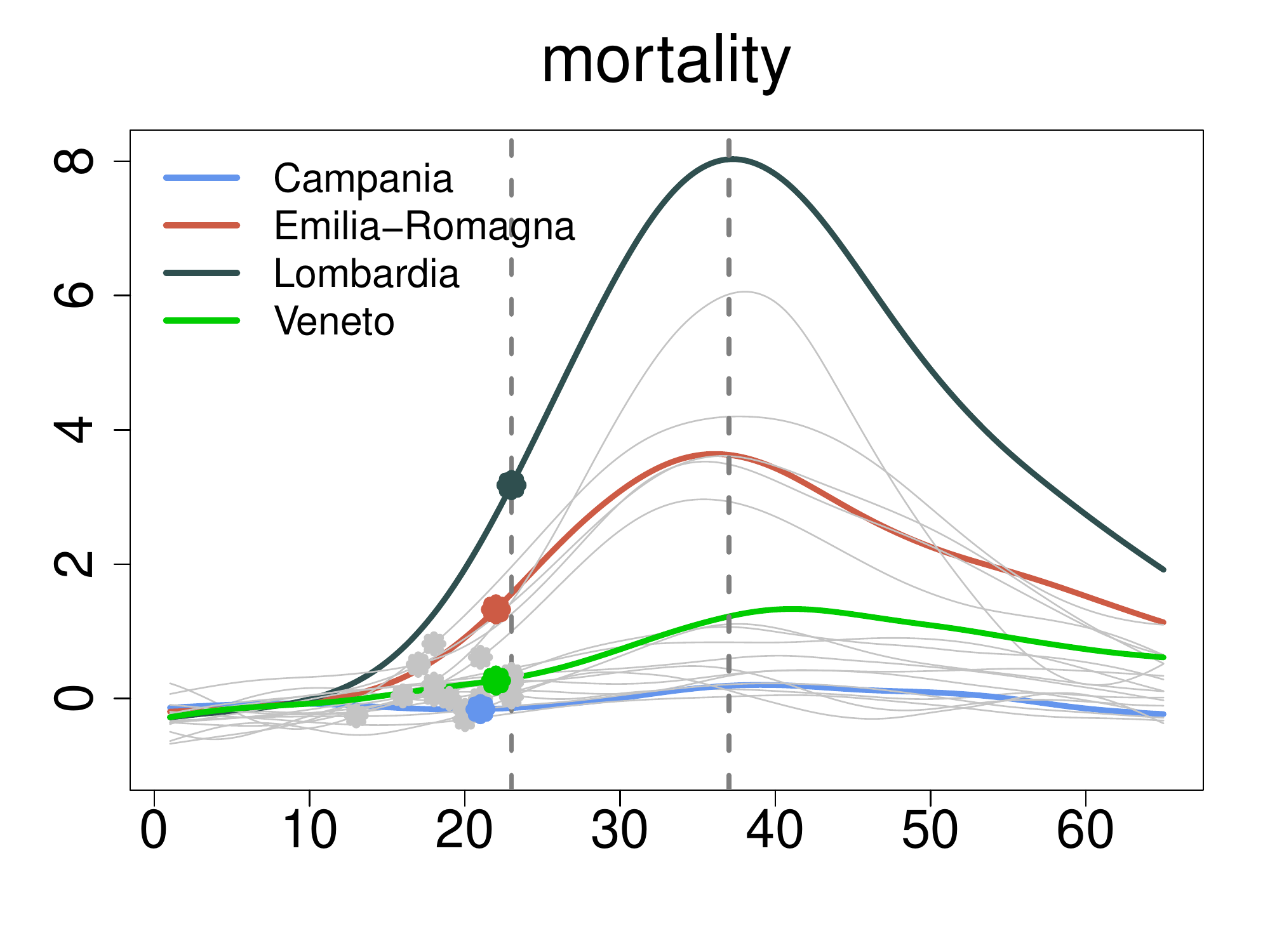}
\hspace{0.1cm}
\includegraphics[width=0.32\linewidth]{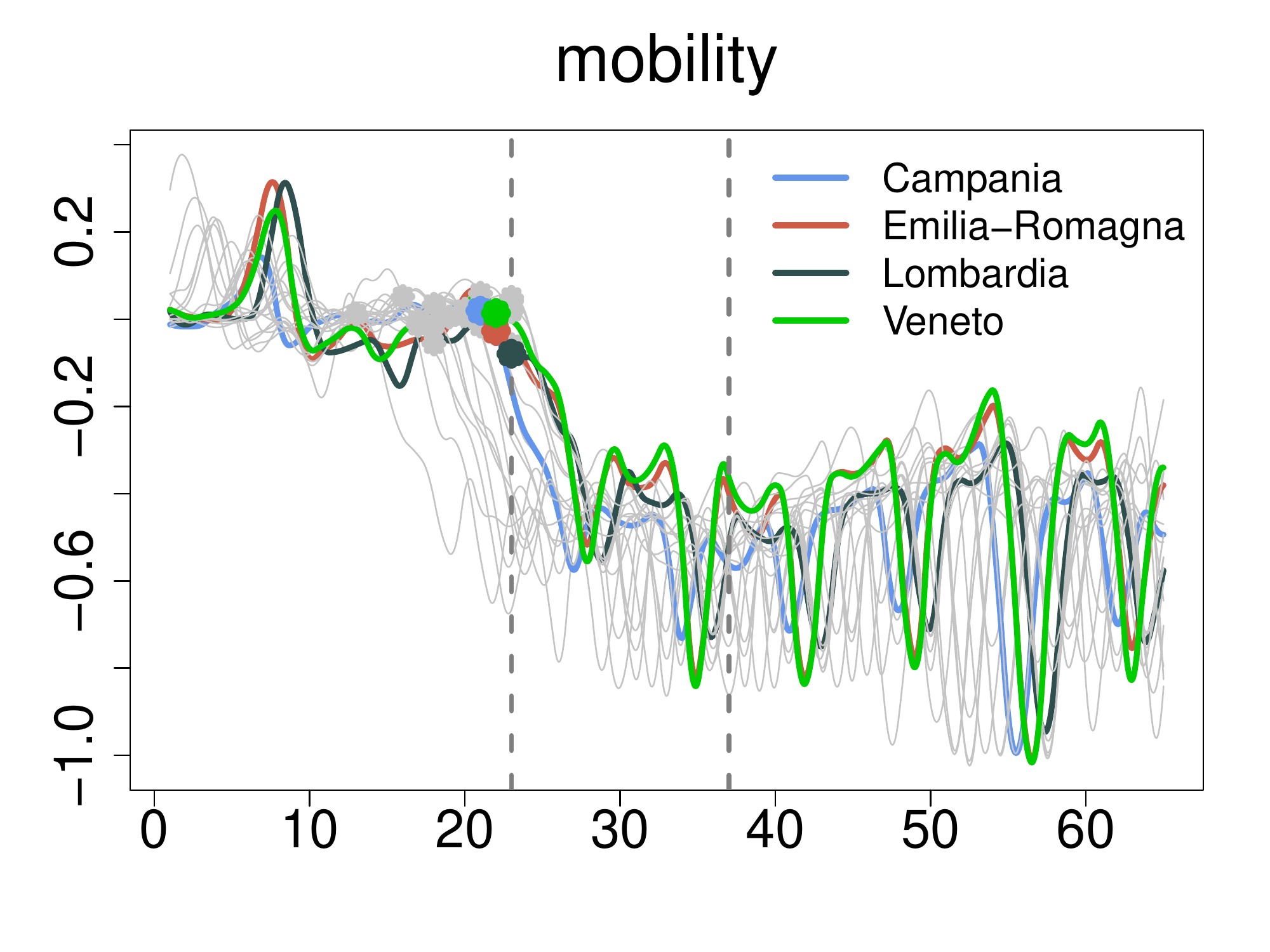}
\hspace{0.1cm}
\includegraphics[width=0.32\linewidth]{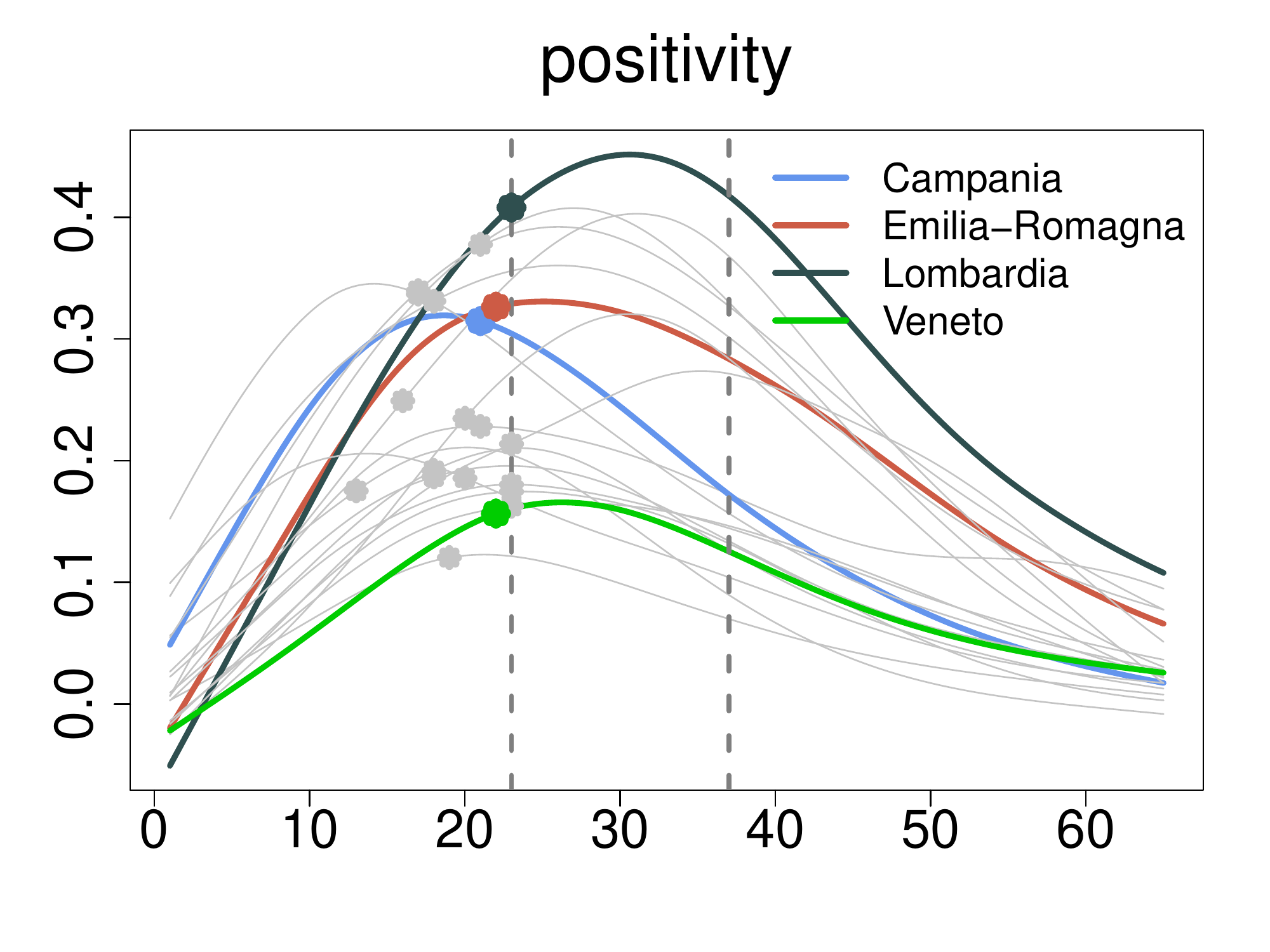}
\vrule width 1pt
\hrule height 1pt
\vspace{0.2cm}  

\begin{center}
{\small \bf shifted curves DPC}
\end{center}
\centering
\vspace{-0.2cm}
\hrule height 1pt
\vrule width 1pt
\includegraphics[width=0.32\linewidth]{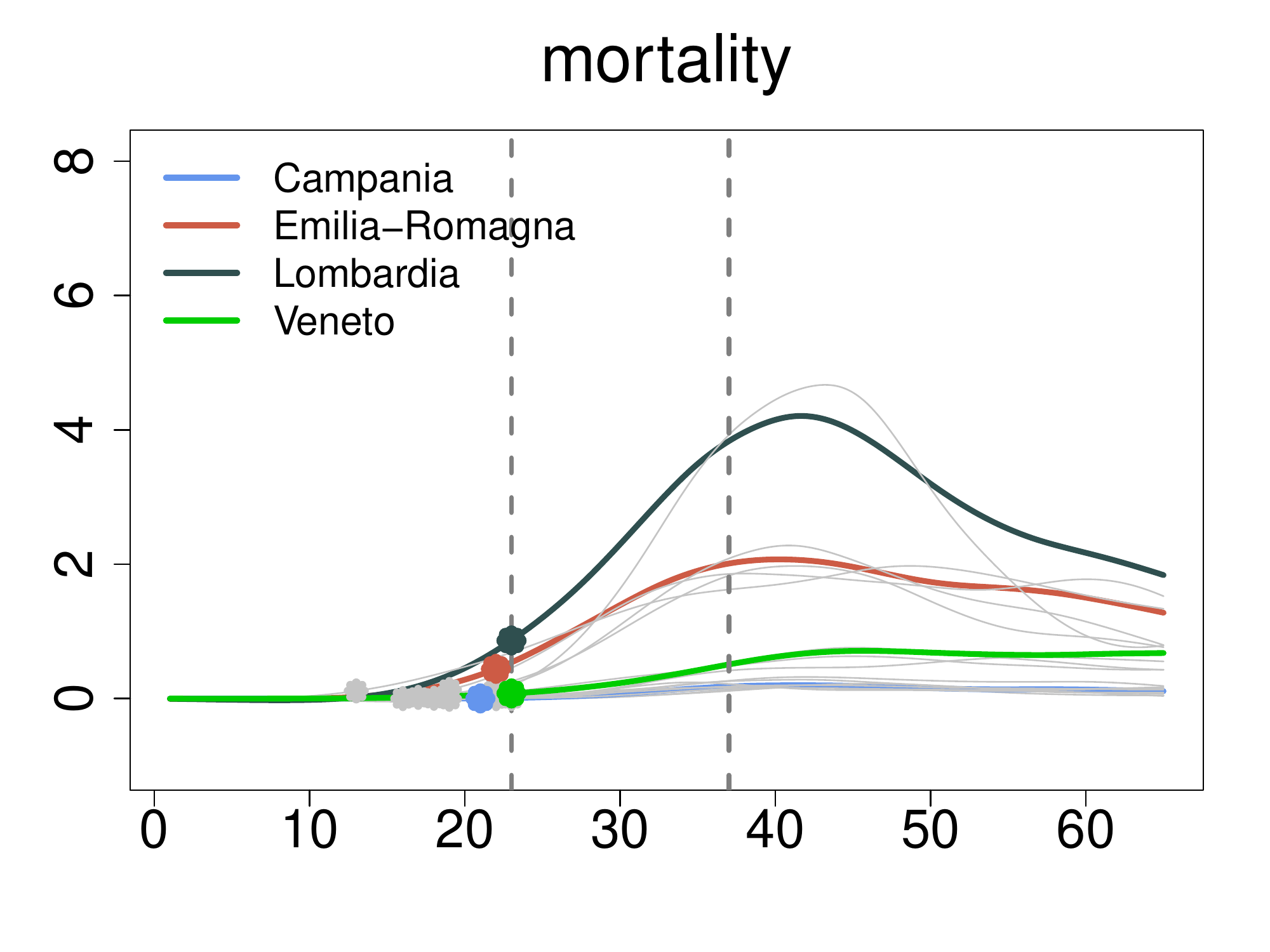}
\hspace{0.1cm}
\includegraphics[width=0.32\linewidth]{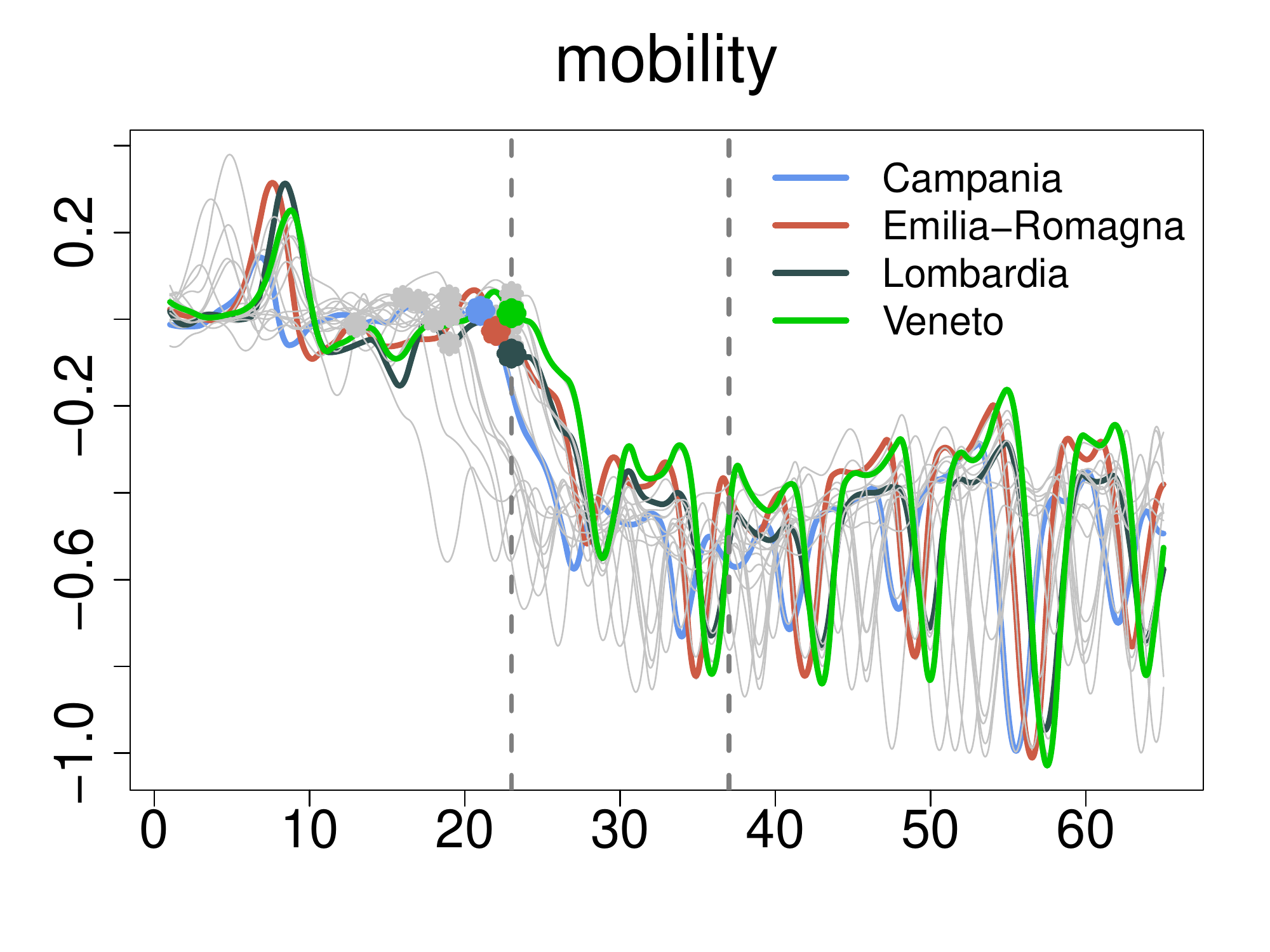}
\hspace{0.1cm}
\includegraphics[width=0.32\linewidth]{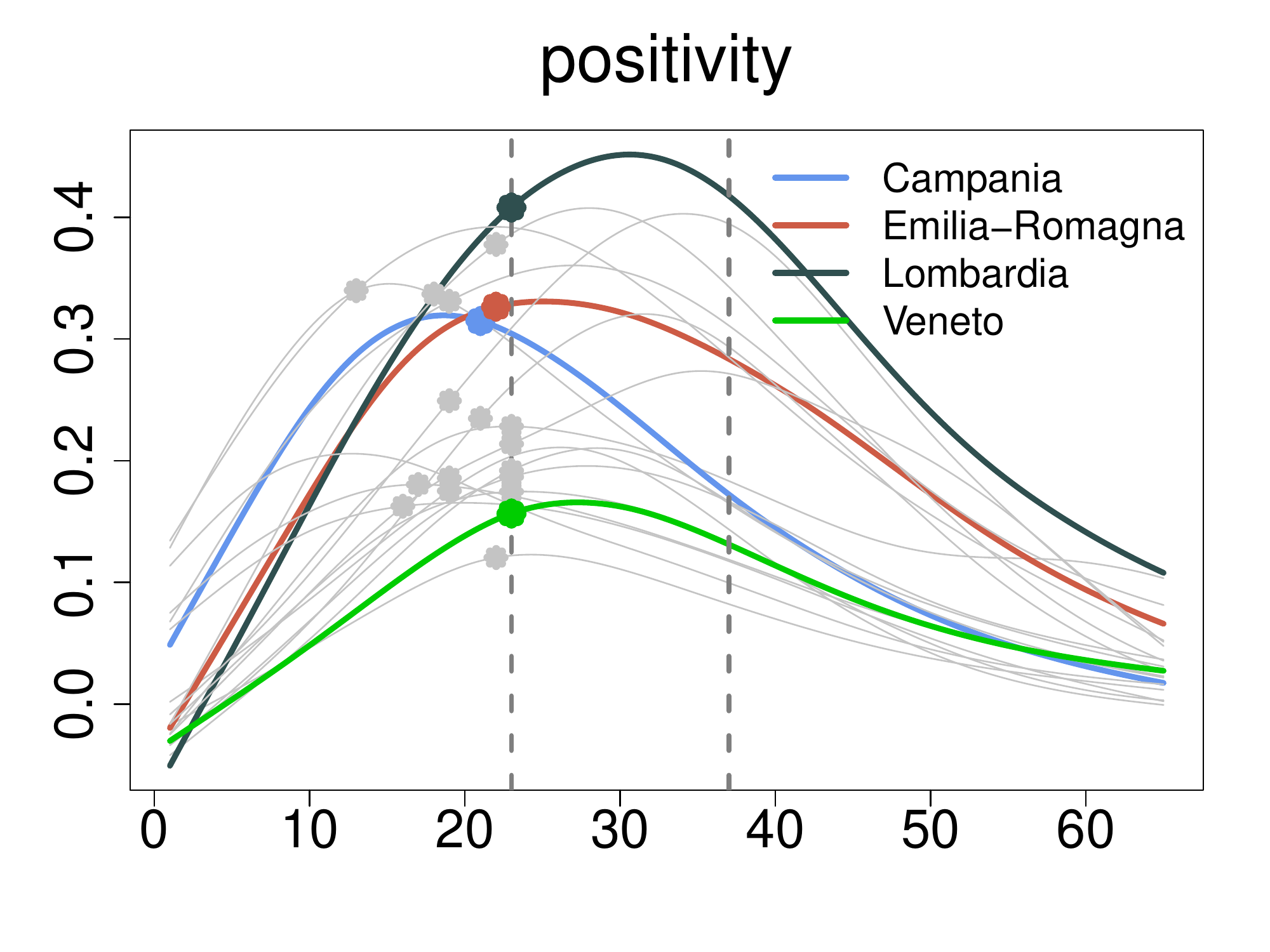}
\vrule width 1pt
\hrule height 1pt
\vspace{0.2cm}  
\caption{ {\bf Shifted curves for ISTAT and DPC.} Mortality (per 100,000 inhabitants), mobility, and positivity curves after the shifts produced by \emph{probKMA} with K=2. Vertical lines mark the dates of the national lock-down (March 9) and the suspension of all non-essential production activities (March 23) without shifts. Stars on the curves mark the lock-down after the region-specific shifts.
}
\label{fig:shifted_curves_istat_dpc}
\end{figure}


\begin{figure}[!tb]
 \centering
    \subfloat[\label{subfig:mob_pos}]{
    \fbox{
    \begin{minipage}[b]{0.36\linewidth}
        \centering
        \includegraphics[width=0.95\linewidth]{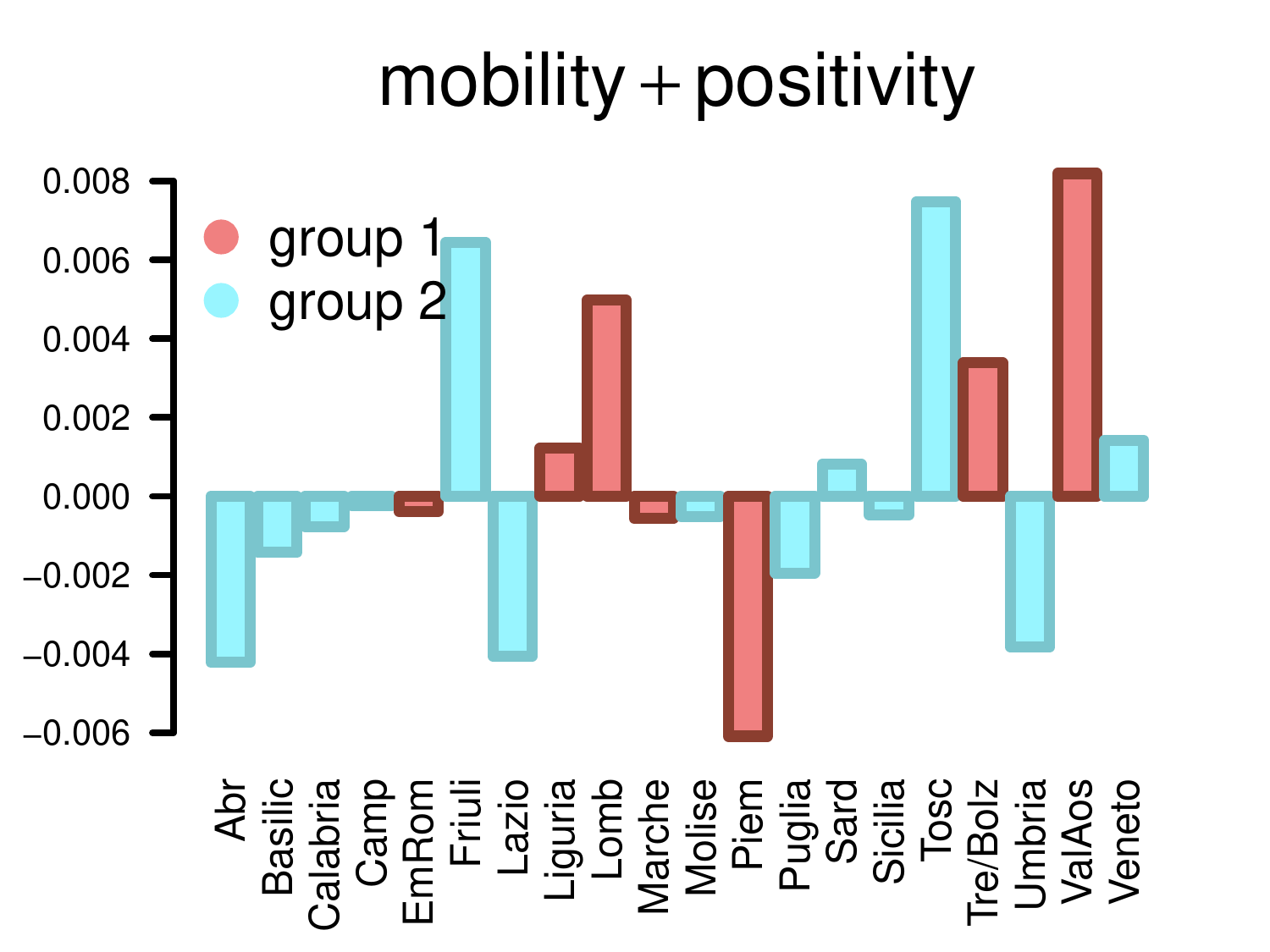}
    \end{minipage}}
    }
    \subfloat[\label{subfig:mob_pos_pc1}]{
    \fbox{
    \begin{minipage}[b]{0.36\linewidth}
        \centering
        \includegraphics[width=0.95\linewidth]{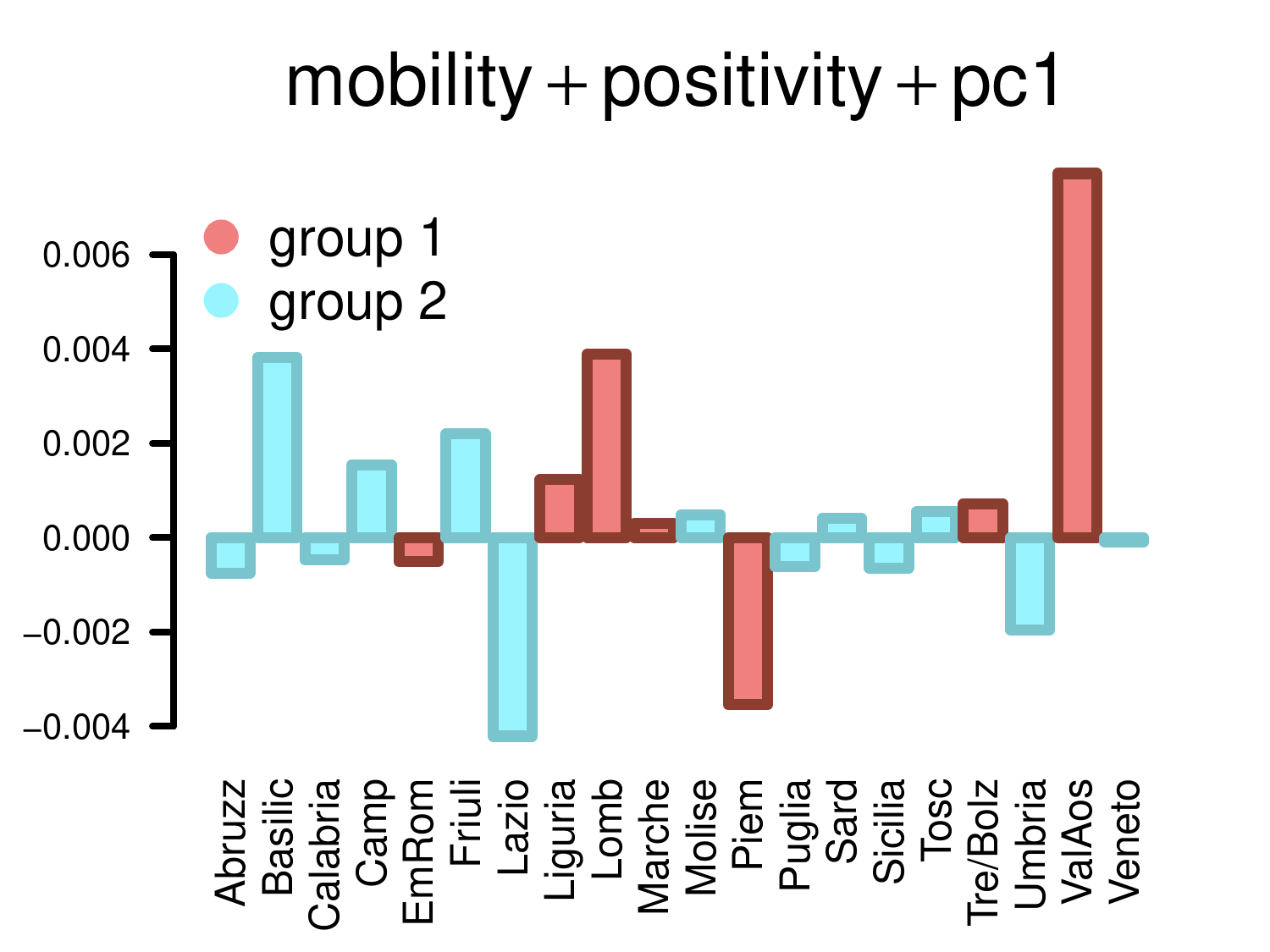}
    \end{minipage}}
    }
\caption{ 
{\bf MAX mortality residuals.} {\bf \protect\subref{subfig:mob_pos}:} residuals of the function-on-function regression of MAX mortality on local mobility and positivity. {\bf \protect\subref{subfig:mob_pos_pc1}:} residuals of the function-on-function regression of MAX mortality on local mobility, positivity, and the first principal component of the top 5 covariates. In both panels curves from Group 1 are in red, and curves from Group 2 are in blue. Residuals with positive sign indicate regions for which the true mortality curve is above the estimated mortality curve. Conversely, residuals with negative sign indicate regions for which the true mortality curve is below the estimated mortality curve.
}
\label{fig:max_res}
\end{figure}

\begin{figure}[!tb]
\begin{center}
{\small \bf mortality $\bm \sim$ mobility} \\
\vspace{0.1cm}
{\small \bf MAX \hspace{5cm} ISTAT \hspace{5cm} DPC}
\end{center}
\centering
\vspace{-0.2cm}
\hrule height 1pt
\vrule width 1pt
\hspace{0.2cm}
\includegraphics[width=0.30\linewidth]{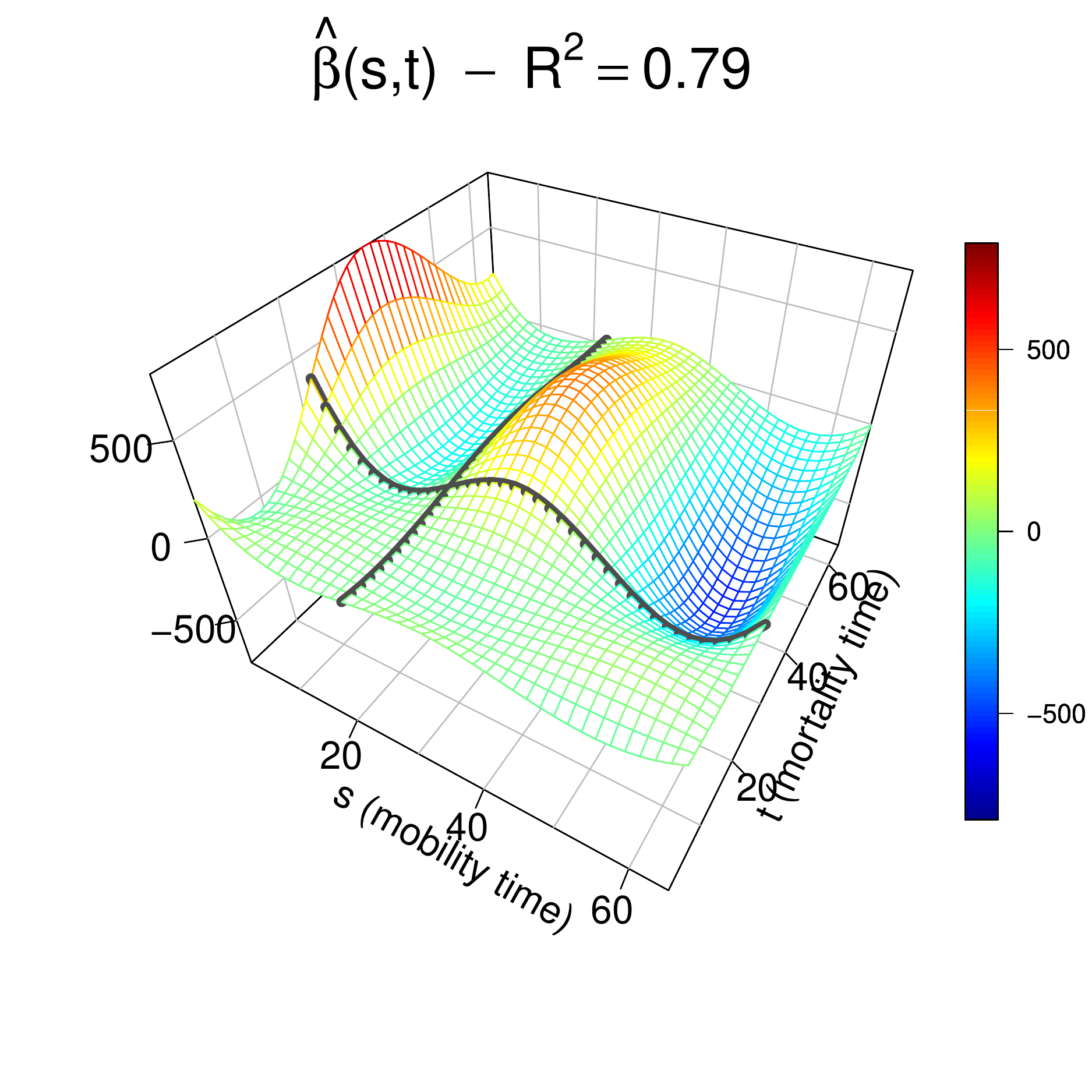}
\hspace{0.3cm}
\vrule width 1pt
\includegraphics[width=0.30\linewidth]{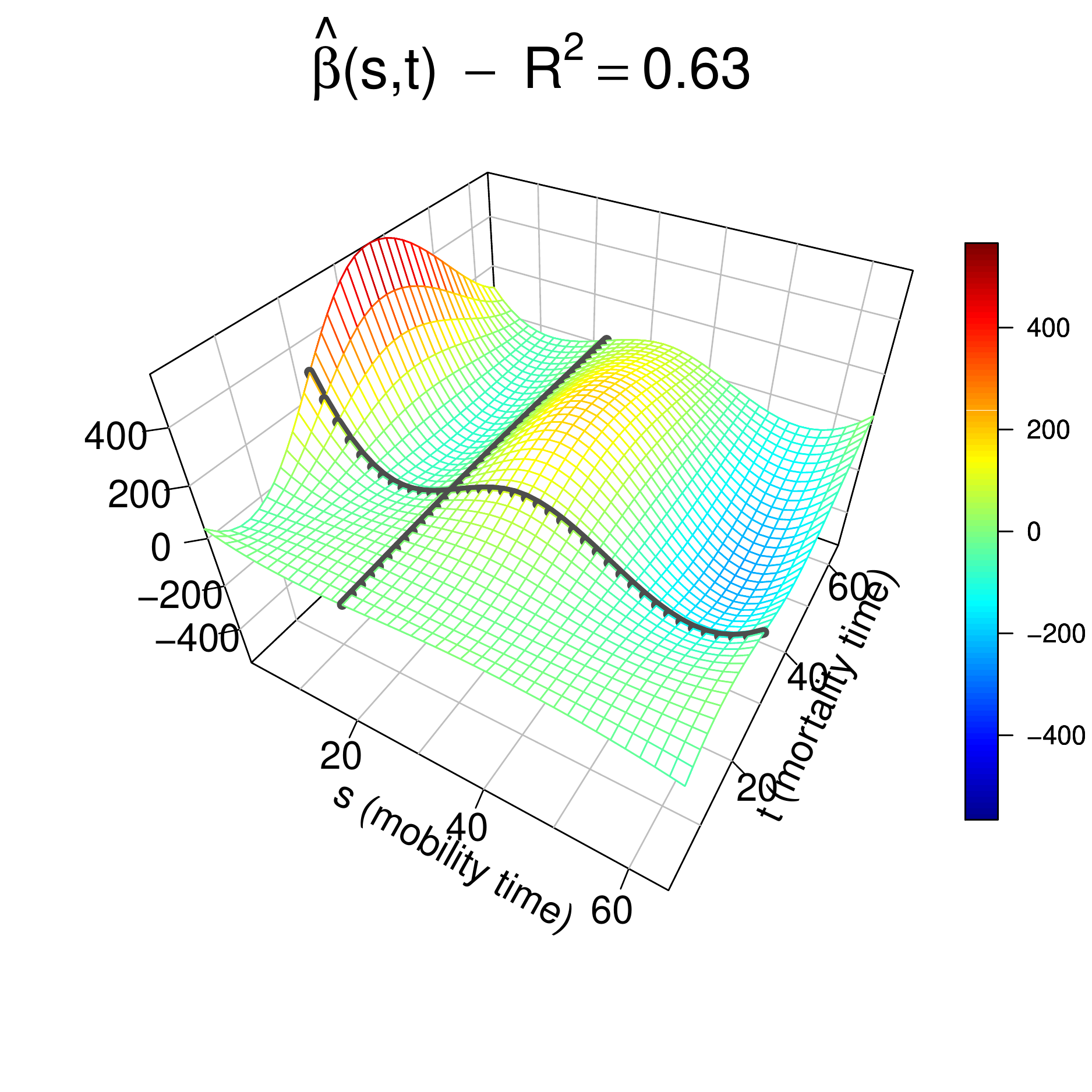}
\hspace{0.3cm}
\vrule width 1pt
\includegraphics[width=0.30\linewidth]{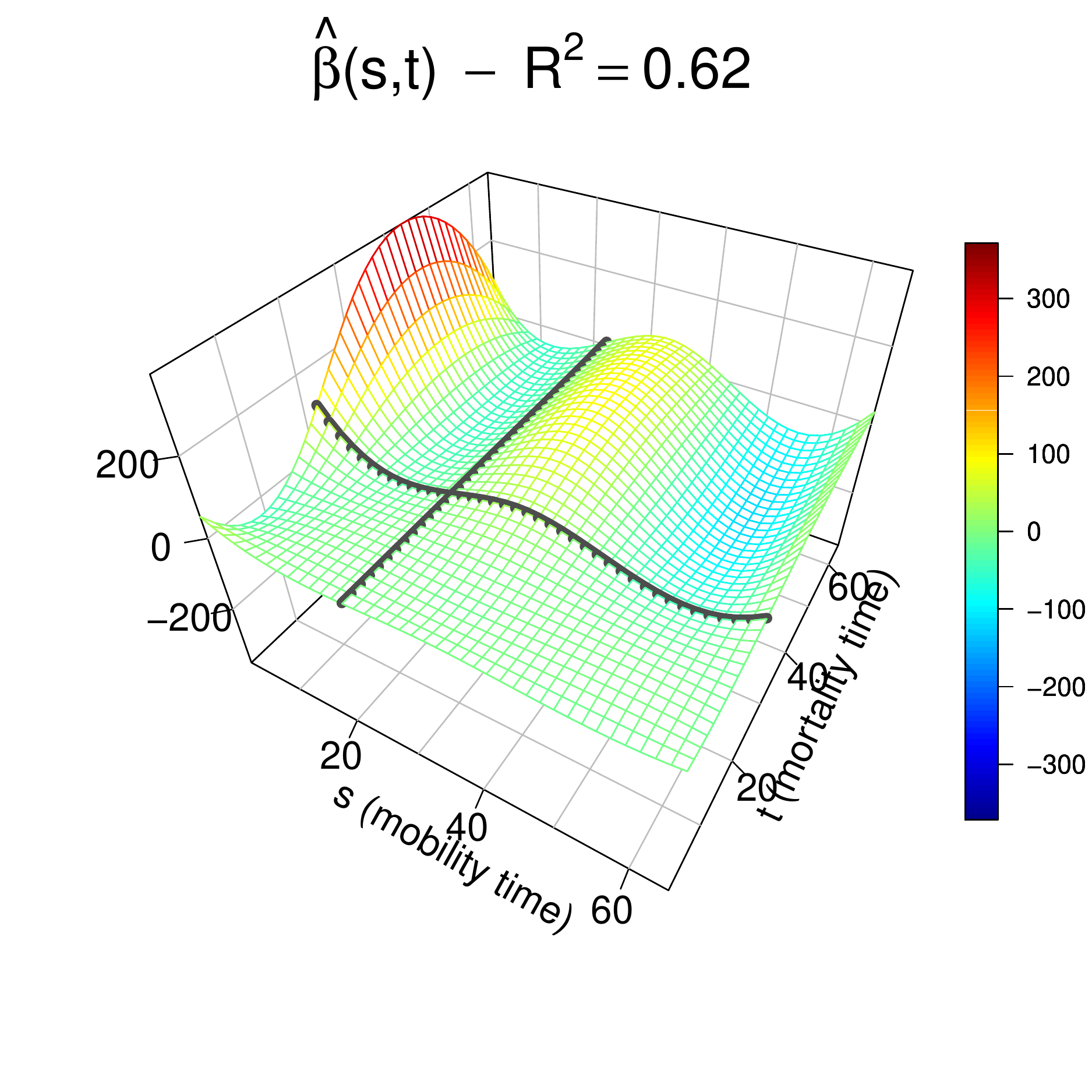}
\hspace{0.2cm}
\vrule width 1pt \\
\vspace{-0.6cm}
\vrule width 1pt
\hspace{0.2cm}
\includegraphics[width=0.3\linewidth]{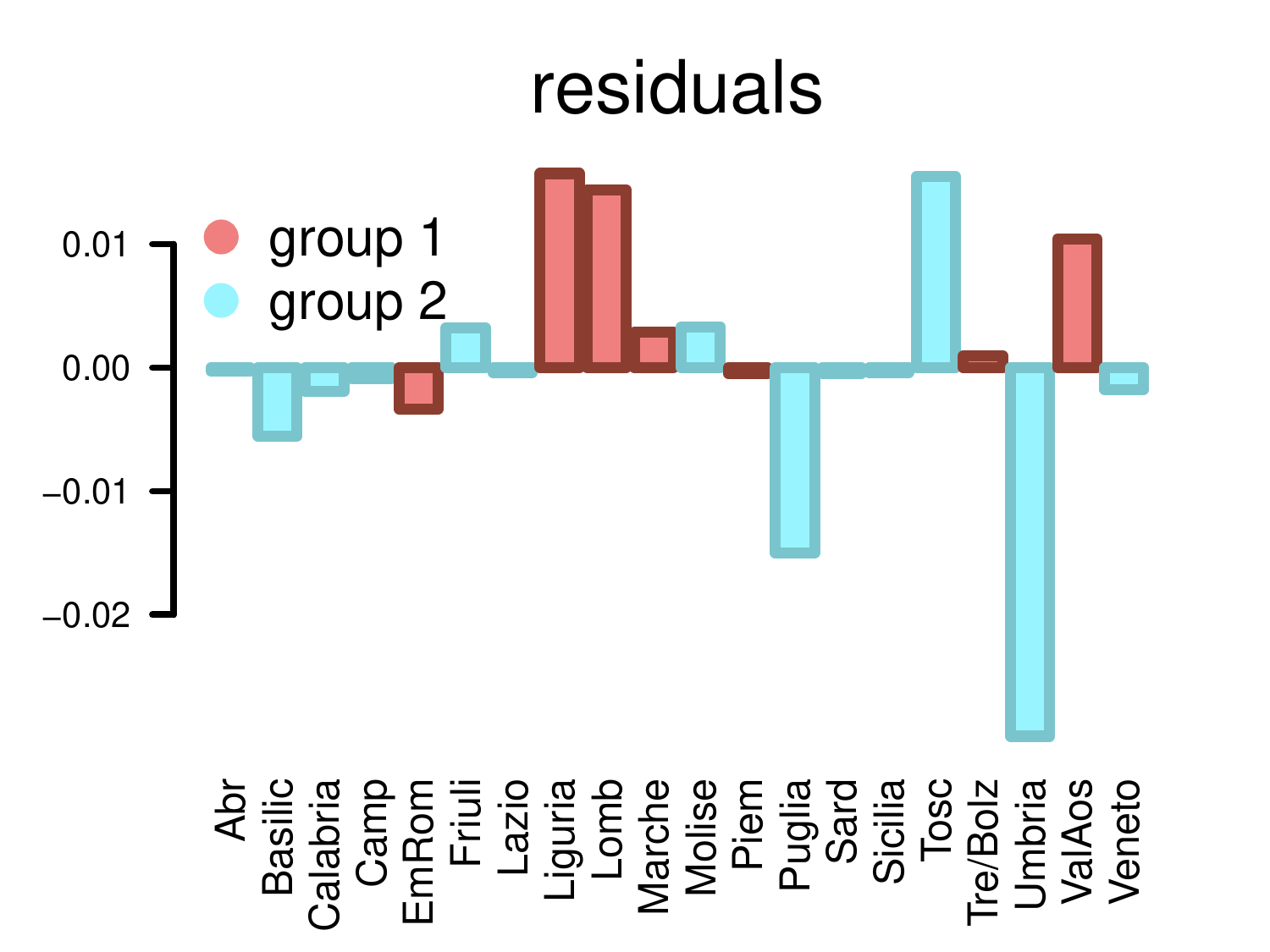}
\hspace{0.3cm}
\vrule width 1pt
\includegraphics[width=0.3\linewidth]{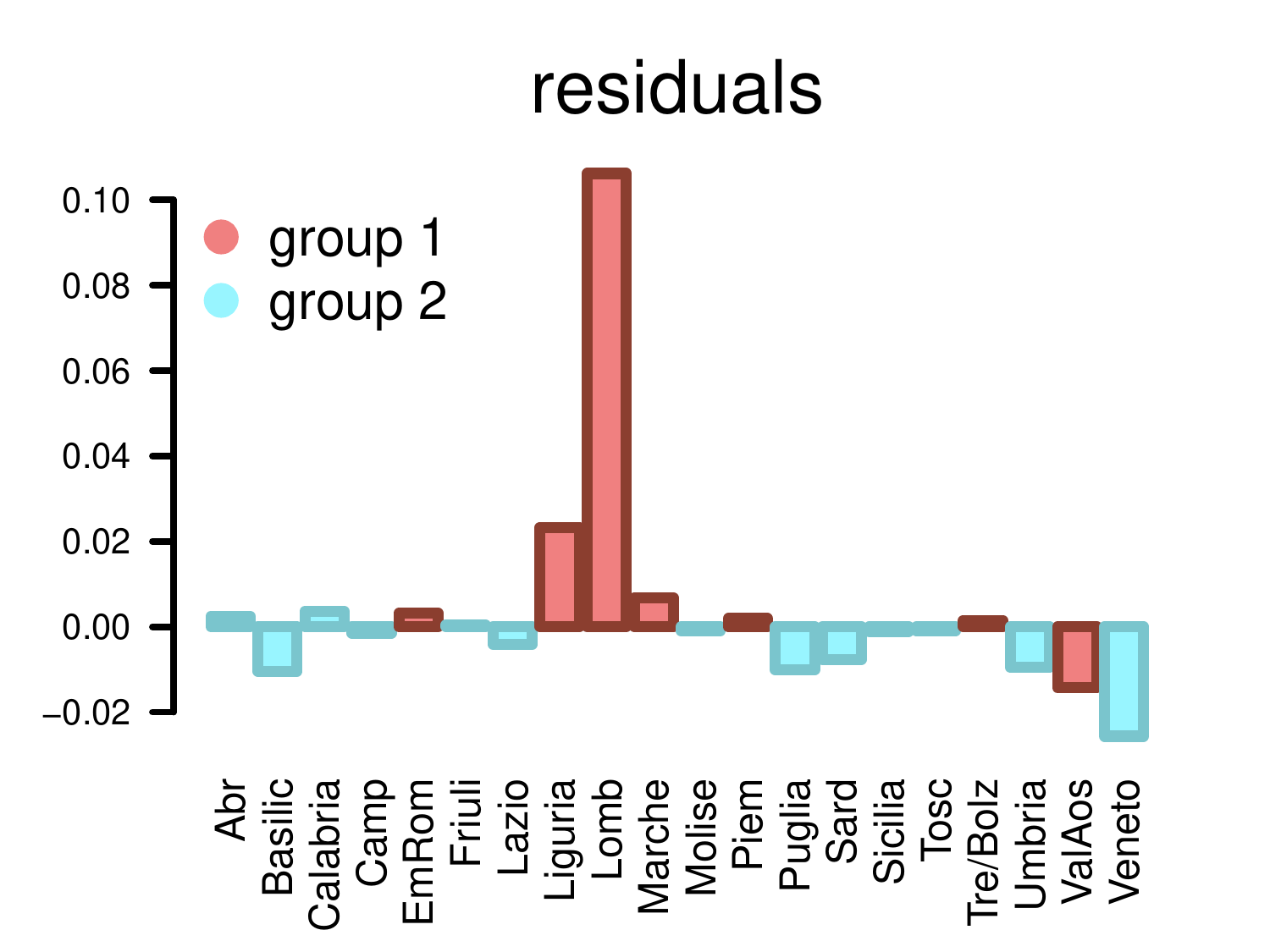}
\hspace{0.3cm}
\vrule width 1pt
\includegraphics[width=0.3\linewidth]{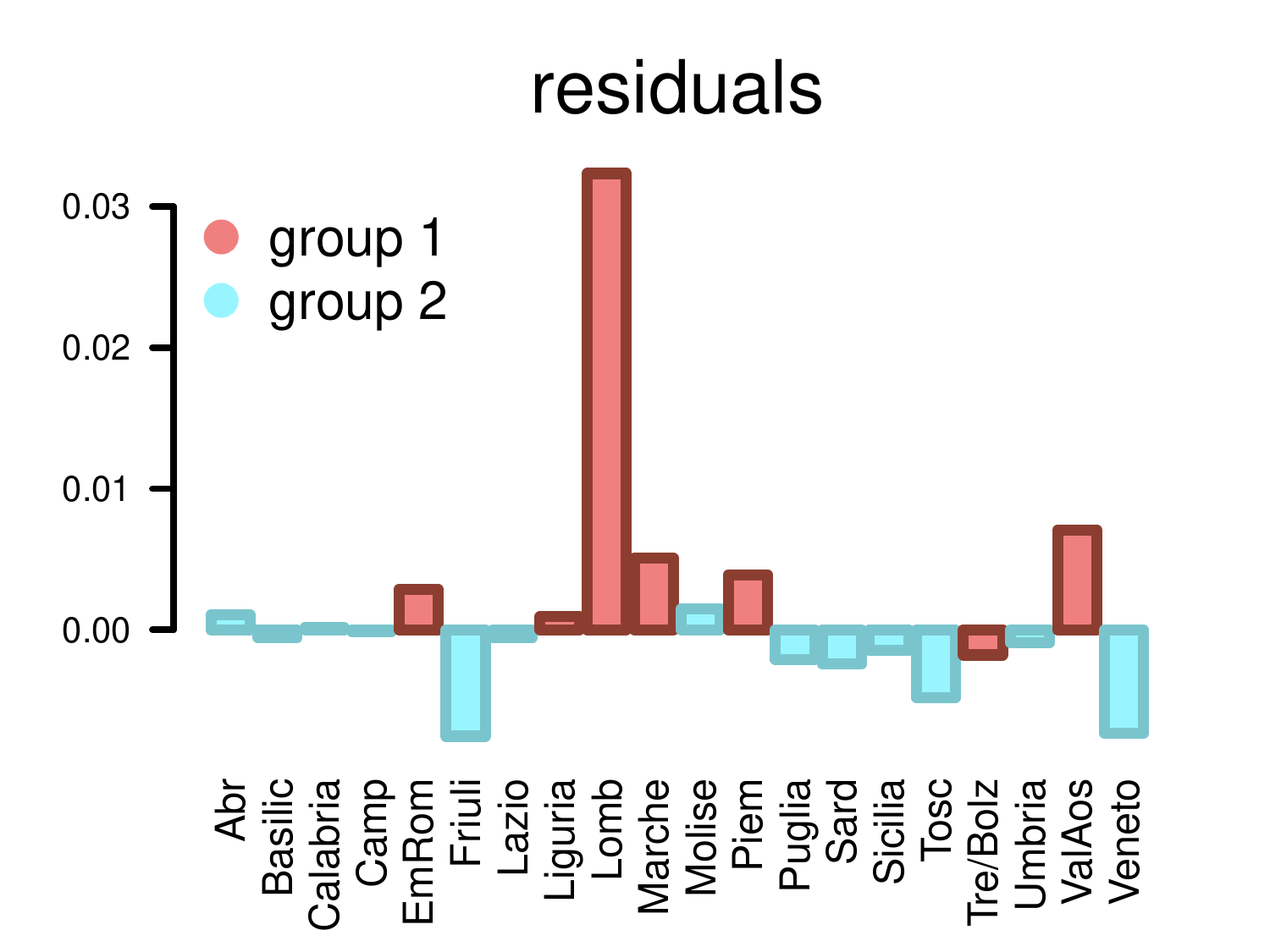}
\hspace{0.2cm}
\vrule width 1pt
\hrule height 1pt
\vspace{0.2cm} 
\caption{ 
{\bf Associating mortality to mobility.}
Results from the function-on-function regression of mortality on local mobility. The top row displays the estimated effect surface (the March 9 date is marked) with respective in-sample $R^2$ (for LOO-CV $R^2$ see Table~\ref{tab:R2}).
The bottom row displays the regression residuals (for barplots interpretation see Fig.~\ref{fig:max_res}).
}
\label{fig:mob_marginal}
\end{figure}

\begin{figure}[!tb]
\begin{center}
{\small \bf mortality $\bm \sim$ positivity} \\
\vspace{0.1cm}
{\small \bf MAX \hspace{5cm} ISTAT \hspace{5cm} DPC}
\end{center}
\vspace{-0.2cm}
\centering
\hrule height 1pt
\vrule width 1pt
\hspace{0.2cm}
\includegraphics[width=0.30\linewidth]{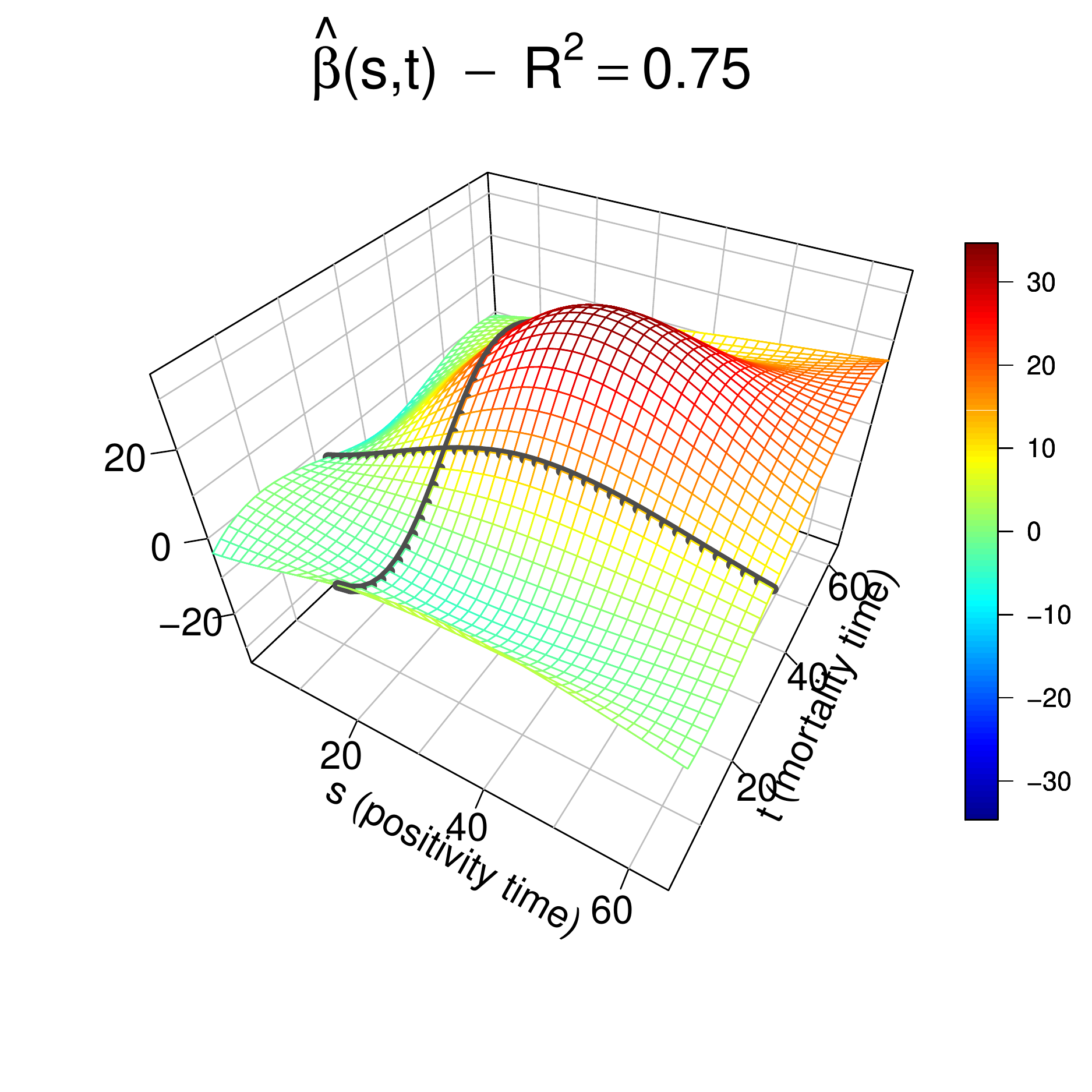}
\hspace{0.3cm}
\vrule width 1pt
\includegraphics[width=0.30\linewidth]{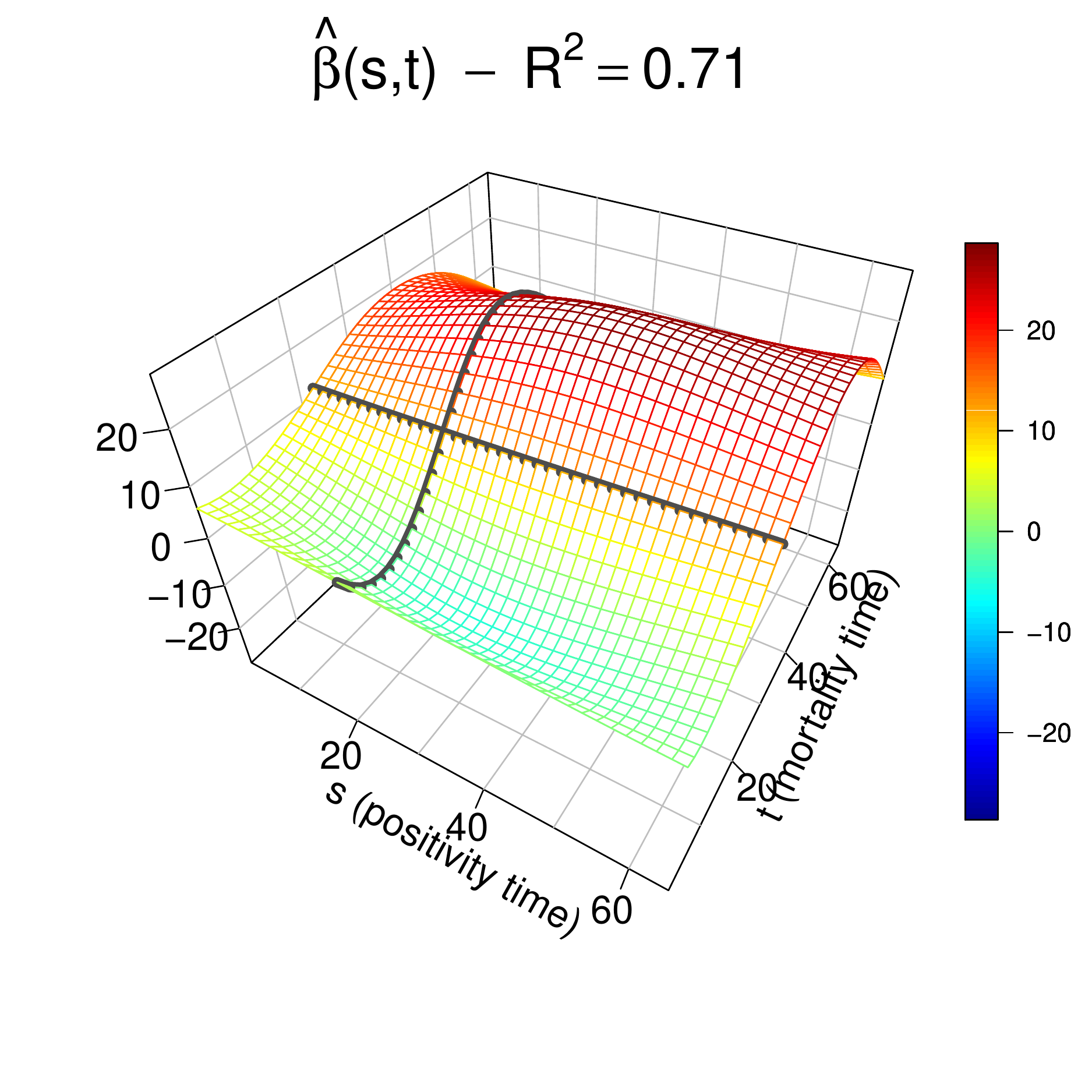}
\hspace{0.3cm}
\vrule width 1pt
\includegraphics[width=0.30\linewidth]{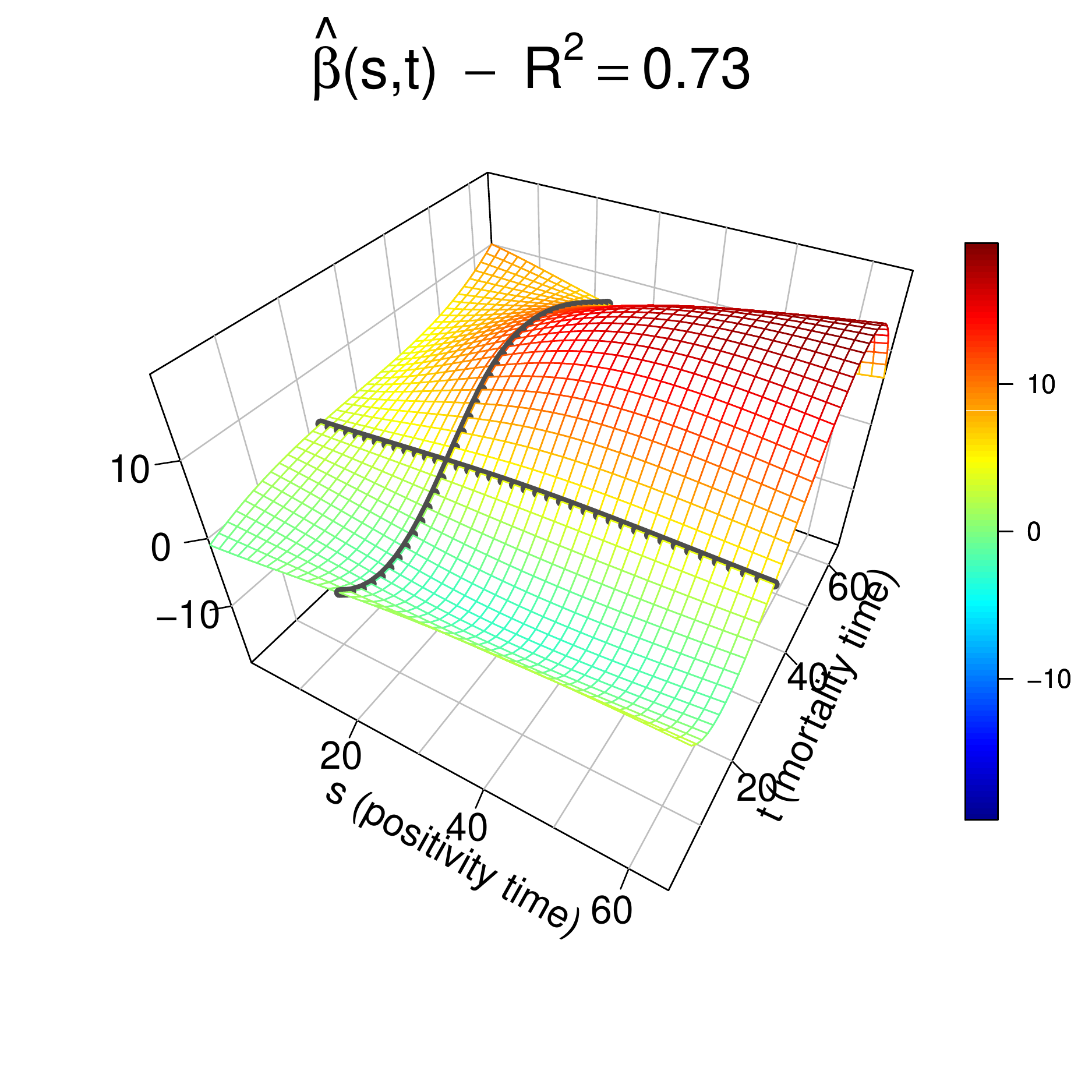}
\hspace{0.2cm}
\vrule width 1pt \\
\vspace{-0.6cm}
\vrule width 1pt
\hspace{0.2cm}
\includegraphics[width=0.3\linewidth]{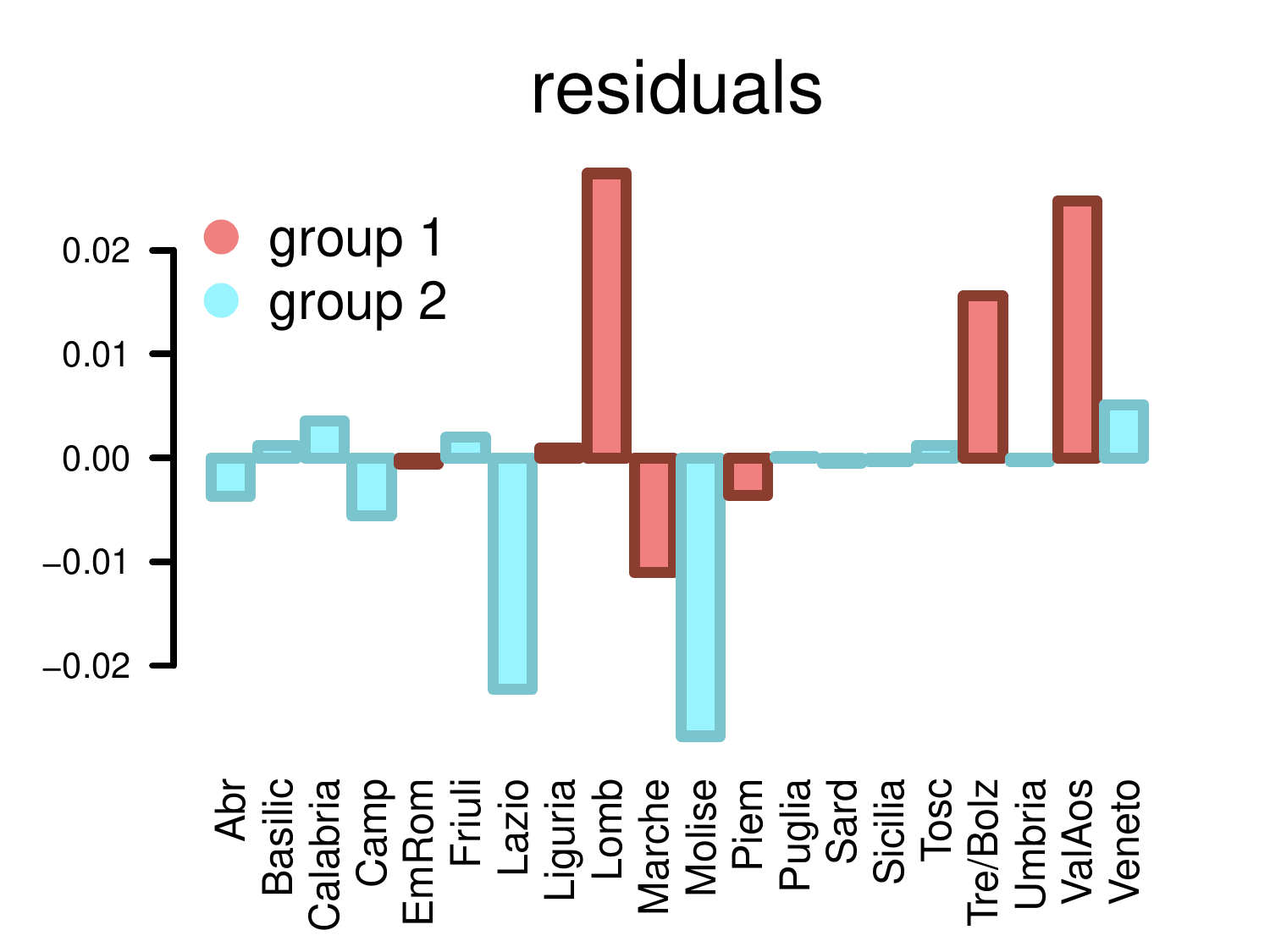}
\hspace{0.3cm}
\vrule width 1pt
\includegraphics[width=0.3\linewidth]{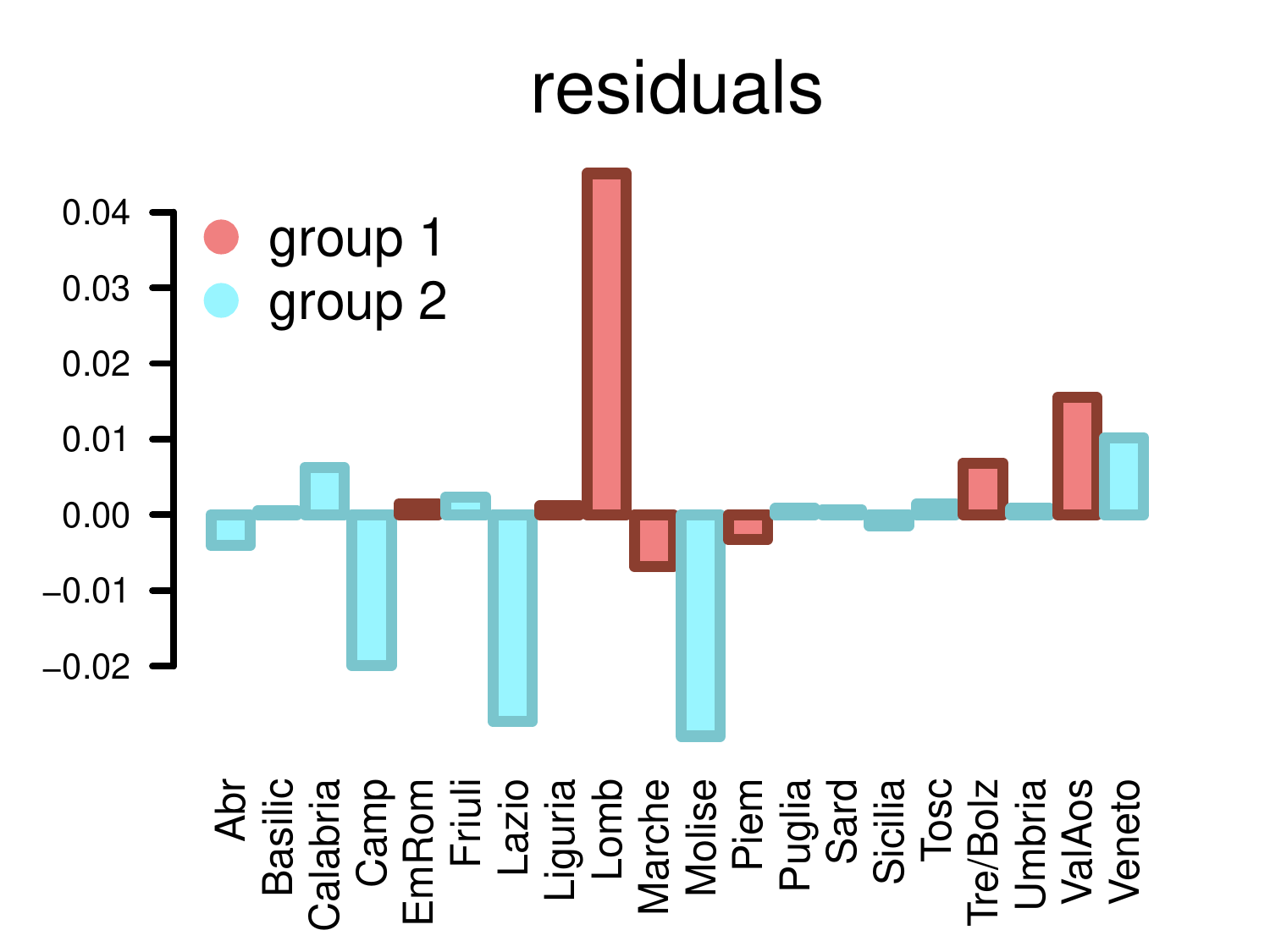}
\hspace{0.3cm}
\vrule width 1pt
\includegraphics[width=0.3\linewidth]{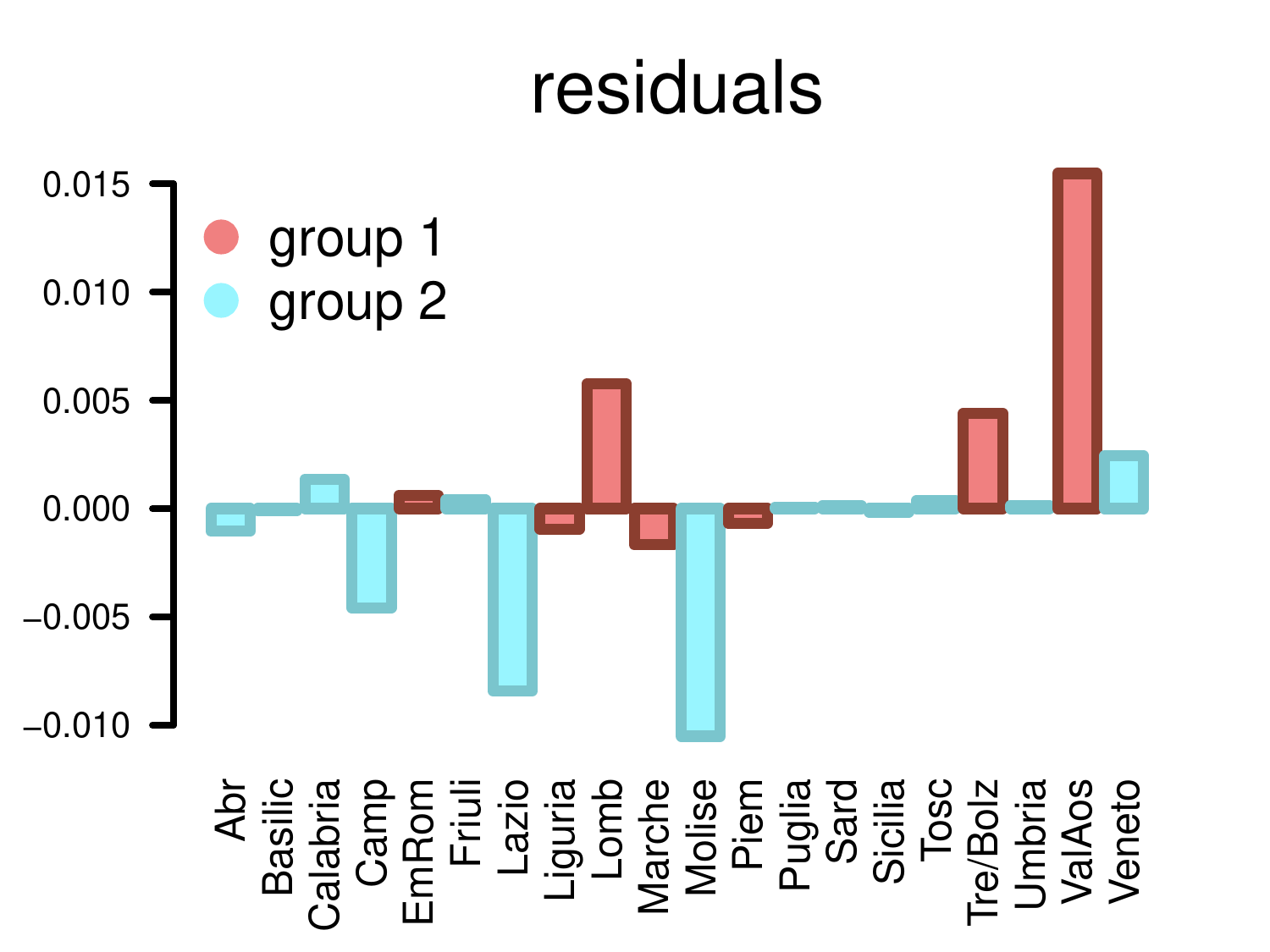}
\hspace{0.2cm}
\vrule width 1pt
\hrule height 1pt
\vspace{0.2cm}   
\caption{
{\bf Associating mortality to positivity.}
Results from the function-on-function regression of mortality on positivity. The top row displays the estimated effect surface (the March 9 date is marked) with respective in-sample $R^2$ (for LOO-CV $R^2$ see Table~\ref{tab:R2}).
The bottom row displays the regression residuals (for barplots interpretation see Fig.~\ref{fig:max_res}).
}
\label{fig:pos_marginal}
\end{figure}

\begin{figure}[!tb]
\begin{center}
{\small \bf mortality $\bm \sim$ mobility  + positivity}
\\
{\small \bf ISTAT \hspace{8cm} DPC}
\end{center}
\centering
\vspace{-0.2cm}
\hrule height 1pt
\vrule width 1pt
\includegraphics[width=0.235\linewidth]{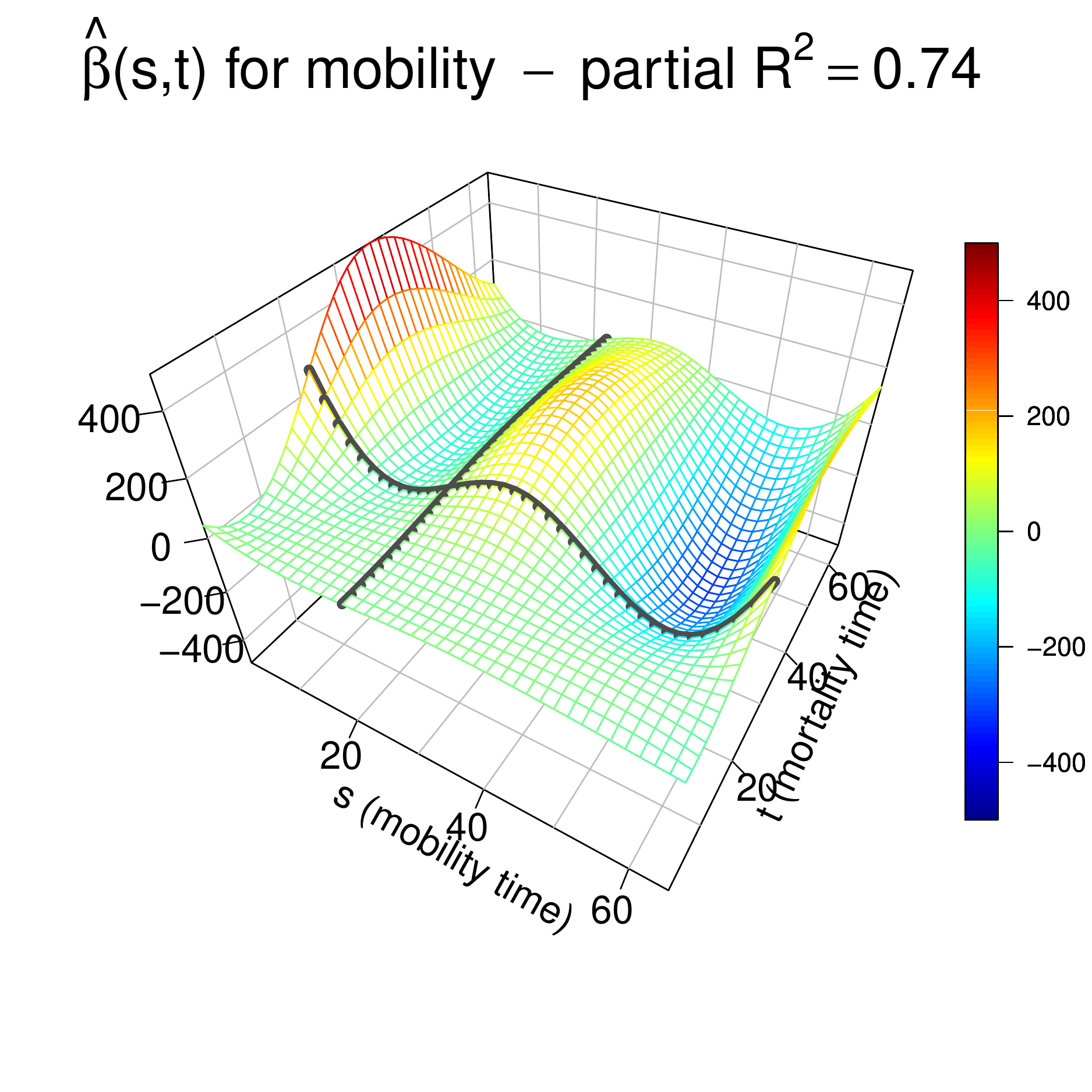}
\includegraphics[width=0.235\linewidth]{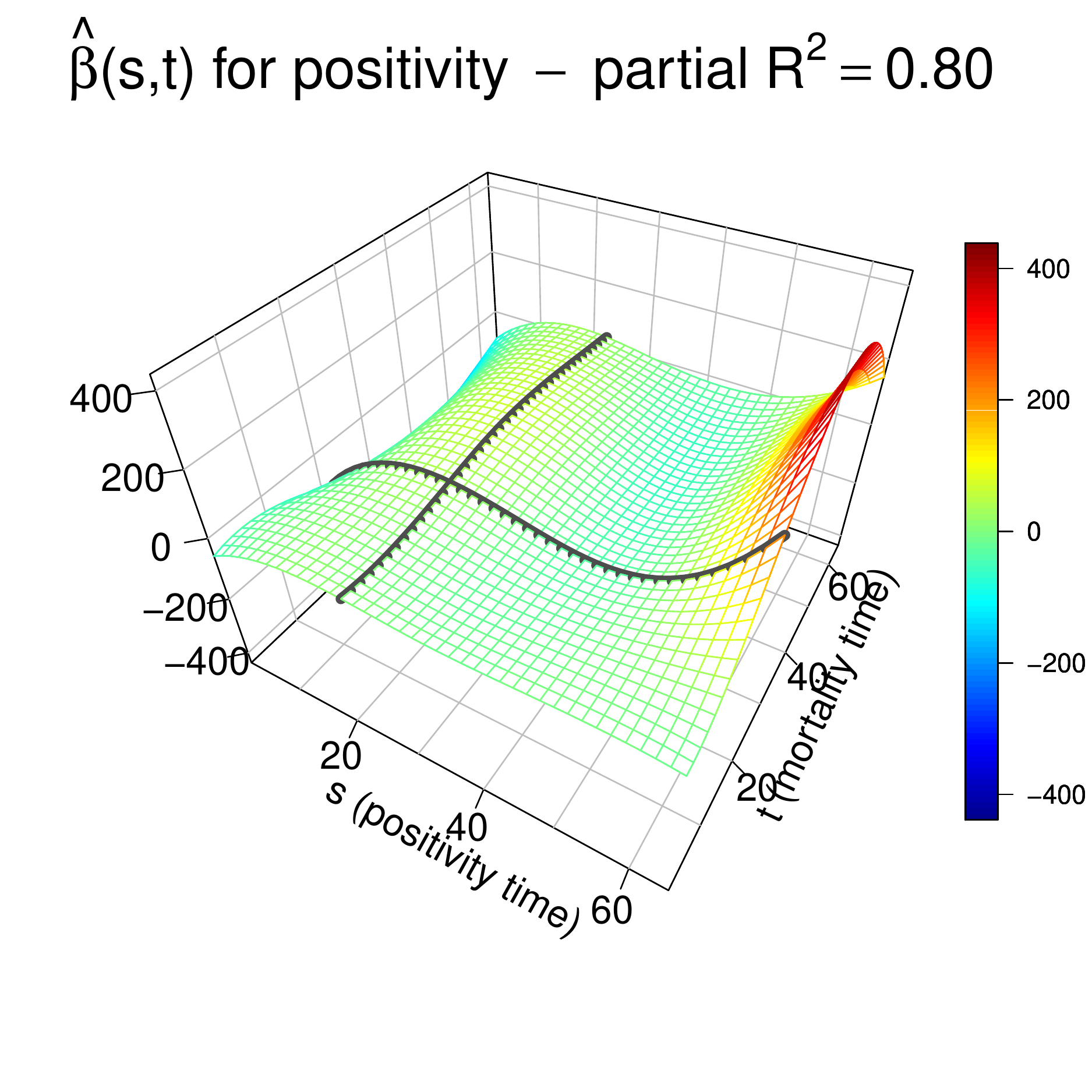}
\hspace{0.2cm}
\vrule width 1pt
\hspace{0.2cm}
\includegraphics[width=0.235\linewidth]{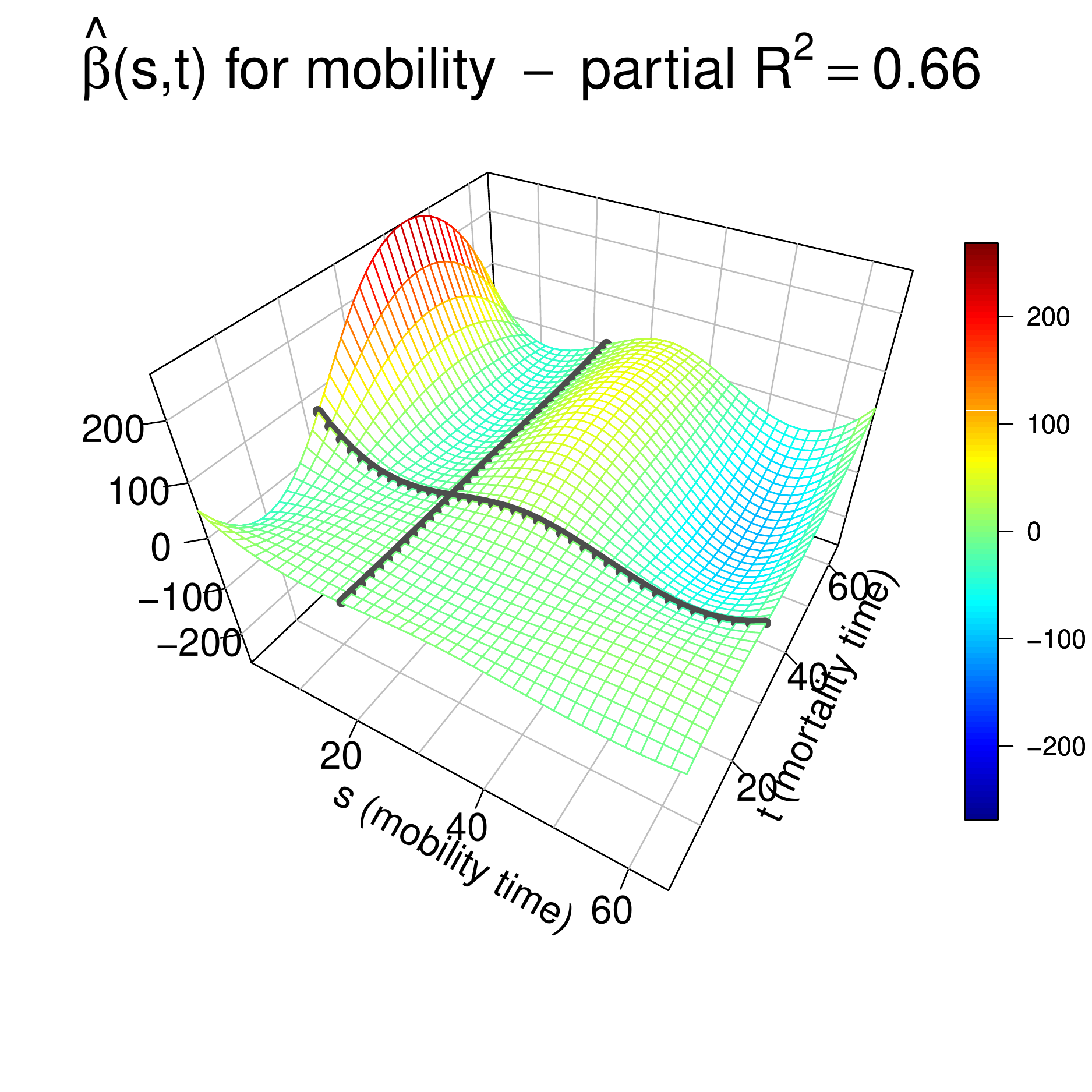}
\includegraphics[width=0.235\linewidth]{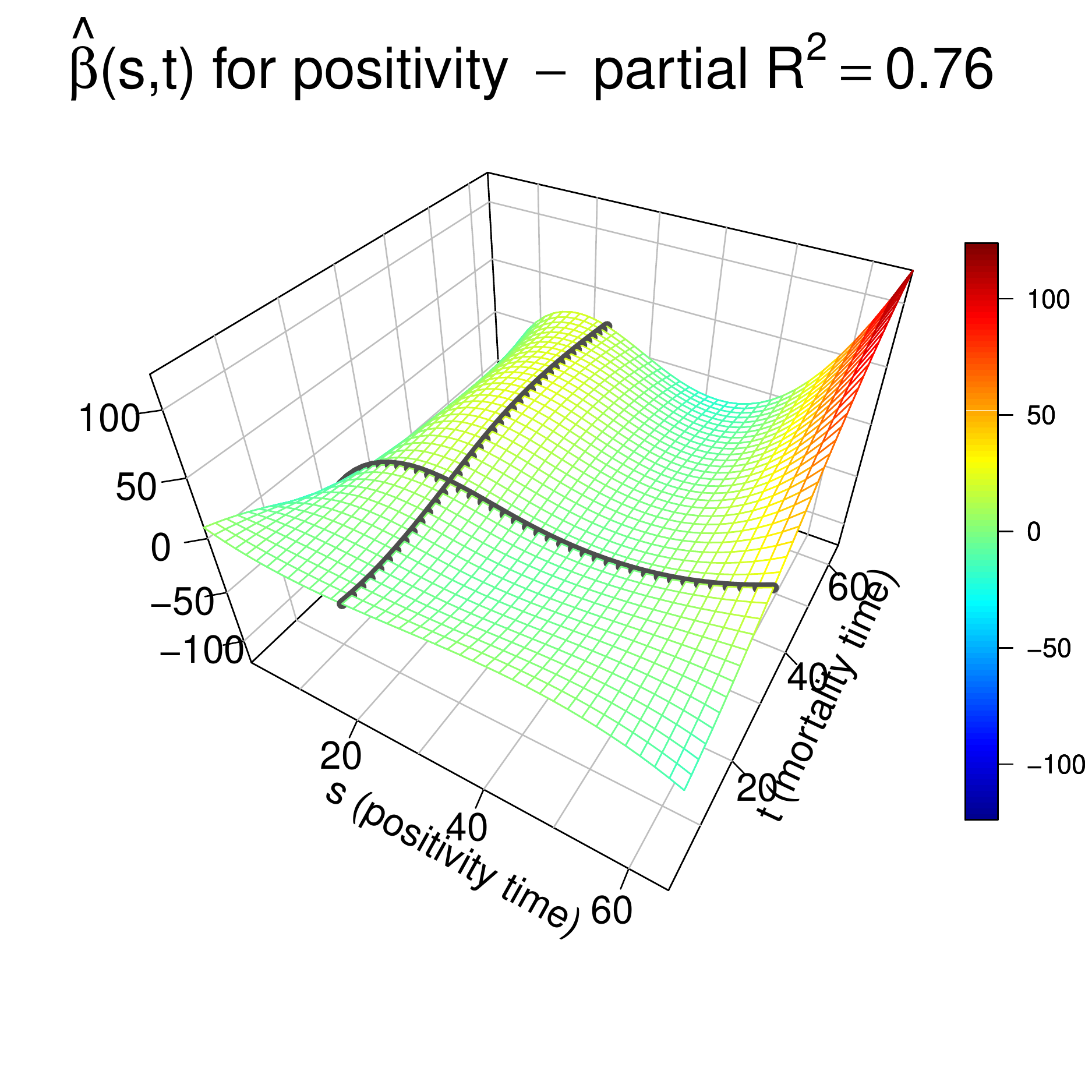}
\vrule width 1pt \\
\vspace{-0.5cm}
\vrule width 1pt
\hspace{1.85cm}
\includegraphics[width=0.27\linewidth]{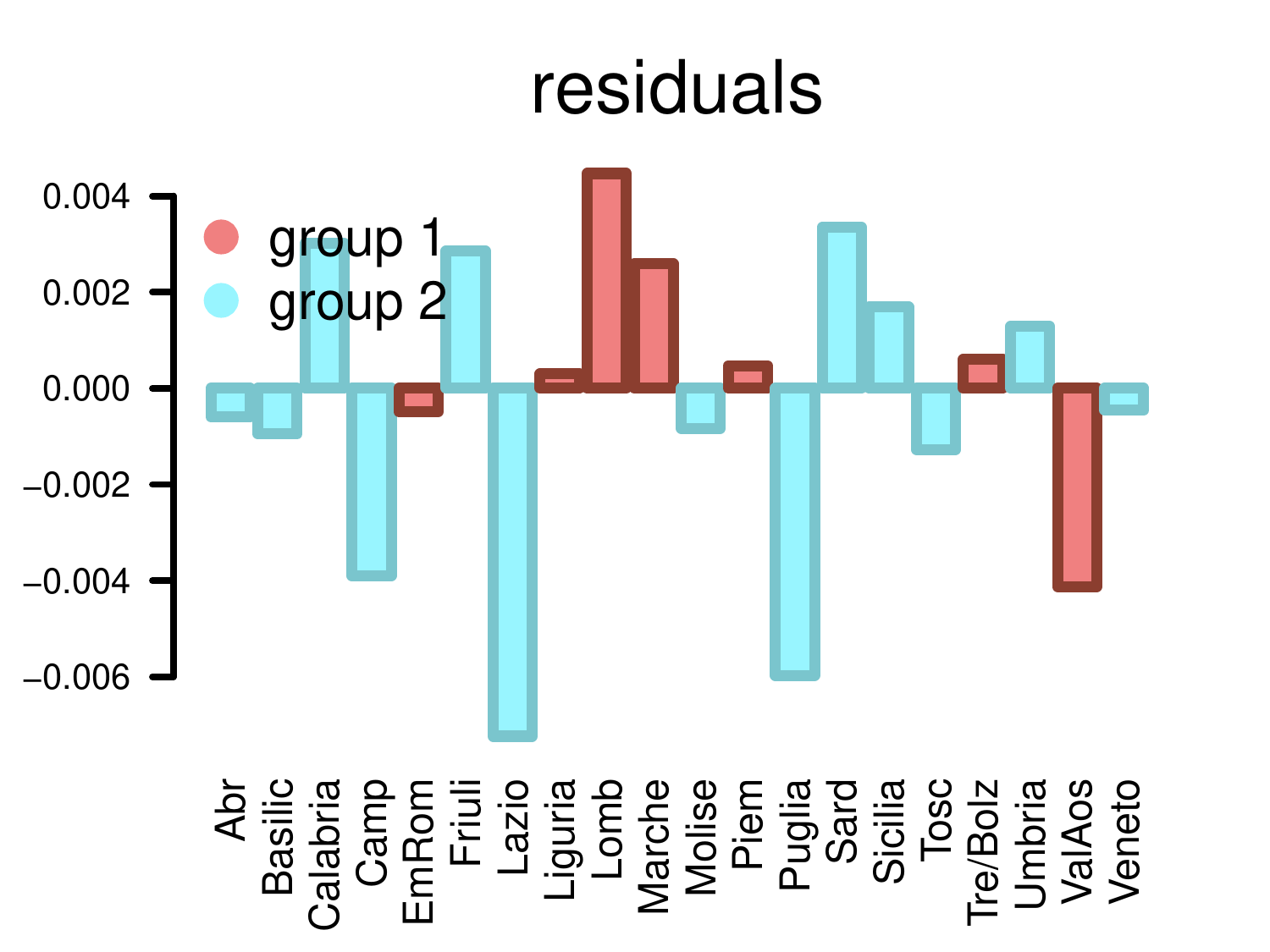}
\hspace{1.87cm}
\vrule width 1pt
\hspace{1.87cm}
\includegraphics[width=0.27\linewidth]{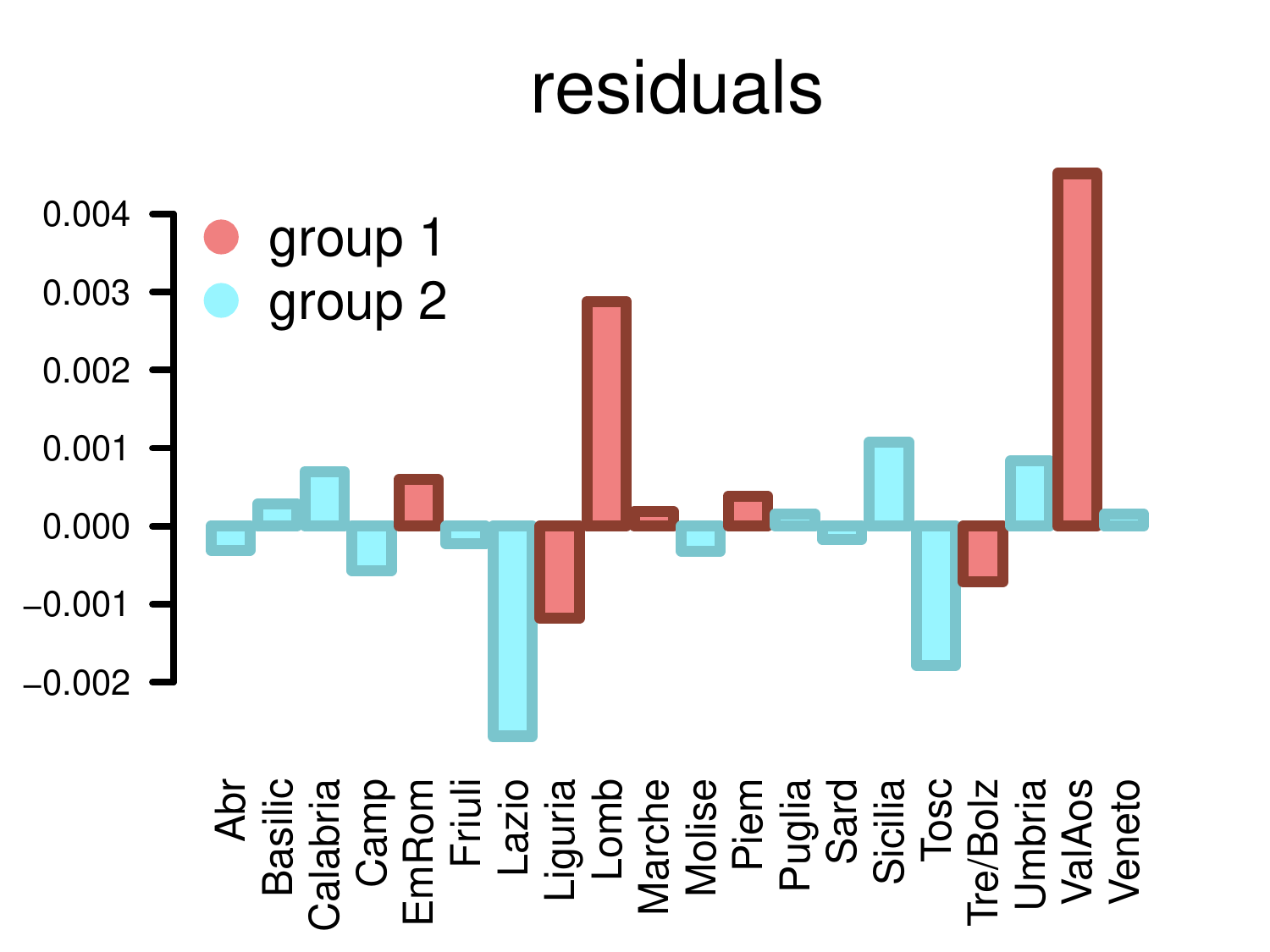}
\hspace{1.85cm}
\vrule width 1pt
\hrule height 1pt
\vspace{0.2cm}  
\caption{ 
{\bf Associating mortality to mobility and positivity - ISTAT and DPC.}
Results from the joint function-on-function regression of ISTAT and DPC mortality on mobility and positivity. The top row shows the estimated effect surfaces (the March 9 date is marked) with respective partial $R^2$ (for in-sample $R^2$ and LOO-CV $R^2$ see Table~\ref{tab:R2}). The bottom row shows the regression residuals (for barplots interpretation see Fig.~\ref{fig:max_res})
}
\label{fig:mob_pos_dpc_istat}
\end{figure}

\begin{figure}[!tb]
\begin{center}
{\small \bf mortality $\bm \sim$ mobility + positivity + reduced PC1}
\\
{\small \bf ISTAT \hspace{8cm} DPC}
\end{center}
\centering
\vspace{-0.2cm}
\hrule height 1pt
\vrule width 1pt
\includegraphics[width=0.235\linewidth]{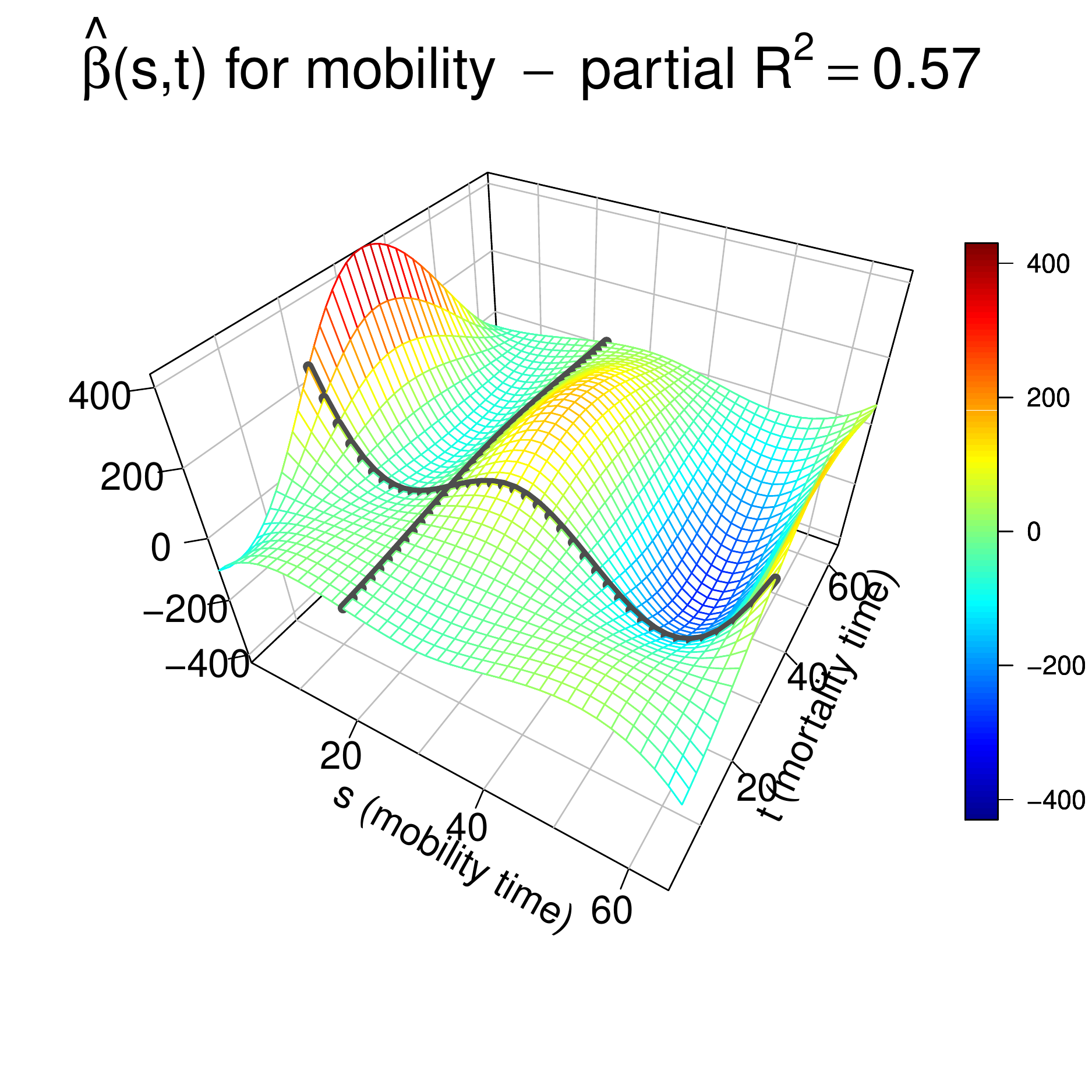}
\includegraphics[width=0.235\linewidth]{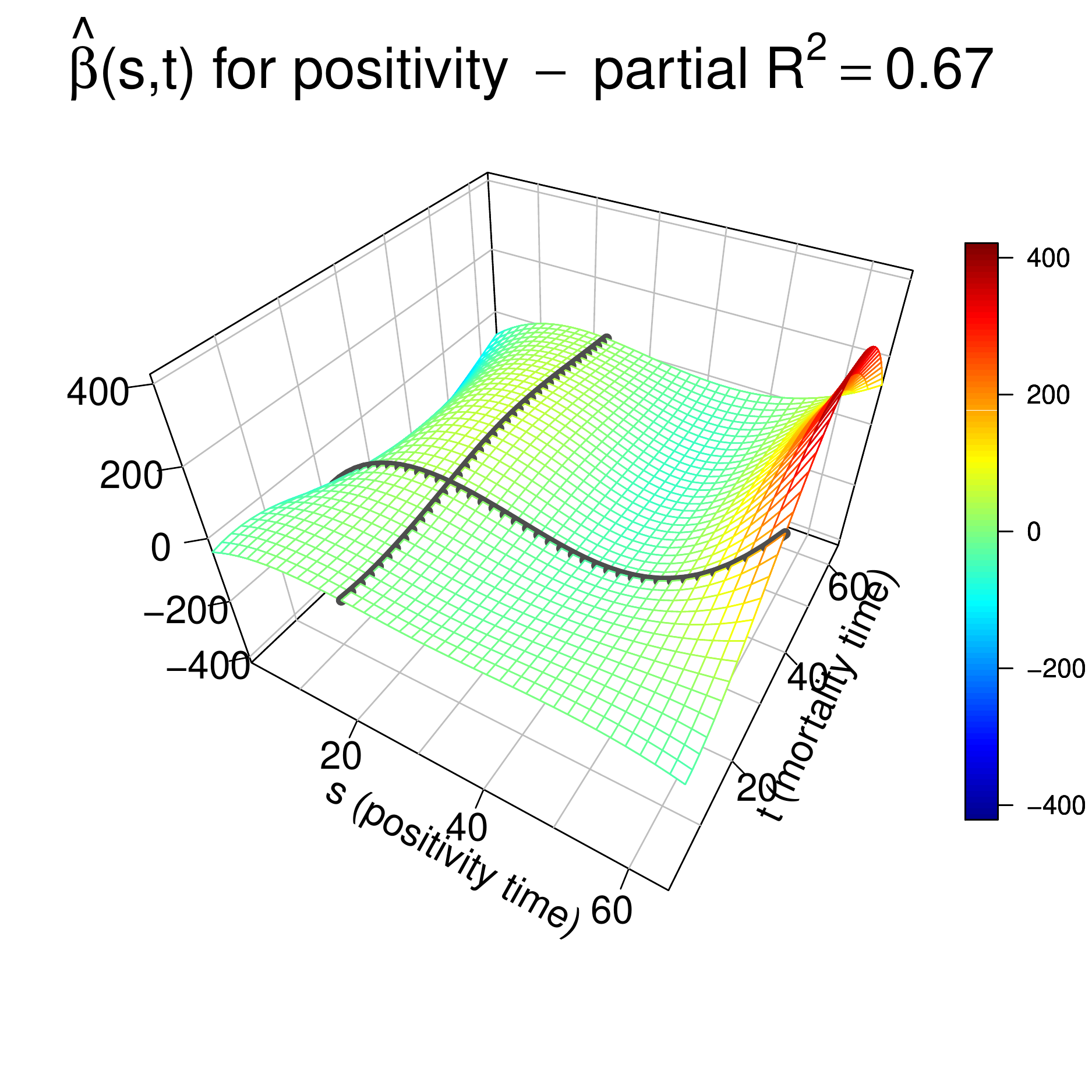}
\hspace{0.2cm}
\vrule width 1pt
\hspace{0.2cm}
\includegraphics[width=0.235\linewidth]{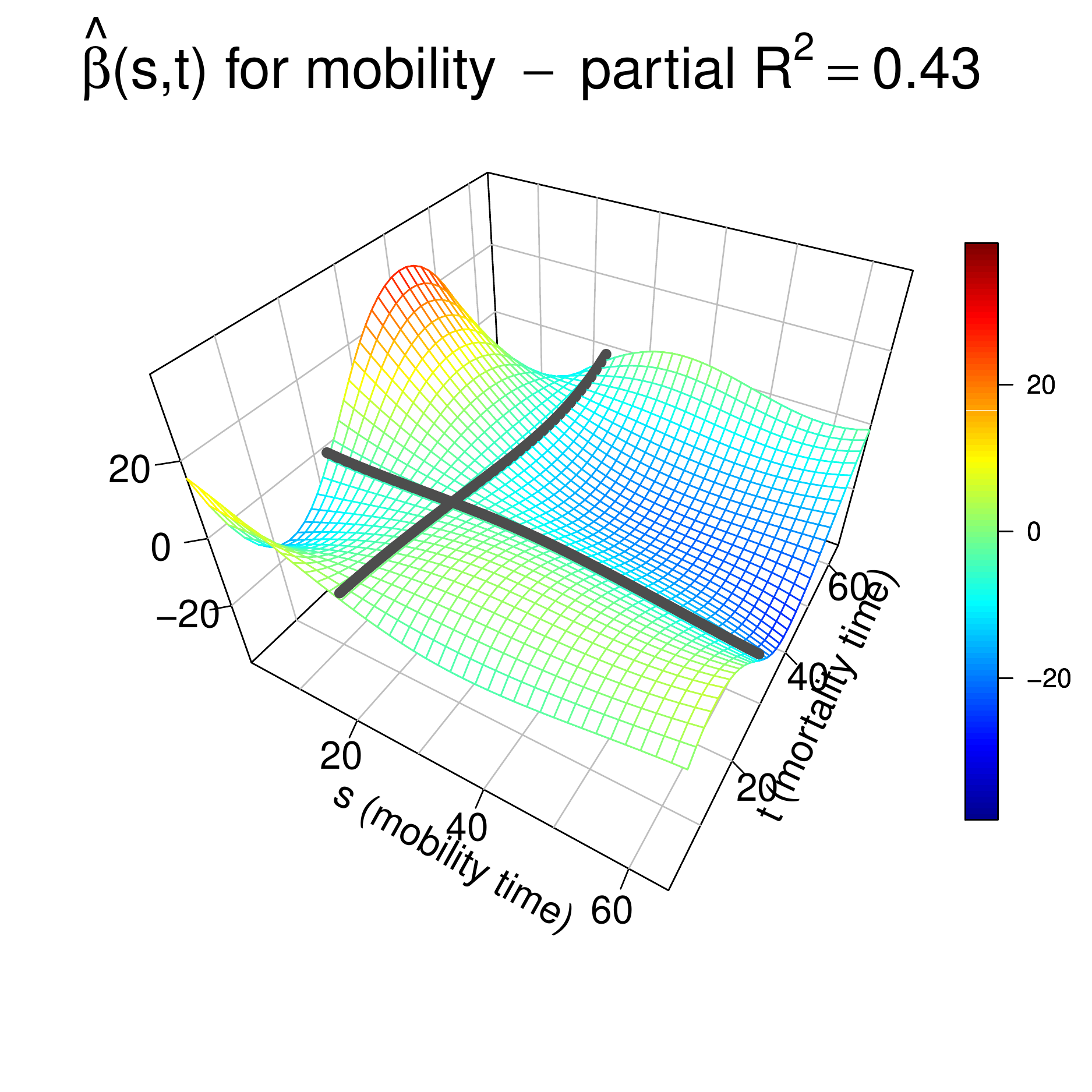}
\includegraphics[width=0.235\linewidth]{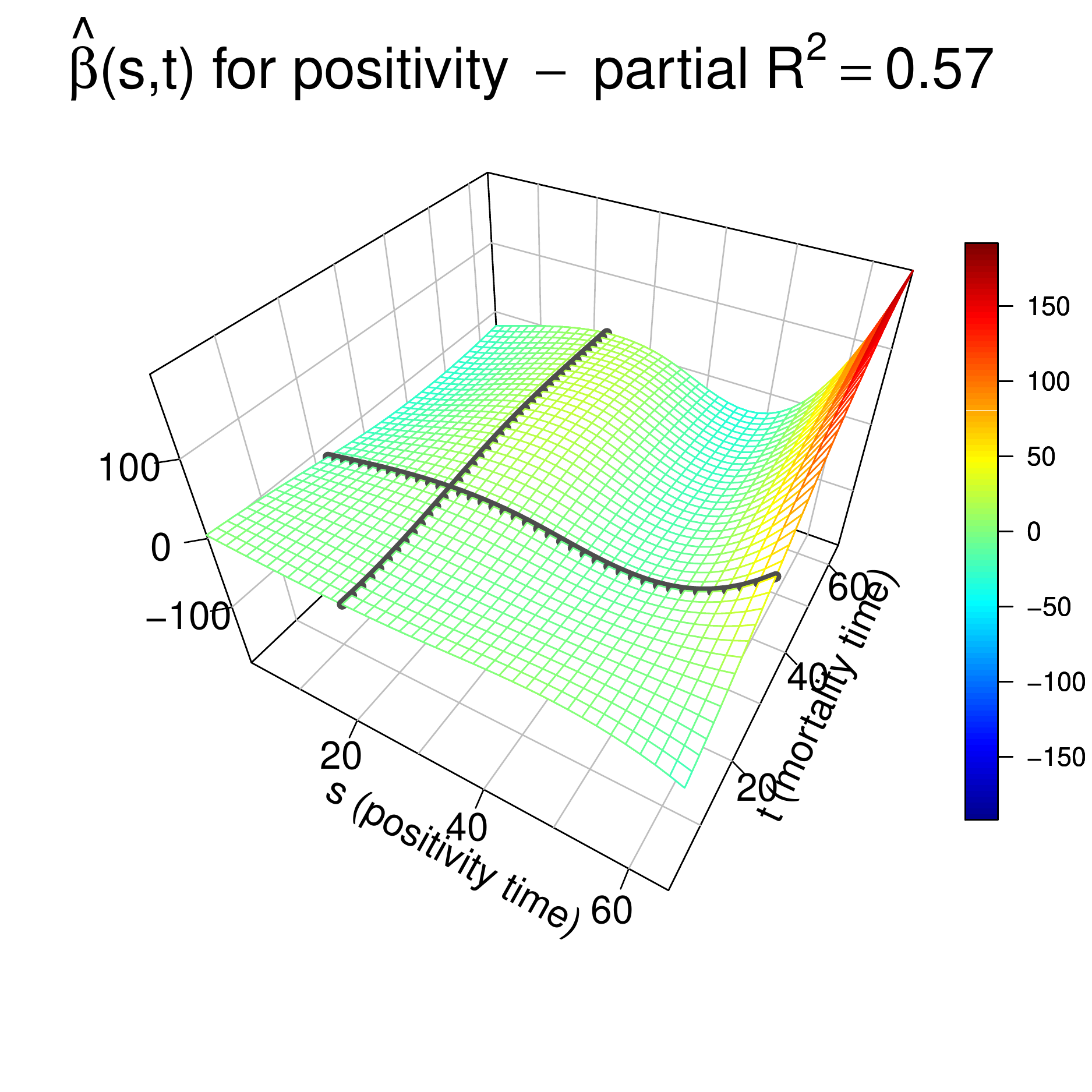}
\vrule width 1pt \\
\vspace{-0.4cm}
\vrule width 1pt
\includegraphics[width=0.235\linewidth]{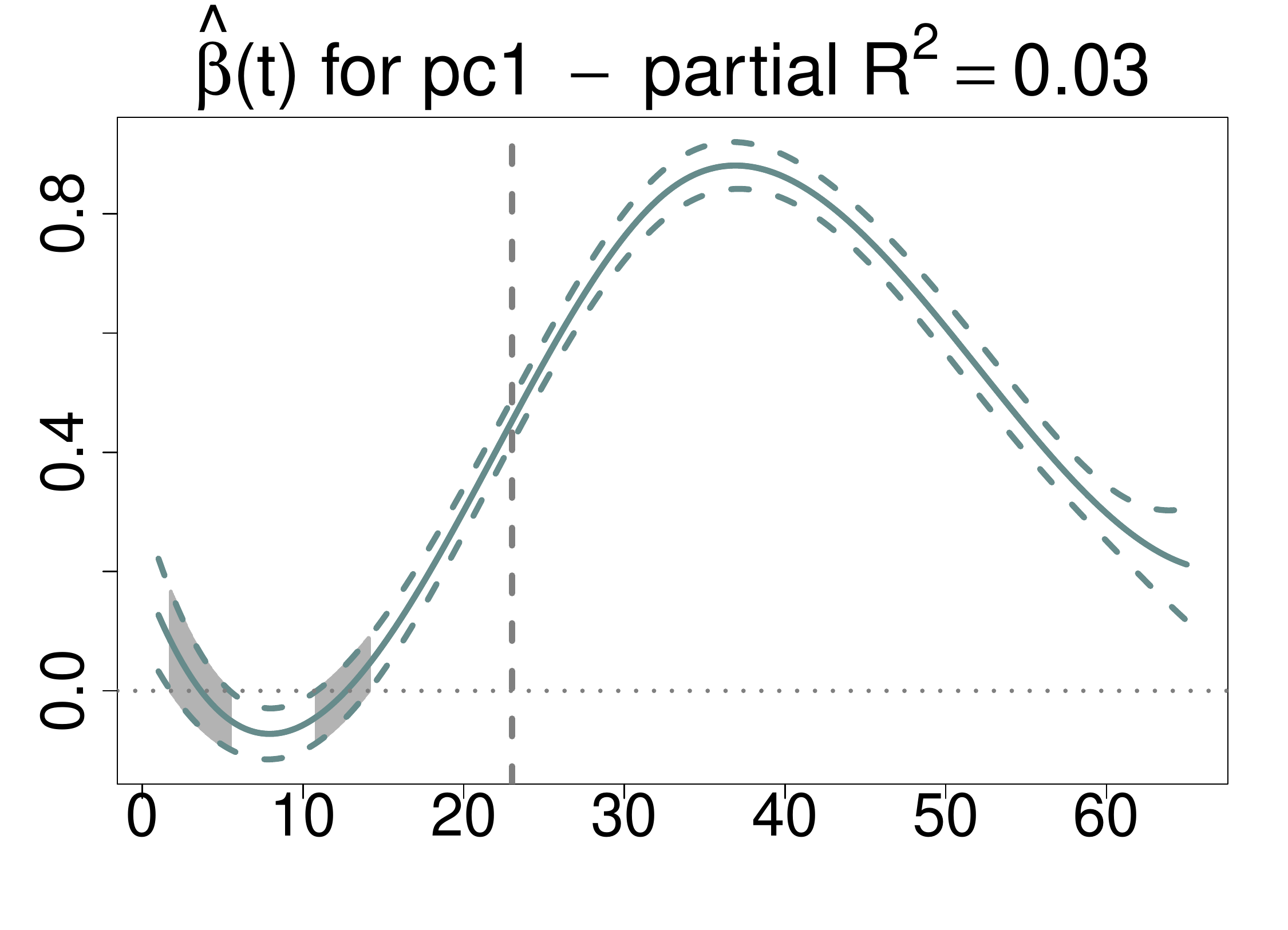}
\includegraphics[width=0.235\linewidth]{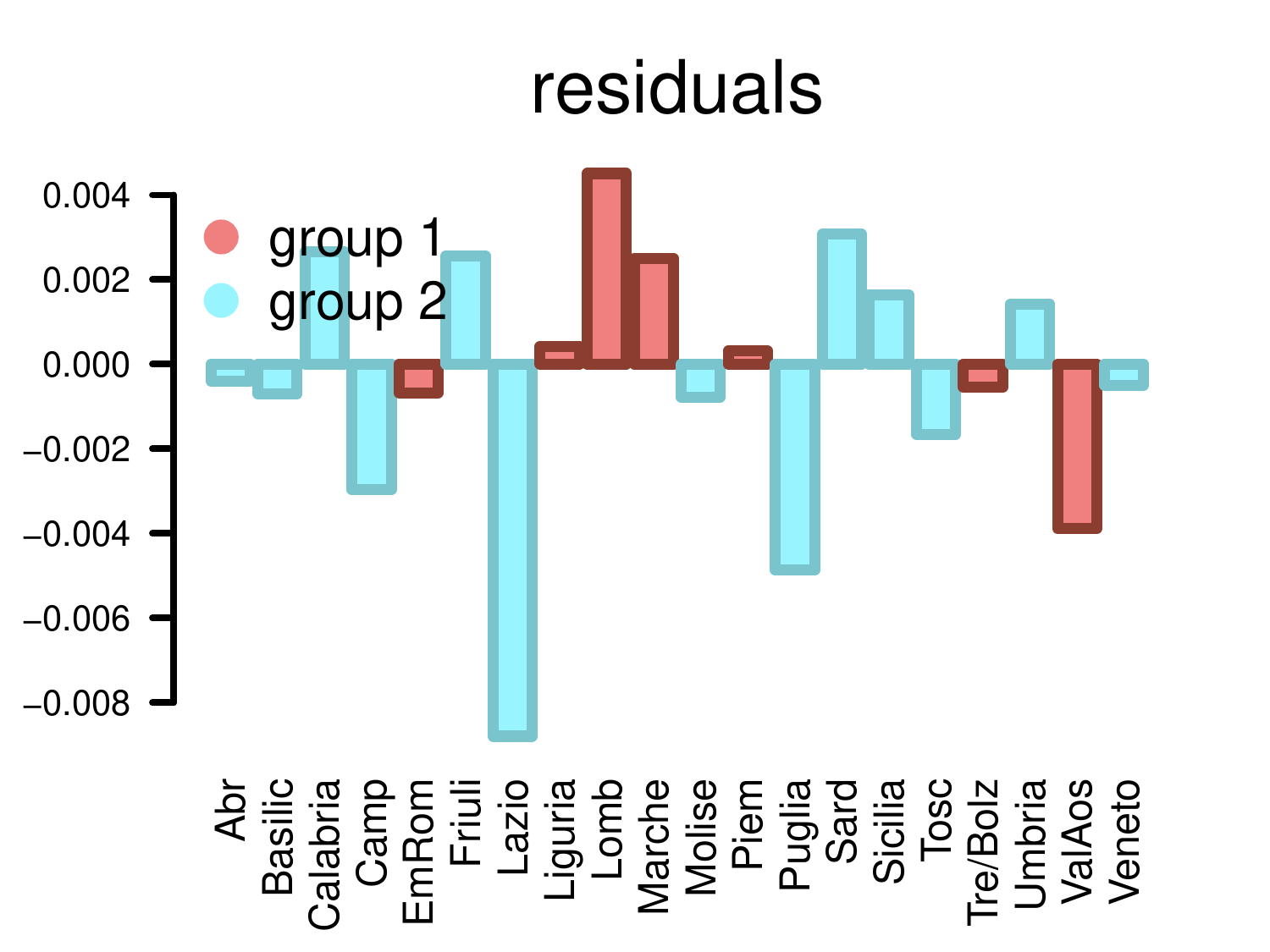}
\hspace{0.2cm}
\vrule width 1pt
\hspace{0.2cm}
\includegraphics[width=0.235\linewidth]{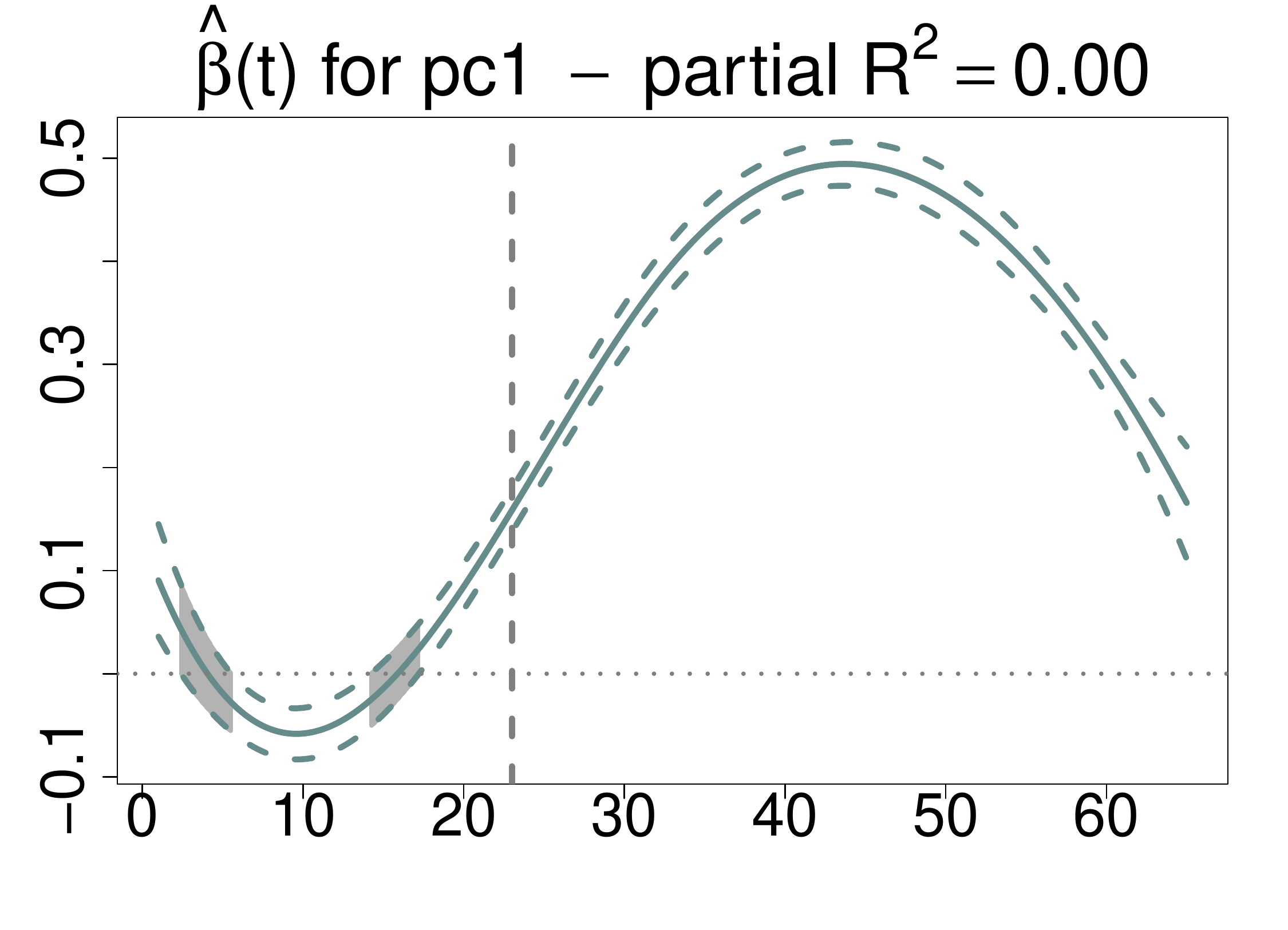}
\includegraphics[width=0.235\linewidth]{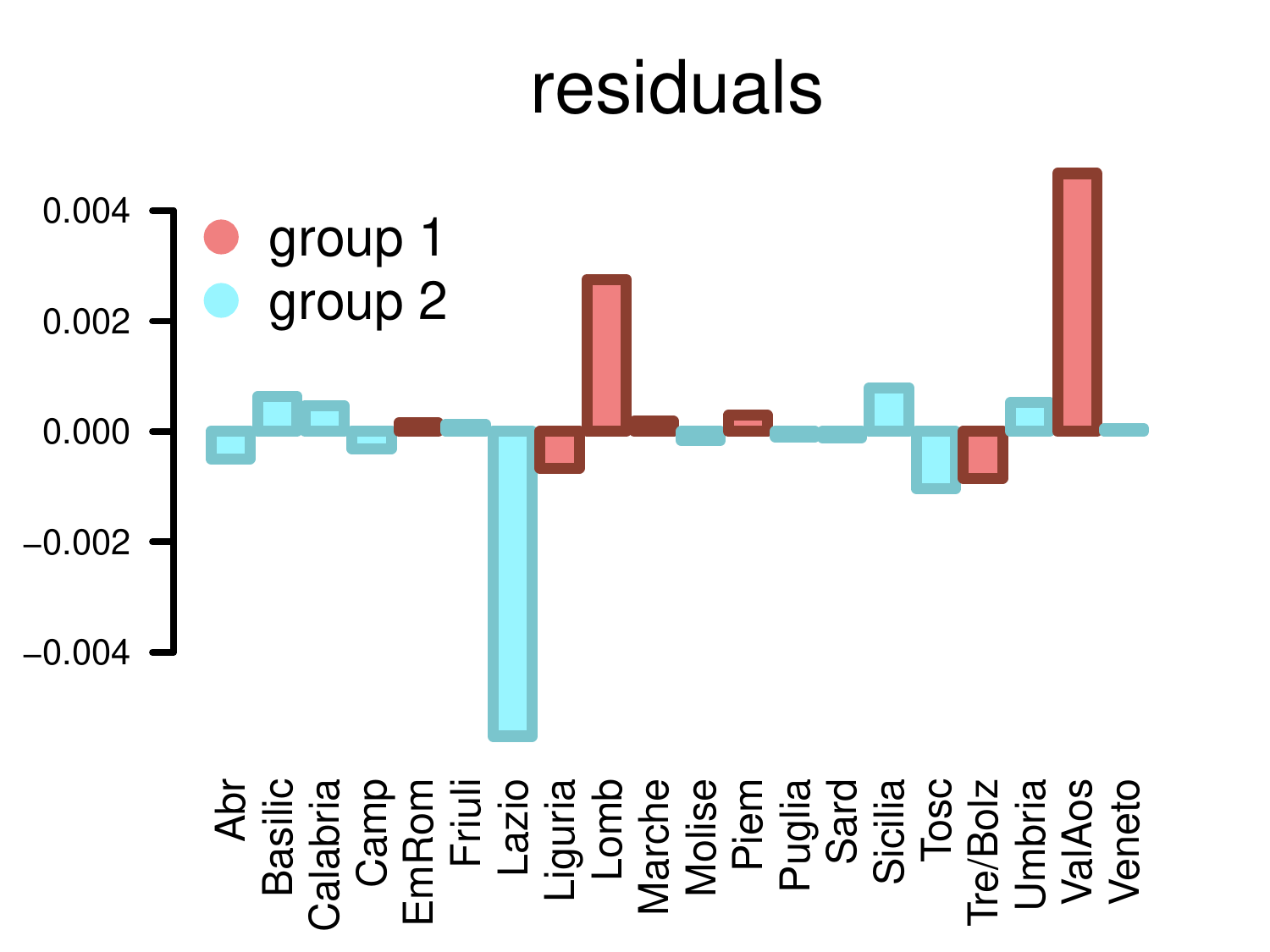}
\vrule width 1pt
\hrule height 1pt
\vspace{0.2cm}  
\caption{ 
{\bf Associating mortality to mobility, positivity and first principal component - ISTAT and DPC.}
Results from the joint function-on-function regression of ISTAT and DPC mortality on mobility, positivity, and the first principal component (pc1) of the top 5 covariates, used as a "summary" control. Each panels shows: the estimated effect surfaces for mobility and positivity and the estimated effect curve for pc1 with respective partial $R^2$ (for in-sample $R^2$ and LOO-CV $R^2$ see Table~\ref{tab:R2}). For interpretating the regression residuals see Fig.~\ref{fig:max_res}.
}
\label{fig:mob_pos_pc1_dpc_istat}
\end{figure}
    
\begin{table}[H]
\centering
\caption{ {\bf Function-on-scalar feature selection.} Top five scalar covariates selected by SsNAL-EN considering as response the MAX, ISTAT, and DPC mortality curves.}
\label{tab:fgen_selection}
\vspace{-0.2cm}
\small
\begin{tabular}{r|r|r|r}
\Xhline{3\arrayrulewidth}
\multicolumn{1}{l|}{} & \multicolumn{1}{c|}{\textbf{MAX}} & \multicolumn{1}{c|}{\textbf{ISTAT}} & \multicolumn{1}{c}{\textbf{DPC}} \\ 
\Xhline{3\arrayrulewidth}
$\bm 1$ & Adults per family doctor & Adults per family doctor & Adults per family doctor \\
$\bm 2$ & Ave.~beds per hospital (whole) & Ave.~students per classroom & Ave.~beds per hospital (whole) \\
$\bm 3$ & Ave.~students per classroom & Ave.~beds per hospital (whole) & Ave.~students per classroom \\
$\bm 4$ & Ave.~members per household & Ave.~employees per firm & Ave.~employees per firm \\
$\bm 5$ & Ave.~employees per firm & Ave.~members per household & Ave.~members per household \\ 
\Xhline{3\arrayrulewidth}
\end{tabular}
\end{table}

\begin{figure}[!tb]
\begin{center}
{\small \bf marginal $\bm \beta (t)$ signs: 1 intercept model}
\end{center}
\centering
\vspace{-0.2cm}
\hrule height 1pt
\vrule width 1pt
\includegraphics[width=0.32\linewidth]{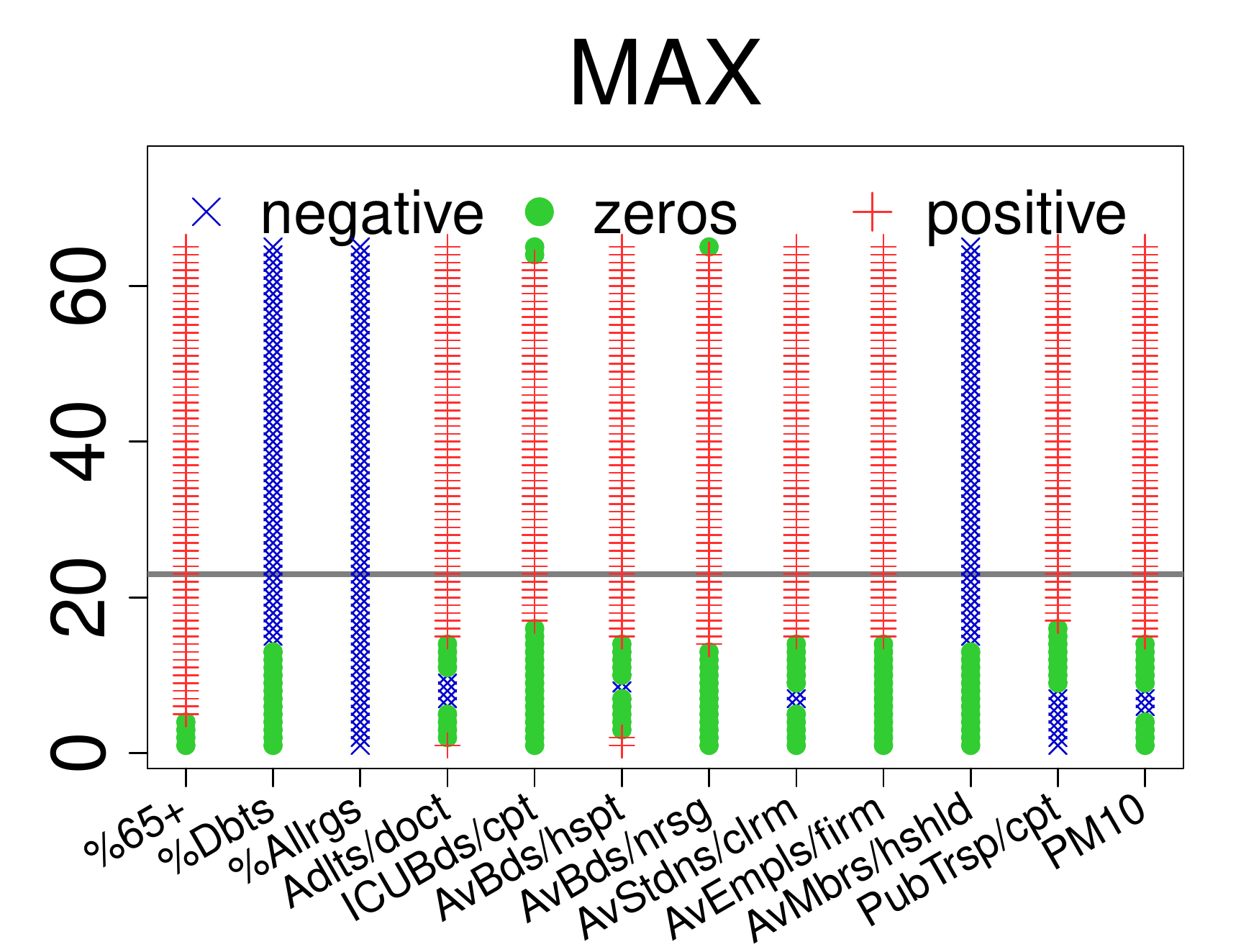}
\hspace{0.1cm}
\includegraphics[width=0.32\linewidth]{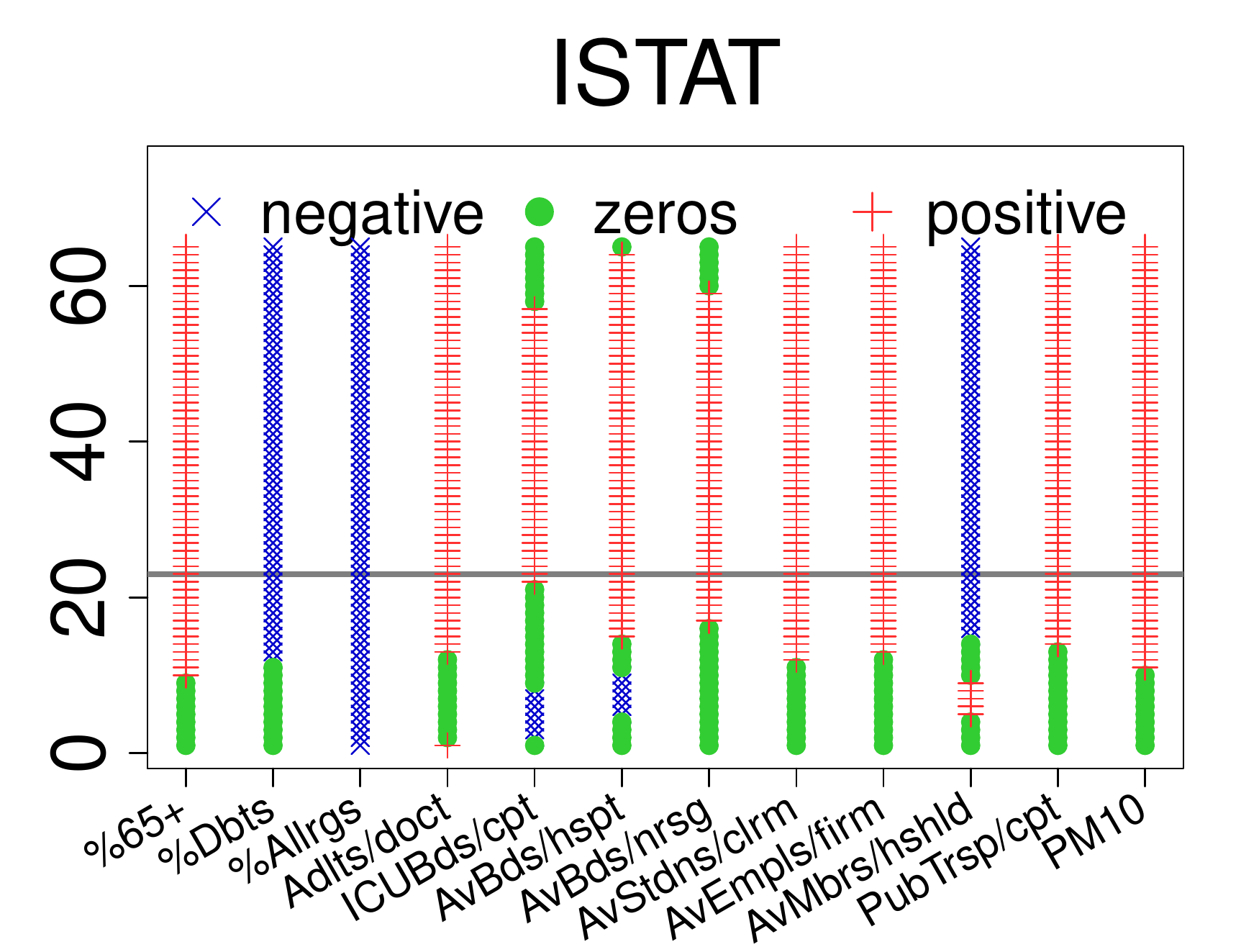}
\hspace{0.1cm}
\includegraphics[width=0.32\linewidth]{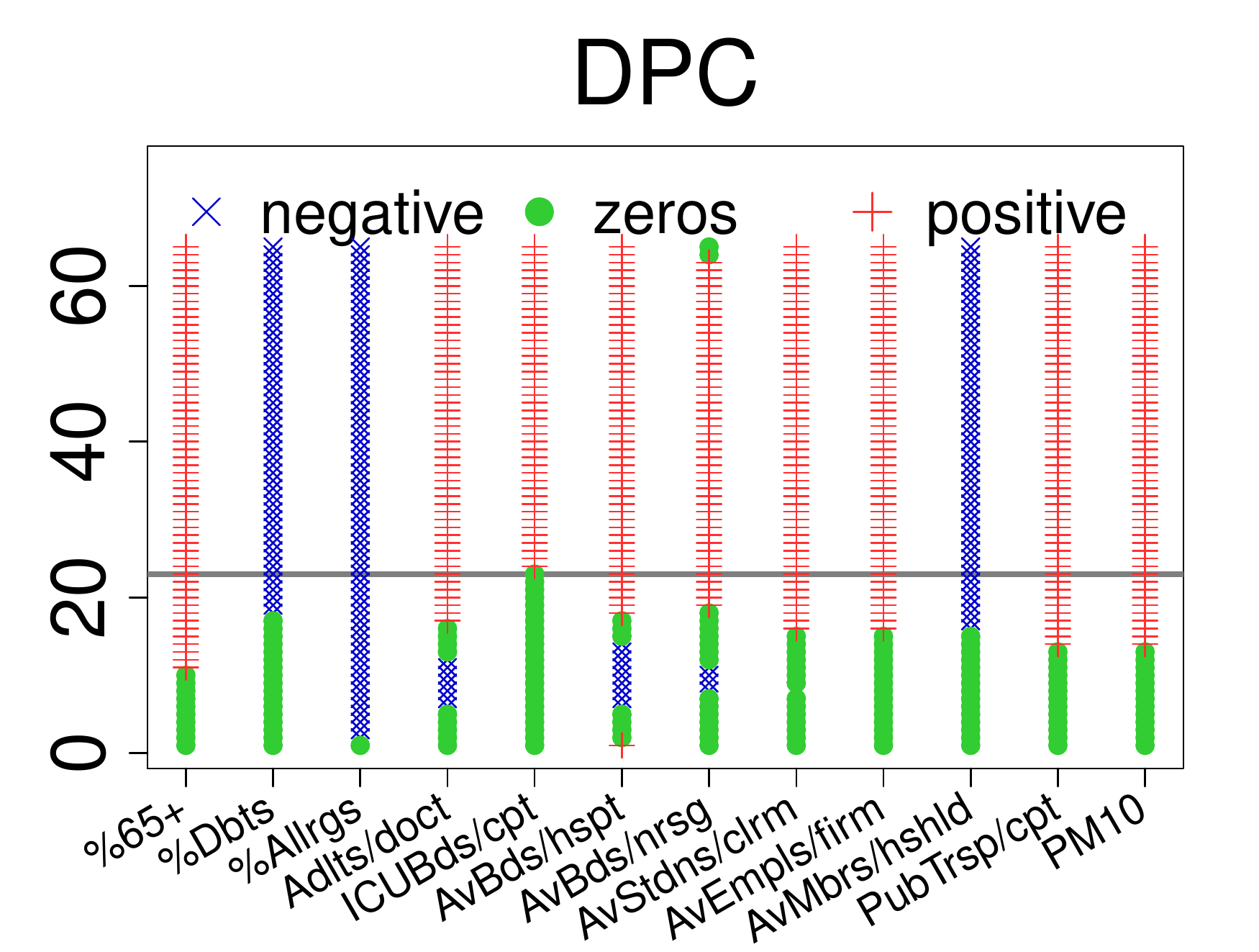}
\vrule width 1pt
\hrule height 1pt
\vspace{0.2cm}  

\begin{center}
{\small \bf marginal $\bm \beta(t)$ signs: 2 intercepts model}
\end{center}
\centering
\vspace{-0.2cm}
\hrule height 1pt
\vrule width 1pt
\includegraphics[width=0.32\linewidth]{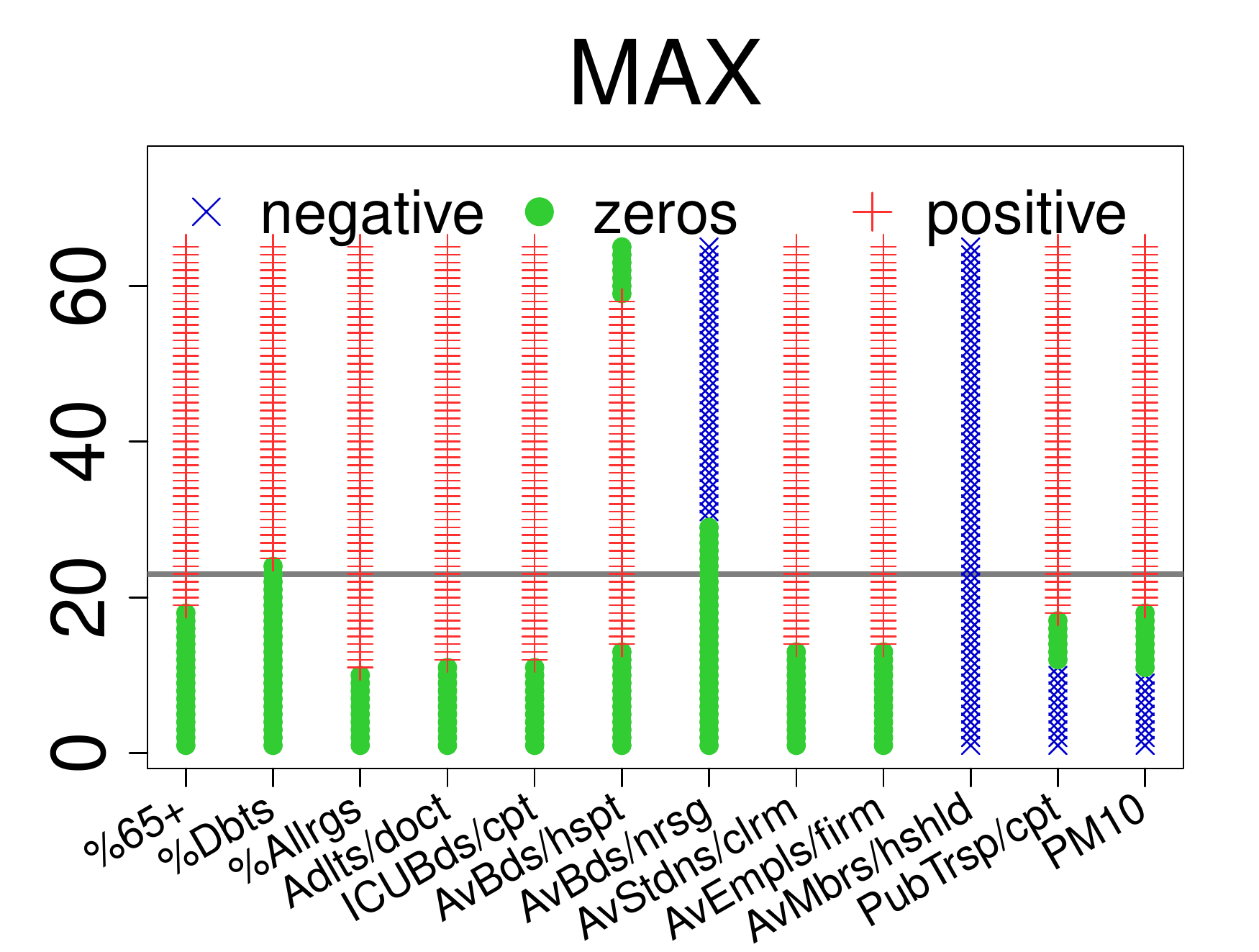}
\hspace{0.1cm}
\includegraphics[width=0.32\linewidth]{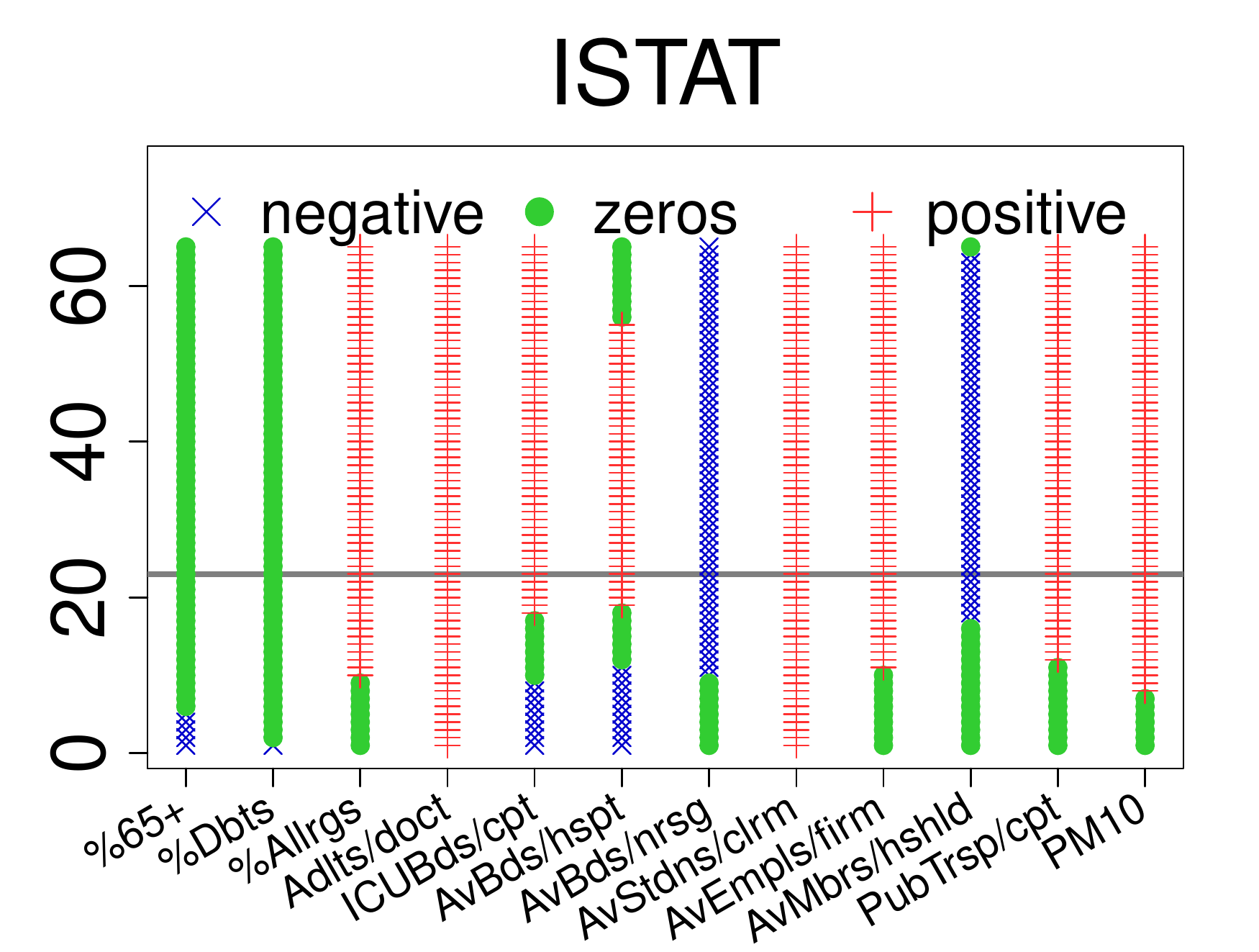}
\hspace{0.1cm}
\includegraphics[width=0.32\linewidth]{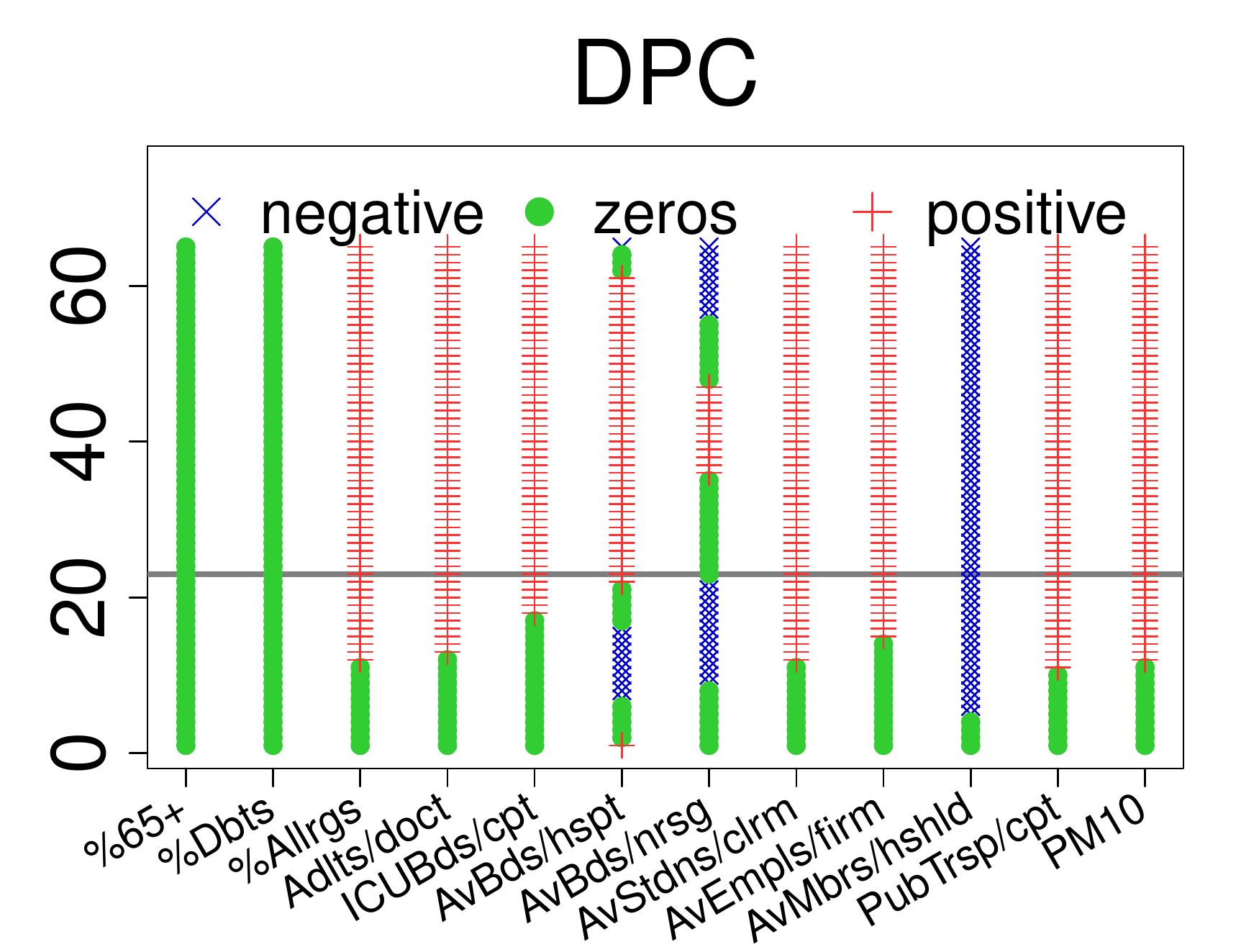}
\vrule width 1pt
\hrule height 1pt
\vspace{0.2cm}  
\caption{ {\bf Marginal function-on-scalar regressions.} Results for marginal function-on-scalar regressions. Mortality curves are regressed against each of the scalar covariates in Table 1. The top-row displays the signs of the effect curves estimated when just one intercept is included in the model. The bottom-row displays the signs of the effect curves estimated when we consider two different intercepts for curves in Group 1 and curves in Group 2. Time is on the vertical axis (the national lock down on March 9, without shift, is marked by a horizontal line). Red, blue and green indicate, respectively, positive, negative and non-significant portions (i.e., where 95\% confidence bands around the estimated effect curve are entirely above, entirely below, or contain 0).
}
\label{fig:beta_signs_1_2_inter}
\end{figure}

\begin{table}[H]
\caption{ {\bf Functional regression models 
(in-sample) $\bm{R^2}$, LOO-CV $\bm{R^2}$ and partial $\bm{R^2}$s.} For each functional linear model which regresses mortality on the covariates listed in the first column, the table reports the (in-sample) $R^2$, the LOO-CV $R^2$ and the partial $R^2$s. 
}
\label{tab:R2}
\hspace{-0.3cm}
\begin{tabular}{r|rrc|rrc|rrc}
\Xhline{3\arrayrulewidth}
\multicolumn{1}{l|}{} & \multicolumn{3}{c|}{\textbf{MAX}} & \multicolumn{3}{c|}{\textbf{ISTAT}} & \multicolumn{3}{c}{\textbf{DPC}} \\ 
\Xhline{3\arrayrulewidth}
\multicolumn{1}{c|}{\textbf{covariates}} & $R^2$ & \multicolumn{1}{l}{LOO-CV $R^2$} & \multicolumn{1}{l|}{partial $R^2$s} & $R^2$ & \multicolumn{1}{l}{LOO-CV $R^2$} & \multicolumn{1}{l|}{partial $R^2$s} & $R^2$ & \multicolumn{1}{l}{LOO-CV $R^2$} & \multicolumn{1}{l}{partial $R^2$s} \\ 
\Xhline{3\arrayrulewidth}
mob & 0.79 & 0.54 & \begin{tabular}[c]{@{}c@{}}$\phantom i$\\ -\\ $\phantom i$\end{tabular} & 0.63 & 0.47 & \begin{tabular}[c]{@{}c@{}}$\phantom i$\\ -\\ $\phantom i$\end{tabular} & 0.62 & 0.33 & \begin{tabular}[c]{@{}c@{}}$\phantom i$\\ -\\ $\phantom i$\end{tabular} \\ \hline
pos & 0.75 & 0.47 & \begin{tabular}[c]{@{}c@{}}$\phantom i$\\ -\\ $\phantom i$\end{tabular} & 0.71 & 0.44 & \begin{tabular}[c]{@{}c@{}}$\phantom i$\\ -\\ $\phantom i$\end{tabular} & 0.73 & 0.47 & \begin{tabular}[c]{@{}c@{}}$\phantom i$\\ -\\ $\phantom i$\end{tabular} \\ \hline
mob + pos & 0.90 & 0.52 & \multicolumn{1}{r|}{\begin{tabular}[c]{@{}r@{}}mob: 0.62\\ pos: 0.53\\ $\phantom{iiii}$ - $\phantom{iiiiii}$\end{tabular}} & 0.93 & 0.64 & \multicolumn{1}{r|}{\begin{tabular}[c]{@{}r@{}}mob: 0.74\\ pos: 0.80\\ $\phantom{iiii}$ - $\phantom{iiiiii}$\end{tabular}} & 0.90 & 0.69 & \multicolumn{1}{r}{\begin{tabular}[c]{@{}r@{}}mob: 0.66\\ pos: 0.76\\ $\phantom{iiii}$ - $\phantom{iiiiii}$\end{tabular}} \\ \hline
mob + pos + pc1 & 0.94 & 0.70 & \multicolumn{1}{r|}{\begin{tabular}[c]{@{}r@{}}mob: 0.66\\ pos: 0.61\\ pc1: 0.39\end{tabular}} & 0.93 & 0.62 & \multicolumn{1}{r|}{\begin{tabular}[c]{@{}r@{}}mob: 0.57\\ pos: 0.67\\ pc1: 0.03\end{tabular}} & 0.94 & 0.68 & \multicolumn{1}{r}{\begin{tabular}[c]{@{}r@{}}mob: 0.43\\ pos: 0.57\\ pc1: 0.00\end{tabular}} \\ 
\Xhline{3\arrayrulewidth}
\end{tabular}
\end{table}

 \begin{figure}[!tb]
    
    \hspace{-1.7cm}
    \begin{minipage}[]{1.15\linewidth}
        {\small \bf \hspace{3.3cm} MAX \hspace{5.8cm} ISTAT \hspace{5.8cm} DPC}
        \vspace*{0.1cm}
        \\
        \vspace*{0.2cm}
        {\small 
        \hspace*{1cm}
        $\bm{\hat \beta(t,s)}$ {\bf mobility} \hspace{1cm} $\bm{\hat \beta(t,s)}$ {\bf positivity}
        \hspace{1.1cm}
        $\bm{\hat \beta(t,s)}$ {\bf mobility} \hspace{1cm} $\bm{\hat \beta(t,s)}$ {\bf positivity}
        \hspace{1.1cm}
        $\bm{\hat \beta(t,s)}$ {\bf mobility} \hspace{1cm} $\bm{\hat \beta(t,s)}$ {\bf positivity}
        }
    \end{minipage}
    \hspace*{-1.7cm}
    \vspace*{0.1cm}
    \begin{minipage}[]{0.1\linewidth}
        {\bf 1}
    \end{minipage}
    \hspace{-1.5cm}
    \fbox{
    \begin{minipage}[]{1.14\linewidth}
        \centering
        \vspace{-0.1cm}
        \hspace{-0.2cm}
        \includegraphics[width=0.163\linewidth]{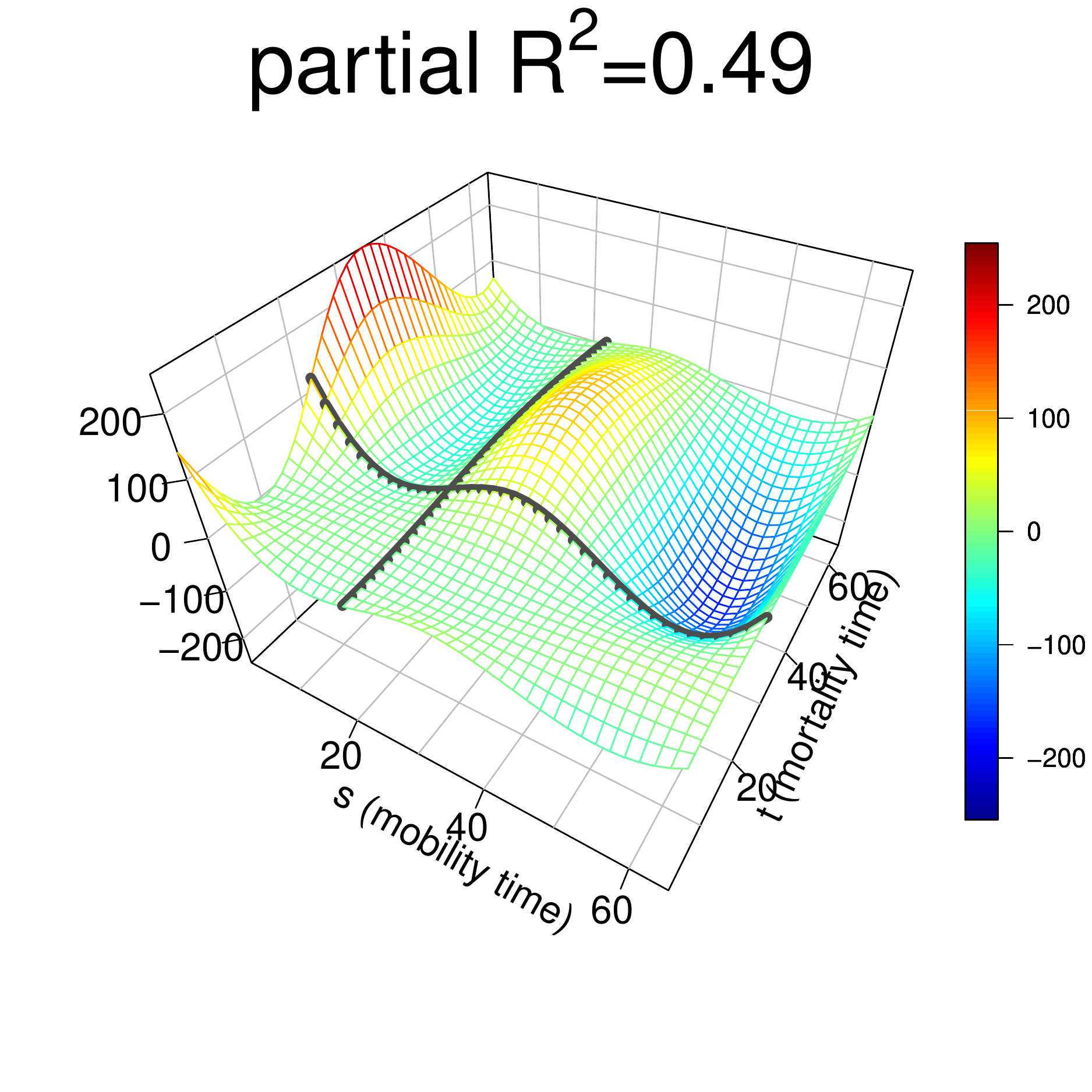}
        \includegraphics[width=0.163\linewidth]{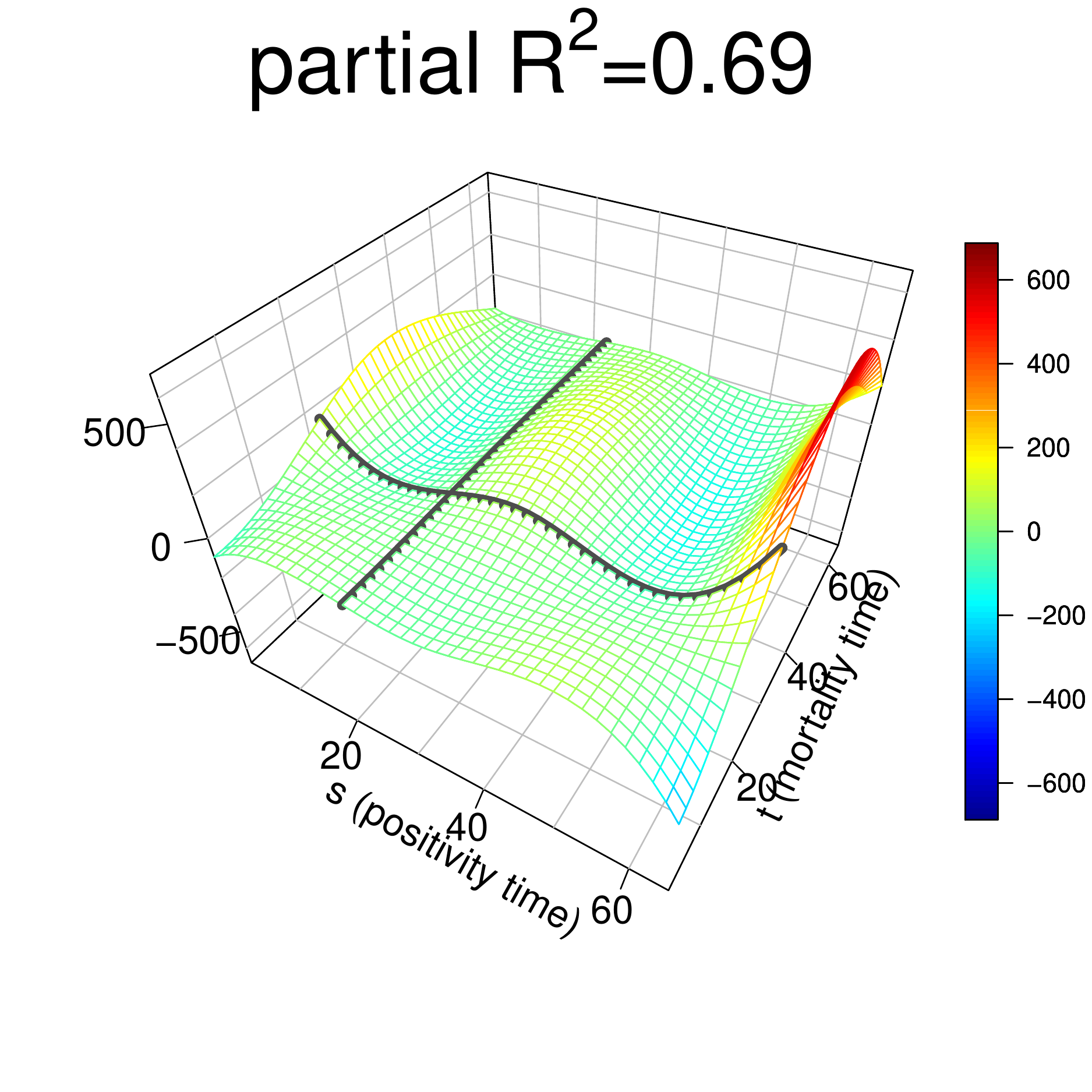}
        \vrule width 1pt
         \includegraphics[width=0.163\linewidth]{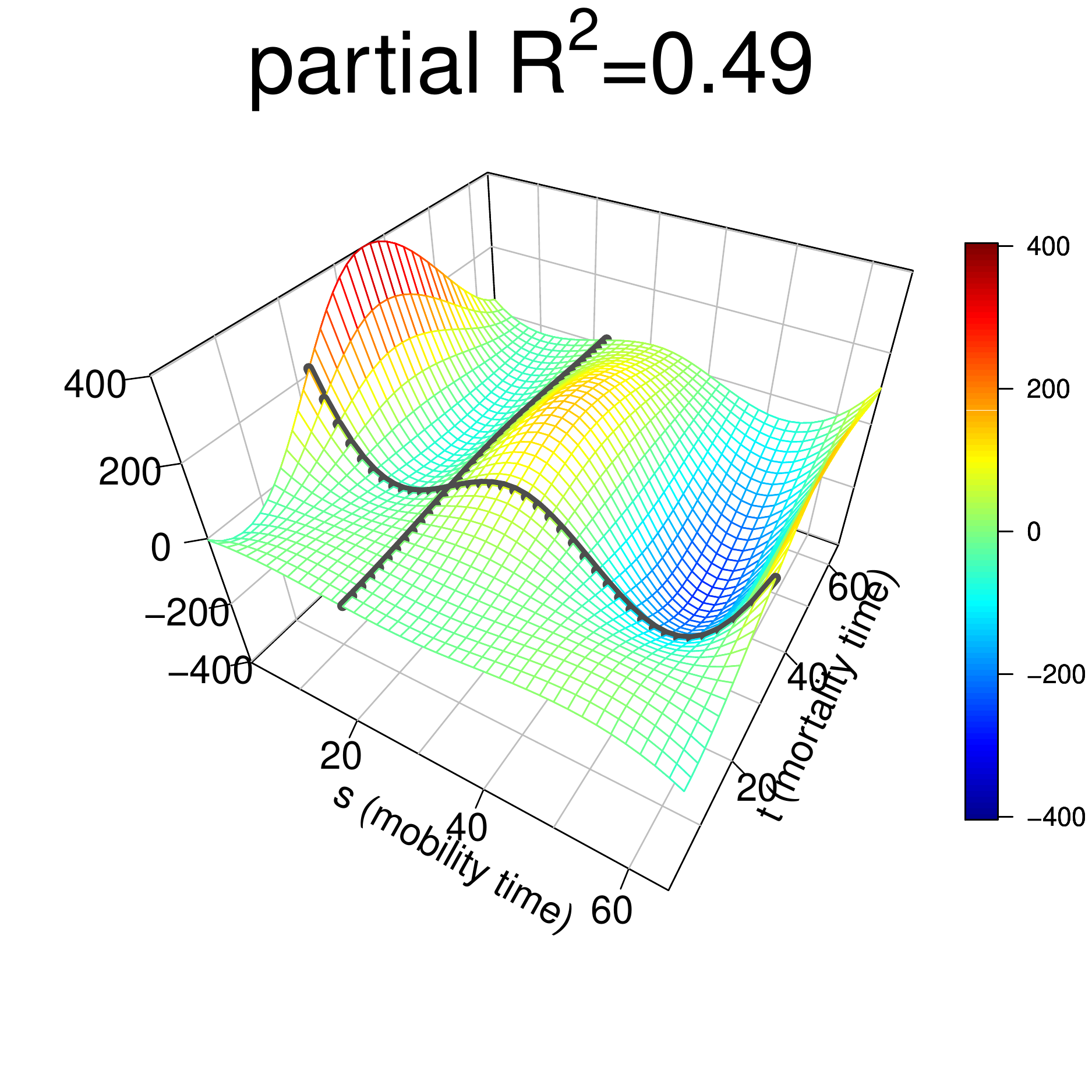}
        \includegraphics[width=0.163\linewidth]{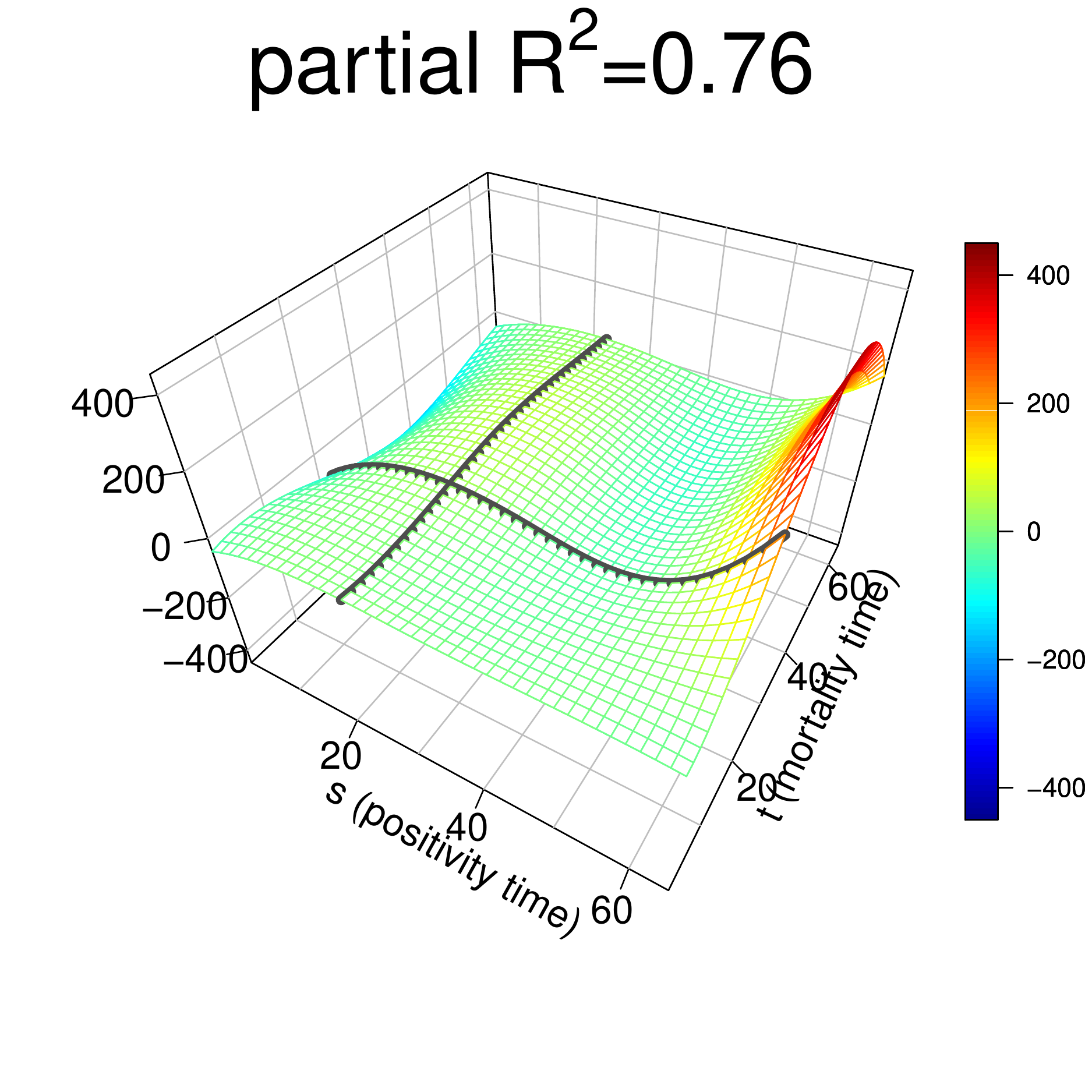}
        \vrule width 1pt
       \includegraphics[width=0.163\linewidth]{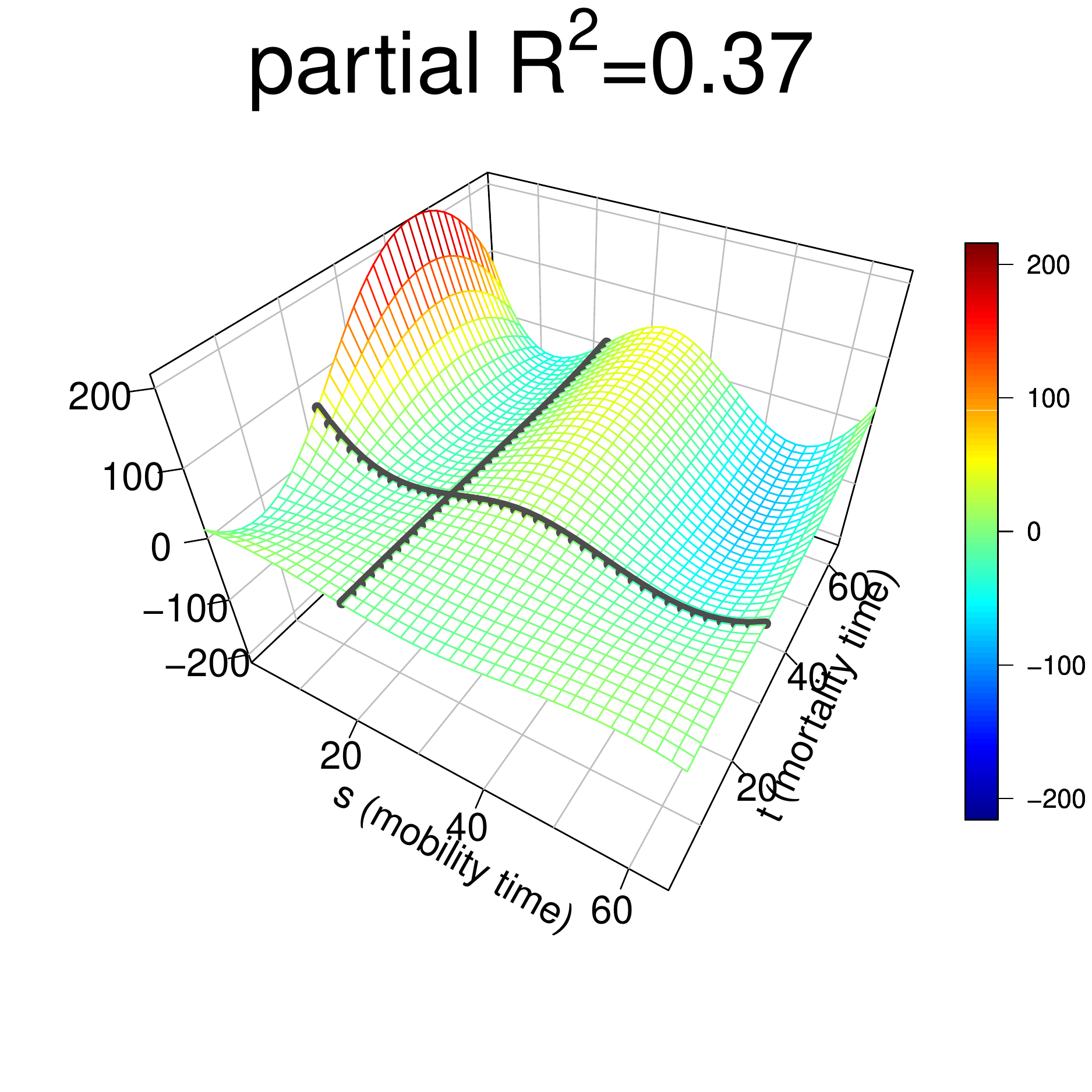}
        \includegraphics[width=0.163\linewidth]{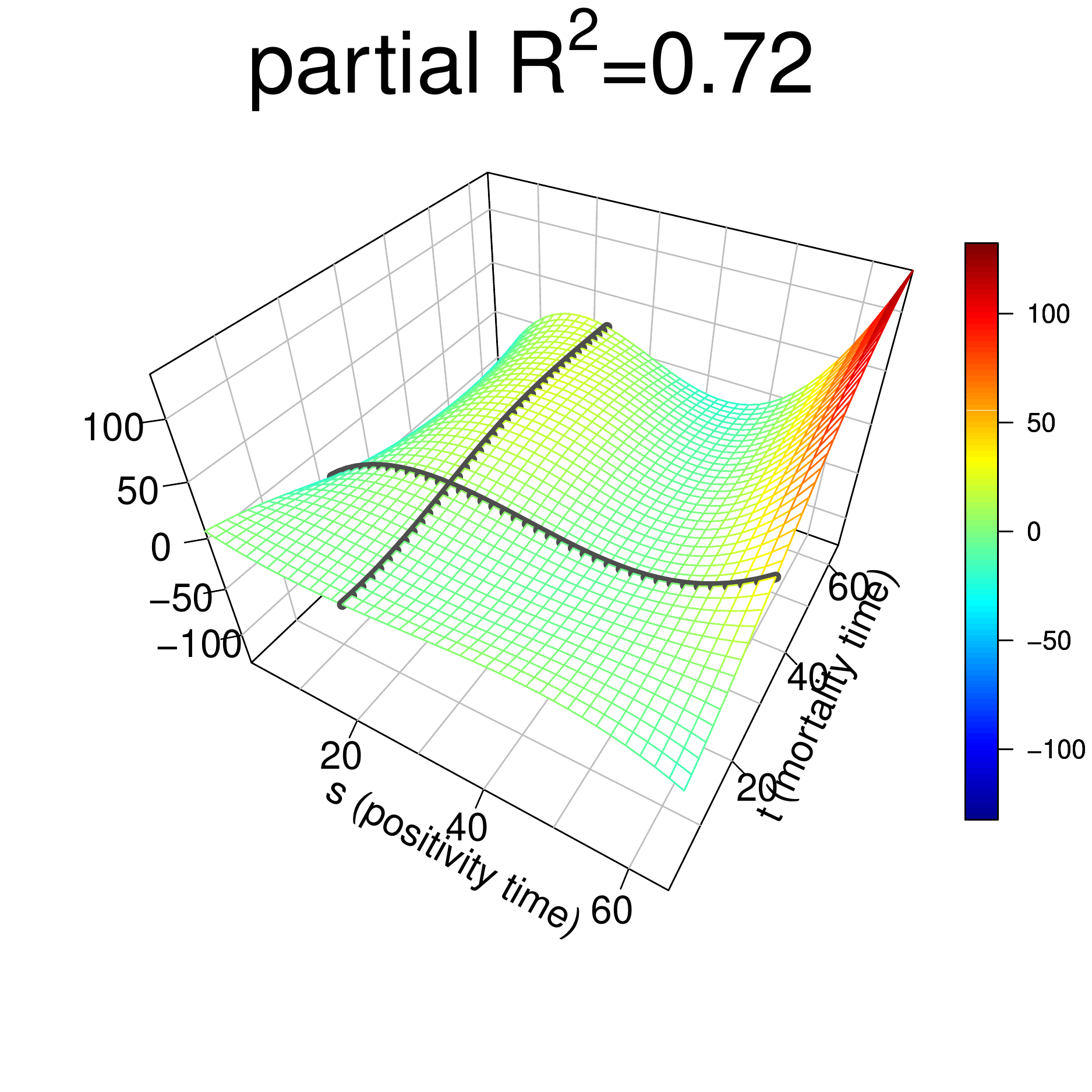} \\
        \vspace{-0.1cm}
    \end{minipage}
    } \hspace{-1.5cm} \\ 
    \hspace*{-1.7cm}
    \vspace*{0.1cm}
    \begin{minipage}[]{0.1\linewidth}
        {\bf 2}
    \end{minipage}
    \hspace{-1.5cm}
    \fbox{
    \begin{minipage}[]{1.14\linewidth}
        \centering
        \vspace{-0.1cm}
        \hspace{-0.2cm}
        \includegraphics[width=0.163\linewidth]{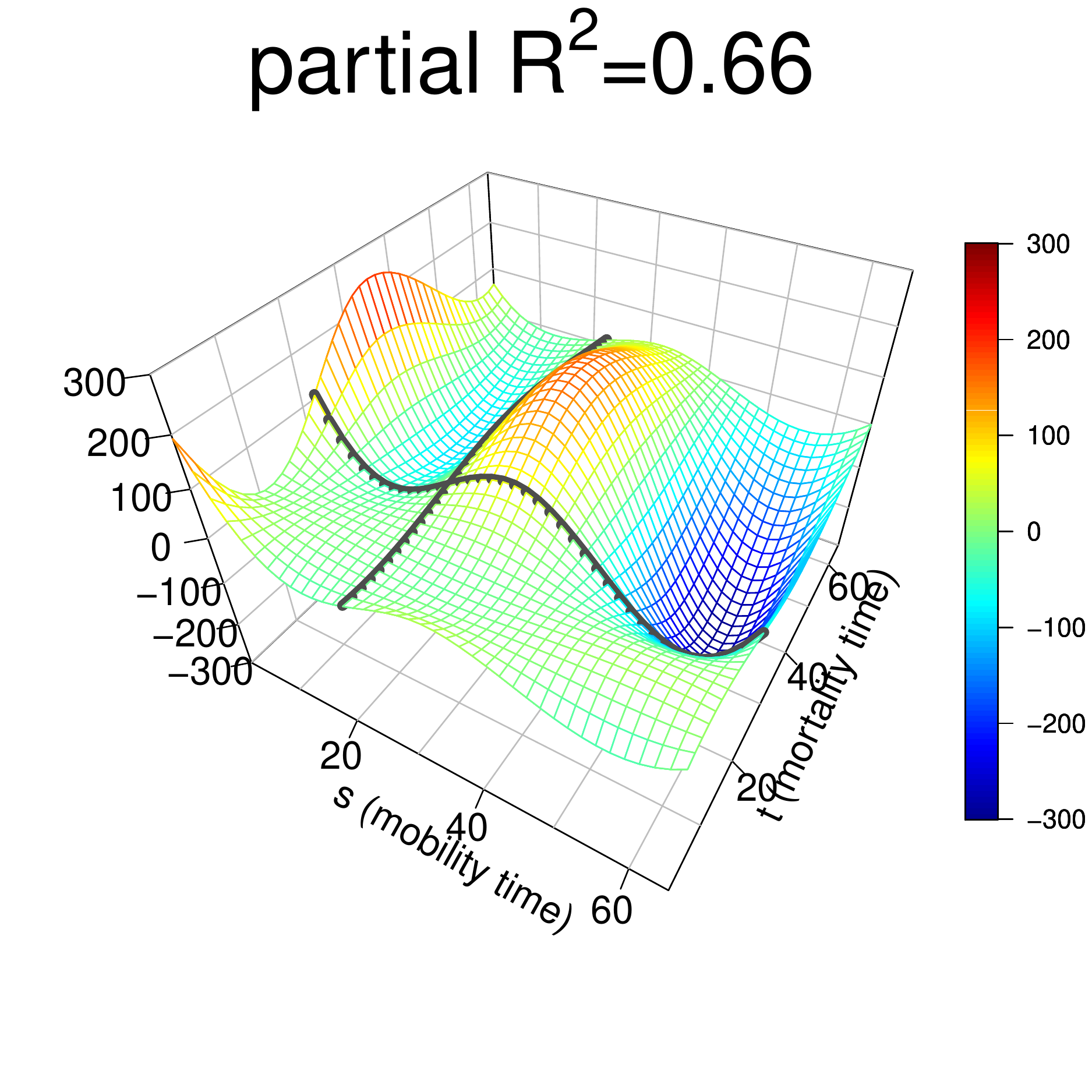}
        \includegraphics[width=0.163\linewidth]{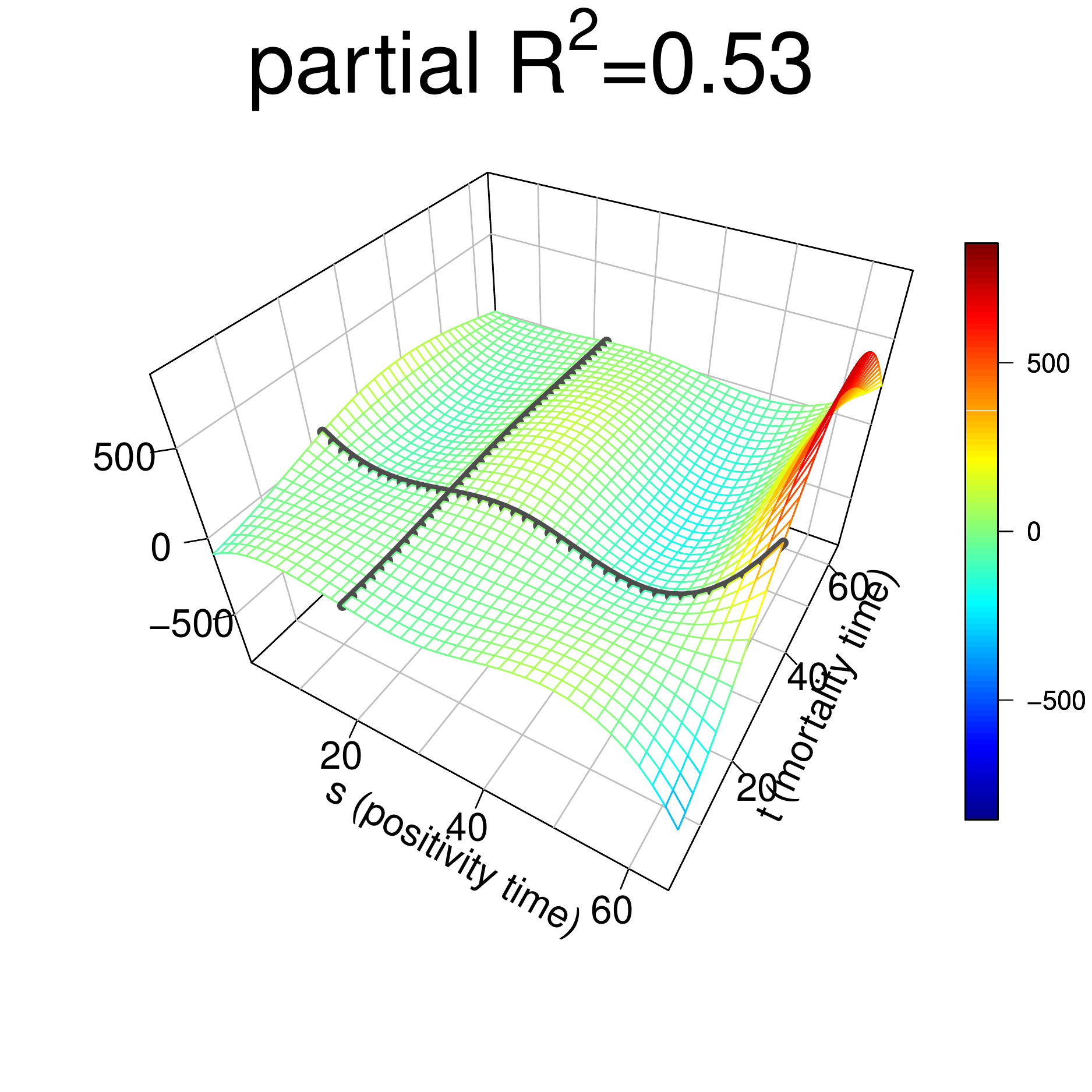}
        \vrule width 1pt
         \includegraphics[width=0.163\linewidth]{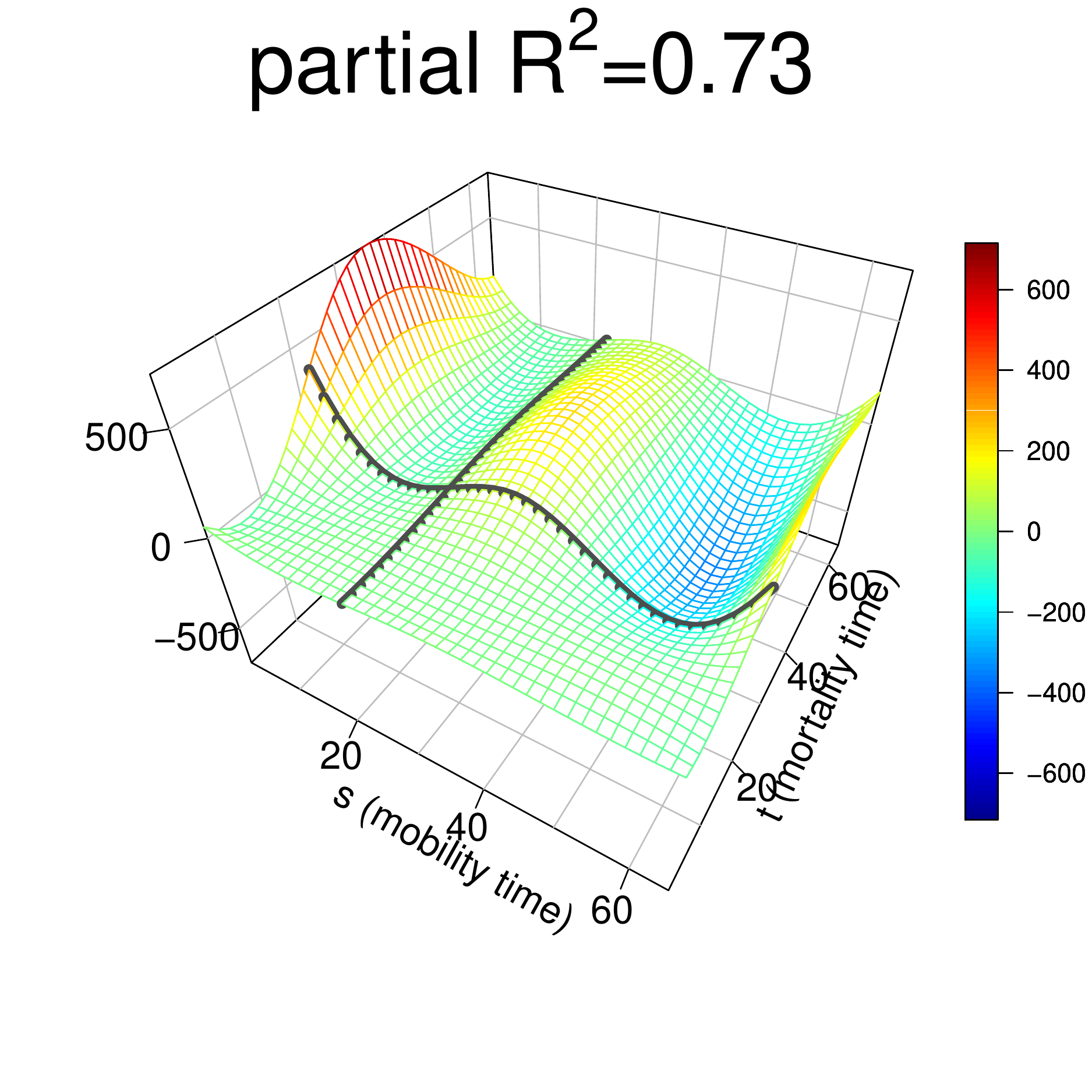}
        \includegraphics[width=0.163\linewidth]{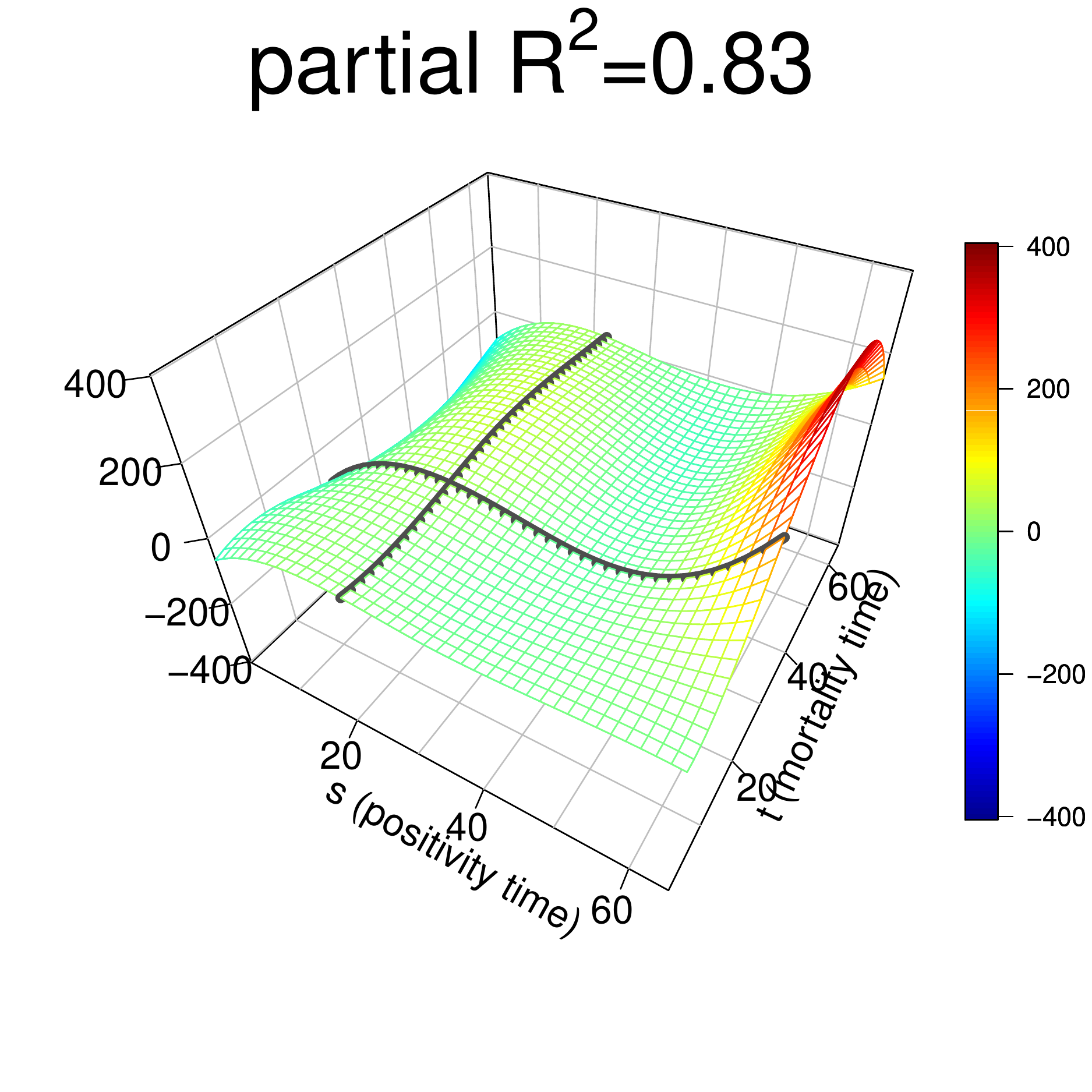}
        \vrule width 1pt
       \includegraphics[width=0.163\linewidth]{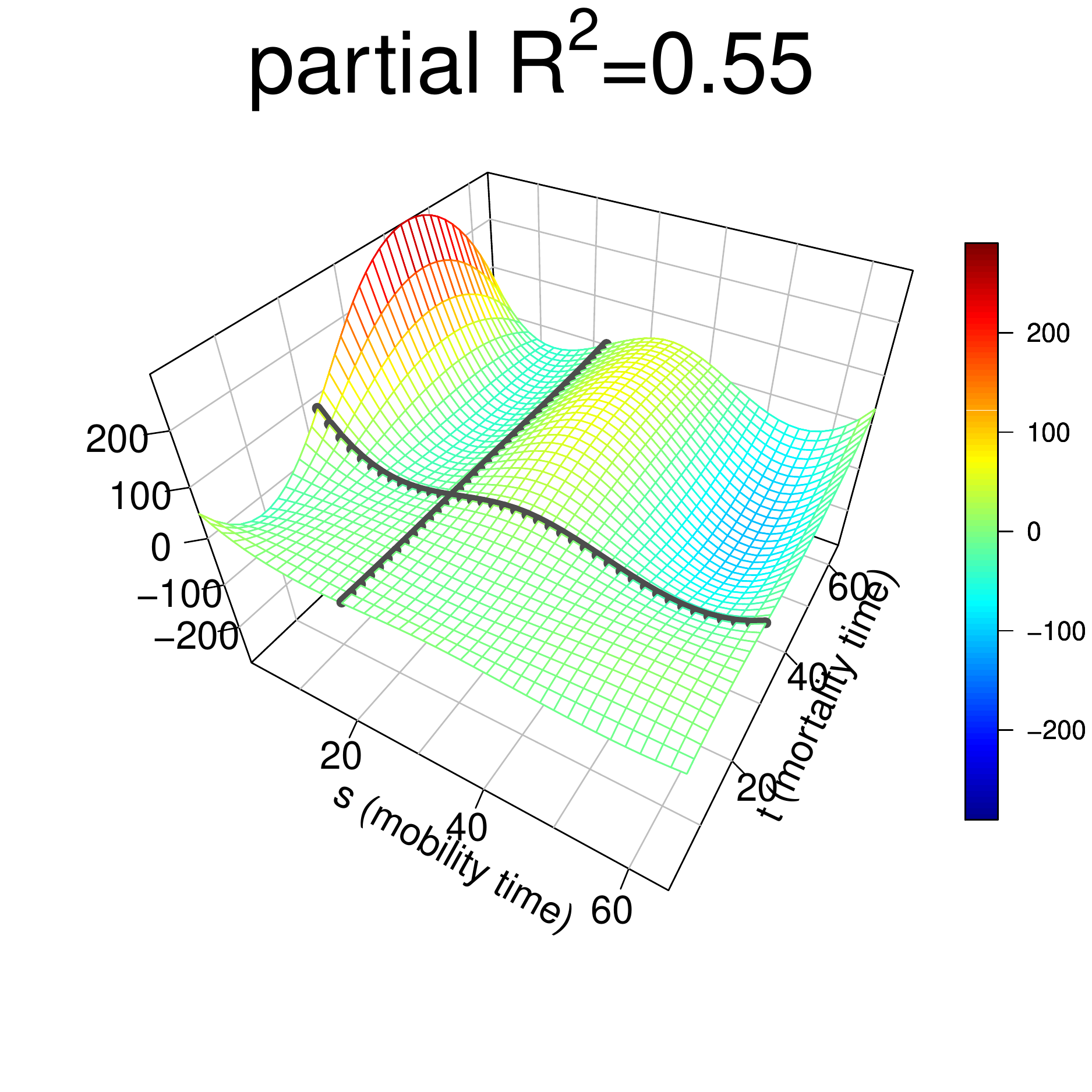}
        \includegraphics[width=0.163\linewidth]{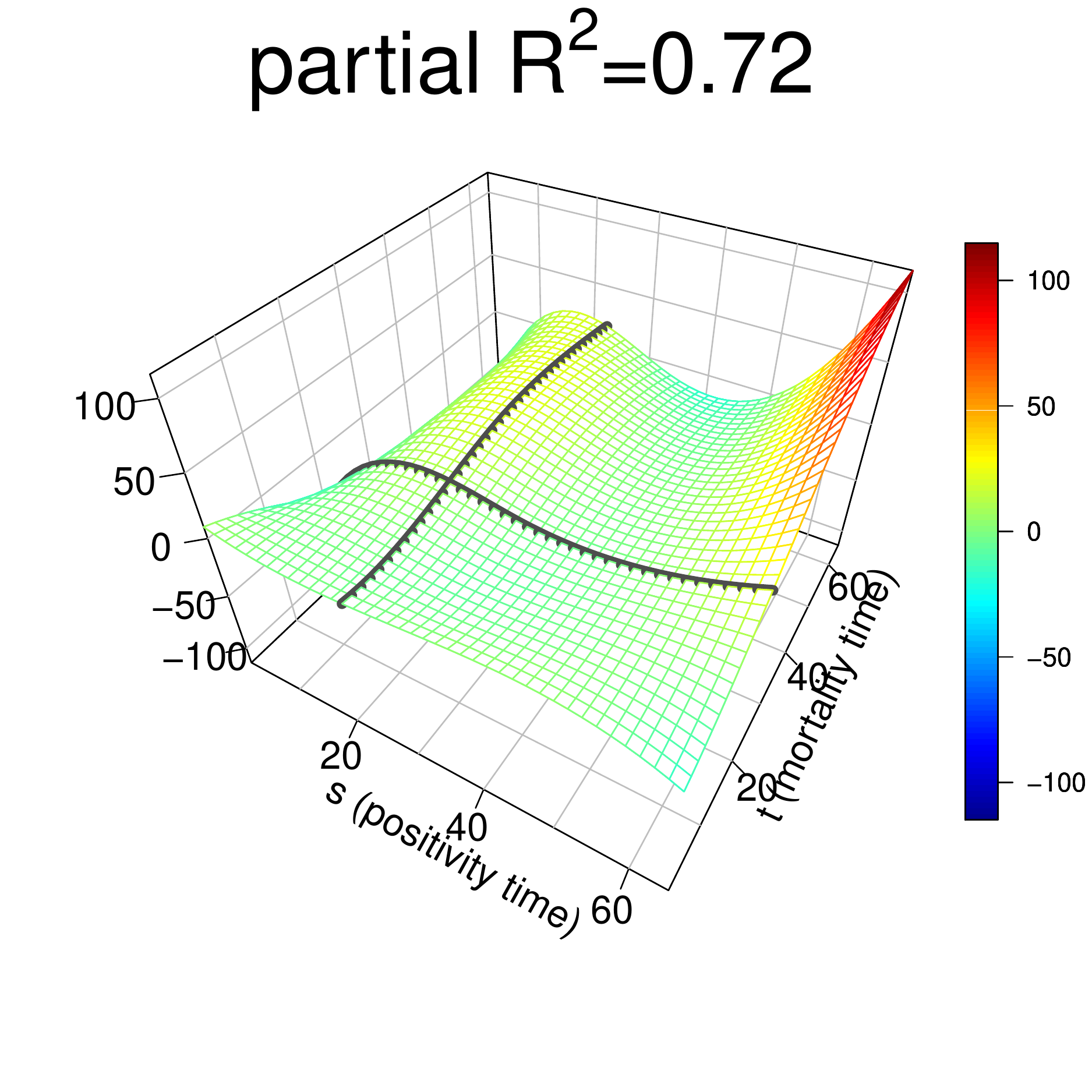} \\
        \vspace{-0.1cm}
    \end{minipage}
    } \\
    \hspace*{-1.7cm}
    \vspace*{0.1cm}
    \begin{minipage}[]{0.1\linewidth}
        {\bf 3}
    \end{minipage}
    \hspace{-1.5cm}
    \fbox{
    \begin{minipage}[]{1.14\linewidth}
        \centering
        \vspace{-0.1cm}
        \hspace{-0.2cm}
        \includegraphics[width=0.163\linewidth]{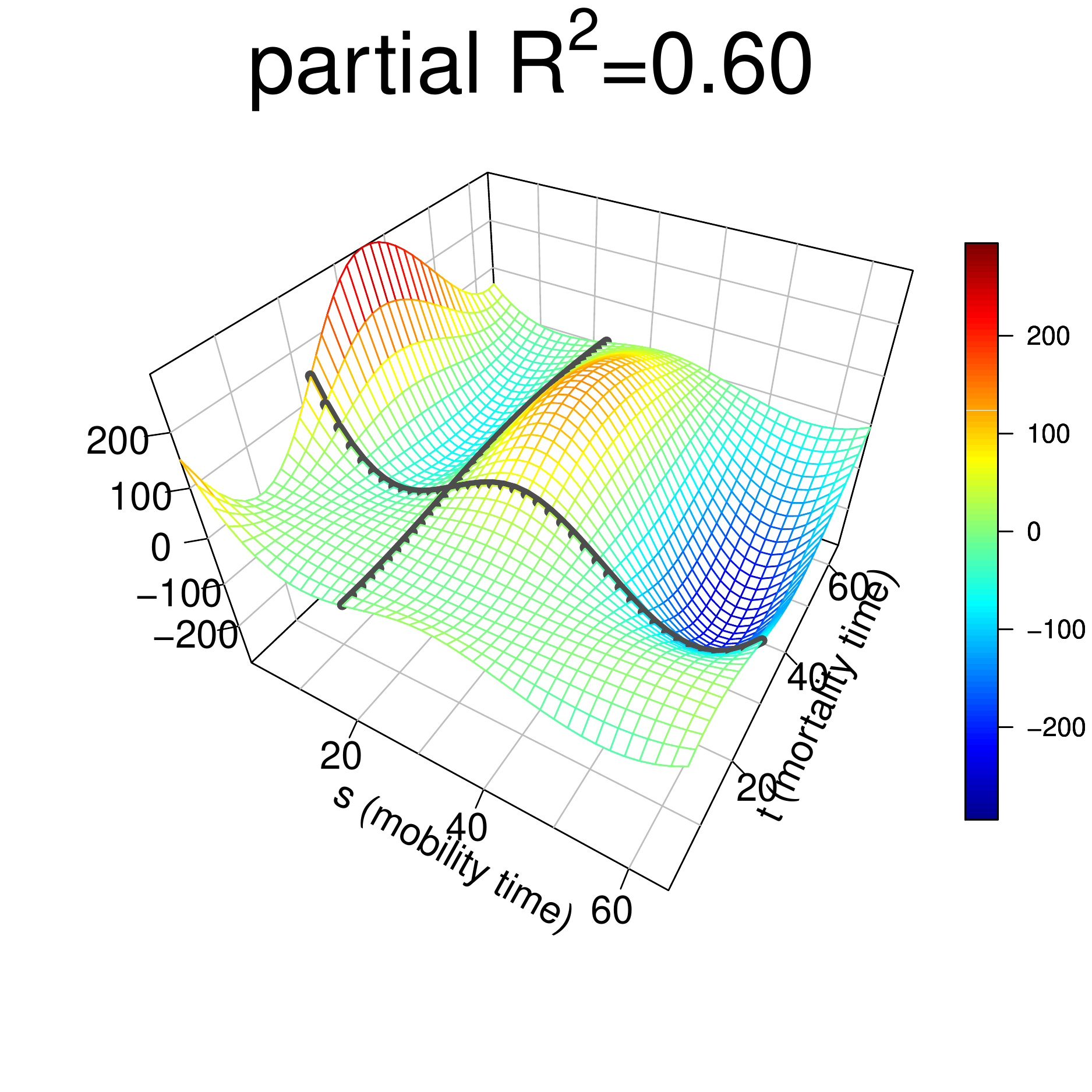}
        \includegraphics[width=0.163\linewidth]{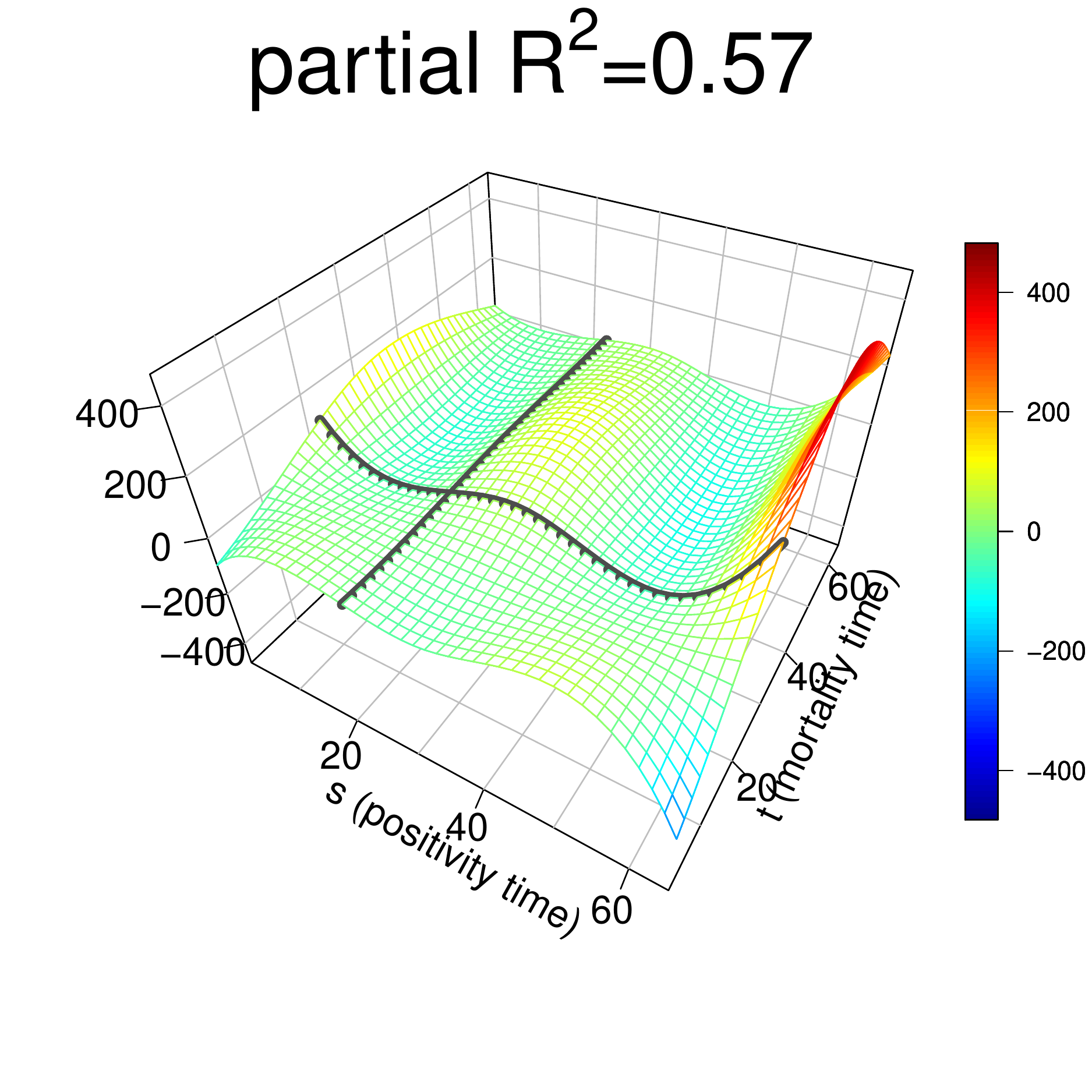}
        \vrule width 1pt
         \includegraphics[width=0.163\linewidth]{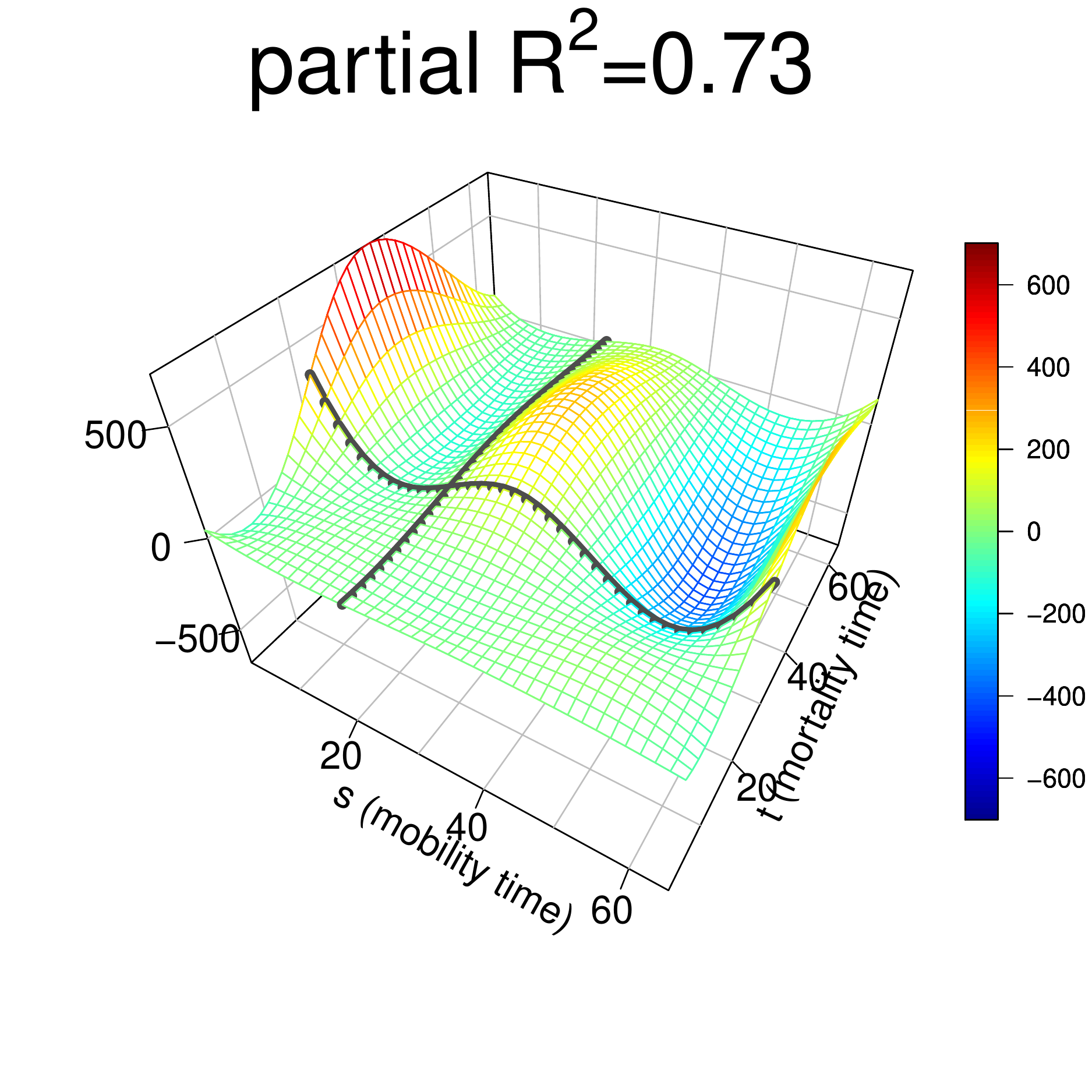}
        \includegraphics[width=0.163\linewidth]{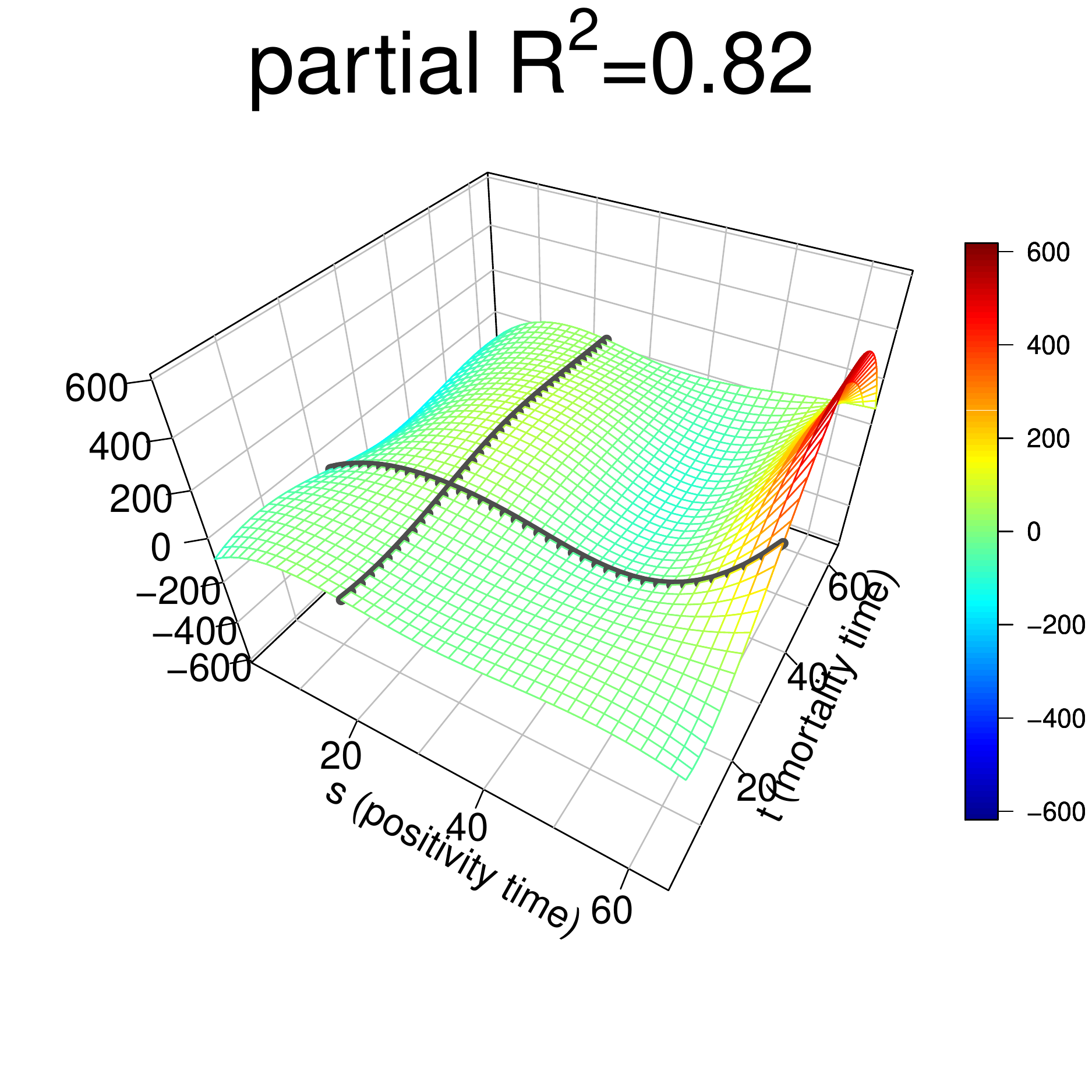}
        \vrule width 1pt
       \includegraphics[width=0.163\linewidth]{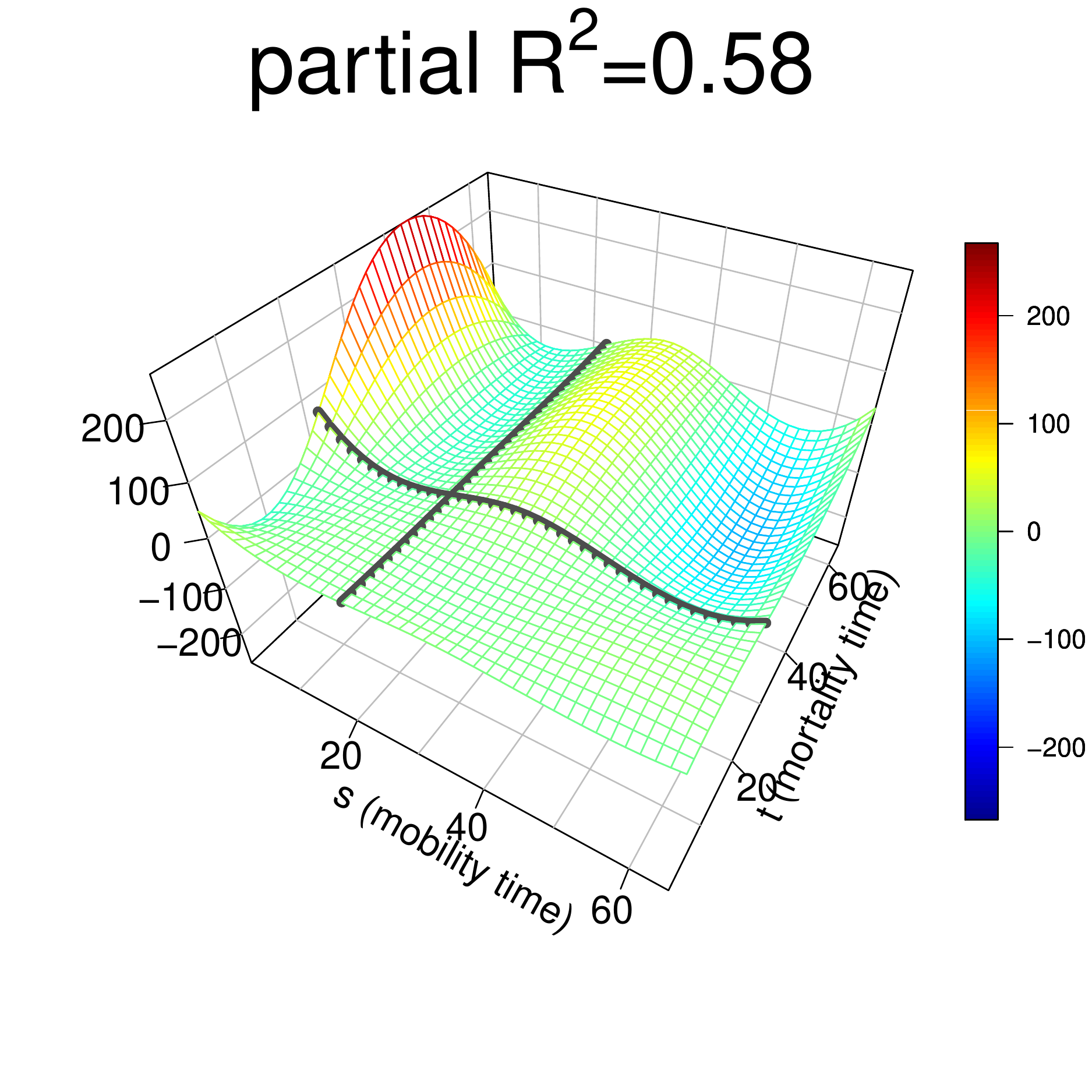}
        \includegraphics[width=0.163\linewidth]{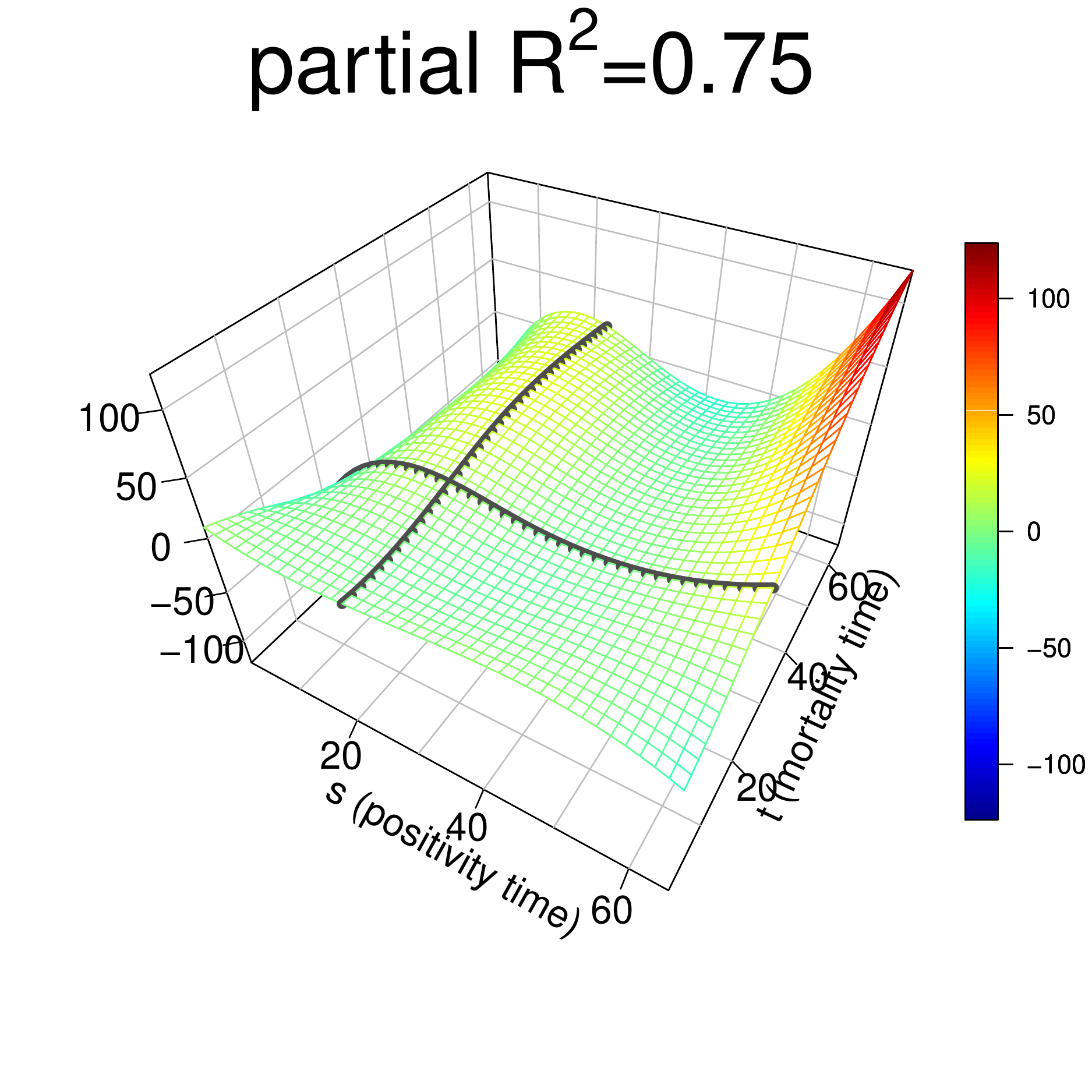} \\
        \vspace{-0.1cm}
    \end{minipage}
    } \\
   \hspace*{-1.7cm}
    \vspace*{0.1cm}
    \begin{minipage}[]{0.1\linewidth}
        {\bf 4}
    \end{minipage}
    \hspace{-1.5cm}
    \fbox{
    \begin{minipage}[]{1.14\linewidth}
        \centering
        \vspace{-0.1cm}
        \hspace{-0.2cm}
        \includegraphics[width=0.163\linewidth]{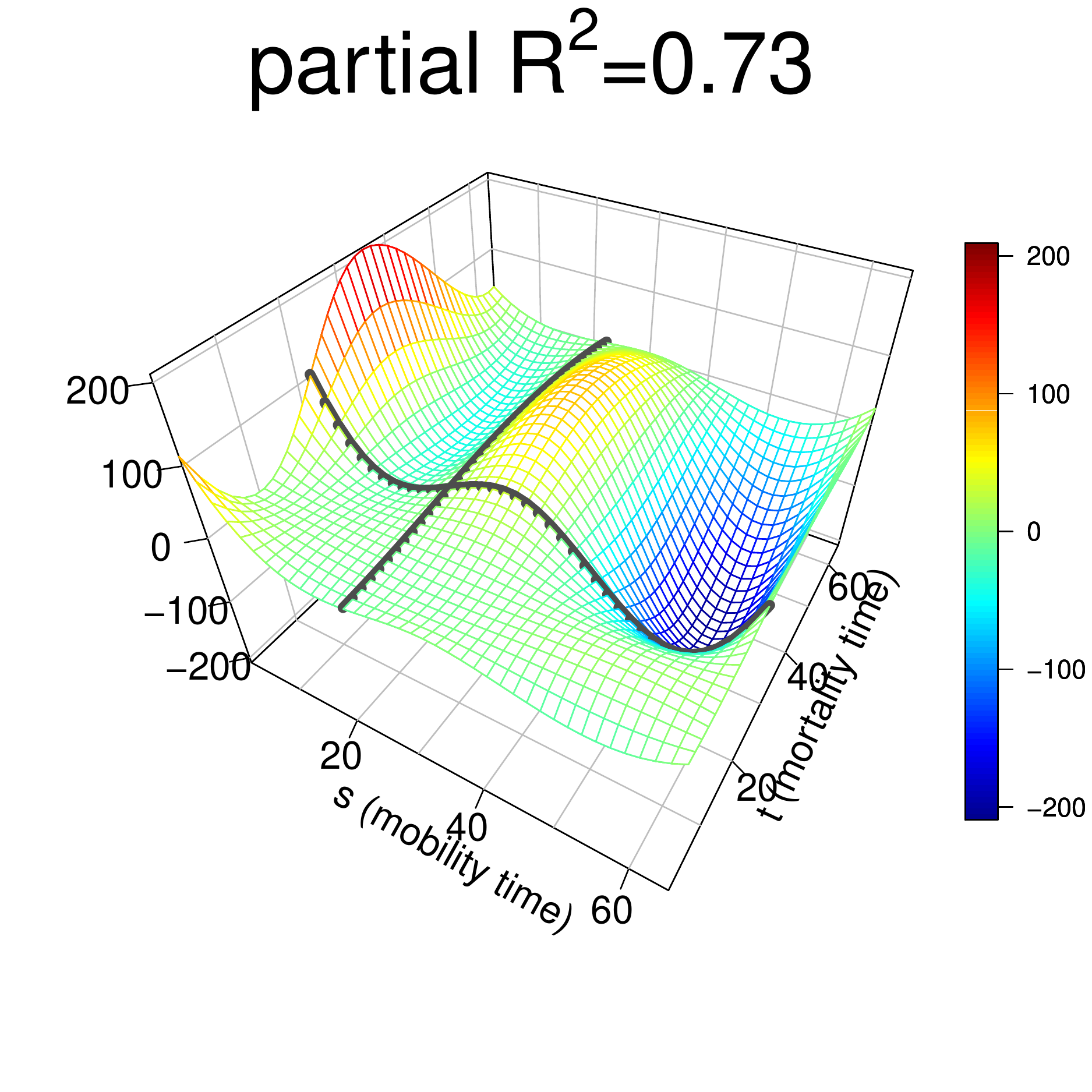}
        \includegraphics[width=0.163\linewidth]{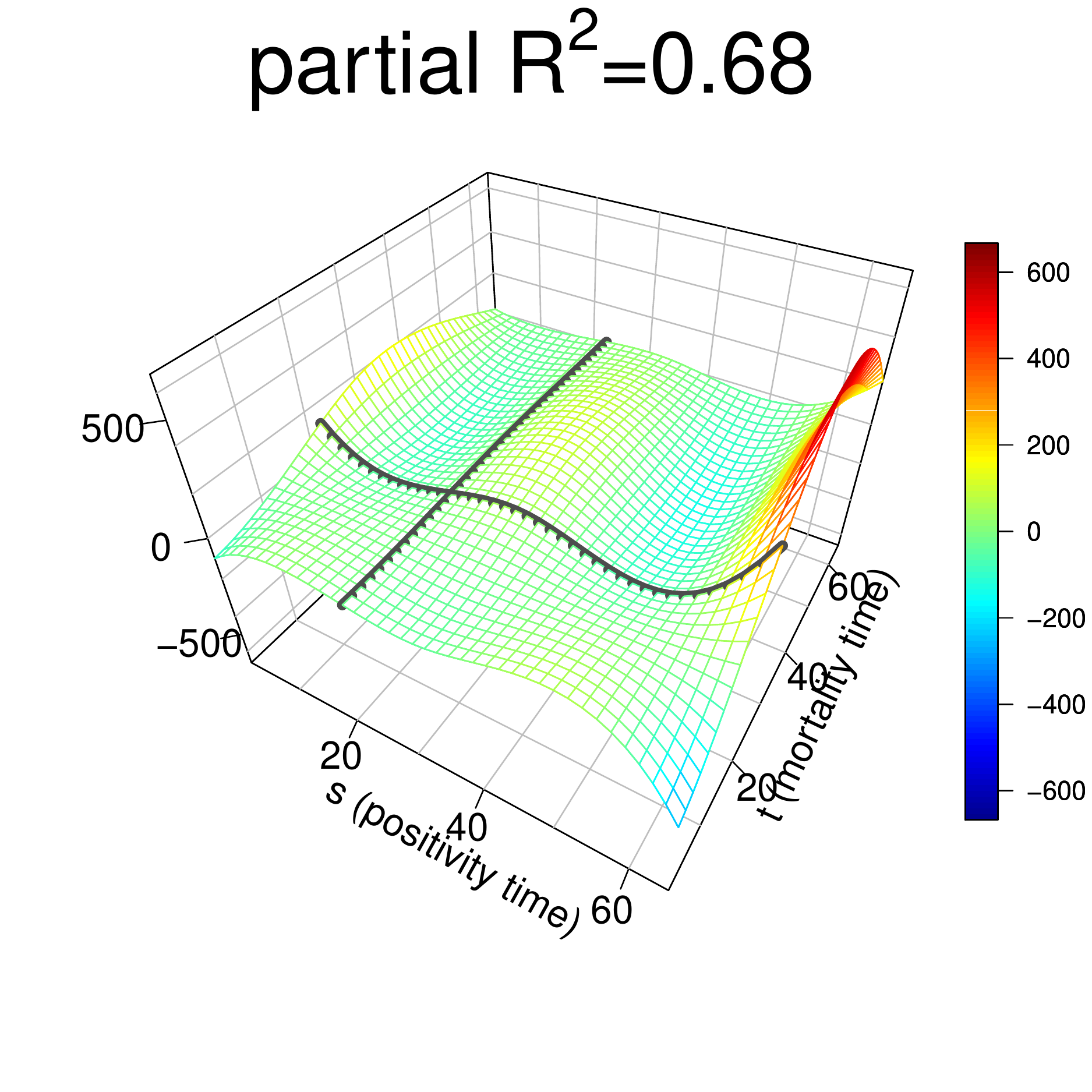}
        \vrule width 1pt
         \includegraphics[width=0.163\linewidth]{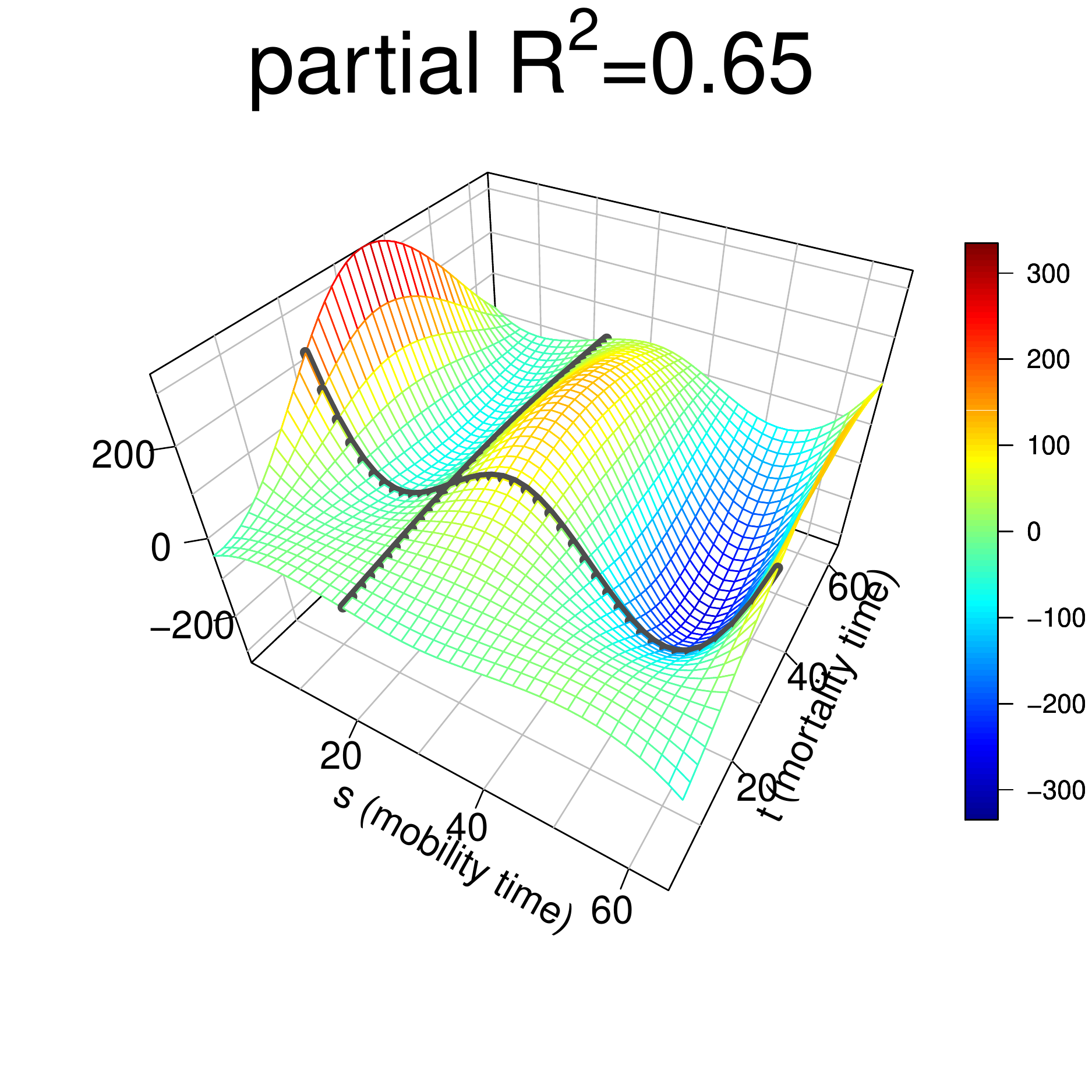}
        \includegraphics[width=0.163\linewidth]{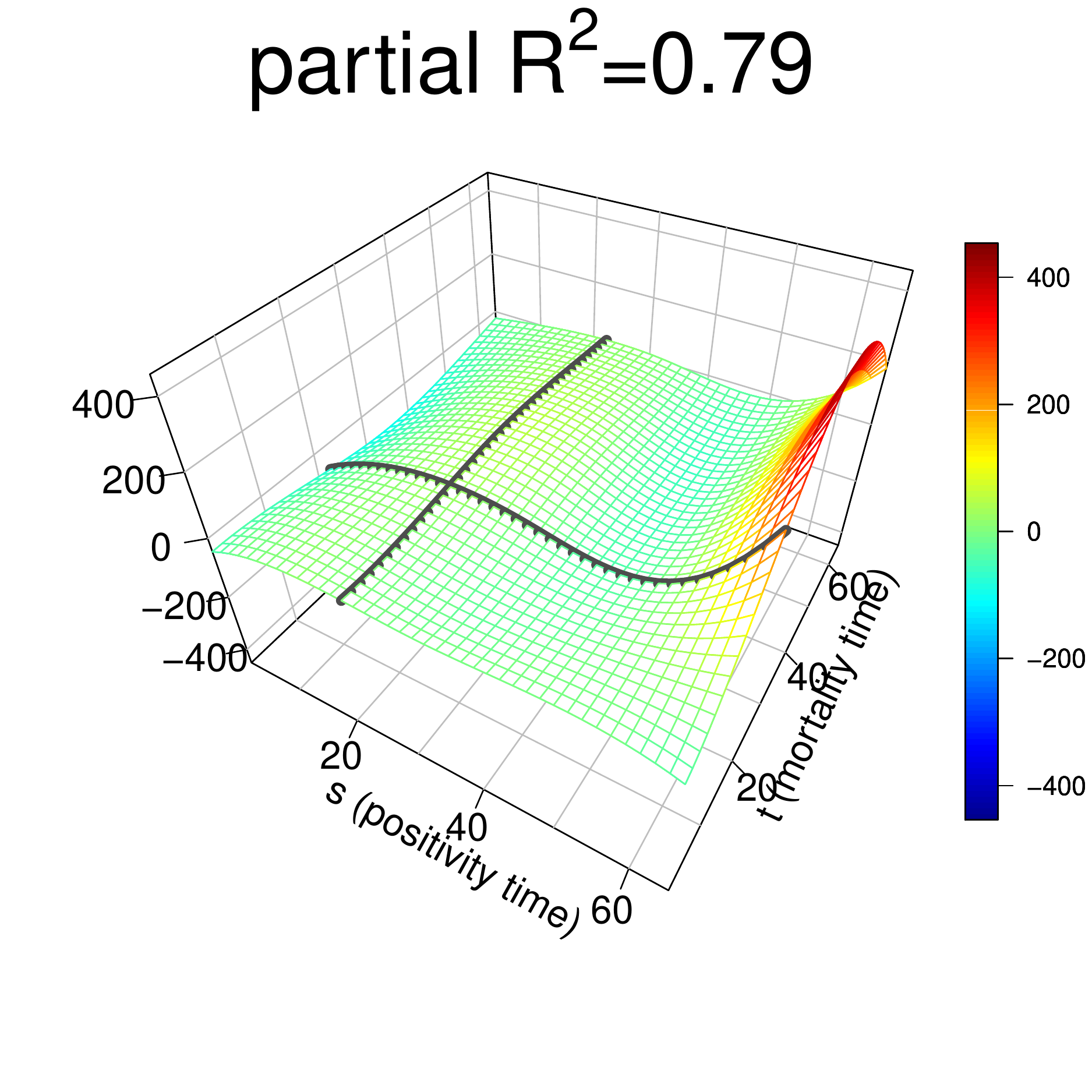}
        \vrule width 1pt
       \includegraphics[width=0.163\linewidth]{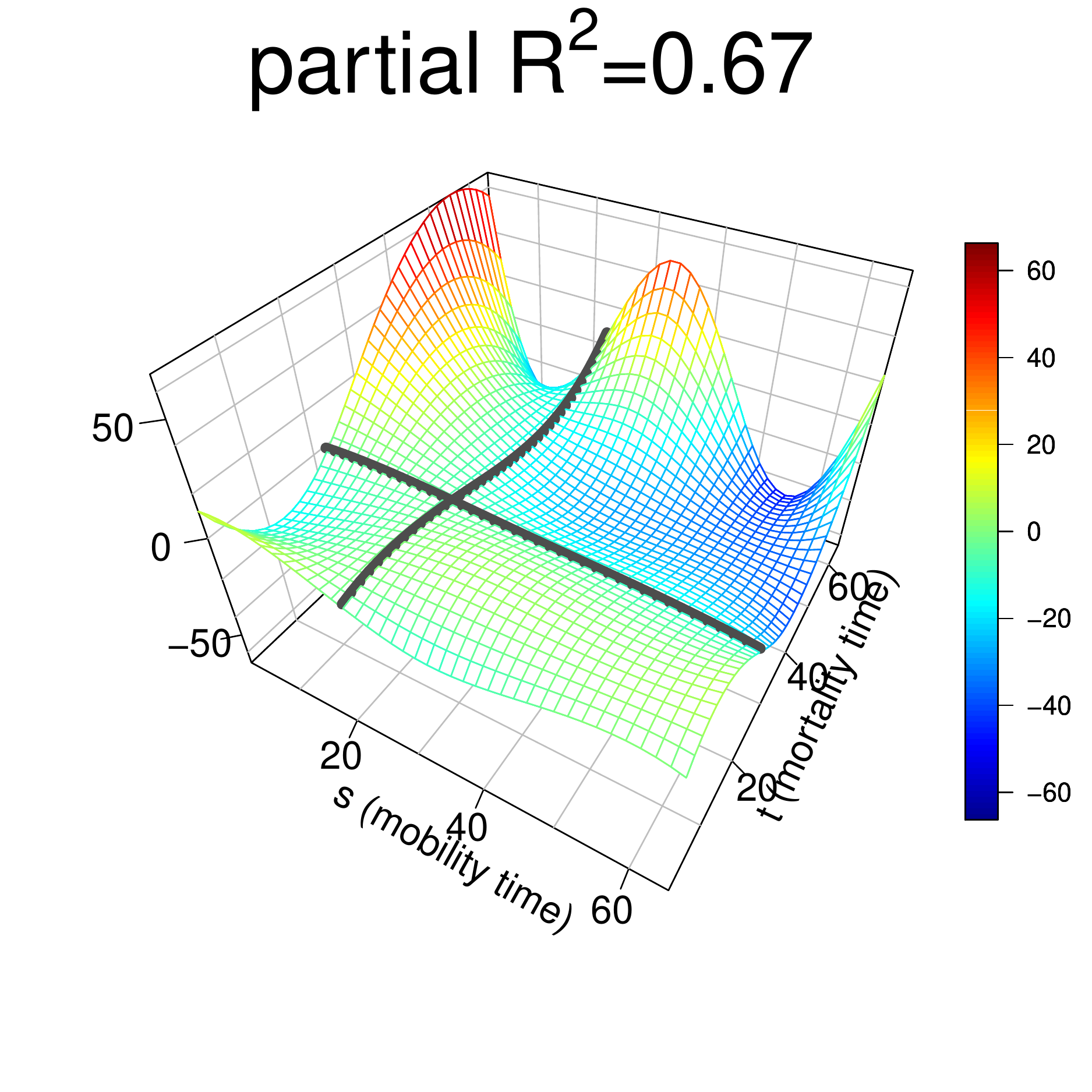}
        \includegraphics[width=0.163\linewidth]{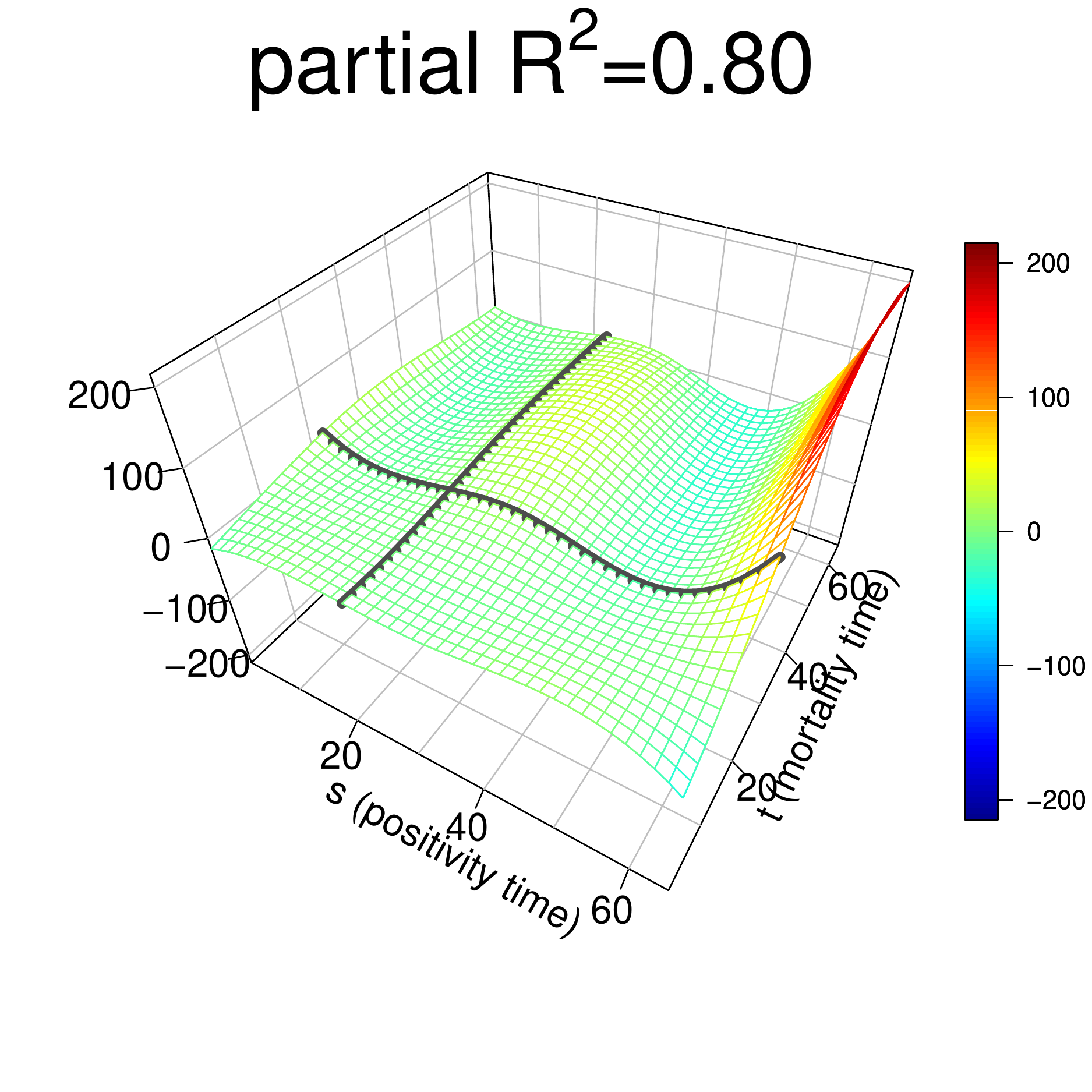} \\
        \vspace{-0.1cm}
    \end{minipage}
    } \\
    \hspace*{-1.7cm}
    \vspace*{0.1cm}
    \begin{minipage}[]{0.1\linewidth}
        {\bf 5}
    \end{minipage}
    \hspace{-1.5cm}
    \fbox{
   \begin{minipage}[]{1.14\linewidth}
        \centering
        \vspace{-0.1cm}
        \hspace{-0.2cm}
        \includegraphics[width=0.163\linewidth]{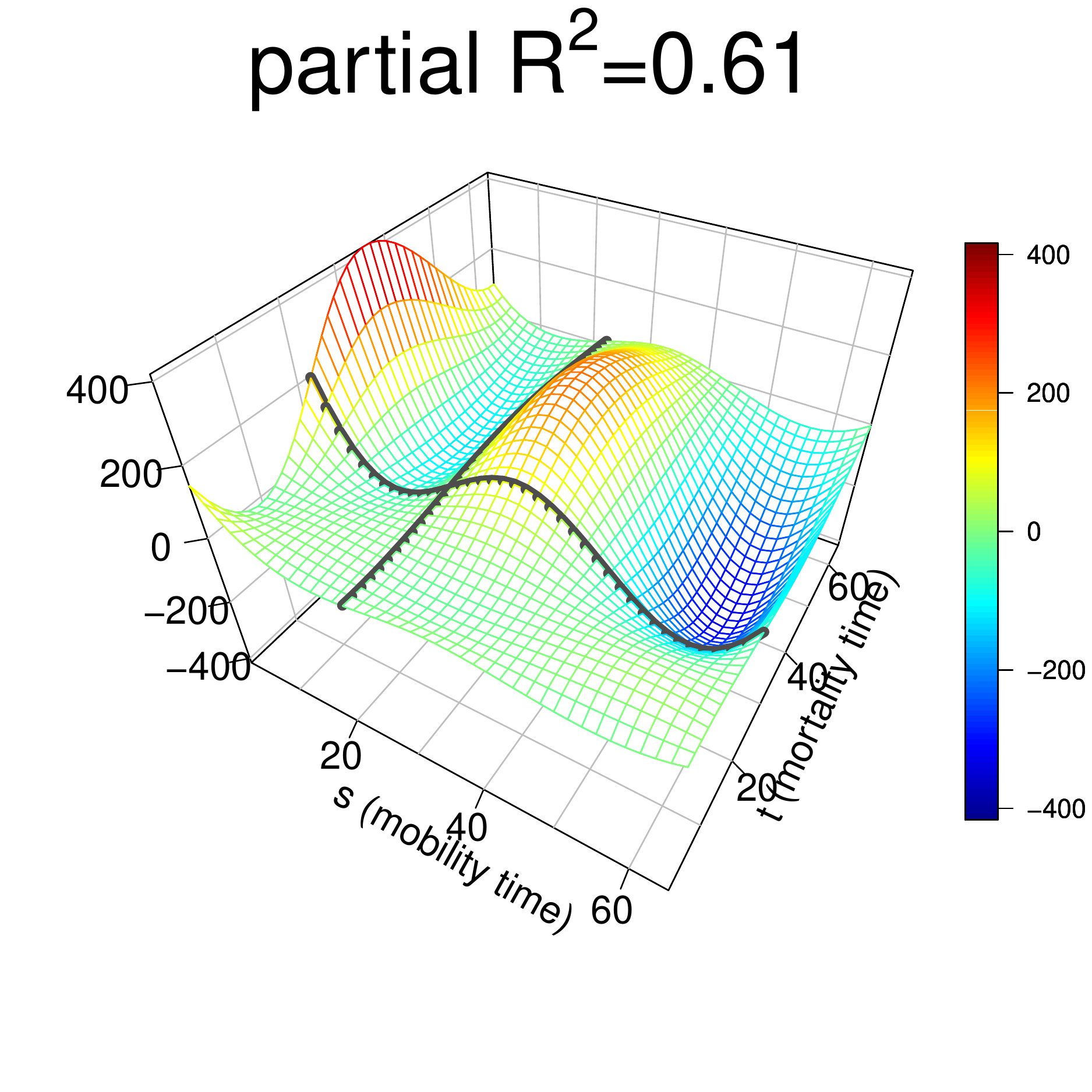}
        \includegraphics[width=0.163\linewidth]{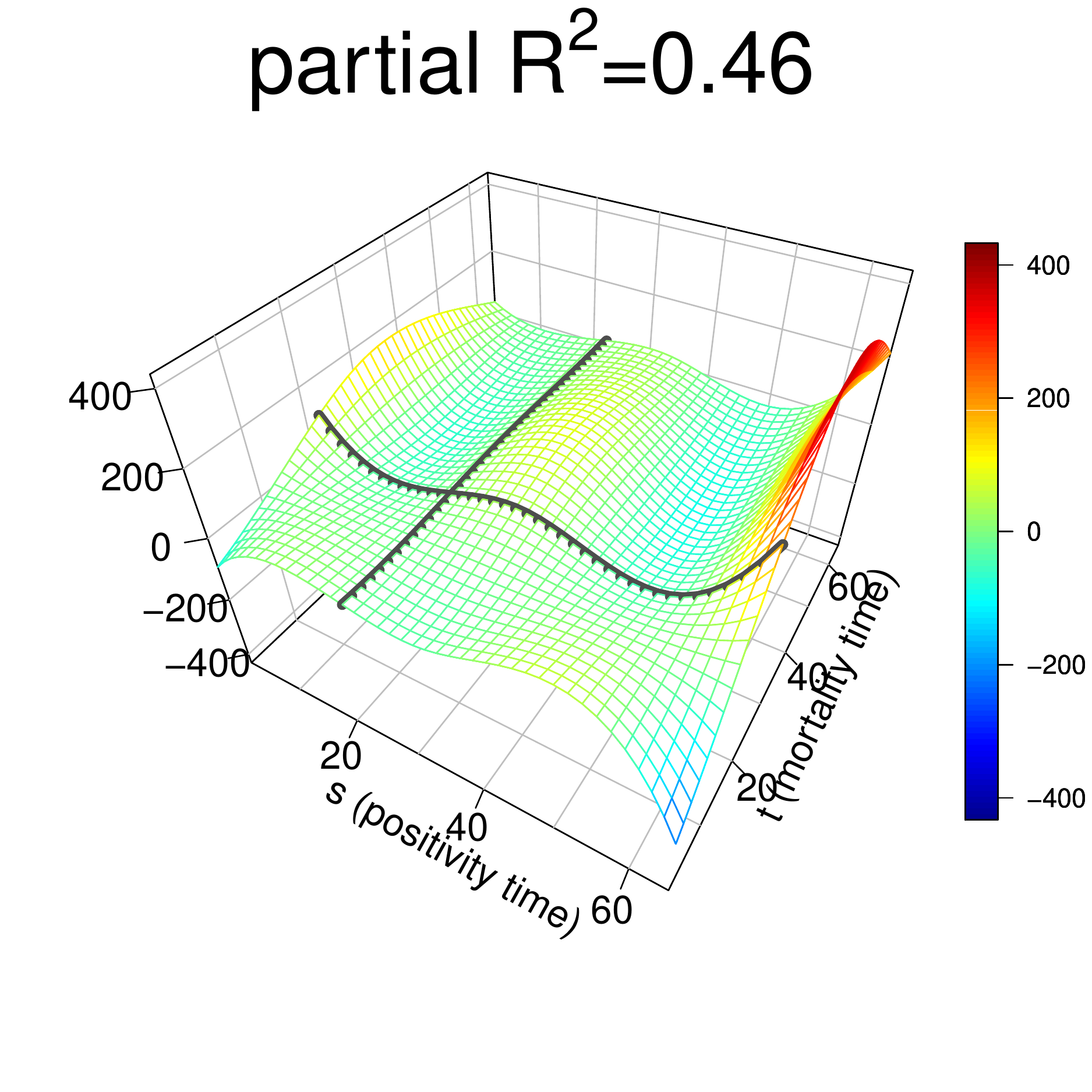}
        \vrule width 1pt
         \includegraphics[width=0.163\linewidth]{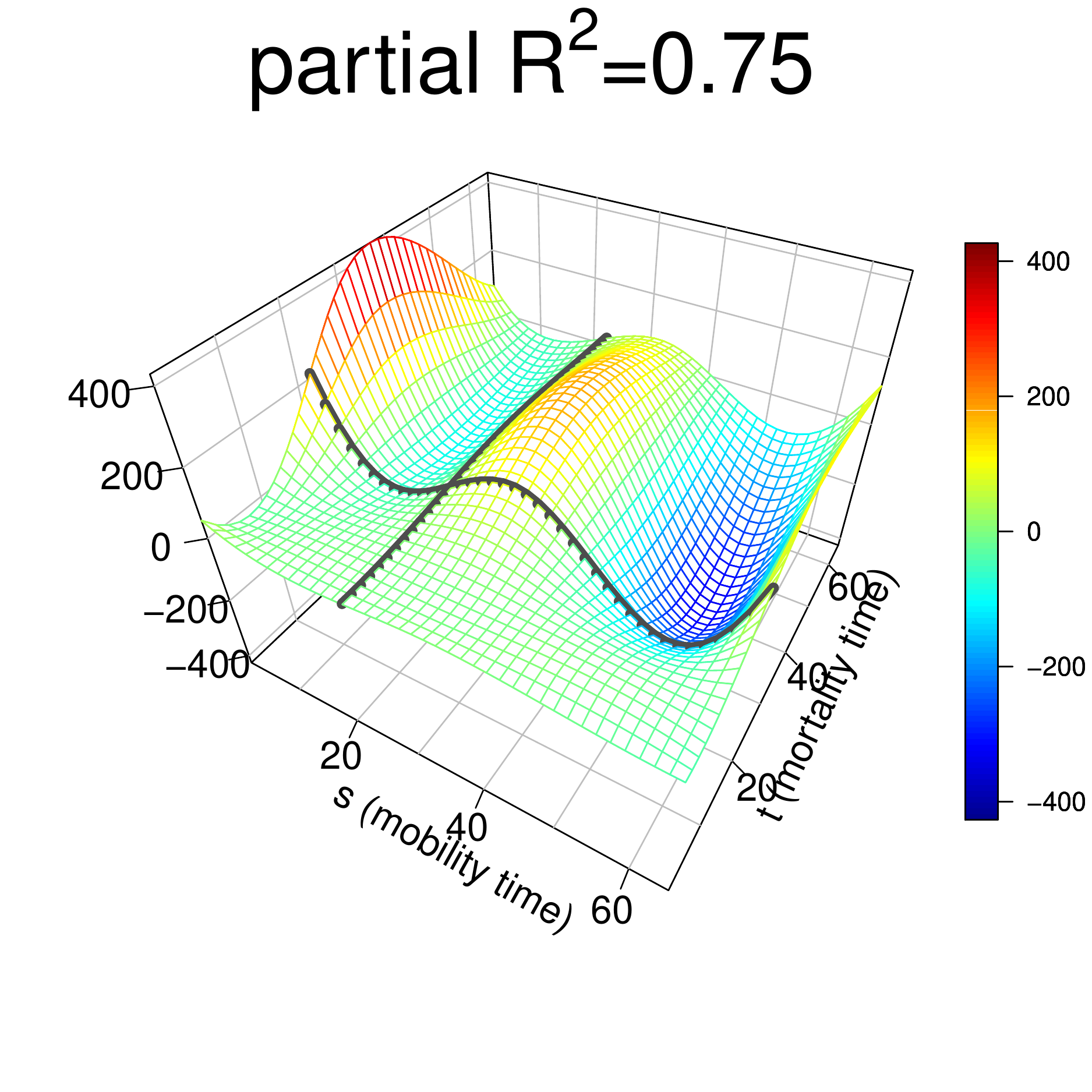}
        \includegraphics[width=0.163\linewidth]{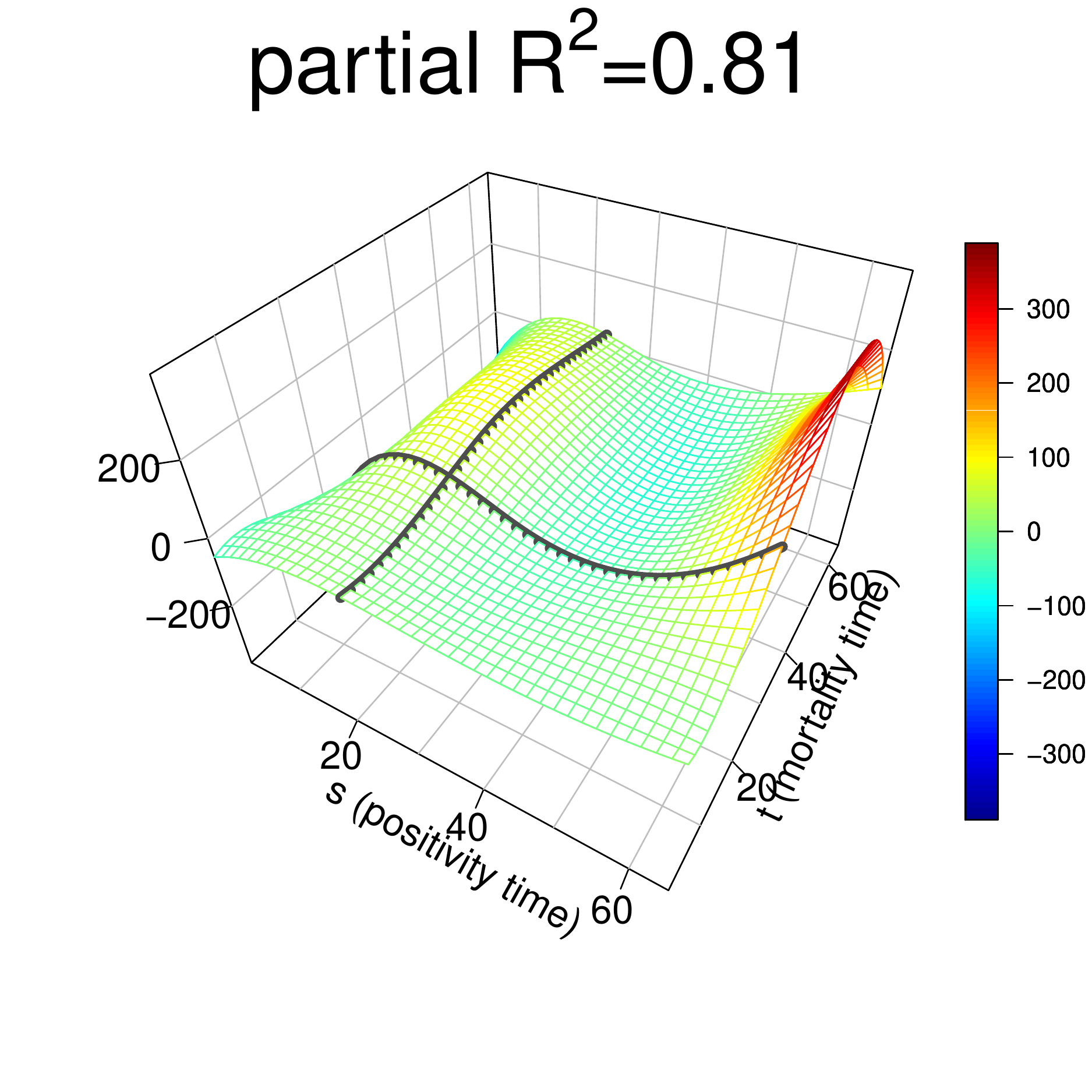}
        \vrule width 1pt
       \includegraphics[width=0.163\linewidth]{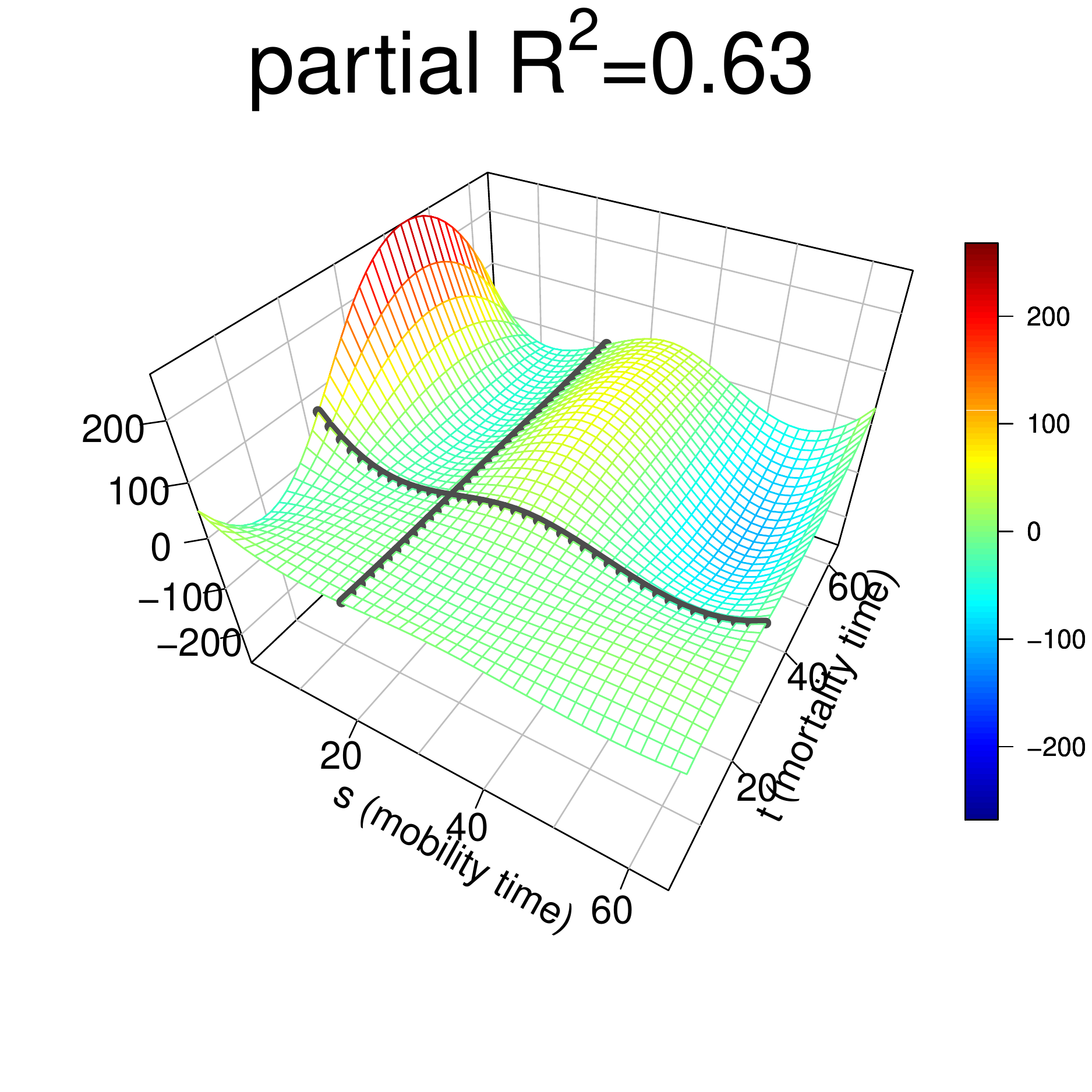}
        \includegraphics[width=0.163\linewidth]{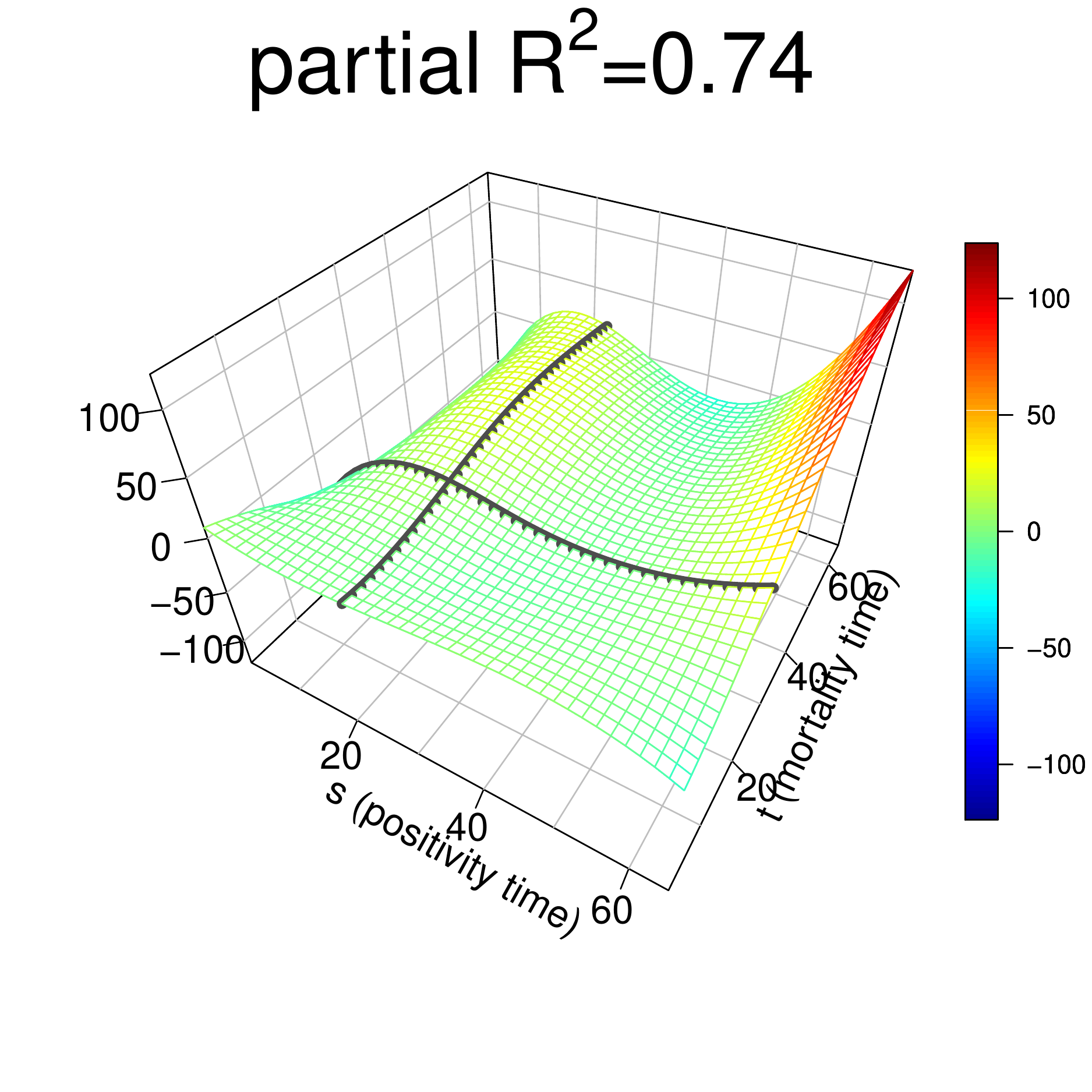} \\
        \vspace{-0.1cm}
    \end{minipage}
    } 
    \caption{
    {\bf Function-on-function regression of mortality on mobility, positivity, and a control scalar covariate.}
    Each row shows some results from the joint function-on-function regression mortality on local mobility, positivity, and one of the top 5 covariates selected by SsNAL-EN, used as control. In particular, we display the estimated effect surfaces for mobility and positivity (the March 9 date, without shift, is marked) with their respective partial $R^2$s. The scalar control covariates associated with each row are the following:
    1:~Adults per family doctor, 2:~Ave.~beds per hospital (whole), 3:~Ave.~students per classroom, 4:~Ave.~employees per firm, 5:~Ave.~members per household}
    \label{fig:mob_pos_controls}
\end{figure}
    

 \begin{table}
 \centering
\caption{List of all scalar covariates considered.}
\label{tab:covariatelis}
\scalebox{0.8}{
\begin{tabularx}{\textwidth}{l|l}
\toprule
\textbf{Covariate} & \textbf{Year and Source} \\
   \midrule
   Resident population, units & 2018, ISTAT\\
   Land area, hectares & 2018, ISTAT \\
   \% population over 65 & 2018, ISTAT \\
   \% population over 70 & 2018, ISTAT \\
   \% population over 80 & 2018, ISTAT \\
   \% population over 85 & 2018, ISTAT \\
   \% male over 18 & 2018, ISTAT\\
   \% female over 18 & 2018, ISTAT \\
   Employees in large supermarket chains, units & 2018, ISTAT \\
   Department stores, units & 2018, ISTAT \\
   Supermarkets, units & 2018, ISTAT \\
   Ipermarkets, units & 2018, ISTAT \\
   Airports, units & 2018, ISTAT \\
   Landed and departed passengers in airports, units & 2018, ISTAT \\
   Landed and departed airplanes in international flights, units & 2018, ISTAT \\
   Healthcare institutes (private and public), units & 2018, Ministry of Health \\
   Public healthcare institutes, units & 2018, Ministry of Health \\
   Days of stay in public and private healthcare institutes & 2015, ISTAT \\
   Days of stay in public healthcare institutes & 2015, ISTAT \\
   Patients in public and private institutes, units & 2015, ISTAT \\
   Patients in public institutes (except for residual psychiatric institutes), units & 2015, ISTAT \\
   Beds in pneumatology in public and private healthcare institutes & 2015, ISTAT \\
   Mechanical lung ventilators, units & 2018, Ministry of Health \\
   People with at least 1 chronic disease, units & 2017, Ministry of Health \\
   People with at least 2 chronic diseases, units & 2017, Ministry of Health \\
   \% people with diabetes & 2017, Ministry of Health \\
   \% people with hypertension & 2017, Ministry of Health \\
   \% people with bronchitis & 2017, Ministry of Health \\
   \% people with osteoporosis & 2017, Ministry of Health \\
   \% people with arthritis & 2017, Ministry of Health \\
   \% people with allergy & 2017, Ministry of Health \\
   \% people with ulcer & 2017, Ministry of Health\\
   Old-age index & 2016, ISTAT \\
   Life expectancy at birth (female) & 2017, ISTAT \\
   Life expectancy at birth (male) & 2017, ISTAT \\
   Active buses per 1000 habitants & 2016, ISTAT \\
   \% children going to school with public transportation & 2016, ISTAT \\
   Public expenditure in healthcare per capita & 2016, ISTAT \\
   Factor risk: alcohol & 2016, ISTAT \\
   Factor risk: smoke & 2016, ISTAT \\
   Factor risk: obesity & 2016, ISTAT \\
   Average household income & 2015, ISTAT \\
   Mobility index (commuting due to work) & 2011, ISTAT \\
   Self-containment index & 2011, ISTAT \\
   Public mobility index & 2011, ISTAT \\
   PM10 & 2017, ISTAT \\
   PM2.5 & 2017, ISTAT \\
   ICU beds per 100K inhabitants & 2018, Ministry of Health \\
   Beds in pneumatology & 2018, Ministry of Health \\
   Additional beds in ICU on April, 10\textsuperscript{th} 2020 & 2020, DPC \\
   Weighted PM10 & 2018, ISTAT \\
   Average students per classroom & 2018, Ministry of Education \\
   Average students per school & 2018, Ministry of Education \\
   Gini index for schools & 2018, Ministry of Education \\
   Average beds per nursing home (ward) & 2018, Ministry of Health \\
   Average beds per nursing home (whole) & 2018, Ministry of Health \\
   Gini index for nursing homes & 2018, Ministry of Health \\
   Average beds per hospital (ward) & 2018, Ministry of Health \\
   Average beds per hospital (whole) & 2018, Ministry of Health \\
   Gini index for hospitals & 2018, Ministry of Health \\
   Average number of employees & 2017, ISTAT \\
   Gini index for firms & 2017, ISTAT \\
   Total number of tests between February 25\textsuperscript{th} and April, 30\textsuperscript{th} 2020 & 2020, DPC \\
   Total number of tests between February 25\textsuperscript{th} and March, 23\textsuperscript{rd} 2020 & 2020, DPC \\
   Total number of tests between March, 23\textsuperscript{rd} and April, 30\textsuperscript{th} 2020 & 2020, DPC \\
   Adults per family doctor & 2017, Ministry of Health \\
   Average members per family & 2018, ASR Lombardia \\
   Public transport rides per capita & 2017, ISTAT \\
   \bottomrule
\end{tabularx}
}
\end{table}

\begin{figure}[h]
    \centering
    \includegraphics[width=1.3\textwidth, angle=90]{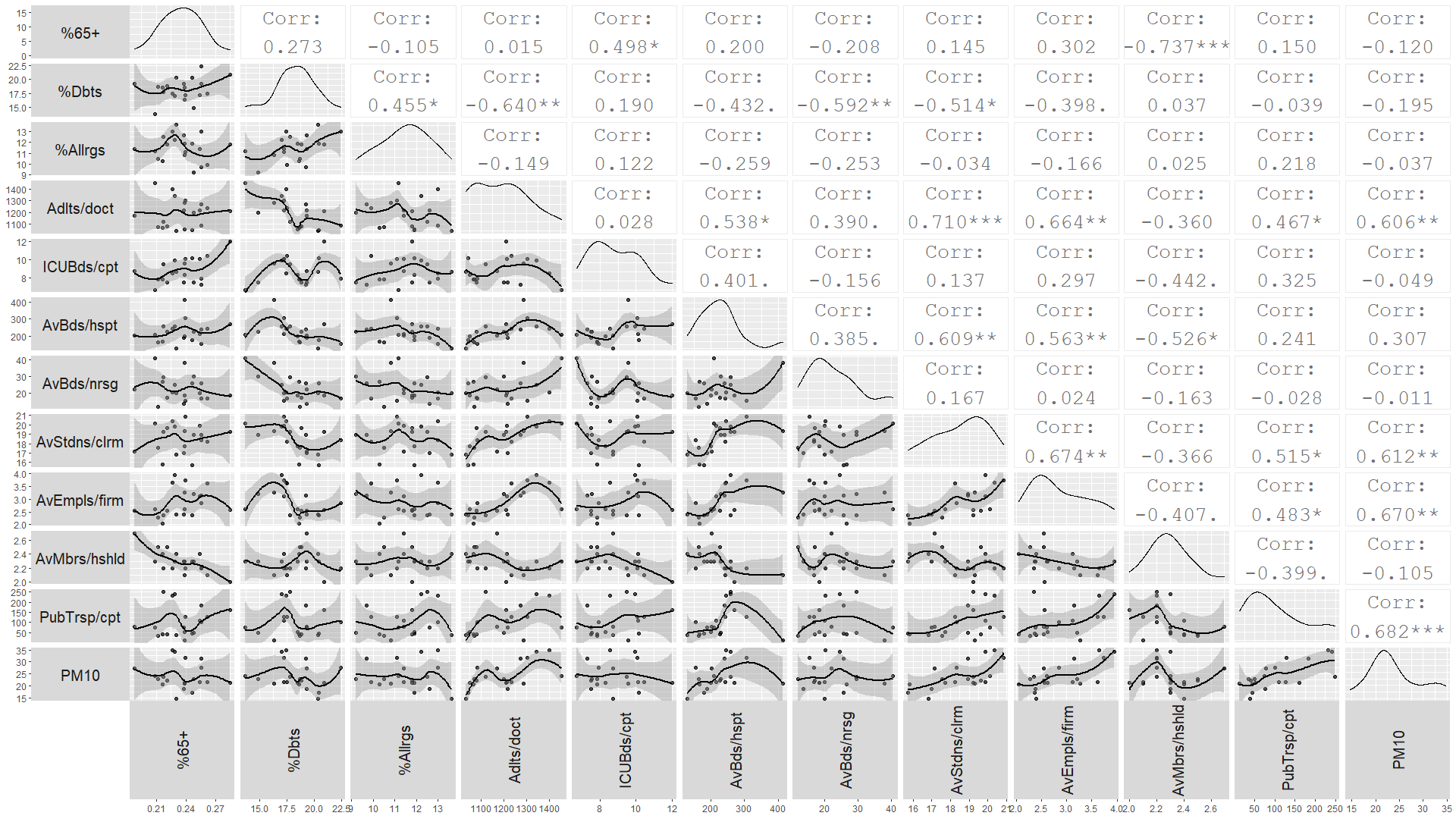}
    \caption{Exploratory matrix containing scatterplots with loess regression, marginal densities and correlation between covariates}
    \label{fig:ggally}
\end{figure}

\begin{table}[h]
\caption{Variance inflation factors (VIF) for the $12$ scalar covariates used in the main analysis.}
\label{tab:vif}
\centering
\begin{tabular}{l|l}
\toprule
\textbf{Covariate} & \textbf{VIF} \\
\midrule
\%65+ & 11.315684\\ 
\%Dbts  & 12.105778\\
\%Allrgs     & 3.933976\\
Adlts/doct  & 4.245479\\
ICUBds/cpt   & 2.892715\\
AvBds/hspt   & 5.012343\\
AvBds/nrsg   & 2.428207\\
AvStdns/clrm & 6.529208\\
AvEmpls/firm & 7.854636\\
AvMbrs/hshld & 6.223056\\
PubTrsp/cpt  & 4.293915\\
PM10  & 12.858811\\
\bottomrule
\end{tabular}
\end{table}

\end{document}